\documentclass{aa}
\usepackage[varg]{txfonts}

\usepackage{natbib}
\usepackage{float}
\usepackage{graphicx}
\usepackage{hyperref}
\usepackage{gensymb}
\usepackage{makecell}
\usepackage{esvect}

\begin{document}

        \title{Exoplanet atmosphere retrievals in 3D using phase curve data with ARCiS: application to WASP-43b}
        
        \titlerunning{Exoplanet atmosphere retrieval of WASP-43b in 3D with ARCiS}
        
        \author{Katy L. Chubb\inst{1,2}\thanks{klc20@st-andrews.ac.uk} and Michiel Min\inst{1}\thanks{m.min@sron.nl}}
        
        \authorrunning{Chubb and Min}
        
        \institute{SRON Netherlands Institute for Space Research, Sorbonnelaan 2, 3584 CA, Utrecht, Netherlands \and Centre for Exoplanet Science, University of St Andrews, North Haugh, St Andrews, UK}

        \abstract {} {Our goal is to create a retrieval framework which encapsulates the three-dimensional (3D) nature of exoplanet atmospheres, and to apply it to observed emission phase curve and transmission spectra of  the `hot Jupiter' exoplanet WASP-43b.}
        {We present our 3D framework, which is freely available as a stand-alone module from GitHub. We use the atmospheric modelling and Bayesian retrieval package ARCiS (ARtful modelling Code for exoplanet Science) to perform a series of eight 3D retrievals on simultaneous transmission (HST/WFC3) and phase-dependent emission (HST/WFC3 and Spitzer/IRAC) observations of WASP-43b as a case study. Via these retrieval setups, we assess how input assumptions affect our retrieval outcomes. In particular we look at constraining equilibrium chemistry vs a free molecular retrieval, the case of no clouds vs parametrised clouds, and using Spitzer phase data that have been reduced from two different literature sources. For the free chemistry retrievals, we retrieve abundances of H$_2$O, CH$_4$, CO, CO$_2$, AlO, and NH$_3$ as a function of phase, with many more species considered for the equilibrium chemistry retrievals.} {We find consistent super-solar C/O (0.6~-~0.9) and super-solar metallicities (1.7~-~2.9~dex) for all retrieval setups that assume equilibrium chemistry. We find that atmospheric heat distribution, hotspot shift ($\approx$15.6$\degree$ vs 4.5$\degree$ for the different Spitzer datasets), and temperature structure are very influenced by the choice of Spitzer emission phase data. We see some trends in molecular abundances as a function of phase, in particular for CH$_4$ and H$_2$O. Comparisons are made with other studies of WASP-43b, including global climate model (GCM) simulations,  available in the literature.} {The parametrised 3D setup we have developed provides a valuable tool to analyse extensive observational datasets such as spectroscopic phase curves. We conclude that further near-future observations with missions such as the James Webb Space Telescope (JWST) and Ariel will greatly improve our understanding of the atmospheres of exoplanets such as WASP-43b. This is particularly evident from the effect that the current phase-dependent Spitzer emission data has on retrieved atmospheres. }

        \maketitle
        
        \section{Introduction}\label{sec:intro}

        The spectra of a star that hosts a transiting planet can offer a wealth of information on that planet's atmosphere, both via transmission spectroscopy during primary transit, and by comparing emission spectra before and during secondary transit. Information about different regions of an exoplanet's atmosphere can be deduced by analysing these complementary spectra. 
        Various `1D' assumptions have been typically employed to model and retrieve parameters from such spectral observations of exoplanet atmospheres. 
1D transmission retrievals, for example, typically assume an isothermal pressure-temperature structure, which is the same at both morning and evening terminator regions of the planet. 1D emission retrievals will often make assumptions such as  the pressure-temperature structure being modelled the same across the entire dayside of the planet, and that molecular abundances are constant throughout the atmosphere.
        These can often be reasonable simplifications, particularly when analysing transmission spectra from a narrow wavelength region that comes from a layer of the atmosphere where the pressure-temperature structure is close to isothermal. The retrieval of emission spectra, which are more sensitive to the pressure-temperature profile than transmission spectra, requires a non-isothermal structure, but this will usually be parametrised~\cite[see, for example,][]{08Hansen,09MaSe,10Guillot.exo,15WaRoTi.taurex,20BaHe}.  One reason for employing such simplifying assumptions is that the computation of many thousands of forward models in a retrieval necessitates each forward model to be computationally efficient. This is particularly useful when analysing multiple datasets in population studies such as~\cite{16SiFoNi,18TsWaZi.exo,18FiHe.exo,18PiMaGa,19WeMaAl}.  
        
        There are a number of works that have explored the biases introduced into atmospheric models and retrievals that assume 1D rather than 3D (or closer to 3D) atmospheres, for example: \cite{16RoWaVe,17BlDoGr,16FeLiFo.exo,20TaPaIr.exo,20PlZiLe,20McGoLe}. 
        These works make clear the requirement to develop modelling tools that take higher dimension effects into account (for example, allowing molecular abundances, pressure-temperature profiles, and cloud coverage to vary as a function of longitude, latitude, and altitude). The development of various such tools, which range from 1.5~D to 3~D effects, has been the focus of studies such as \cite{19IrPaTa.wasp43b,19CaLeSe,20ChAl.phase,20FeLiFo.phase,20BeRaBr,21CuKeCo.exo,21LeWaPr,21FaZiPl,21MaLe}. 
        These studies build on early examples of work into developing 3D transmission spectra models, such as \cite{10FoShSh,10BuRaSp,12DoAgBu}.

        WASP-43b is a very well-studied ``hot Jupiter'' exoplanet~\citep[e.g.][]{18LoKr.exo,17GaMa.exo,17KoShTa.wasp43b,17KeCo.wasp43b,20GaJe,20FeLiFo.phase,21LuStMa,21GaPrPa}. It has a rich variety of available observational data, including an emission phase curve~\citep{14StDeLi.wasp43b,17StLiBe.wasp43b} and a transmission spectrum that has been found to contain strong evidence of molecular signatures~\citep[e.g.][]{14KrBeDe.wasp43b,18FiHe.exo,20ChMiKa.wasp43b}. The close-in giant exoplanet is assumed to be tidally locked, so information on atmospheric variability across the planet's surface can be inferred by utilising the phase curve data to generate spectra at different phases during the planet's orbit.  
        This makes WASP-43b an ideal candidate for detailed studies of atmospheric circulation models, and for testing retrieval models that extend beyond the traditional 1D formalism, such as the one presented in this work. It has already been the target for such retrievals that go beyond the traditional 1D formalism for retrieving properties of exoplanet atmospheres, for example: \cite{19IrPaTa.wasp43b,20FeLiFo.phase,20ChAl.phase,21CuKeCo.exo}.
        The presence of strong equatorial jets has been suggested by previous studies, such as \cite{15KaShFo.wasp43b}, largely due to the strong day--night temperature contrast~\citep{14StDeLi.wasp43b,15KaShFo.wasp43b,17GaMa.exo,19IrPaTa.wasp43b}. 
        Another recent study from \cite{20CaBaMo.exo} has also investigated the potential presence of deep atmospheric wind jets in WASP-43b, which are linked to retrograde (westwards) airflow to explain an apparent small hotspot shift and a large day-night temperature gradient. 
        
        The publicly available transmission and emission phase data for WASP-43b are primarily a result of observations by the Wide Field Camera 3 (WFC3) instrument on the Hubble Space Telescope (HST) and the Infrared Array Camera (IRAC) instrument on board the Spitzer Space Telescope (Spitzer). HST/WFC3 was used to observe three full-orbit phase curves, three primary transits, and two secondary eclipses (proposal ID 13467, PI: Jacob Bean~\citep{13Bean.hst}). Spitzer/IRAC was used to observe three broadband photometric phase curves of WASP-43b (proposal IDs 10169 and 11001, PI: Kevin Stevenson~\citep{17StLiBe.wasp43b}). The light curve fitting of the transmission data was first carried out by \cite{14KrBeDe.wasp43b} and later, independently, by \cite{18TsWaZi.exo}. We consider the former in this work. The Spitzer IRAC emission phase curve data measured at 3.6~$\mu$m and 4.5~$\mu$m by \cite{17StLiBe.wasp43b} were later reduced independently by \cite{18MeMaDe.wasp43b,19MoDaDi.wasp43b,20MaSt.wasp43b,21BeDaCo.wasp43b}. 
        Previous analyses of the WFC3 data by \cite{14KrBeDe.wasp43b} and \cite{17StLiBe.wasp43b} have found evidence of H$_2$O in both transmission and emission, with \cite{17StLiBe.wasp43b}  deducing the presence of CO and/or CO$_2$ based on the Spitzer data points. \cite{19WeLoMe.wasp43b}  also find  evidence of H$_2$O in transmission. CH$_4$ was found to vary in emission with phase by \cite{17StLiBe.wasp43b}, with some caution on the derived abundances demonstrated by \cite{16FeLiFo.exo}. \cite{20ChMiKa.wasp43b} recently found evidence for AlO in the transmission spectra of WASP-43b, based on a systematic search for evidence of all species that have available opacity data with spectral features in the HST/WFC3 wavelength region.

        The main aim of the present study is to create a retrieval framework that encapsulates the 3D nature of an exoplanet's atmosphere. We have developed and implemented such a framework into the atmospheric modelling and Bayesian retrieval package ARCiS~(ARtful modelling Code for exoplanet Science)~\citep{20MiOrCh.arcis,18OrMi.arcis}. We utilise available observed emission phase curve data and transmission spectra of WASP-43b. We perform a series of retrievals on these spectra in order to test various input assumptions on retrieval outcomes.
        
        This paper is structured as follows. Section~\ref{sec:arcis} outlines the theory behind our 3D retrieval model. This model is applied to WASP-43b using retrieval code ARCiS, as outlined in Section~\ref{sec:wasp43b}. A total of eight different retrieval setups are detailed here, the results of which are outlined in Section~\ref{sec:results} (and corresponding figures in Appendix~\ref{sec:appendix_retrieval} and the supplementary information document associated with this work\footnote{\url{https://doi.org/10.5281/zenodo.6325489}}). Discussions on various aspects of these results are also included in Section~\ref{sec:results}. Finally, our conclusions and future outlook are presented in Section~\ref{sec:conclusions}.

        \section{3D model for use in retrievals: theory}\label{sec:arcis}
        
        ARCiS is an atmospheric modelling and Bayesian retrieval package. General details can be found in \cite{20MiOrCh.arcis} and \cite{18OrMi.arcis}. ARCiS can handle simultaneous retrieval of transmission and emission spectra, including phase-dependent spectra. A benchmark of the emission spectra, based on that of \cite{2017ApJ...850..150B}, can be found in Appendix~\ref{sec:emis_benchmark}. We have recently updated ARCiS to allow for a 3D characterisation of exoplanet atmospheres during this simultaneous retrieval, as outlined in this work. There is no limit to the amount of observational data that can be included in such a retrieval, although the computational time does scale with the number of observed data points. Although the 3D setup implemented in ARCiS is used in this work, we also provide a general routine for the 3D setup used\footnote{\url{https://github.com/michielmin/DiffuseBeta}}. This is with the intention that it can be implemented into other general retrieval codes. 
        
        In the 3D retrieval setup, we have to model how the incoming stellar flux on the dayside of the planet is spread throughout the atmosphere. Here we model this using diffusion and horizontal winds. In general the normalised incoming stellar flux at the top of the atmosphere 
        as a function of longitude $\Lambda$ and latitude $\Phi$ is
        \begin{equation}\label{eq:source_day}
                \beta_\star = \cos{\Lambda}\cos{\Phi}
        \end{equation}
        on the dayside, and 
        \begin{equation}\label{eq:source_night}
                \beta_\star = 0
        \end{equation}
        on the nightside. For a static atmosphere (with no diffusion or winds), this allows us to compute the temperature structure at each location. We know this is unphysical. The atmosphere is not static and part of the energy incoming on the dayside is transported to the nightside. We therefore define a new parameter $\beta$, which specifies how the energy incident from the host star, $\beta_\star$, is spread throughout the atmosphere. Again, $\beta$ is a function of longitude $\Lambda$ and latitude $\Phi$. We model this spread of energy by a diffusion equation on $\beta$:
        \begin{equation}\label{eq:diff1}
                \nabla \cdot \vv{v}\beta - K_{\Lambda\Phi}\nabla^2 \beta = S
        \end{equation}
        Here, $\beta$ now specifies how the energy incident from the central star is spread through the atmosphere, and is found as a function of longitude $\Lambda$ and latitude $\Phi$. For simplicity we convert to a unit sphere and dimensionless velocity and diffusion coefficients. The diffusion is computed over this unit sphere assuming the velocity vector $\vv{v}$ is in the longitudinal direction. It is known from global circulation models (GCMs) that these longitudinal winds are strong around the equator and much weaker at the poles \citep[see e.g.][]{15KaShFo.wasp43b,20CaBaMo.exo}. Therefore, we also use a coefficient $n_\phi$ to specify the latitudinal extent of the winds: $\vv{v}=V_\Lambda\cos^{n_\phi}{\Phi}$. In spherical coordinates, Eq.~\ref{eq:diff1} is then given by:
        \begin{equation}\label{eq:diff2}
                \frac{V_{\Lambda}\cos^{n_\phi}{\Phi}}{\cos{\Phi}}\frac{\partial \beta}{\partial \Lambda} - \frac{K_{\Lambda\Phi}}{\cos{\Phi}}\frac{\partial}{\partial \Phi} \left( \cos{\Phi}\frac{\partial \beta}{\partial \Phi}\right) - \frac{K_{\Lambda\Phi}}{\cos^2{\Phi}}\frac{\partial^2 \beta}{\partial \Lambda^2} = S
        ,\end{equation}
        where the source term is given by
        \begin{equation}\label{eq:source_day2}
                S = f_{\rm day}\beta_\star
        \end{equation}
        and where $f_{\rm day}$ is the efficiency with which the planet absorbs the stellar radiation (i.e. in principle $f_{\rm day}=1-A_b$, where $A_b$ is the planetary bond albedo). 
        
        To aid the interpretation of the variables in our equations, we substitute a few parameters with those that are better tailored to the interpretation of observations. First we introduce a new parameter $f_{v} = \frac{V_{\Lambda}}{K_{\Lambda\Phi}}$ as the ratio of the equatorial wind speed over the horizontal diffusion parameter. 
        Eq.~\ref{eq:diff2} then becomes:
        \begin{equation}\label{eq:diff3}
                f_{v}\frac{\cos^{n_\phi}{\Phi}}{\cos{\Phi}}\frac{\partial \beta}{\partial \Lambda} - \frac{1}{\cos{\Phi}}\frac{\partial}{\partial \Phi} \left( \cos{\Phi}\frac{\partial \beta}{\partial \Phi}\right) - \frac{1}{\cos^2{\Phi}}\frac{\partial^2 \beta}{\partial \Lambda^2} = \frac{S}{K_{\Lambda\Phi}}
.        \end{equation}
        Finally, we define a parameter $f_{red}$, which is the integrated value of $\beta$ on the nightside over the integrated value of $\beta$ on the dayside. This gives a measure of the heat distribution across the atmosphere. For a static atmosphere, we have $f_{red}=0$, while for a fully mixed atmosphere we have $f_{red}=1$. We iteratively adjust the value of $K_{\Lambda\Phi}$ to reach the given value of $f_{red}$. Numerically this is rather easy since with all other parameters fixed $f_{red}$ is always changing in the same direction as $K_{\Lambda\Phi}$.
        
\begin{figure*}[!tp]
                \centerline{\resizebox{\hsize}{!}{\includegraphics{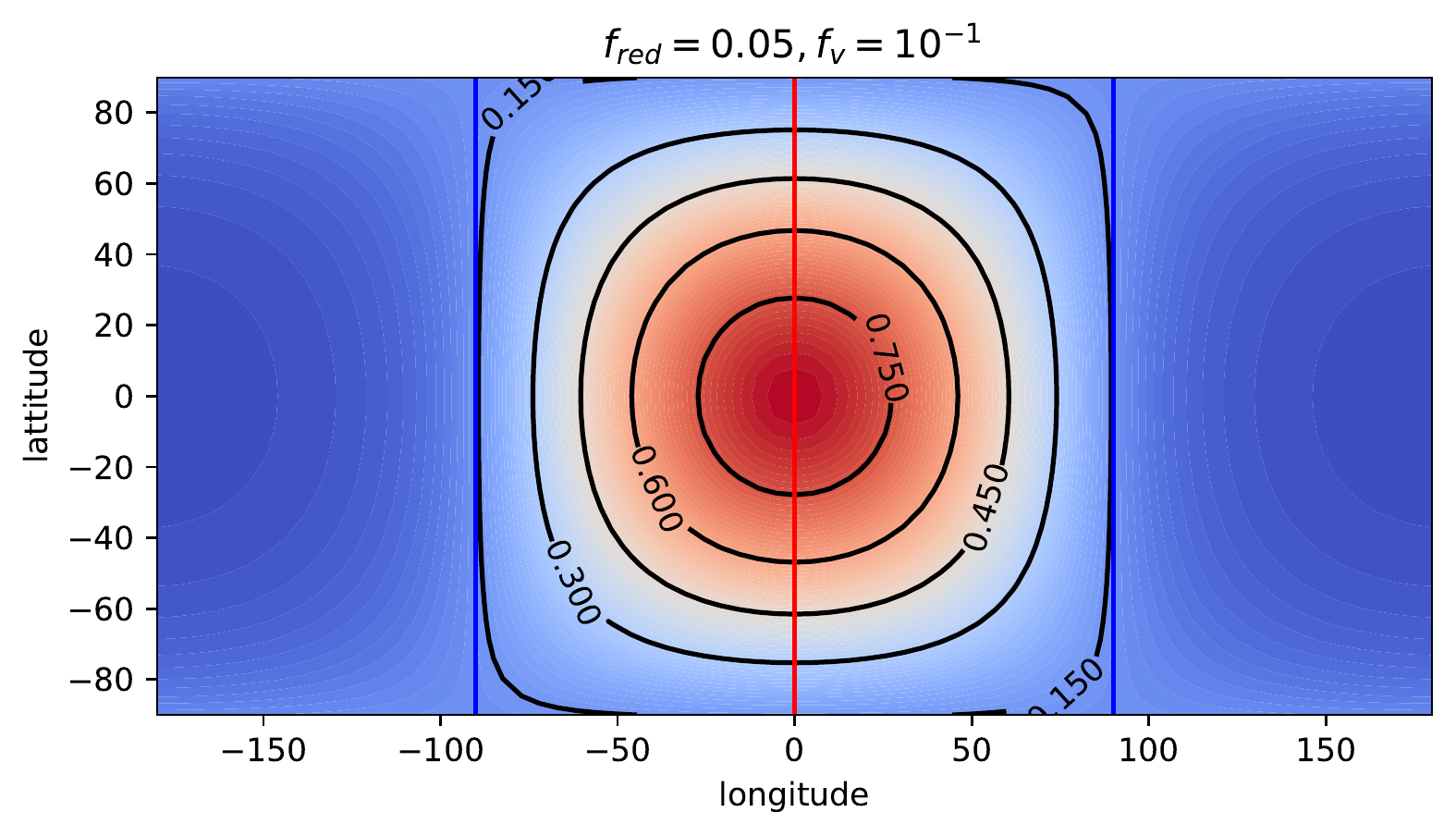}\includegraphics{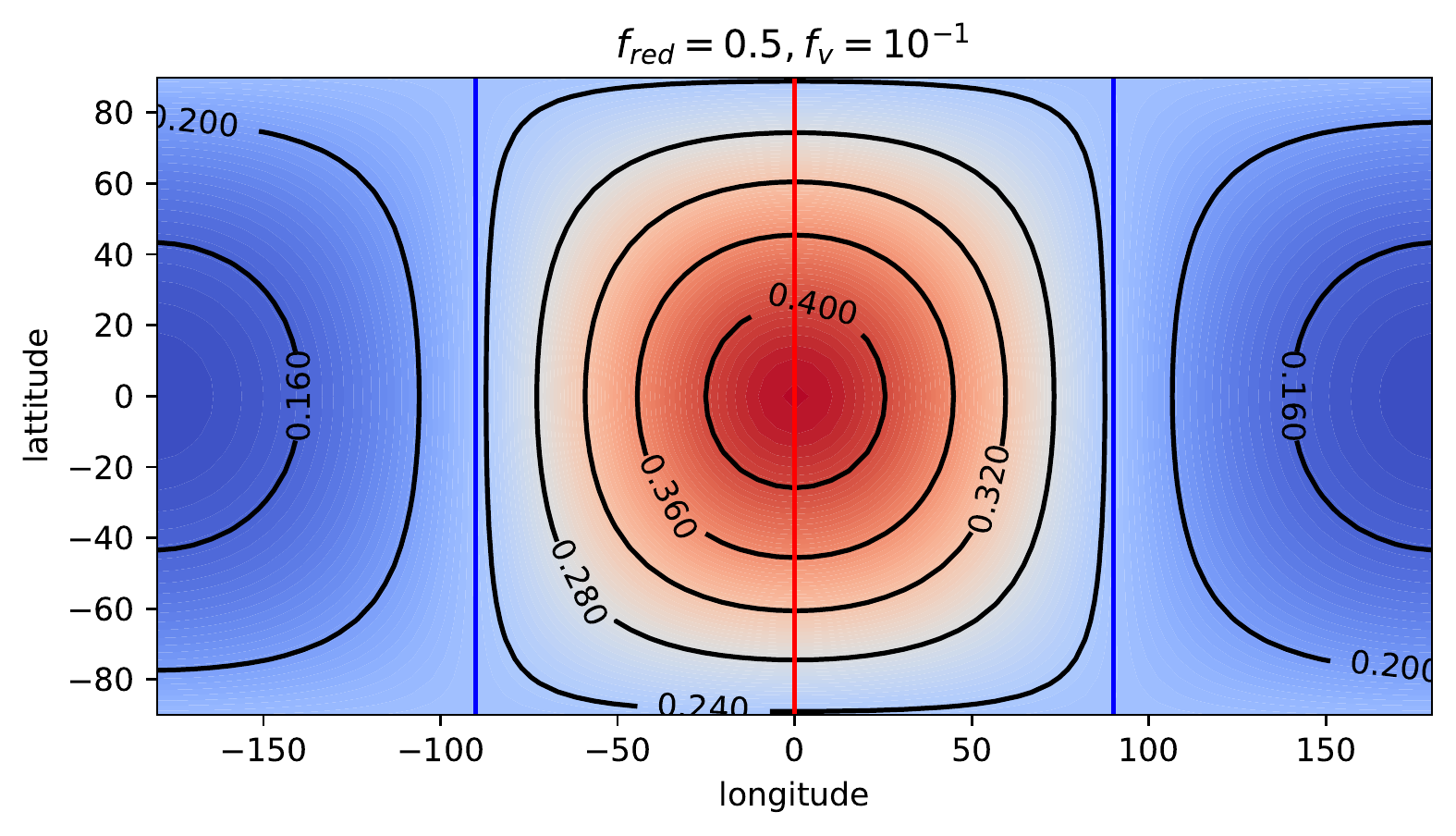}}}
                \centerline{\resizebox{\hsize}{!}{\includegraphics{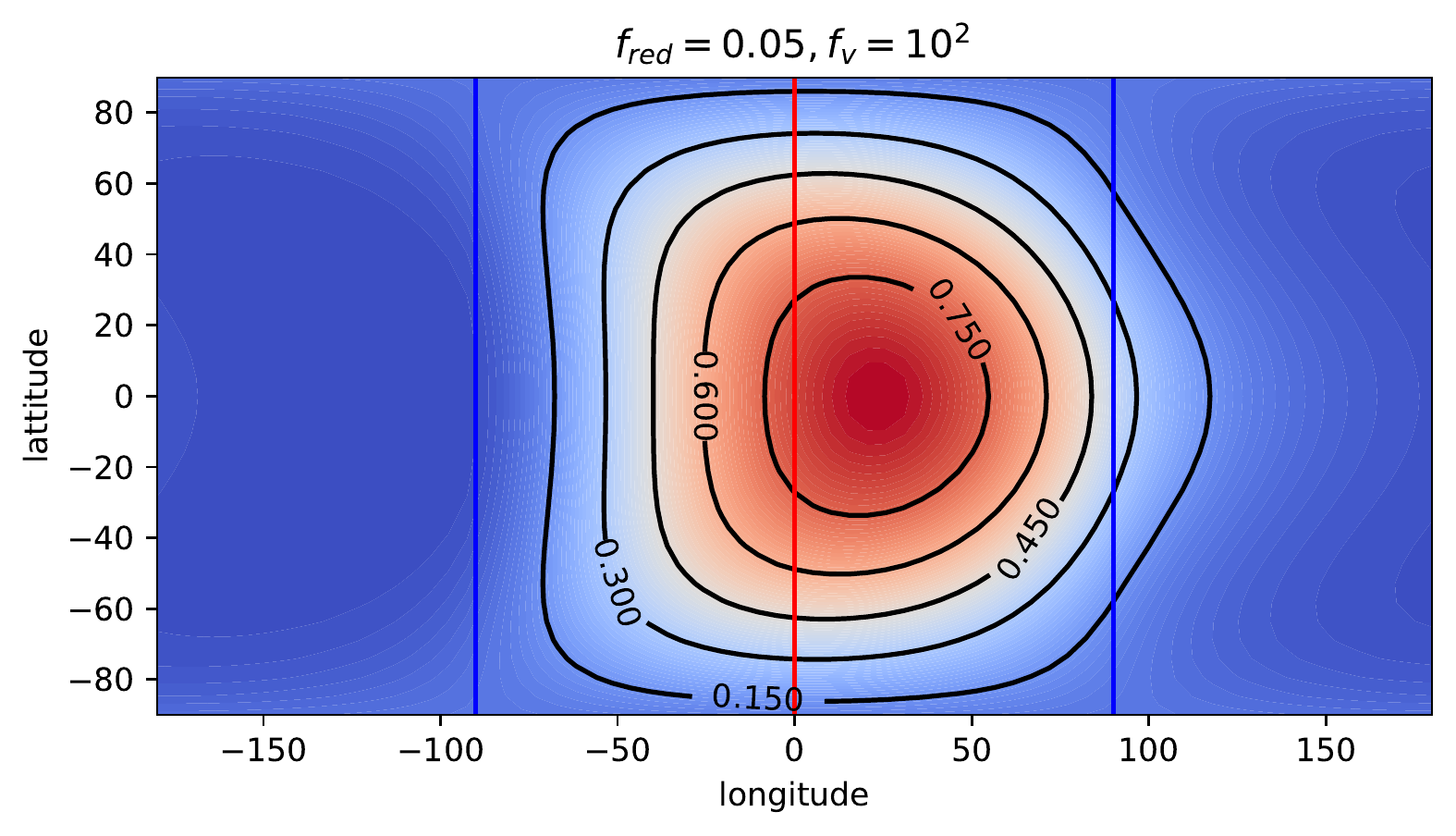}\includegraphics{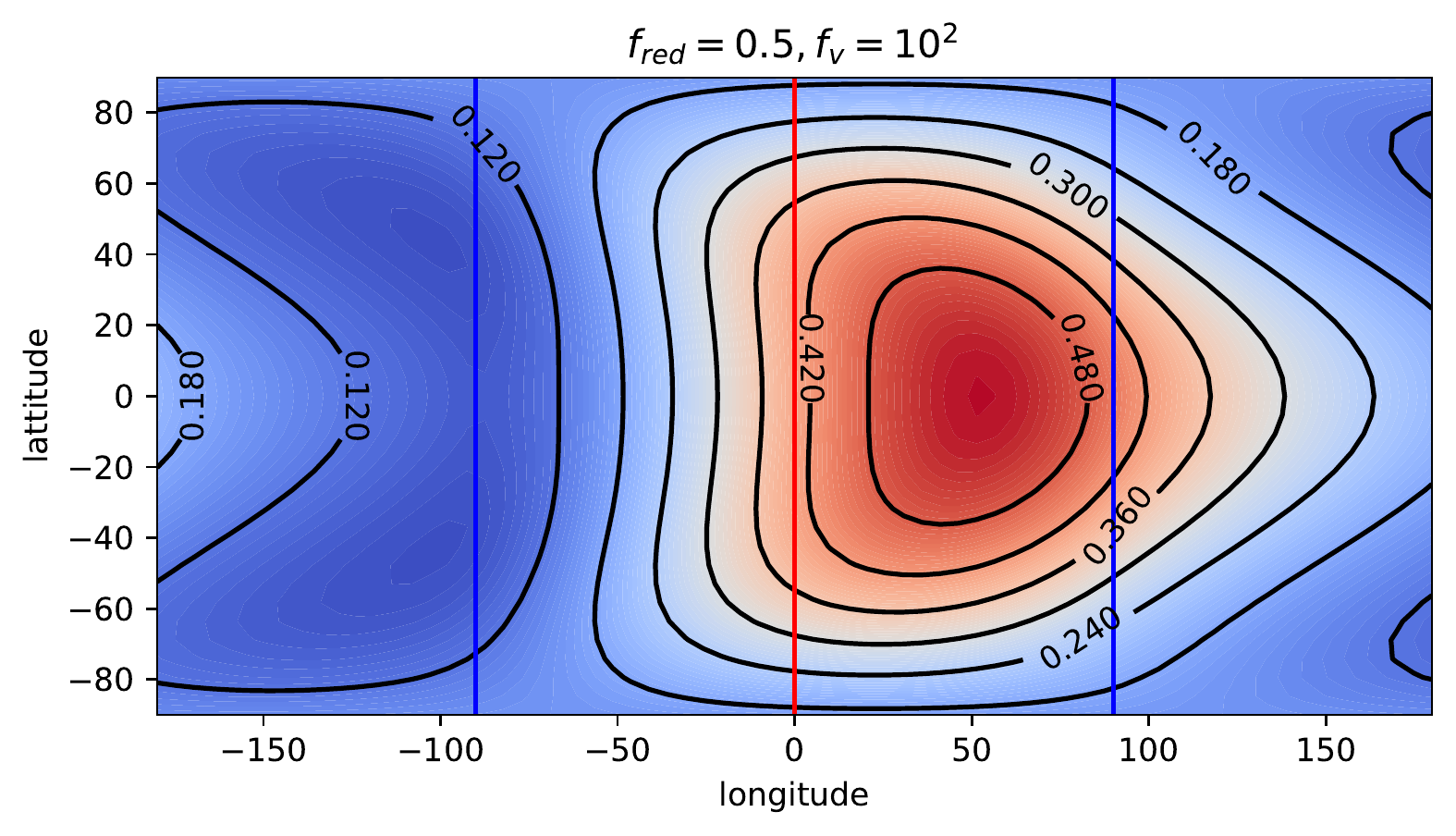}}}
                \caption{Example structures of $\beta$ as a function of latitude and longitude. The top row shows two examples for a pure diffusive case (no significant longitudinal wind), while the lower panels show two examples for significant longitudinal winds. The red vertical lines indicate the position of the longitude 0, while the blue vertical lines indicate the location of the terminator. For all four cases, we take $f_{\rm day}=1$, $n_\Phi=3$, where $f_{\rm day}$ and $n_\Phi$ are the efficiency with which the planet absorbs incoming stellar radiation, and the coefficient specifying the latitudinal extent of the longitudinal winds, respectively.}
                \label{fig:betamaps}
        \end{figure*}

        To summarise, our base 3D setup (before including other free parameters related to molecular abundance or temperature profile) consists of only four parameters: $(f_{red}, f_v, f_{\rm day},$ and $n_\Phi)$. For these parameters we iteratively solve for $K_{\Lambda\Phi}$ using Eq.~\ref{eq:diff3} and solve for $\beta$ at each location on the planet's globe. Next, at each $\Lambda, \Phi$ coordinate, we compute a 1D temperature structure. We allow some of the parameters in our model (or their logarithm) to vary linearly with $\beta$, while other parameters are kept fixed over the entire planet. This has the advantage that for each value of $\beta$, we only have to compute a single 1D structure and position it at different latitudinal and longitudinal locations. Example structures of $\beta$ as a function of longitude and latitude are shown in Fig.~\ref{fig:betamaps}.
        
        In the retrievals presented in this paper, we use ray tracing to compute the outgoing flux from the parametrised pressure-temperature structure.
        Ray tracing is done on a 3D structure, with no assumptions on the geometry. This means that effects such as limb darkening are automatically taken into account. In addition, effects of the planet becoming aspherical because of the larger vertical extend of hotter regions, are also automatically computed. For both scattering and temperature computations, where non-Monte Carlo, ray-tracing techniques are used, we achieve good agreement with literature benchmarks, as can be seen in Appendix~\ref{sec:emis_benchmark}.
        
        \subsection{Pressure-temperature profiles}\label{sec:PT_par}
        
        In principle, knowing $\beta$ (where $\beta$ specifies how the energy incident from the central star is spread through the atmosphere; see Eq.~\ref{eq:diff1}) at each location allows the pressure temperature structure to be computed from radiative transfer. However, for a retrieval model this is very computationally demanding. Therefore, here we use the pressure-temperature profile parametrisation of \cite{10Guillot.exo} to model the atmospheric structure at a given longitude and latitude. It is possible to have a completely free temperature profile with our setup; however, this introduces many extra free parameters into the retrieval.  We instead opt here for the Guillot profile, which requires fewer free parameters but is still relatively flexible. In the future, when more data is available from missions such as the
James Webb Space Telescope (JWST), we may be able to constrain the free pressure-temperature profile also. For the Guillot parametrisation, the temperature $T$ at a given atmospheric layer, defined by the optical depth $\tau$,  is given by:
        \begin{equation}\label{eq:beta}
                T^4 = \frac{3T_{\rm int}^4}{4}\left[\frac{2}{3}+ \tau\right] + \frac{3T_{\rm irr}^4}{4}\beta\left[\frac{2}{3} + \frac{1}{\gamma\sqrt{3}}+\left(\frac{\gamma}{\sqrt{3}} - \frac{1}{\gamma\sqrt{3}}\right){\rm e}^{-\gamma\tau\sqrt{3}}\right]
        \end{equation}
        Here, $\gamma = \frac{\kappa_{\rm v}}{\kappa_{\rm IR}}$  (i.e. the ratio of  mean optical to infra-red opacities). $\tau$ is the optical depth at IR wavelengths, related to pressure $P$ by gravity $g$ and the mean infra-red opacity $\kappa_{\rm IR}$:
        \begin{equation}
                \tau = \frac{\kappa_{\rm IR}P}{g}
        \end{equation}
        These equations were derived for a static atmosphere where the stellar flux received by the planet at the top of the atmosphere is given  by $\frac{\sigma T_{\rm irr}^4\beta_\star}{4\pi}$. The irradiation from the host star is computed using $T_{\rm irr} = T_{\star}\sqrt{R_{\star}/d_p}$, where $T_{\star}$ is the stellar temperature, $R_{\star}$ the stellar radius, and $d_p$ the distance between the planet and host star. The internal heat flux which is emitted by the planet is given by  $\sigma T_{\rm int}^4$ , where $\sigma$ is the Stefan-Boltzmann constant. We apply this equation also for our non-static atmosphere by using $\beta$ instead of $\beta_\star$ at each longitude and latitude (see the text surrounding Equation~\ref{eq:diff2}). We therefore compute a pressure-temperature profile for each of these grid points using our $\Lambda-$ and $\Phi-$ dependent value of $\beta$ together with Eq.~\ref{eq:beta}. Using this setup, the full 3D temperature structure is determined by seven parameters: $(f_{red}, f_v, f_{\rm day}, n_\Phi, T_{\rm int}, \kappa_{\rm IR,}$ and, $\gamma)$.
        
        \subsection{Retrieval setup}\label{sec:ret_intro}
        
        We are using the Bayesian retrieval package ARCiS~\citep{20MiOrCh.arcis,18OrMi.arcis}, which utilises the {\sc Multinest} \citep{08FeHo.multi,09FeGaHo.multi,13FeHoCa.multi} Monte Carlo nested sampling algorithm, to sample the specified parameter space for the region of maximum likelihood. We use 500 live points, an evidence tolerance of 0.5, and a sampling efficiency of 0.4 in our retrievals. The parameters and associated priors relevant for the present study are given in Table~\ref{t:ret_pars}. A benchmark of the transmission  spectra computed by ARCiS can be found in \cite{18OrMi.arcis}, and a benchmark for the emission spectra, based on that of \cite{2017ApJ...850..150B}, can be found in Appendix~\ref{sec:emis_benchmark}.
        
        Above, we described the 3D setup of the forward model. Some of the parameters in Table~\ref{t:ret_pars} are global, while some others  (or their logarithmic values) are varied linearly with $\beta$ (the parameter which determines how incoming stellar radiation is spread throughout the atmosphere; see Equation~\ref{eq:diff2}). 
        For these parameters which are found as a function of $\beta$, we have a \textit{night} value, occurring at the lowest value of $\beta$, and a \textit{day} value, corresponding to the highest value of $\beta$. The values in between these two extremes are varied linearly.
                For example, in the free molecular abundance retrievals, two free parameters exist (day and night abundances) and the longitudinal abundances are computed by interpolating with respect to $\beta$.
                In the equilibrium case, the metallicity, [Z], and carbon-to-oxygen ratio, C/O, are conserved globally, but the abundances change longitudinally with the local temperature.
                The cloud parameters (we use a simple cloud layer and allow the pressure-top level, cloud opacity, and cloud albedo to vary; see Section~\ref{sec:clouds_vs_no}), are also retrieved using two free parameters. Similar to the free molecular abundances, these are related to the minimum and maximum values of  $\beta$ (night and day), with the values in between computed by interpolating linearly with respect to $\beta$. All other retrieval parameters (see the section titled \textit{Parameters included in all setups}  in Table~\ref{t:ret_pars}) are retrieved globally (i.e. there is only one free parameter). The pressure-temperature profiles, computed using Eq.~\ref{eq:beta}, depend on the defined parameters used to describe the Guillot profile, and on $\beta$, which leads to different temperature structures as a function of longitude and latitude.

        We compute the phase-averaged temperature-profiles for each specified phase, which are better linked to the observed data points and can be directly compared to the results of other studies and the outputs of GCMs. The phase-averaged outputs are centred about the latitudinal line on the atmosphere with the same value in degrees. Outputs given as a function of longitude give the exact value at that point on the atmosphere, whereas outputs given as a function of phase give the average value integrated over the visible disk of the planet, centred on the given latitude (see Section~\ref{sec:results} for all outputs). The parameters retrieved as free parameters in the 3D ARCiS retrieval are given in Table~\ref{t:ret_pars}, with a short description of each. Further details on the molecules included in the equilibrium chemistry and free molecular retrieval setups are given in Sections~\ref{sec:eq_chem} and~\ref{sec:free_ret}. 
        
        
        \begin{table*}[!tp]
                \caption{Description of the parameters retrieved by ARCiS using the different 3D setups in this study, as described in the text. }
                \label{t:ret_pars} 
                \centering  
                \begin{tabular}{lcc}
                        \hline
                        \hline
                        \rule{0pt}{3ex}Name & Description & Priors \\
                        \hline
                        \multicolumn{3}{l}{\rule{0pt}{3ex}Parameters included in all setups}\\
                        \rule{0pt}{3ex}$R_p$ &  \makecell{\rule{0pt}{3ex} Planet radius} & 0.94 to 1.13 $R_J$\\
                        \rule{0pt}{3ex}$f_{\rm day}$ &  \makecell{\rule{0pt}{3ex} Proportion of incoming stellar flux absorbed  \\ by the planet's dayside atmosphere. $f_{\rm day} = 1 - {\rm A_b}^a$} & 0 (no flux absorbed) to 1 (all flux absorbed)\\
                        \rule{0pt}{3ex}$f_{\rm red}$ & \makecell{\rule{0pt}{3ex}Heat redistribution factor. Integrated energy on the nightside \\ divided by integrated energy on the dayside} & \makecell{0 (no heat redistributed from day to night) \\ to 1 (all heat redistributed from day to night)} \\
                        \rule{0pt}{3ex}log($f_{v}$) & \makecell{\rule{0pt}{3ex}Base-10 logarithm of the ratio of equatorial wind speed $V_{\rm \Lambda}$  \\ over horizontal diffusion parameter ${K_{\rm \Lambda\Phi}}^b$} &  -1 for negligible winds to 8 for strong winds \\
                        \rule{0pt}{3ex}$n_{\phi}$ & \makecell{\rule{0pt}{3ex} Exponent determining the latitudinal extent of  \\ equatorial winds, given by $\cos^{n_\phi}{\Phi}$} & \makecell{\rule{0pt}{3ex}3 for far-reaching winds to \\ 10 for a narrow band of wind}\\
                        \rule{0pt}{3ex}T$_{\rm int}$ &  \makecell{\rule{0pt}{3ex} Temperature at an optical depth $\tau$~=~$\frac{2}{3}$ as caused by \\ internal heat from the planet} & \makecell{\rule{0pt}{3ex} 10 to 3000~K}\\
                        \rule{0pt}{3ex}log($\gamma$) &  \makecell{\rule{0pt}{3ex}Base-10 logarithm of the ratio between \\ optical and IR opacities i.e. $\gamma = \frac{\kappa_{\rm v}}{\kappa_{\rm IR}}$} & \makecell{\rule{0pt}{3ex} -2 to 2}\\
                        \rule{0pt}{3ex}log($\kappa_{\rm IR}$) &  \makecell{\rule{0pt}{3ex} Base-10 logarithm of the mean IR opacity} & \makecell{\rule{0pt}{3ex} $\kappa_{\rm IR}$ varies from $10^{-4}$ to $10^4$ cm$^2$/g}\\
                        \rule{0pt}{3ex}log($g_p$) &  \makecell{\rule{0pt}{3ex} Base-10 logarithm of the planet surface gravity$^c$} & \makecell{\rule{0pt}{3ex} 3.56 to 3.83 (with $g$ given in cgs units)}\\[2mm]
                        \Xhline{2\arrayrulewidth}
                        \multicolumn{3}{l}{\rule{0pt}{3ex}Parameters only included in setups with clouds } \\
                        \rule{0pt}{3ex}log($P_{\rm cloud}$) &  \makecell{\rule{0pt}{3ex} Cloud top pressure (function of phase)} & \makecell{\rule{0pt}{3ex} $P_{\rm cloud}$~=~10$^{-3}$~bar for high-altitude clouds to \\ 10$^{3}$~bar for  deep-atmosphere clouds}\\
                        \rule{0pt}{3ex}log($\kappa_{\rm cloud}$)&  \makecell{\rule{0pt}{3ex} A measure of cloud opacity (function of phase)} & \makecell{\rule{0pt}{3ex} $\kappa_{\rm cloud}$~=~10$^{-7}$ cm$^2$/g for transparent clouds to \\ 10$^{2}$ cm$^2$/g for very opaque clouds}\\
                        \rule{0pt}{3ex}$A_{\rm cloud}$ &  \makecell{\rule{0pt}{3ex} A measure of cloud albedo (function of phase)} & \makecell{\rule{0pt}{3ex} 0 to 1}\\[2mm]
                        \Xhline{2\arrayrulewidth}
                        \multicolumn{3}{l}{\rule{0pt}{3ex}Parameters only included in free molecular retrieval setups} \\
                        \rule{0pt}{3ex}log(VMR(H$_2$O)) &  \makecell{\rule{0pt}{3ex} Base-10 logarithm of VMR of H$_2$O as a function of phase} & \makecell{\rule{0pt}{3ex} 10$^{-12}$ for negligible to 1 for very abundant}\\
                        \rule{0pt}{3ex}log(VMR(CO)) &  \makecell{\rule{0pt}{3ex} Base-10 logarithm of VMR of CO as a function of phase} & \makecell{\rule{0pt}{3ex} 10$^{-12}$ for negligible to 1 for very abundant}\\
                        \rule{0pt}{3ex}log(VMR(CO$_2$)) &  \makecell{\rule{0pt}{3ex} Base-10 logarithm of VMR of CO$_2$ as a function of phase} & \makecell{\rule{0pt}{3ex} 10$^{-12}$ for negligible to 1 for very abundant }\\
                        \rule{0pt}{3ex}log(VMR(CH$_4$)) &  \makecell{\rule{0pt}{3ex} Base-10 logarithm of VMR of CH$_4$ as a function of phase} & \makecell{\rule{0pt}{3ex} 10$^{-12}$ for negligible to 1 for very abundant}\\
                        \rule{0pt}{3ex}log(VMR(NH$_3$)) &  \makecell{\rule{0pt}{3ex} Base-10 logarithm of VMR of NH$_3$ as a function of phase} & \makecell{\rule{0pt}{3ex} 10$^{-12}$ for negligible to 1 for very abundant }\\
                        \rule{0pt}{3ex}log(VMR(AlO)) &  \makecell{\rule{0pt}{3ex} Base-10 logarithm of VMR of AlO as a function of phase} & \makecell{\rule{0pt}{3ex} 10$^{-12}$ for negligible to 1 for very abundant }\\[2mm]
                        \Xhline{2\arrayrulewidth}
                        \multicolumn{3}{l}{\rule{0pt}{3ex}Parameters only included in equilibrium chemistry retrieval setups} \\
                        \rule{0pt}{3ex}C/O &  \makecell{\rule{0pt}{3ex} Atmospheric carbon-to-oxygen ratio (global)} & \makecell{\rule{0pt}{3ex} 0.1 (sub-solar) to 1.5 (super-solar)}\\
                        \rule{0pt}{3ex}[Z] &  \makecell{\rule{0pt}{3ex} Atmospheric metallicity (global) } & \makecell{\rule{0pt}{3ex} -3 (sub-solar) to 3 (super-solar) dex}\\[2mm]
                        \hline
                        \hline
                \end{tabular}
                \rule{0pt}{0.2ex}
                \flushleft{\textit{$^a$: $A_b$ is the bond albedo.}}
                \flushleft{\textit{$^b$: The value of $K_{\rm \Lambda\Phi}$ is adjusted until the given value of $f_{\rm red}$ is reached.}}
                \flushleft{\textit{$^c$: We use a Gaussian prior on the mass based on radial velocity measurements of WASP-43b~\citep{12GiTrFo.wasp43b}. We compute the mass based on $R_p$ and log($g_p$) and then place the prior on this derived parameter.}}
                
        \end{table*}
        
        
        \section{Case study: 3D retrievals of WASP-43b}\label{sec:wasp43b}

        WASP-43b was discovered by \cite{11HeAnCo.wasp43b} around an active K7V star, with deduced planetary parameters from radial velocity measurements and transit observations of 2.034~$\pm$~0.052~$M_J$ and 1.036~$\pm$~0.019~$R_J$~\citep{12GiTrFo.wasp43b}. 
        It is assumed to be tidally locked. Therefore, the emission phase observations from \cite{14StDeLi.wasp43b,17StLiBe.wasp43b} can be used to essentially map the atmosphere across the planetary surface.    ARCiS allows a combination of emission, phase, and transmission spectra to be included simultaneously in the retrieval. The data we included in this case study of WASP-43b is summarised in Table~\ref{t:obs_data}. We follow the convention that the morning terminator corresponds to 270$^{\circ}$ or 0.75 phase, and the evening terminator to 90$^{\circ}$ or 0.25 phase.  We choose not to use Spitzer transmission data points in this study, as there may be potential issues with combining data from different instruments (see, for example, \cite{20YiChEd.exo}). Nevertheless, as will be shown by our results, and as pointed out in previous studies, the Spitzer emission data are very important in order to distinguish between different interpretations of the atmospheric composition and dynamics. We thus include the Spitzer data from the emission phase curves along with the HST/WFC3 emission phase curve data. Having observations with more extensive wavelength coverage taken from a single instrument, such as JWST, will greatly aid future retrievals.

        \begin{table*}[!tp]
                \caption{Observed data used in the WASP-43b retrieval setups described in this work. All phase data from HST are from \cite{14StDeLi.wasp43b}. The phase data observed using Spitzer are from \cite{17StLiBe.wasp43b} and were independently reduced by two different sources: \cite{17StLiBe.wasp43b} or \cite{18MeMaDe.wasp43b}. We use slightly different phases in this work for these two cases, as detailed in the footnote of this table. Transmission HST data are from \cite{14KrBeDe.wasp43b}. HST WFC3 refers to the Hubble Space Telescope Wide Field Camera 3, and IRAC to the Infrared Array Camera on board Spitzer.  
                }
                \label{t:obs_data} 
                \centering  
                \begin{tabular}{l|l|l|l}
                        \hline\hline
                        \rule{0pt}{3ex}Observation type & Phase ($^{\circ}$)         & Centred on    & Instrument \\
                        \hline
                        \rule{0pt}{3ex}Phase & 45  &  Nightside &  HST WFC3 / Spitzer IRAC\\
                        Phase & 67.5$^{a}$ &  Nightside &  HST WFC3 / Spitzer IRAC\\
                        Phase & 90$^{c}$ &  Evening &  HST WFC3 / Spitzer IRAC\\
                        Phase & 112.5$^{a}$ &  Dayside &  HST WFC3 / Spitzer IRAC\\
                        Phase & 135$^{b}$ &  Dayside &  HST WFC3 / Spitzer IRAC\\
                        Phase & 157.5 &  Dayside &  HST WFC3 / Spitzer IRAC\\
                        Phase & 180 &  Dayside &  HST WFC3 / Spitzer IRAC\\
                        Phase & 202.5 &  Dayside &  HST WFC3 / Spitzer IRAC\\
                        Phase & 225$^{b}$ &  Dayside &  HST WFC3 / Spitzer IRAC\\
                        Phase & 247.5$^{a}$ &  Dayside &  HST WFC3 / Spitzer IRAC\\
                        Phase & 270 &  Morning &  HST WFC3 / Spitzer IRAC\\
                        Phase & 315 &  Nightside &  HST WFC3 / Spitzer IRAC\\
                        Transmission & - & Terminators (combined) &  HST WFC3 \\
                        \hline \hline
                \end{tabular}
                \rule{0pt}{0.2ex}
                \flushleft{\textit{$^a$: This phase was used in retrieval setups using the \cite{18MeMaDe.wasp43b} Spitzer points only.}}
                \flushleft{\textit{$^b$: This phase was used in retrieval setups using the \cite{17StLiBe.wasp43b} Spitzer points only.}}
                \flushleft{\textit{$^c$: This phase considers HST WFC3 data only for the retrieval cases using the \cite{18MeMaDe.wasp43b} Spitzer points.}}
                
        \end{table*}

        As mentioned in Section~\ref{sec:ret_intro}, we are using Bayesian retrieval package ARCiS~\citep{20MiOrCh.arcis,18OrMi.arcis}, which uses the {\sc Multinest} \citep{08FeHo.multi,09FeGaHo.multi,13FeHoCa.multi} algorithm to sample the specified parameter space for the region of maximum likelihood.
        In order to fully explore the effects of including different physical parameters in a retrieval (in particular, including parametrised clouds, or allowing free vs equilibrium chemistry), and also the effects of using different analyses of the Spitzer phase curve emission data, we performed a series of various setups of retrievals, as described in Table~\ref{t:ret_setups}.
        The physical parameters of the WASP-43 system that we used in all retrieval setups are given in Table~\ref{t:fixed_pars}. Below, we explain briefly the differences between these eight retrieval setups. We included isotropic scattering from stellar and planetary flux in all retrieval setups.
        
        \subsection{Clouds vs no clouds}\label{sec:clouds_vs_no}
        
        Clouds are predicted to be present on WASP-43b, particularly on the nightside~\citep{19HeGrSa.wasp43b,21HeLeSa.clouds}. We therefore performed a set of retrievals with clouds included in a parametrised way. We used a simple cloud setup, for which we introduced 3 parameters to describe a cloud or haze layer; $P_{\rm cloud}$ (cloud-top pressure, ranging from 10$^{-3}$~bar for high-altitude clouds or hazes to 10$^{3}$~bar for  deep-atmosphere clouds), $\kappa_{\rm cloud}$ (a measure of cloud opacity from 10$^{-7}$ cm$^2$/g for transparent clouds to 10$^{2}$ cm$^2$/g for very opaque clouds), and $A_{\rm cloud}$ (cloud albedo, ranging from 0 for completely unreflective clouds to 1 for fully reflective clouds). As outlined in Section~\ref{sec:arcis}, this cloud setup introduces six more free parameters in the retrieval setups that include clouds than in those that don't. This is because we introduced values of each of the three parameters that correspond to both minimum and maximum values of $\beta$ (i.e. the hottest dayside and the coolest nightside parts of the planet), and then allowed the parameters to vary linearly in between. These six extra free parameters inevitably increased the computation time for our retrievals. One way to make these computations more feasible was to explore the settings and modes available in the {\sc Multinest} algorithm used in ARCiS to sample parameter space. We thus used the constant efficiency mode of {\sc Multinest}, which noticeably improved the efficiency. A short description of this is given below in Section~\ref{sec:const_eff}.
        
        \subsection{Computational demands: constant efficiency mode for cloud setups}\label{sec:const_eff}

        The constant efficiency mode of {\sc Multinest} can vastly speed up retrievals computations~\citep{09FeGaHo.multi,13FeHoCa.multi}. We tested this for our 3D retrieval setup and found significant speed improvements. To put into context the increased computational effort of the 3D retrieval setup in comparison to a standard 1D retrieval using ARCiS: a standard 1D retrieval (of combined emission and transmission spectra) computes approximately 5 models per second (on a MacBook, 2 GHz Quad-Core Intel Core i5, 16 GB memory), whereas the 3D model computes approximately 1 model per second. A retrieval of just transmission spectra using a 1D setup, with no scattering included, can compute around 33 models per second. 
        \cite{09FeGaHo.multi} note that although {\sc Multinest}'s constant efficiency mode could potentially miss sampling some regions of parameter space, it may nevertheless produce reasonably accurate posterior distributions for parameter estimation purposes. The evidence estimates in this mode, however, should not be relied upon. We therefore do not use these values as a way to judge between our different retrieval setups, such as in \cite{08Trotta.stats}.
        In order to test whether {\sc Multinest} was missing degenerate solutions in this mode, we ran a test by setting up the same retrieval as setup D (with clouds, equilibrium chemistry, and \cite{17StLiBe.wasp43b} Spitzer analysis; see Table~\ref{t:ret_setups}), but using full efficiency instead of constant efficiency. We refer to this as retrieval setup I. The number of model calls is over ten times larger for the full efficiency mode in comparison to the equivalent constant efficiency mode (comparing retrieval setup D at 60,000 and retrieval setup I at 700,000). This had a big impact on computational time. We did find some differences between some of the retrieved parameters of the two retrievals, but many were within $\sigma$ error bounds (see the posterior plots contained in the supplementary information document of this work. This document also contains a comparison of results from setups D and I). The main difference that we note is that the parameters retrieved in the constant efficiency case appear to be better constrained than those in the full-efficiency case. This has a noticeable effect on the pressure-temperature profiles (see Section~\ref{sec:heat_redis}). We conclude this is artificial, and it should be kept in mind when viewing our results that we do not expect them to be as well constrained as they often appear in the cases where clouds are included in the setups.

        \begin{table*}[!tp]
                \caption{Description of the different retrieval setups presented in this work. We are considering setups: (i) with and without clouds, (ii) free chemistry vs equilibrium chemistry, and (iii) using different sets of Spitzer points. The retrievals with clouds included were run in the constant efficiency mode of {\sc Multinest}, and those without clouds in the full efficiency mode of {\sc Multinest}, with the exception of setup I. See text of Section~\ref{sec:const_eff} for more information. }
                \label{t:ret_setups} 
                \centering  
                \begin{tabular}{llll}
                        \hline
                        \hline
                        \rule{0pt}{3ex}Setup & Clouds? & Chemistry & Spitzer points \\[2mm]
                        \hline
                        \multicolumn{4}{l}{\rule{0pt}{3ex}Retrievals using Stevenson Spitzer data:}\\
                        \rule{0pt}{3ex} A & No & Free & \cite{17StLiBe.wasp43b}\\
                        \rule{0pt}{3ex} B & No & Equilibrium & \cite{17StLiBe.wasp43b}\\
                        \rule{0pt}{3ex} C & Yes & Free & \cite{17StLiBe.wasp43b}\\
                        \rule{0pt}{3ex} D & Yes & Equilibrium & \cite{17StLiBe.wasp43b}\\
                        [2mm]
                        \hline
                        \multicolumn{4}{l}{\rule{0pt}{3ex}Retrievals using Mendon{\c{c}}a Spitzer data:}\\
                        \rule{0pt}{3ex} E & No & Free & \cite{18MeMaDe.wasp43b}\\
                        \rule{0pt}{3ex} F & No & Equilibrium & \cite{18MeMaDe.wasp43b}\\
                        \rule{0pt}{3ex} G & Yes & Free & \cite{18MeMaDe.wasp43b}\\
                        \rule{0pt}{3ex} H & Yes & Equilibrium & \cite{18MeMaDe.wasp43b}\\
                        [2mm]
                        \hline
                        \multicolumn{4}{l}{\rule{0pt}{3ex}Retrieval including clouds but using the full efficiency mode of  {\sc Multinest}:}\\
                        \rule{0pt}{3ex} I & Yes & Equilibrium & \cite{17StLiBe.wasp43b}\\
                        \hline
                        \hline
                \end{tabular}
        \end{table*}

        \begin{table*}[!tp]
                \caption{Fixed parameters used in the ARCiS retrievals of WASP-43b.  }
                \label{t:fixed_pars} 
                \centering  
                \begin{tabular}{l|l|l}
                        \hline\hline
                        \rule{0pt}{3ex}Parameter        &       Value   & Description     \\
                        \hline
                        \rule{0pt}{3ex}$T_{*}$ ($K$) & 4520  $\pm$ 120&   Stellar temperature$^1$ \\
                        $R_{*}$ ($R_\odot$) & 0.667$\pm$ 0.01  &  Stellar radius$^1$ \\
                        $M_{p}$ ($M_J$) & 2.034 $\pm$ 0.051  &  Planetary mass$^1$  \\
                        $M_{*}$ ($M_\odot$) & 0.717 $\pm$ 0.025  &  Stellar mass$^1$ \\
                        H$_2$ / He  & 0.17 &  (H$_2$ / He) ratio\\
                        n$_{P_{\rm layers}}$ & 100  & Number of pressure layers  \\
                        log(P$_{\rm layers}$ (Pa))& -5 \dots +6 & Range of pressure layers  \\
                        CIA (H$_2$-H$_2$), (H$_2$-He) & HITRAN &  Collision induced absorption$^2$  \\
                        \hline \hline
                \end{tabular}
                {\flushleft
                        $^1$ \cite{12GiTrFo.wasp43b};
                        $^2$ \cite{HITRAN_2016,01BoJo.cia}
                }
        \end{table*}

        \subsection{Equilibrium chemistry}\label{sec:eq_chem}
        
        In the equilibrium chemistry retrievals, the molecular abundances are computed assuming equilibrium chemistry using the GGchem code~\citep{18WoHeHu.exo}, which is fully integrated into ARCiS. Two parameters are used to parametrise the elemental abundances: carbon-to-oxygen ratio, C/O, and metallicity, [Z]. Opacities were all computed in ARCiS format as part of the ExoMolOP database~\citep{20ChRoAl.exo}, and are publicaly available on the ExoMol~\citep{jt804} website\footnote{\url{http://www.exomol.com/data/data-types/opacity/}}. Opacity formats for other retrieval codes are also available from ExoMolOP: TauREx3~\citep{15WaTiRo.taurex,15WaRoTi.taurex,19AlChWa.taurex}, NEMESIS~\citep{NEMESIS}, and petitRADTRANS~\citep{19MoWaBo.petitRADTRANS}. We include the following set of species, given along with the source of the line-list used for the corresponding opacities: H$_2$O~\citep{jt734}, CO$_2$~\citep{20YuMeFr.co2}, CH$_4$~\citep{jt698}, CO~\citep{15LiGoRo.CO},
        AlO~\citep{jt598},
        NH$_3$~\citep{jt771},
        H$^-$~\citep{88John},
        SiO~\citep{ExoMol_SiO},
        FeH~\citep{10WeReSe.FeH},
        TiO~\citep{ExoMol_TiO},
        VO~\citep{jt644},
        SH~\citep{19GoYuTe},
        H$_2$S~\citep{ExoMol_H2S},
        C$_2$H$_2$~\citep{ExoMol_C2H2},
        CN~\citep{14BrRaWe.CN},
        CP~\citep{14RaBrWe.CP},
        H$_2$CO~\citep{jt597},
        H$_2$O$_2$~\citep{jt638},
        K~\citep{16AlSpKi.broad,19AlSpLe.broad,NISTWebsite},
        Na~\citep{19AlSpLe.broad,NISTWebsite},
        OH~\citep{16BrBeWe.OH,18YoBeHo.OH,MOLLIST},
        PH$_3$~\citep{jt592},
        and ScH~\citep{jt599}. There are other species available from ExoMolOP, such as \cite{14YoYuLo.PN,jt732,jt729}, but we limit the species included in our chemistry retrievals to those listed above. This is because combining the k-tables of multiple species has a significant effect on computing time for emission spectra in particular. The number of molecular species included is less of an issue for the constrained chemistry case than the free molecular retrieval, where in the latter the number of free parameters also has a significant effect on computational time. 
                We therefore made a selection of which species to include based on those that are expected to be there, and those with notable absorption features in the wavelength region able to be constrained by our data. We would consider including additional species in the future if we found that they could affect our resulting C/O.
        
        For a C/O above solar, we adjust the oxygen abundance and keep carbon at solar abundances, while for a C/O below solar, we adjust the carbon abundance and keep oxygen at solar abundances. This simulates the removal of these elements during the formation or evolution of the planet. 
        After the relative abundances of all elements heavier than He are computed using this C/O, the H and He abundances are then scaled using the metallicity parameter [Z]: 
        \begin{equation}
                [Z] = \log_{10}\left[ \left(\frac{\sum_{i\ne \rm H,He}X_i}{\sum_{i=\rm H,He}X_i}\right)\cdot \left(\frac{\sum_{i= \rm H,He}X_i}{\sum_{i\ne \rm H,He}X_i}\right)_{\rm Solar}\right].
        \end{equation}
        Here, $X_i$ is the number density of element $i$. 
        
        \subsection{Free molecular chemistry}\label{sec:free_ret}
        For the free molecular retrievals, we include the base set of molecules (H$_2$O, CO, CO$_2$, CH$_4$, and NH$_3$) which are typically included in previous studies of WASP-43b \cite[for example,][]{17StLiBe.wasp43b,19IrPaTa.wasp43b,14KrBeDe.wasp43b}. We also include AlO due to the studies of \cite{20ChMiKa.wasp43b}, who found some potential evidence for AlO in the \cite{14KrBeDe.wasp43b} transmission spectra of WASP-43b. 
        As with the equilibrium chemistry setups, all opacity data were formatted for ARCiS as part of the publicly available ExoMolOP database~\citep{20ChRoAl.exo}. The species 
        used in the free molecular retrieval setup are as follows: H$_2$O,
        CH$_4$,
        CO,
        CO$_2$,
        AlO, and NH$_3$.
        All line lists used are the same as those listed in Section~\ref{sec:eq_chem}.

        \subsection{Using different analyses of the Spitzer phase curve}\label{sec:spitz_phase}
        
        Although we do not consider them all here, there have been multiple reductions of the same observed Spitzer phase curves of WASP-43b; for both 3.6~$\mu$m and 4.5~$\mu$m (programmes 10169 and 11001, PI: Kevin Stevenson). These include the two which are used in this study: \cite{17StLiBe.wasp43b} and \cite{18MeMaDe.wasp43b}. Subsequent reductions have been published by \cite{19MoDaDi.wasp43b},        \cite{20MaSt.wasp43b}, and \cite{21BeDaCo.wasp43b}. 
        
        There are some disagreements between these different analyses. For example,        \cite{19MoDaDi.wasp43b} found higher nightside temperatures and smaller hotspot offsets than \cite{17StLiBe.wasp43b}.  The Spitzer points from the latter are used in a number of retrieval studies of WASP-43b, including \cite{19IrPaTa.wasp43b} and \cite{21ChAlEd.wasp43b}. We therefore chose to use them in this work in order to aid comparisons. We also use those from \cite{18MeMaDe.wasp43b} that represent spectra where the apparent day-night temperature difference is less pronounced, in order to observe the effect on our retrieval results. We do not consider the other reductions of the Spitzer phase curves that are mentioned in the present study, but we note that this choice can have an effect on the retrieval results, as will be illustrated in Section~\ref{sec:results}.

        \section{Results and discussion of 3D retrievals of WASP-43b}\label{sec:results}
        As previously mentioned, we performed eight different retrievals using slightly different setups, as detailed in Table~\ref{t:ret_setups}. This allowed us to explore the parameter space and the effects that different prior assumptions have on retrieval results. 
        Table~\ref{t:ret_pars_results} gives the values of a selection of retrieved parameters found from the different retrieval setups, with parameters all described in Table~\ref{t:ret_pars}. Further results can be found in Appendix~\ref{sec:append1}, as detailed below.  We also include all corner plots in the supplementary information of this work. A reminder of the eight retrieval setups is given in the footnote of Table~\ref{t:ret_pars_results}. We also include a ninth setup, labelled I, which is the same as setup D but using the full efficiency mode of  {\sc Multinest} (see Section~\ref{sec:const_eff}). 
        
        Here, we focus on some key findings. In particular, we highlight the retrieved parameters that appear consistent between different setups and could therefore be considered more robust. We also discuss degeneracies and covariance between different parameters. Various figures are presented for these eight different outputs, as detailed in the text of this section. These can be found in the appendix to this paper, with corner plots in a separate supplementary information document.

        
        \begin{table*}[!tp]
                \caption{Retrieved parameters found using different retrieval setups. See Table~\ref{t:ret_pars} for definitions of the parameters. The different retrieval setups are explained in the footnote to this table. The error bounds on $R_p$ are all in the region of 1~$\times$~10$^{-4}$ (in units of Jupiter radii). 
                }
                \label{t:ret_pars_results} 
                \centering  
                \begin{tabular}{lccccccccccc}
                        \hline
                        \hline
                        \rule{0pt}{3ex}No.$^{*}$        &               log($f_{v}$)            &               $f_{red}$                       &               $f_{\rm day}$                   &               log($\gamma$)           &               log($\kappa_{\rm IR}$)                   &               $T_{\rm int}$   &               $R_p$   &               log($g_p$)              &               $n_{\phi}$                      &               C/O             &               [Z]             \\
                        \rule{0pt}{3ex} &                       &                               &                                       &                       &                       &                (K)             &                       &                       &                       &                       &                       \\[2mm]
                        \hline
                        \multicolumn{12}{l}{\rule{0pt}{3ex}Retrievals using Stevenson Spitzer data:}\\
                        \rule{0pt}{3ex} A       &       $       5.35    ^{+     1.69    }_{     -1.69   }$      &       $       0.004   ^{+     0.002   }_{-    0.001   }$      &       $       0.28    ^{+     0.07    }_{     -0.07   }$      &       $       -1.24   ^{+     0.12    }_{     -0.13   }$      &       $       -0.49   ^{+     0.31    }_{     -0.38   }$      &       $       54      ^{+     3       }_{-    3       }$      &               1.05                                            &       $       3.78    ^{+     0.03    }_{     -0.05   }$      &       $       4.68    ^{+     2.75    }_{     -1.11   }$      &                                                       N/A     &                                                       N/A     \\
                        \rule{0pt}{3ex} B       &       $       5.49    ^{+     1.64    }_{     -1.99   }$      &       $       0.006   ^{+     0.002   }_{-    0.002   }$      &       $       0.09    ^{+     0.04    }_{     -0.02   }$      &       $       -1.28   ^{+     0.17    }_{     -0.14   }$      &       $       0.22    ^{+     0.21    }_{     -0.25   }$      &       $       355     ^{+     1       }_{-    1       }$      &               1.05                                            &       $       3.76    ^{+     0.05    }_{     -0.1    }$      &       $       8.65    ^{+     0.96    }_{     -1.94   }$      &       $       0.88    ^{+     0.01    }_{     -0.02   }$      &       $       1.71    ^{+     0.10    }_{     -0.08   }$      \\
                        \rule{0pt}{3ex} C       &       $       5.72    ^{+     1.03    }_{     -1.27   }$      &       $       0.008   ^{+     0.002   }_{-    0.001   }$      &       $       0.47    ^{+     0.02    }_{     -0.02   }$      &       $       -0.88   ^{+     0.02    }_{     -0.02   }$      &       $       -1.52   ^{+     0.13    }_{     -0.09   }$      &       $       41      ^{+     2       }_{-    2       }$      &               1.05                                            &       $       3.7     ^{+     0.09    }_{     -0.06   }$      &       $       7.05    ^{+     1.29    }_{     -1.74   }$      &                                                       N/A     &                                                       N/A     \\
                        \rule{0pt}{3ex} D       &       $       4.73    ^{+     0.97    }_{     -0.82   }$      &       $       0.004   ^{+     0.001   }_{-    0.001   }$      &       $       0.26    ^{+     0.02    }_{     -0.02   }$      &       $       -0.89   ^{+     0.03    }_{     -0.04   }$      &       $       -0.31   ^{+     0.09    }_{     -0.09   }$      &       $       490     ^{+     1       }_{-    1       }$      &               1.05                                            &       $       3.78    ^{+     0.02    }_{     -0.03   }$      &       $       4.83    ^{+     2.09    }_{     -0.95   }$      &       $       0.88    ^{+     0.01    }_{     0.01    }$      &       $       1.89    ^{+     0.12    }_{     -0.06   }$      \\
                        [2mm]
                        \hline                                          
                        \multicolumn{12}{l}{\rule{0pt}{3ex}Retrievals using Mendon{\c{c}}a  Spitzer data:}\\                                                                                                                                                                                                                                                                                                                                                                                
                        \rule{0pt}{3ex} E       &       $       0.15    ^{+     0.19    }_{     -0.23   }$      &       $       0.585   ^{+     0.094   }_{-    0.094   }$      &       $       0.12    ^{+     0.04    }_{     -0.03   }$      &       $       -1.56   ^{+     0.14    }_{     -0.15   }$      &       $       -0.88   ^{+     0.18    }_{     -0.16   }$      &       $       50      ^{+     3       }_{-    3       }$      &               1.05                                            &       $       3.74    ^{+     0.06    }_{     -0.08   }$      &       $       6.56    ^{+     2.18    }_{     -2.29   }$      &                                                       N/A     &                                                       N/A     \\
                        \rule{0pt}{3ex} F       &       $       5.19    ^{+     1.82    }_{     -1.86   }$      &       $       0.001   ^{+     0.001   }_{-    0.000   }$      &       $       0.32    ^{+     0.03    }_{     -0.04   }$      &       $       -0.68   ^{+     0.04    }_{     -0.06   }$      &       $       0.44    ^{+     0.17    }_{     -0.12   }$      &       $       398     ^{+     1       }_{-    1       }$      &               1.06                                            &       $       3.65    ^{+     0.1     }_{     -0.06   }$      &       $       6.63    ^{+     2.38    }_{     -2.27   }$      &       $       0.65    ^{+     0.03    }_{     -0.04   }$      &       $       2.82    ^{+     0.14    }_{     -0.31   }$      \\
                        \rule{0pt}{3ex} G       &       $       0.76    ^{+     0.11    }_{     -0.13   }$      &       $       0.086   ^{+     0.012   }_{-    0.010   }$      &       $       0.58    ^{+     0.02    }_{     -0.01   }$      &       $       -0.76   ^{+     0.03    }_{     -0.02   }$      &       $       -1.74   ^{+     0.10    }_{     -0.08   }$      &       $       65      ^{+     3       }_{-    3       }$      &               1.05                                            &       $       3.75    ^{+     0.04    }_{     -0.05   }$      &       $       7.95    ^{+     1.0     }_{     -1.3    }$      &                                                       N/A     &                                                       N/A     \\
                        \rule{0pt}{3ex} H       &       $       0.83    ^{+     0.07    }_{     -0.08   }$      &       $       0.050   ^{+     0.019   }_{-    0.006   }$      &       $       0.51    ^{+     0.02    }_{     -0.02   }$      &       $       -0.96   ^{+     0.03    }_{     -0.02   }$      &       $       -0.45   ^{+     0.09    }_{     -0.11   }$      &       $       17      ^{+     2       }_{-    1       }$      &               1.06                                            &       $       3.77    ^{+     0.03    }_{     -0.05   }$      &       $       5.23    ^{+     1.21    }_{     -1.07   }$      &       $       0.87    ^{+     0.02    }_{     -0.02   }$      &       $       2.90    ^{+     0.06    }_{     -0.09   }$      \\
                        [2mm]
                        \hline                                                                  
                        \multicolumn{12}{l}{\rule{0pt}{3ex}Full efficiency for equilibrium chemistry using Stevenson Spitzer data and including clouds:}\\                                                                                                                                                                                                                                                                                                                                                                                                                                                                 \rule{0pt}{3ex} I       &       $       5.51    ^{+     1.54    }_{     -1.64   }$      &       $       0.004   ^{+     0.003   }_{-    0.005   }$      &       $       0.42    ^{+     0.06    }_{     -0.1    }$      &       $       -1.78   ^{+     0.24    }_{     0.16    }$      &       $       -0.57   ^{+     0.18    }_{     -0.18   }$      &       $       56      ^{+     3       }_{-    3       }$      &               1.05                                            &       $       3.74    ^{+     0.06    }_{     -0.08   }$      &       $       4.61    ^{+     1.95    }_{     -1.09   }$      &       $       0.90    ^{+     0.01    }_{     -0.02   }$      &       $       2.37    ^{+     0.30    }_{     -0.42   }$      \\[2mm]
                        \hline
                        \hline
                \end{tabular}
                \rule{0pt}{0.2ex}
                \flushleft{$^{*}$\textit{Stevenson Spitzer data:}}
                \flushleft{\textit{A: Free chemistry, no clouds, Stevenson Spitzer data}}
                \flushleft{\textit{B: Equilibrium chemistry, no clouds, Stevenson Spitzer data}}
                \flushleft{\textit{C: Free chemistry, with clouds, Stevenson Spitzer data}}
                \flushleft{\textit{D: Equilibrium chemistry, with clouds, Stevenson Spitzer data}}
                \flushleft{\textit{Mendon{\c{c}}a Spitzer data:}}
                \flushleft{\textit{E: Free chemistry, no clouds, Mendon{\c{c}}a Spitzer data}}
                \flushleft{\textit{F: Equilibrium chemistry, no clouds, Mendon{\c{c}}a Spitzer data}}
                \flushleft{\textit{G: Free chemistry, with clouds, Mendon{\c{c}}a Spitzer data}}
                \flushleft{\textit{H: Equilibrium chemistry, with clouds, Mendon{\c{c}}a Spitzer data}}
                \flushleft{\textit{Clouds but full efficiency:}}
                \flushleft{\textit{I: Equilibrium chemistry, with clouds, Stevenson Spitzer data (full efficiency)}}
                
        \end{table*}
        
        
        \subsection{C/O and metallicity}
        
        Knowledge of atmospheric  C/O and [Z] can have implications for inferring information about planet formation history~\cite[see, for example,][]{11MiFo,16MoBoMo,16ObEd,17MaBiJo,20ShHeIk,22ShHeIk,21KhMiDe,19CrDiAl,21TuCoDa}.
        It can be seen from Table~\ref{t:ret_pars_results} that both C/O and [Z] are consistently retrieved to be super-solar for all the equilibrium chemistry retrieval setups. Super-solar values are consistent with those found in the full retrieval by \cite{21ChAlEd.wasp43b}, who find $Z$~=~$1.81^{+0.19}_{-0.17}$ and C/O~=~$0.68^{+0.11}_{-0.12}$ in the scenario where they allow a completely free temperature profile. This temperature profile setup is different to our parametrised Guillot profile (see Section~\ref{sec:PT_par}). Similar to the present study,  \cite{21ChAlEd.wasp43b} test various retrieval setups when performing simultaneous phase curve retrievals of  WASP-43b observations. While they retrieve super-solar metallicity for all runs, they find the C/O to be more dependent on the exact setup and, in particular, the thermal profile used. \cite{15KaShFo.wasp43b} find that an atmospheric metallicity that is 5~$\times$~solar (without TiO/VO) is the best match to the WASP-43b observations.
        \cite{19IrPaTa.wasp43b} use optimal estimation for their 2.5D retrieval of WASP-43b, but recommend a more extensive Bayesian sampling of parameter spaces, such as nested sampling, as is used in the present work. Based on their high retrieved abundances of CO and CH$_4$, they infer a C/O of $\sim$~0.91, which is similar to the values found in our equilibrium chemistry retrievals. They use the Spitzer emission data from \cite{17StLiBe.wasp43b}. \cite{19IrPaTa.wasp43b} also point out that  any errors in the offset between HST and Spitzer data  could lead to erroneous molecular abundances, which could then have an effect on both the C/O and metallicity. 
        In the context of planet formation theory, a high metallicity typically indicates that  the planet was formed further out from the star and migrated inwards. However, this is also generally assumed to be linked to a low C/O because of the assumed oxygen-rich nature of the accreted solids that leads to the high metallicity~\citep{21KhMiDe,21TuCoDa}. A combination of high metallicity and a high C/O is therefore potentially unexpected, based on predictions from planet formation theory.  A high C/O, however, could also be an indication of cloud formation, which typically de-oxidises the gas phase \citep{19Helling}.

        \subsection{Retrieved spectra at different phases}\label{sec:ret_phase}
        
\begin{figure*}[!tp]
                \centering
                \includegraphics[width=0.4\textwidth]{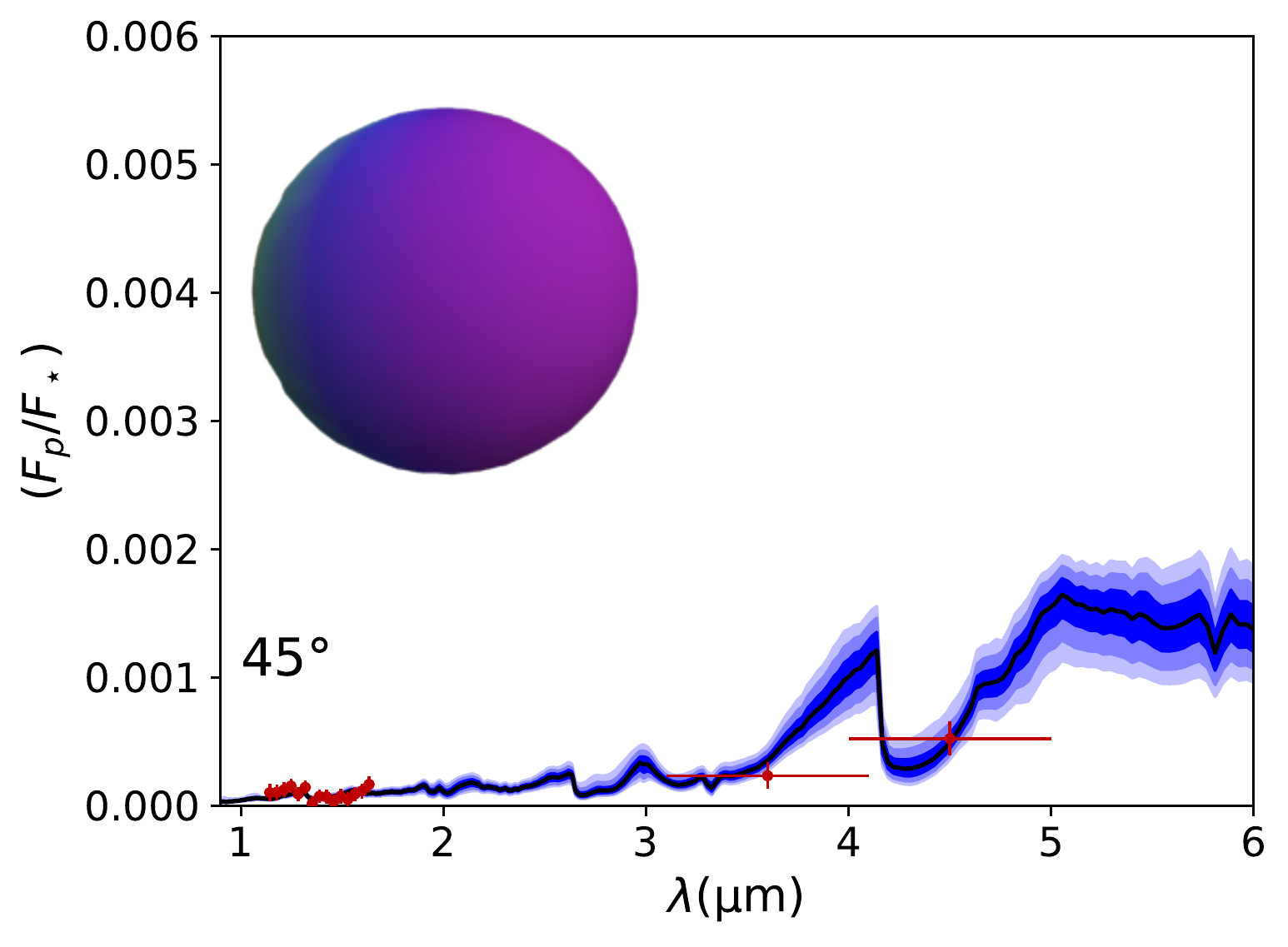}
                \includegraphics[width=0.4\textwidth]{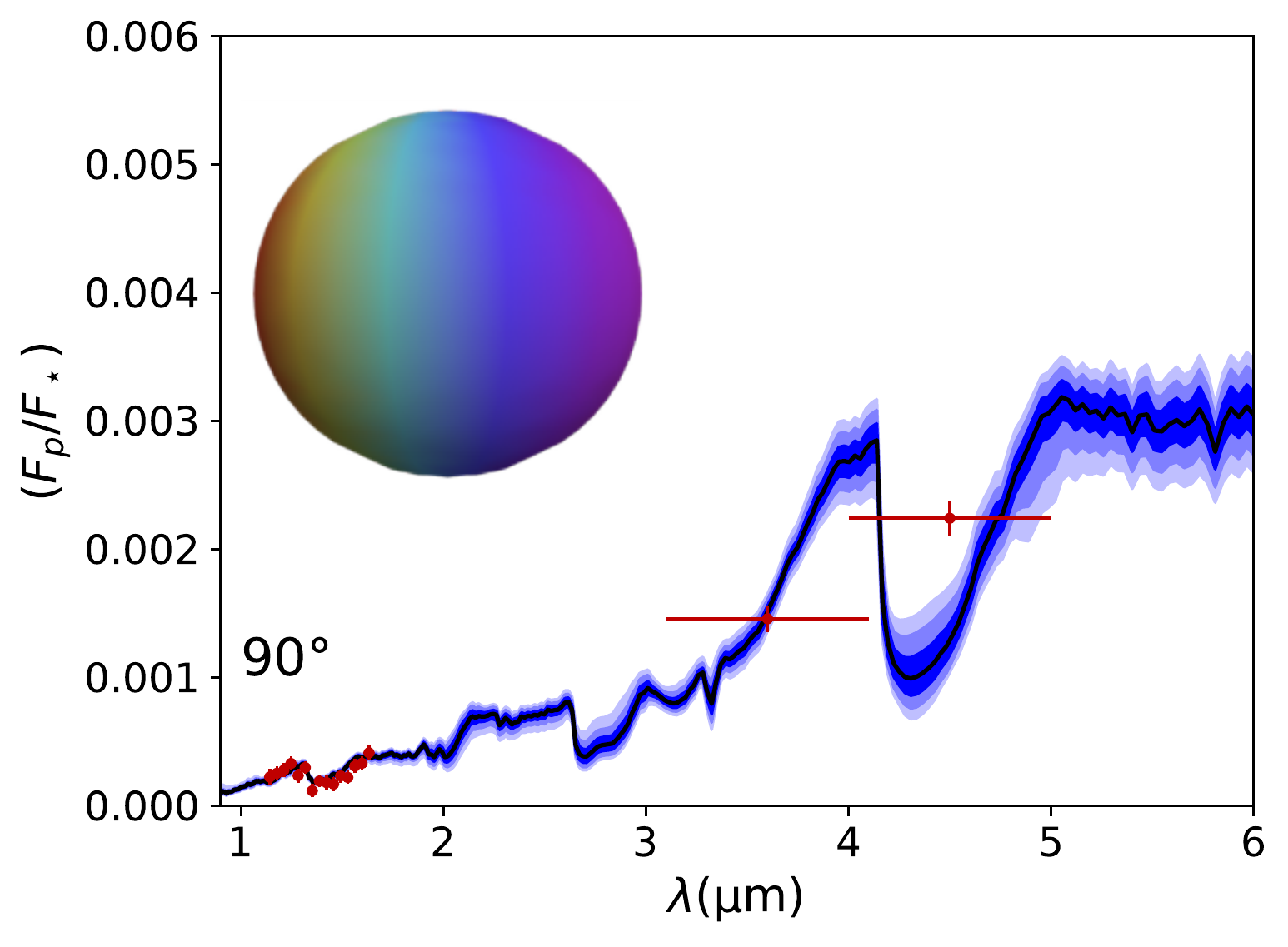}
                \includegraphics[width=0.4\textwidth]{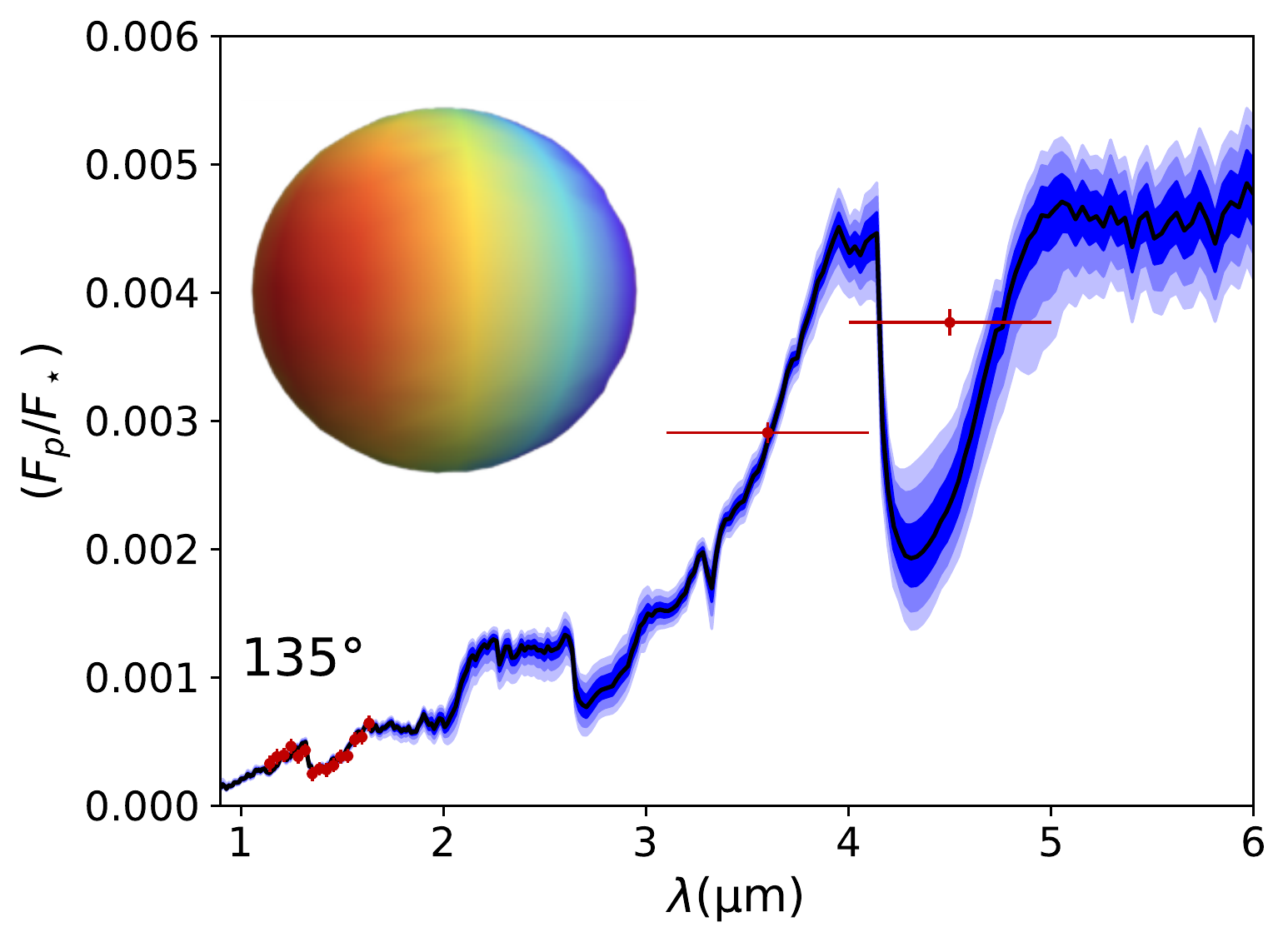}
                \includegraphics[width=0.4\textwidth]{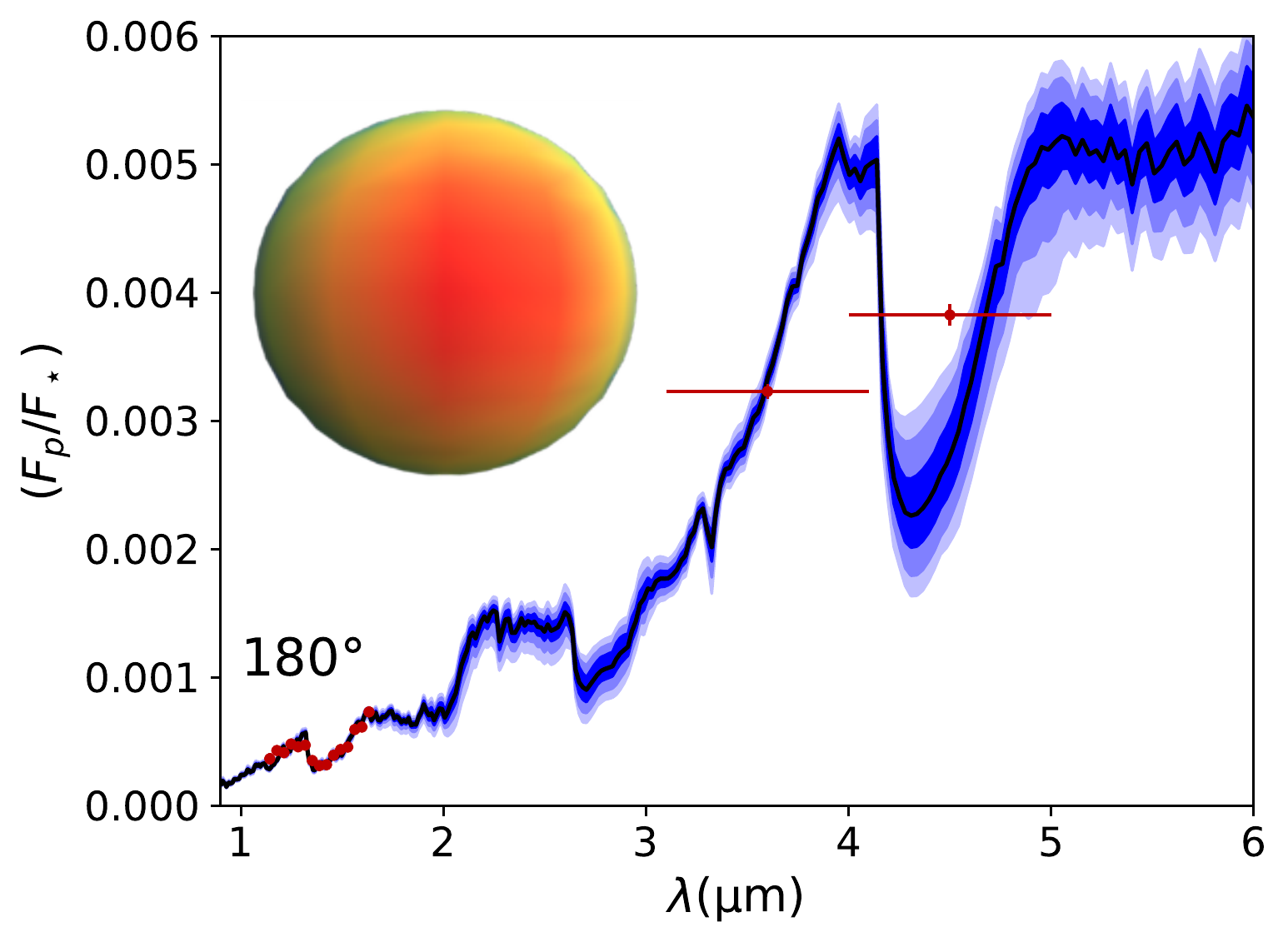}
                \includegraphics[width=0.4\textwidth]{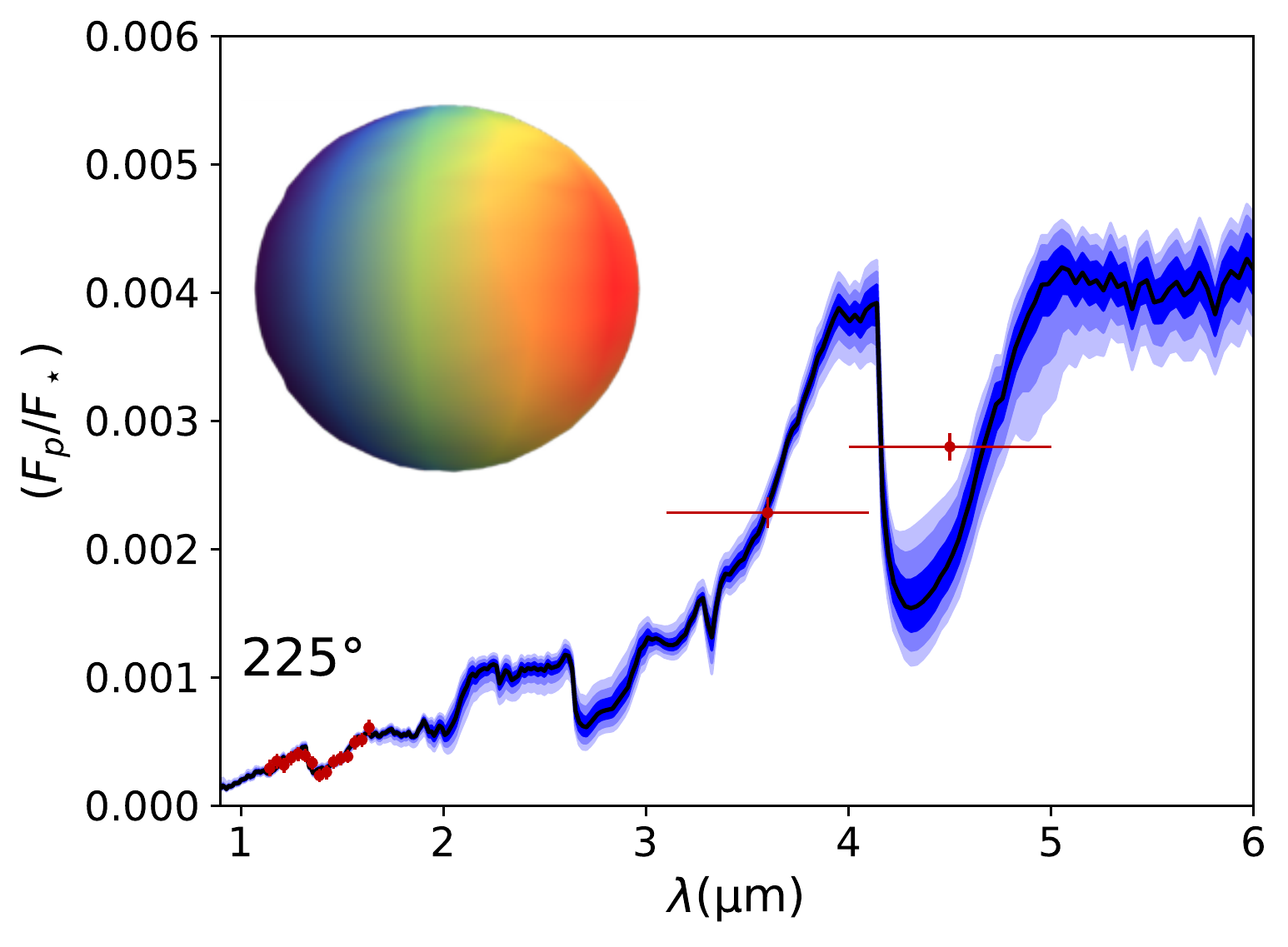}
                \includegraphics[width=0.4\textwidth]{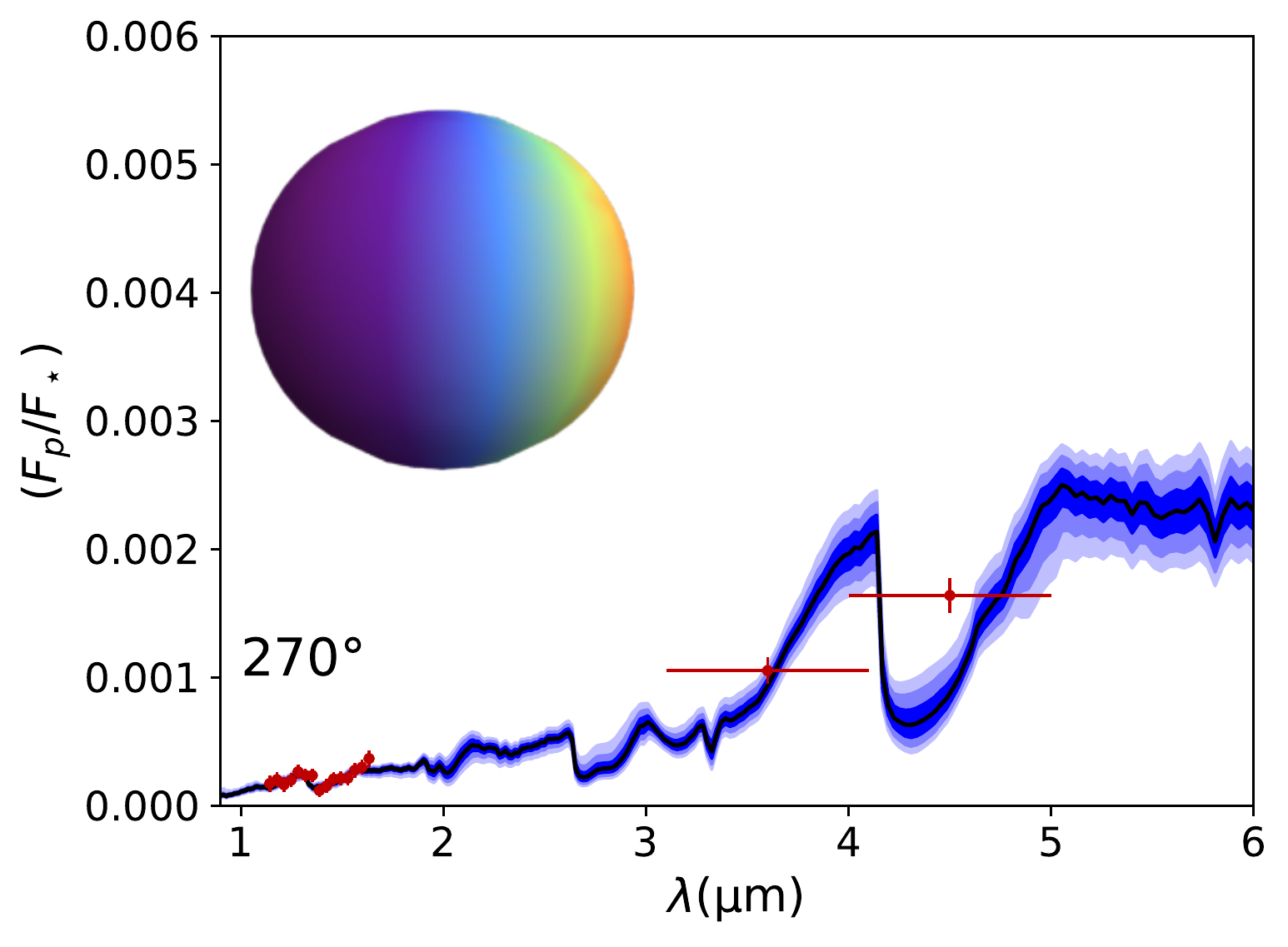}
                \caption{Emission spectra retrieved at different phases plotted with the observed data for each phase (as given in the bottom left of each panel) for retrieval setup A (no clouds, free chemistry, and \cite{17StLiBe.wasp43b} Spitzer analysis). The left of each panel shows the heat distribution over the surface of the planet, as computed by ARCiS during the retrieval. 
                }\label{fig:phase_spectra_A}
        \end{figure*}

        Figure~\ref{fig:phase_spectra_A} shows the observed emission data for different phases plotted alongside the retrieved spectra from ARCiS for retrieval setup A. The shading represents the 1, 2, and 3 $\sigma$ bounds. On the left of each panel is the corresponding heat distribution for each phase, with the cooler nightside in purple and the dayside hotspot in red. 
        Figures~\ref{fig:spectra_3phase} and~\ref{fig:spectra_3phase_zoomed} 
        show retrieved best-fit spectra at three different phases (90$\degree$ (evening), 180$\degree$ (dayside), 270$\degree$ (morning)) for each of the eight different retrieval setups, as detailed in each panel.

        It can be seen that there is some obvious asymmetry in the terminators from the observed data, with one terminator apparently hotter, and therefore emitting more flux than the other. This is particularly apparent for the setups A-D which use the Spitzer data from \cite{17StLiBe.wasp43b}. Setups E-H which use the Spitzer data from \cite{18MeMaDe.wasp43b} do not have the same extent of asymmetry. It is apparent that all retrieved spectra fit the observed data reasonably well, with no drastic outliers.

        \subsection{Retrieved transmission spectra}\label{sec:ret_trans}

        Figure~\ref{fig:spectra_trans} shows the retrieved and observed transmission spectra for all eight of our retrieval setups, as detailed in each panel. Retrieval setup I (which is the same as setup D, but in the full efficiency mode of {\sc Multinest}) is also included. Our 3D model, which we use for all retrievals simultaneously, retrieves both the emission phase spectra presented in Section~\ref{sec:ret_phase} and the transmission spectra shown in Figure~\ref{fig:spectra_trans}, with equal weight assigned to all observed data points. The fit is generally reasonable, although it does not show the absorption features associated with AlO as was found in our previous retrieval of the transmission spectra only~\citep{20ChMiKa.wasp43b}. The flatter transmission spectra retrieved in setups F and H (equilibrium chemistry, using Spitzer data from \cite{18MeMaDe.wasp43b}, without and with clouds included, respectively) is likely due to the high metallicities retrieved for both these cases, as given in Table~\ref{t:ret_pars_results}. Such high metallicities are associated with heavier atmospheres, which thus reduce the scale height of the atmosphere and flatten absorption features. The absorption features in the emission spectra are not affected in the same way as they are more controlled by the temperature structure of the atmosphere. It should also be noted that setup H (including clouds) uses the constant efficiency mode of {\sc Multinest} (see Section ~\ref{sec:const_eff}), and so the posterior distributions given in the corresponding plots are more constrained than they should be.

        \subsection{Atmospheric heat distribution}\label{sec:heat_redis}
        
        Figure~\ref{fig:contour_plots} gives a 2D projection (as a function of longitude and latitude) of heat distribution, with contours representing different values of  $\beta$ (where $\beta$ is a function of longitude and latitude, and gives the heat due to stellar radiation at the top of the atmosphere after diffusion has been taken into account; see Section~\ref{sec:arcis}). 
        This value of $\beta$ can be used in Equation~\ref{eq:beta}, along with the retrieved Guillot parameters, to find the  pressure-temperature profile at a given longitude and latitude point of the atmosphere, and at a given altitude (as determined by atmospheric pressure). This is illustrated in Figures~\ref{fig:contour_plots_temp_0.01}~and~\ref{fig:contour_plots_temp_0.12}, which give the temperature as a function of longitude and latitude for a cut of the atmosphere at a pressure layer of 0.01bar and 0.12 bar, respectively. Some comparison can be made to outputs from GCMs of WASP-43b such as \cite{15KaShFo.wasp43b}, \cite{18MeTsMa.wasp43b}, and \cite{20CaBaMo.exo}. 
        See for example Figure~2~(d)~and~(h) of  \cite{18MeTsMa.wasp43b}, which show horizontal temperature maps  at a pressure-layer cut of 0.01bar for atmospheres which do include and do not include a thermal inversion in the upper atmosphere, respectively. It should be noted that the longitude axis is from 0 to 360$\degree$, whereas ours is from -180 to +180$\degree$. 0.01~bar is the pressure layer where their jet reaches maximum strength. A comparison can be made to our Figure~\ref{fig:contour_plots_temp_0.01}, which gives our retrieved temperature maps at a pressure layer of 0.01~bar. For the case where a thermal inversion is included in the GCM of \cite{18MeTsMa.wasp43b} (labelled G=2.0), 
        their model appears qualitatively to match best to our retrieval setup F (equilibrium chemistry, no clouds, and Spitzer data from \cite{18MeTsMa.wasp43b}).  For the GCM of \cite{18MeTsMa.wasp43b} without a thermal inversion (labelled G=0.5), the temperature map appears most similar to the results of our retrieval setup G (free chemistry, with clouds, and Spitzer data from \cite{18MeTsMa.wasp43b}; see Figure~\ref{fig:contour_plots_temp_0.01}).
        
        Figure~3 of  \cite{15KaShFo.wasp43b} gives the longitude-latitude temperature cut at various pressure levels for 1~$\times$~solar  metallicity and for 5~$\times$~solar in their Figure~7. The temperature cuts at a pressure layer of 10.63~mbar ($\approx$~0.01~bar ) can be compared to our Figure~\ref{fig:contour_plots_temp_0.01}, which gives temperature maps at a pressure cut of 0.01~bar. The results from our retrieval setups B and D (i.e. equilibrium chemistry setups with Spitzer data from \cite{17StLiBe.wasp43b}, without and with clouds, respectively) appear most similar to both  the 1~$\times$~solar  and 5~$\times$~solar  GCM models of \cite{15KaShFo.wasp43b}.  \cite{15KaShFo.wasp43b} assume local equilibrium chemistry (accounting for condensate rainout, the same as in the present work) in their models, and ignore clouds and hazes. Their models could only be compared to HST/WFC3 data from \cite{14StDeLi.wasp43b} at the time they were published as the Spitzer data of \cite{17StLiBe.wasp43b} had not been observed. Our results indicate that their GCM models offer a good match to our retrieval results using \cite{17StLiBe.wasp43b} Spitzer data and assuming equilibrium chemistry.

        Figure~2 (Ib) of \cite{20CaBaMo.exo} gives the longitude-latitude temperature map as a result of their GCM at a cut of 0.12~bar, which can be compared to our retrieved temperature maps cut at the same pressure in Figure~\ref{fig:contour_plots_temp_0.12}. The results of our retrieval setup C (i.e. free chemistry, with clouds, and using Spitzer data from \cite{17StLiBe.wasp43b}), match the GCM of \cite{20CaBaMo.exo} most closely at this pressure layer. Beneath this pressure layer, the GCM models of \cite{20CaBaMo.exo} predict retrograde equatorial wind jets. 
        The deep circulation framework of all these GCMs are far more complex than the wind parametrisation used in this work, and the models of \cite{20CaBaMo.exo} show potential retrograde equatorial wind flow due to deep atmospheric wind jets in WASP-43b. Other variabilities and dynamical processes such as storms are expected to have impacts on the atmosphere of hot Jupiter exoplanets~\citep[see e.g.][]{03ChMeHa,21ChSkTh,21SkCh}. 
        Our retrieved spectra generally fit very well to the observed data, demonstrating that the retrieval model outlined in this work is well placed to deal with observations of WASP-43b using HST/WFC3 and Spitzer/IRAC. We expect it to also be well placed to deal with observations from the recently launched JWST, which will become available in the near future. This will be explored in future work.
        This, as well as the fact that we show very reasonable (and in some cases very good) matches to the GCM model outputs of \cite{15KaShFo.wasp43b}, \cite{18MeTsMa.wasp43b}, and \cite{20CaBaMo.exo}, is promising and gives weight to the ability of the GCMs to predict atmospheric conditions on exoplanets such as WASP-43b.

        We note that the coefficient $n_\phi$ which we use to specify the latitudinal extent of the winds is hard to constrain, because the observed data we are retrieving from is averaged across all latitudes. 
        
        One of the parameters we retrieve, $f_{\rm red}$, gives an indication of how heat is distributed from the day to the nightside of the planet (see Table~\ref{t:ret_pars}; $f_{\rm red}$  can vary between 0 (no heat redistributed from day to night) to 1 (all heat redistributed from day to night)). \cite{17StLiBe.wasp43b} derive a related parameter, $F$, to represent this heat redistribution: 
        \begin{equation}
                F = \frac{1}{2}(1+f_{\rm red}), 
        \end{equation}
        with $F$ varying between 0.5 (low heat redistribution) and 1 (high heat redistribution). 
        \cite{17StLiBe.wasp43b} find a value of $F$~=~0.501$^{+ 0.005}_{-0.001}$ 
        by integrating their retrieved model spectra. This corresponds to a value of $f_{red}$~=~0.002$^{+                0.010}_{-0.002}$, which can be compared to the values of $f_{red}$ found in our eight different retrieval setups, as can be found in Table~\ref{t:ret_pars_results}. In general our retrieved values of $F$ agree very well, particularly for the setups which use the Spitzer data from \cite{17StLiBe.wasp43b}, as may be expected. See Section~\ref{sec:spitz} for further discussion on the heat distribution parameter $f_{\rm red}$ and potential degeneracies with other parameters.

        \cite{19KeCoDa.wasp43b} find a heat recirculation factor (which varies from 0 for no recirculation to 1 for perfect heat recirculation) of 0.27$\pm$0.07 for WASP-43b. They infer a dayside temperature of 1664$\pm$69~K and a relatively hot nightside temperature of 984$\pm$67~K; this hotter nightside is more consistent with the results we get from the retrieval setups which use the \cite{18MeMaDe.wasp43b} Spitzer data. \cite{19KeCoDa.wasp43b} also use the \cite{18MeMaDe.wasp43b} analysis of the Spitzer phase curve in their study.

        \subsection{Influence of Spitzer phase curve data reduction}\label{sec:spitz}

        As mentioned in Section~\ref{sec:spitz_phase}, there are various different reductions of the Spitzer phase curves available in the literature. We find the choice to influence our interpretations of the data through our 3D retrievals.  This can be seen in Table~\ref{t:ret_pars_results} and as described in Section~\ref{sec:heat_redis}.
        
        In general, $f_{\rm red}$, the flux redistribution factor, is very low or zero for the retrievals using the Spitzer data from \cite{17StLiBe.wasp43b}. It is typically slightly higher for the retrievals that instead use the \cite{18MeMaDe.wasp43b} Spitzer data. There is a relation here with $f_{v}$, with a higher value of $f_{v}$ representing a scenario with a higher ratio of wind speed to diffusion (i.e. a greater hotspot shift).  It appears that using the \cite{17StLiBe.wasp43b} Spitzer data results in a greater hotspot shift (which represents a greater asymmetry between the East and West of the planet), but overall very little heat distributed across from the dayside to the nightside. Using the \cite{18MeMaDe.wasp43b} Spitzer data, however, generally results in a very small hotspot shift, but a larger amount of heat distributed across from the dayside to the nightside. This is evident when looking at the spectra (e.g. see Figure~\ref{fig:spectra_3phase}), where the difference in flux is not so extreme between night and day when using the \cite{18MeMaDe.wasp43b} Spitzer data in comparison to the \cite{17StLiBe.wasp43b} Spitzer data. The outlier here is 
        the retrieval setup where Spitzer data is from \cite{18MeMaDe.wasp43b}, equilibrium chemistry is enforced, and no clouds are included (setup F). It is unclear exactly why this is, but this hints that there are degeneracies in the interpretation of the observed data used in this study with our model. We note that due to data availability, we do not include Spitzer data at 90$\degree$
        (i.e centred on the evening terminator) for the Mendon{\c{c}}a retrievals, and only include data from HST/WFC3 at this phase (see Table~\ref{t:obs_data}). However, due to the way the 3D retrieval is setup, and the fact that we use Spitzer data for phases either side of 90$\degree$ (67.5$\degree$ and 112.5$\degree$), we do not expect this to influence our findings significantly.

        \subsection{Hotspot shift}
        
        All panels of Figures~\ref{fig:contour_plots}, \ref{fig:contour_plots_temp_0.01}, and \ref{fig:contour_plots_temp_0.12} show a hotspot shifted slightly away from the middle of the dayside. 
        The value of this hotspot shift for the different retrieval setups is given in Table~\ref{t:hotspot_shifts}, which can be compared to values found in previous studies, which are also given in Table~\ref{t:hotspot_shifts}. The caveat is that these hotspot shifts were all derived based on the 4.5$\mu$m Spitzer data only, which could give limited information. The values determined in the present work are based on our full 3D retrievals, which include effects due to the geometry of the planet. Nevertheless, a comparison is still of interest. In general there is a greater shift in hotspot for those retrievals that use 
        Spitzer data from \cite{17StLiBe.wasp43b} as opposed to Spitzer data from \cite{18MeMaDe.wasp43b}. This could be linked to the higher asymmetry  between the morning and evening terminator regions in the spectra 
        of  \cite{17StLiBe.wasp43b} (see Figure~\ref{fig:spectra_3phase}).  In general, our values broadly agree with those found in the other studies noted in Table~\ref{t:hotspot_shifts}. However, we note that those studies do only consider the 4.5$\mu$m Spitzer data to infer the hotspot shifts given in Table~\ref{t:hotspot_shifts}.

        \begin{table*}[!tp]
                \caption{Hotspot offsets for WASP-43b based on various analyses of  the WASP-43b phase curve in the literature, based on the 4.5$\mu$m Spitzer point only (as published by \cite{20MaSt.wasp43b} and \cite{21BeDaCo.wasp43b}). The hotspot shifts found as a results of our 3D retrievals are given underneath these literature values. 
                }
                \label{t:hotspot_shifts} 
                \centering  
                \begin{tabular}{lc}
                        \hline
                        \hline
                        \rule{0pt}{3ex}Analysis & Hotspot offset ($\degree$) \\
                        \hline
                        \rule{0pt}{3ex}\cite{17StLiBe.wasp43b}$^{a}$ &  \makecell{\rule{0pt}{3ex} 21.1$\pm$1.8} \\
                        \rule{0pt}{3ex}\cite{18MeMaDe.wasp43b}$^{b}$ &  \makecell{\rule{0pt}{3ex} 12$\pm$3.0} \\
                        \rule{0pt}{3ex}\cite{19MoDaDi.wasp43b} &  \makecell{\rule{0pt}{3ex} 11.3$\pm$2.1} \\
                        \rule{0pt}{3ex}\cite{20MaSt.wasp43b} &  \makecell{\rule{0pt}{3ex} 20.6$\pm$2.0} \\
                        \rule{0pt}{3ex}\cite{21BeDaCo.wasp43b} &  \makecell{\rule{0pt}{3ex} 20.4$\pm$3.6} \\
                        \hline
                        \hline
                        \multicolumn{2}{l}{\rule{0pt}{3ex}Retrieval results from this study:}\\                             
                        \multicolumn{2}{l}{\rule{0pt}{3ex}Using Stevenson Spitzer data:}\\        
                        \rule{0pt}{3ex} A       &       $       13.5    ^{+     3.7     }_{-    0.7     }$      \\
                        \rule{0pt}{3ex} B       &       $       17.3    ^{+     1.8     }_{-    2.3     }$      \\
                        \rule{0pt}{3ex} C       &       $       18.3    ^{+     2.0     }_{-    1.1     }$      \\
                        \rule{0pt}{3ex} D       &       $       13.2    ^{+     2.1     }_{-    1.2     }$      \\
                        \multicolumn{2}{l}{\rule{0pt}{3ex}Using Mendon{\c{c}}a Spitzer data:}\\        
                        \rule{0pt}{3ex} E       &       $       2.8     ^{+     1.4     }_{-    1.0     }$      \\
                        \rule{0pt}{3ex} F       &       $       7.4     ^{+     2.6     }_{-    1.6     }$      \\
                        \rule{0pt}{3ex} G       &       $       4.1     ^{+     0.8     }_{-    0.8     }$      \\
                        \rule{0pt}{3ex} H       &       $       3.8     ^{+     1.2     }_{-    0.6     }$      \\
                \end{tabular}
                \rule{0pt}{0.2ex}
                \flushleft{$^{a}$\textit{used in this study for setups A-D, and I}}
                \flushleft{$^{b}$\textit{used in this study for setups E-H}}
        \end{table*}

        \subsection{Variation of molecular abundances across the atmosphere}

        Figures \ref{fig:AlO_COratio}~-~\ref{fig:NH3_COratio} give the variation of molecular abundances as a function of phase for each of the eight different retrieval setups and six different molecules that are included in the free molecular retrievals. These are the volume mixing ratio (VMR) abundances which are averaged over the visible part of the disk, centred at each of the given phases, and taken at a pressure-layer cut of 0.18~bar at the equator. 
        These  phase-averaged plots allow us to compare the molecular abundances found in the free molecular retrieval case with those found in the corresponding equilibrium chemistry case; see Figures~\ref{fig:AlO_COratio}~-~\ref{fig:NH3_COratio}. The same data used as presented for these six species are available for each species included in the equilibrium chemistry setup (see Section~\ref{sec:eq_chem}), but we only present here those which can be compared to the six species which are included in the free molecular retrieval setups (i.e. AlO, CH$_4$, CO, CO$_2$, H$_2$O, and NH$_3$).  
        The plots of Figures~\ref{fig:CH4_COratio}~-~\ref{fig:H2O_COratio} (which correspond to CH$_4$, CO, CO$_2$, and H$_2$O, respectively) can be compared to Figure~7 of \cite{17StLiBe.wasp43b}. We generally find a similar pattern for these four species that we can compare to \cite{17StLiBe.wasp43b}, as described below.  We also discuss briefly the similarities between our eight different retrievals, and highlight where the same trend is seen in all or most retrieval setups. 
        
        \subsubsection{AlO}
        
        As can be seen in Figure~\ref{fig:AlO_COratio}, AlO is predicted by all the equilibrium chemistry retrievals to be most abundant on the dayside, where the atmosphere is hottest. Although we find this same trend in two of the free chemistry retrievals, we also find it to be unconstrained or not very abundant in the other two free chemistry retrievals.  
        A potential indication for AlO was found in WASP-43b's transmission spectra by \cite{20ChMiKa.wasp43b}.
                We note that if there really is AlO present at one or more of the terminators, then we would expect to also see it on the dayside of the planet. It is possible that the emission data is not precise enough to allow for a strong determination of its presence. In any case, observing a wider wavelength region would really help to confirm whether this species is present in the atmosphere of WASP-43b, and if it is present, to place tighter constraints on its abundance.

        \subsubsection{CH$_4$}
        
        We find a very consistent trend of CH$_4$ abundance as a function of phase, particularly in the retrievals cases which use the \cite{17StLiBe.wasp43b} Spitzer data (see Figure~\ref{fig:CH4_COratio}). We find CH$_4$ to vary between $\sim$$10^{-4}$ on the dayside and $\sim$$10^{-3}$ on the nightside. This can be compared to Figure~7 of \cite{17StLiBe.wasp43b}, where a higher abundance is also found on the nightside than on the dayside, but with different overall abundances. The abundance trend as a function of phase for the four retrievals that use the  \cite{17StLiBe.wasp43b} Spitzer data are particularly consistent. 
        
        \subsubsection{CO and CO$_2$}
        
        We retrieve relatively high abundances of both CO and CO$_2$ (see Figures~\ref{fig:CO_COratio} and~\ref{fig:CO2_COratio}). \cite{17StLiBe.wasp43b} combine these two species together. They also find a relatively consistent abundance across day and night, at a VMR of around $10^{-4}$. This is not too dissimilar from our findings. In particular, our equilibrium chemistry retrievals find an extremely high abundance of CO across all phases. As pointed out by \cite{19IrPaTa.wasp43b}, any errors in the offset between HST and Spitzer data  could lead to erroneous molecular abundances; in particular of CO. Although there is likely to be some degeneracy between these species, which makes it potentially hard to distinguish between them with our given dataset, we choose to keep them separate when performing our retrievals. Both species have a very strong absorption feature at 4.5$\mu$m, but CO$_2$ also has some weaker features in the HST/WFC3 region.

        \subsubsection{H$_2$O}
        
        We find the same trend with all of our retrieval setups (with one exception; see Figure~\ref{fig:H2O_COratio}): a higher abundance of H$_2$O on the dayside than on the nightside. This appears for both the equilibrium and free chemistry retrievals. This trend is also found by the 1D set of  retrievals from \cite{17StLiBe.wasp43b}. We note that the apparent tight constraints on H$_2$O abundance for the free retrievals that include clouds are artificial, due to the constant efficiency mode of {\sc Multinest} (see Section~\ref{sec:const_eff}).
        
        \subsubsection{NH$_3$}
        
        Although NH$_3$ is predicted to be relatively abundant by the equilibrium chemistry retrievals (see Figure~\ref{fig:NH3_COratio}), it remains unconstrained by our free chemistry retrievals. A wider wavelength region is likely needed in order to properly constrain the abundances of this species, if it is present in the observable atmospheric region.

        \subsection{Temperature-pressure profiles}
        
        As previously mentioned, we also retrieve pressure-temperature profiles for each phase, averaged over the visible disk of the planet. Figure~\ref{fig:TP_4phase} gives these for the dayside (180$\degree$), nightside (315$\degree$), morning (270$\degree$) and evening (90$\degree$) terminators, with each panel giving the temperature profiles retrieved as a result of each of our different retrieval setups. 
        
        We use our 3D retrievals to find the temperature structure at each of the points specified on the grid of 36 longitude and 18 latitude points. We then average for each specified phase in order to facilitate comparisons with other works. In Figure~\ref{fig:TP_4phase} we present four phases only: 90$\degree$ (evening), 180$\degree$ (dayside), 270$\degree$ (morning), and 315$\degree$ (nightside). The shading of the plots represents the 1, 2, and 3 $\sigma$ regions. These plots can also be compared to other works such as those from the 1D retrievals of \cite{17StLiBe.wasp43b} (e.g. see their Figure~6). We can also compare to GCM outputs, for example see the first panel of Figure~2 of \cite{19HeGrSa.wasp43b}, which presents the results of 3D GCM models of WASP-43b~\citep{18PaLiBe.exo}.
        The horizontal lines added to the plots given in Figure~\ref{fig:TP_4phase} represent the  approximate region outside which the temperature cannot be constrained from the observed data. It is important to note that the temperature may appear to be artificially well constrained, in particular for the setups which include clouds, largely due to the mode of {\sc Multinest} used to make some of the computations feasible (see Section ~\ref{sec:const_eff}), but likely also because of the shape of the parametrised profile. We also include the results of retrieval setup I as part of Figure~\ref{fig:TP_4phase}, which is the same as setup D but using the full efficiency as opposed to the constant efficiency mode of {\sc Multinest} (see Section ~\ref{sec:const_eff}). It is very apparent, by comparing these two panels, that the retrieval using the constant efficiency mode (setup D) is too artificially well constrained compared to using the full efficiency mode (setup I). 

        There is a link here to the retrieved parameter T$_{\rm int}$ (see Table~\ref{t:ret_pars}), which in the absence of stellar radiation (i.e. due to internal planet heat) gives the atmospheric temperature at an optical depth of $\tau$~=~$\frac{2}{3}$.
        It can be seen from the posterior distributions contained in the supplementary information that this parameter is well constrained as having a relatively high value of $\log_{10}$~(T$_{\rm int}$)~=~2.69~$\pm$~0.02, whereas  the same parameter in retrieval setup I remains unconstrained. T$_{\rm int}$ appears in Eq.~\ref{eq:beta}, which determines the temperature at a given atmospheric layer in the \cite{10Guillot.exo} temperature-pressure parametrisation that we use. A similar trend in terms of how constrained a parameter is is seen for other parameters, including log($\gamma$) and log($\kappa_{IR}$), which also determine the pressure-temperature structure. 
        
        \cite{21ChAlEd.wasp43b} point out that the Guillot profile (which we use in this work; see Section~\ref{sec:PT_par}) may not represent the nightside atmospheric structure of the planet so well, and so they test for allowing the use of a free profile in their retrievals of WASP-43b. This has also been explored in other studies such as \cite{17BlDoGr,09MaSe}. Our retrieval setup allows for the use of different pressure-temperature profiles, such as a free temperature profile, but this would come with an increase in computational time. Allowing for different pressure-temperature profiles while performing our 3D analysis is something we will investigate further in the future, particularly for the nightside of the planet. When using a Guillot profile, \cite{21ChAlEd.wasp43b} retrieve a non-inverted temperature profile and a chemistry structure similar to that of \cite{20FeLiFo.phase}. However, when allowing for a free temperature profile, as opposed to the parametrised Guillot profile, which we also use in the present study, they find that the temperature structure (at least on the dayside and at hotspot regions), is consistent with a thermal inversion.

        \subsection{Clouds}\label{sec:clouds}
        
        Figure~\ref{fig:cloud_all} gives the retrieved parameters that represent our cloud setup for each of the four retrievals for which clouds are included, as detailed in each panel (see Section~\ref{sec:clouds_vs_no} for details). We retrieve P$_{\rm cloud}$ (cloud-top pressure), $\kappa_{\rm cloud}$ (cloud opacity), and $A_{\rm cloud}$ (cloud albedo). For each we retrieve the values corresponding to the minimum and maximum values of $\beta$, and allow them to vary linearly in between (see Section~\ref{sec:clouds_vs_no}). 
        In general, most of the cloud parameters in all of the retrievals that include clouds are not well constrained, particularly the albedo. This is perhaps unsurprising, considering the narrow wavelength coverage of the observations we are retrieving, and  particularly because we do not include any data in the optical region. We find decent fits to the data without clouds included, which could indicate their inclusion is not necessary to fit the observed data. Nevertheless, clouds are expected throughout the atmosphere \cite[see, for example,][]{19HeGrSa.wasp43b,21HeLeSa.clouds}.
        In particular, the detailed cloud and haze models of WASP-43b computed by \cite{19HeGrSa.wasp43b} predict relatively transparent haze 
        at very high altitudes on the dayside of WASP-43b, and more opaque clouds 
        on the nightside. 
        It could therefore be that the current observed data does not contain enough information to retrieve information on these clouds predicted to be present.
        In order to properly explore such a trend in a retrieval, we would not only need data that covers a wider wavelength region, but we would also ideally use more detailed cloud formation models than those used in the present work. There is a cloud formation model in ARCiS~\citep{18OrMi.arcis,20MiOrCh.arcis} that we did not use for the current study in order to make the retrievals more computationally feasible. This model allows the C/O to vary locally throughout the atmosphere, with an explicit link to cloud formation throughout the atmosphere. 
        This more sophisticated approach to treating the presence of clouds in the atmosphere, including allowing for patchy clouds, has obvious benefits, with the familiar caveat of increased computational time.
        
        \cite{21FrMaSt.wasp43b} analyse reflected light observations in the optical, taken using the HST WFC3/UVIS instruments.  They do not detect clouds on the dayside of the planet within the pressure layers probed (P~>~1~bar). 
        They rule out the presence of a high-altitude, bright, uniform cloud layer, but they cannot rule out the presence of patchy clouds.          \cite{19IrPaTa.wasp43b} retrieve a colder nightside of WASP-43b using their 2.5D-setup than predicted from GCMs. They assume from their retrievals of WASP-43b that there is a thick cloud on the nightside with a cloud-top pressure~$<$~0.2~bar.

        \subsection{Albedo}
        
        In theory we should be able to compare our retrieved bond albedo values, indicating the proportion of energy reflected as opposed to absorbed, with those deduced from previous studies by 1~-~$f_{\rm day}$. However, we do not compute the temperature structure of the atmosphere self-consistently, which would take into account the absorption properties of the atmospheric species and their retrieved abundances. ARCiS does have the capability to do this, but including it in these 3D retrievals is currently computationally unfeasible. 
        
        The values of bond albedo of WASP-43b found in the literature are generally much lower than the values which we can deduce from our retrievals using $1-f_{day}$ (see Table~\ref{t:ret_pars_results}). 
        For example, \cite{19KeCoDa.wasp43b} find a bond albedo for WASP-43b of 0.22$^{+0.13}_{-0.12}$, and \cite{14StDeLi.wasp43b} derive a bond albedo of 0.078 – 0.262 based on computed day and night fluxes.

        \section{Conclusions and future outlook}\label{sec:conclusions}
        
        In this work we have outlined the general theory behind a 3D retrieval setup for use in characterising exoplanet atmospheres. This setup models incoming radiation from the host star and how it spreads throughout the planetary atmosphere via diffusion and longitudinal winds, which is used to find the vertical pressure-temperature profile on a grid of longitude-latitude points. The routine is provided as a stand-alone code, which can be freely downloaded via GitHub\footnote{\url{https://github.com/michielmin/DiffuseBeta}}. Here, we have implemented it into the Bayesian retrieval package ARCiS and run a series of 3D retrievals on spectral observations of WASP-43b as a demonstration. In running eight different retrieval setups, we are able to compare a number of different input assumptions; in particular, constraining equilibrium chemistry vs a free molecular retrieval, testing the case of no clouds vs parametrised clouds, and using Spitzer phase data which has been reduced from two different sources (\cite{17StLiBe.wasp43b} and \cite{18MeMaDe.wasp43b}). We give some examples of how results from our 3D retrievals can be compared to GCM simulations, with indications of some good agreements. We hope to expand on this in the future, in a more quantitative way. 
        
        In summary, we find a consistent super-solar metallicity and super-solar C/O for all retrieval setups which assume equilibrium chemistry (see Table~\ref{t:ret_pars_results} for a summary of retrieved parameters). We find that all eight retrieval setups fit the observed spectra reasonably well, with no drastic outliers. This indicates that there are degeneracies in how the 3D setup can fit the observed data. We find that the retrievals that use observed phase-dependent emission Spitzer data from \cite{17StLiBe.wasp43b} have a low heat redistribution factor, while the value is retrieved to be higher when using Spitzer data from \cite{18MeMaDe.wasp43b}. There is a hotspot shift found in all retrievals, with a higher shift for those using the \cite{17StLiBe.wasp43b} Spitzer data than for those using the \cite{18MeMaDe.wasp43b} Spitzer data. The \cite{18MeMaDe.wasp43b} Spitzer dataset apparently requires a more diffusion-based solution rather than strong equatorial winds, which are more favoured by the \cite{17StLiBe.wasp43b} Spitzer dataset. The trend of molecular abundances varying across the atmosphere is most consistent in the case of CH$_4$, where is it found to be higher on the nightside and decreasing on the dayside in all retrievals. Generally H$_2$O is found to consistently be lower on the nightside and increasing on the dayside, with one exception; AlO is generally expected to be most abundant on the dayside, as found in all equilibrium chemistry retrievals and in two of the free molecular abundance retrievals. The AlO abundance in the other two free retrievals remains unconstrained. NH$_3$ is also predicted to be relatively abundant if assuming equilibrium chemistry, but is unconstrained in all the free molecular retrievals, most likely due to the observed data wavelength coverage. CO and CO$_2$ are generally found to be present in high abundances for all retrievals,  although again we find they are less constrained for the free molecular  abundance retrievals. We find the temperature-pressure profiles to be particularly influenced by the mode of {\sc Multinest} used (see Section~\ref{sec:const_eff}). Artificially well-constrained parameters lead to seemingly very well-constrained profiles. 
        T$_{\rm int}$ is in general hotter for all retrievals where equilibrium chemistry is assumed. Although such values of  T$_{\rm int}$ are feasible based on the theoretical predictions of~\cite{19ThGaFo} for an exoplanet such as WASP-43b,
        we note that setup I has a lower value than setup D (both setups are identical, including clouds, and assume equilibrium chemistry, but they use different modes of {\sc Multinest}). It can be seen from the respective posterior plots in the supplementary information document of this work that this is due to T$_{\rm int}$ being unconstrained in setup I, but artificially constrained in setup D. We therefore advise caution when interpreting such results.  Our retrieved cloud parameters remain unconstrained for all retrievals where clouds were included. We therefore conclude that clouds cannot be inferred from the currently available observed data using our retrieval setup with a relatively simple cloud setup (we allow for clouds to vary linearly from day to night, but do not allow for patchy clouds, for example). This can be improved by more data and a more sophisticated cloud setup. We do not infer that the atmosphere is cloud-free from our results, but perhaps our cloud-free retrievals can be considered more reliable in the current study due to the mode of {\sc Multinest} used. 
        
        It was mentioned a few times in this work that knowledge of the atmosphere of WASP-43b would be significantly improved by more data observed over a wider wavelength range. We find that the choice of phase-dependent Spitzer emission data has a big impact on the retrieved parameters, particularly those relating to heat distribution. More observations across this wavelength region will be vital to improve our understanding of the atmosphere of WASP-43b. Fortunately, observations of WASP-43b are expected to be obtained as a result of several JWST Cycle 1 guaranteed time observations (GTO) and Early Release Science (ERS) programmes. 
        Amongst these, JWST will observe a phase curve of WASP-43b in the 5~-~12~$\mu$m region using the MIRI LRS instrument, as detailed in \cite{20VaPaBl.exo} and also in ~\cite{18BeSt.exo}. This will also be complemented by another phase curve observation by JWST, using the NIRSpec instrument to cover the region around 3~-~5~$\mu$m (GTO programme 1224, PI: S. Birkmann). The increased wavelength coverage that JWST will observe is expected to really help in breaking degeneracies and constraining our retrieved parameters. Ideally these proposed observations will be complemented by observations of WASP-43b at shorter wavelengths, including the visible, which will particularly help with constraining cloud parameters and temperature-pressure profiles through the ratio of mean optical to infra-red opacity. Observations over a large wavelength region using one instrument will also deal with any potential calibration issues between different instruments (in particular between Spitzer and HST in this work). This will greatly help in interpreting the physical properties of the atmosphere. We demonstrate that when such observations become available in the near future, we have a powerful tool to help interpret the observations and compare them to current theoretical predictions and properties inferred from currently available observations. 
        
        Some studies predict that there are disequilibrium chemistry effects affecting the atmosphere, and therefore affecting the observed spectra of WASP-43b \cite[see, for example,][]{18MeTsMa.wasp43b}. Although we ran a set of free chemistry retrievals that allowed for disequilibrium chemistry effects, it would be of interest in the future to treat these effects in a parametrised way by including the disequilibrium chemistry parametrisation that has recently been included in ARCiS~\citep{21KaMi.arcis}. However, there are currently only a limited number of species included in this disequilibrium chemistry scheme. Aluminium-bearing species such as AlO are currently not included, largely due to the availability of suitable data. This is something that could be explored further in the future.

        \section*{Supplementary material}
        
        The supplementary information document associated with this work, which includes extra figures, is available from \url{https://doi.org/10.5281/zenodo.6325489}.

        \section*{Acknowledgements}
        
        This project has received funding from the European Union's Horizon 2020 Research and Innovation Programme, under Grant Agreement 776403, ExoplANETS-A.

        \bibliographystyle{aa}
        \bibliography{3D_phases}
        
        \onecolumn
        
        \begin{appendix}\label{sec:append1}
                
                \section{Figures of retrieval results}\label{sec:appendix_retrieval}

                \begin{figure}[H]
                        \centering
                        \includegraphics[width=0.4\textwidth]{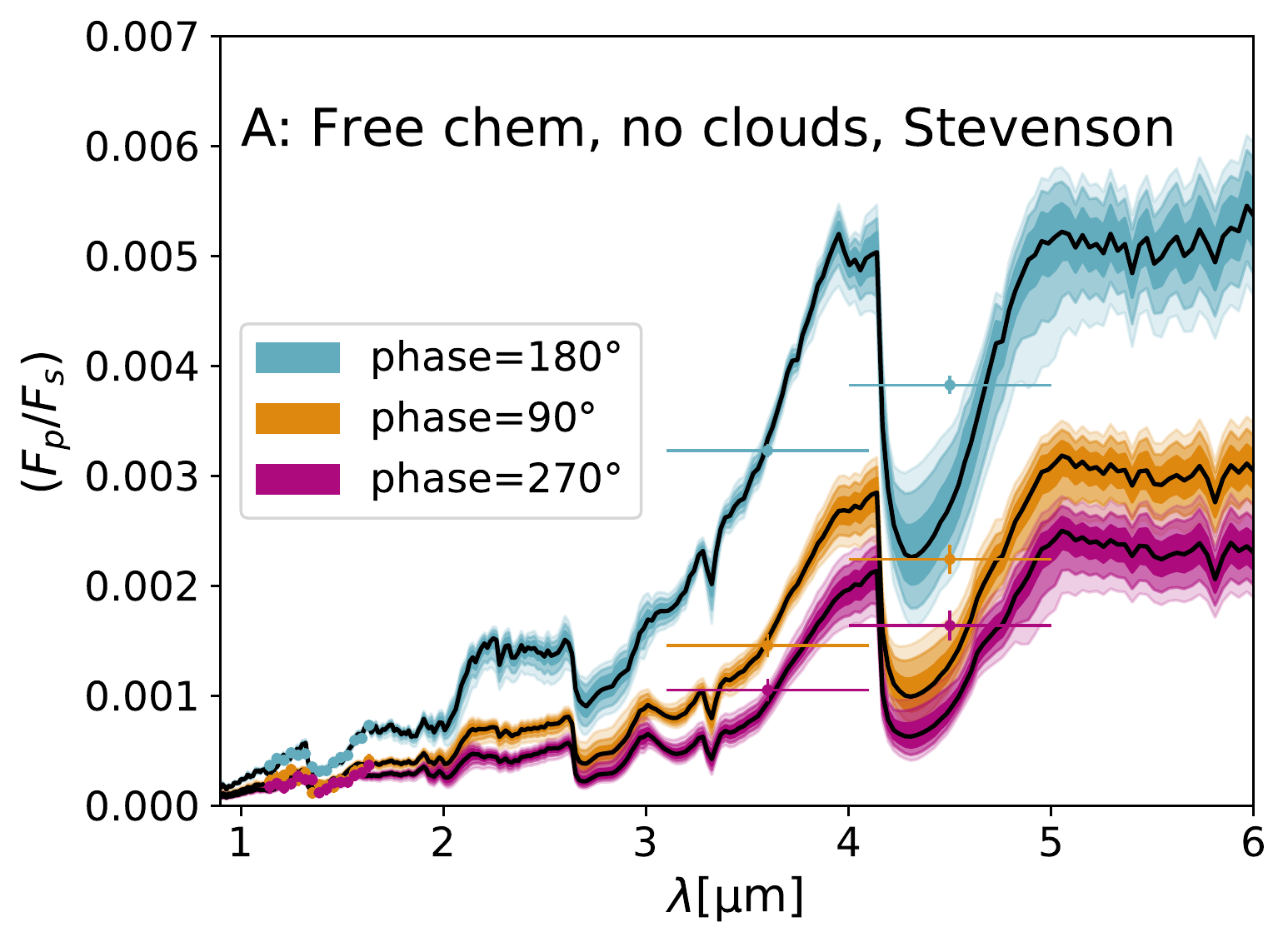}
                        \includegraphics[width=0.4\textwidth]{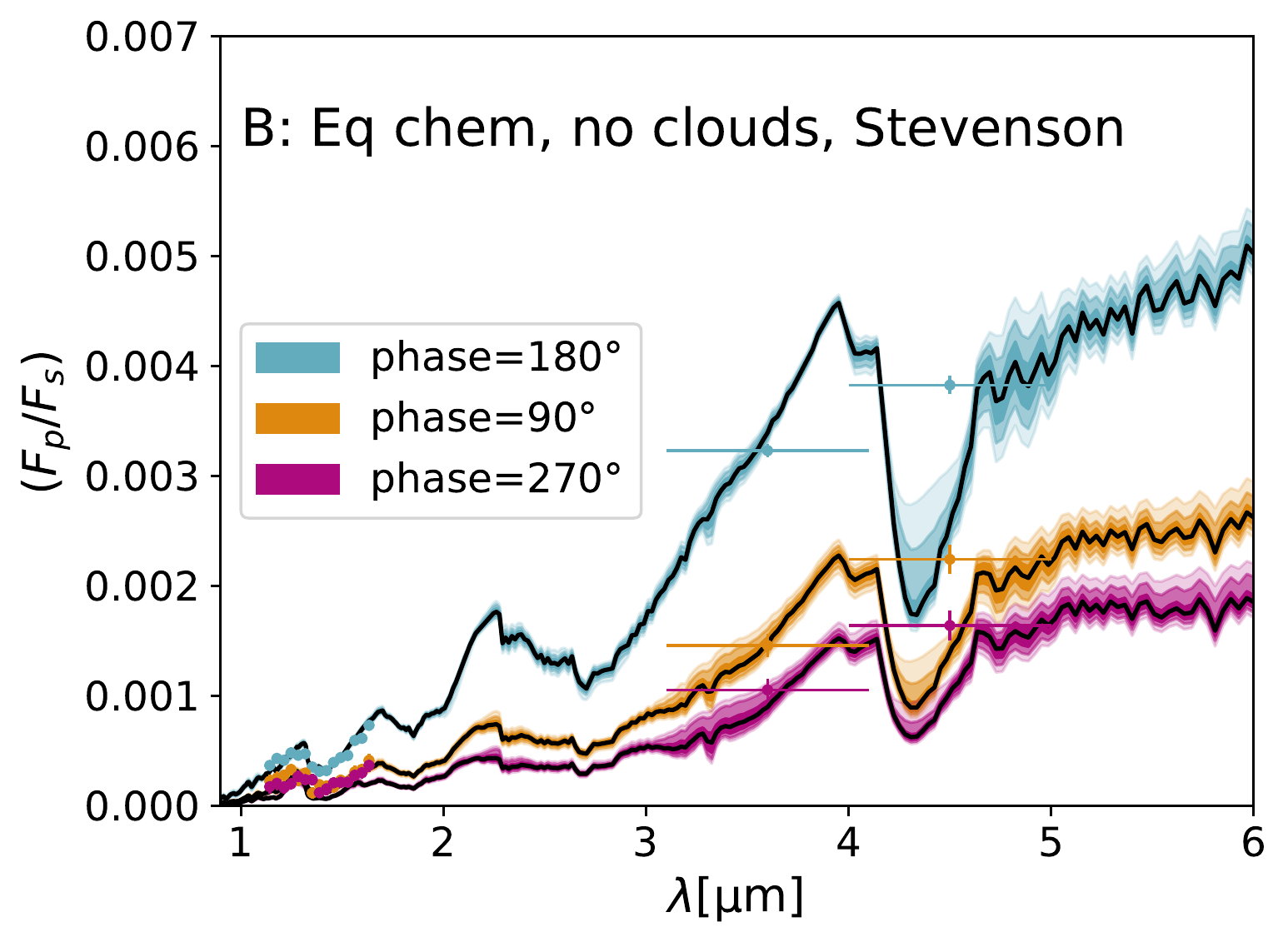}
                        \includegraphics[width=0.4\textwidth]{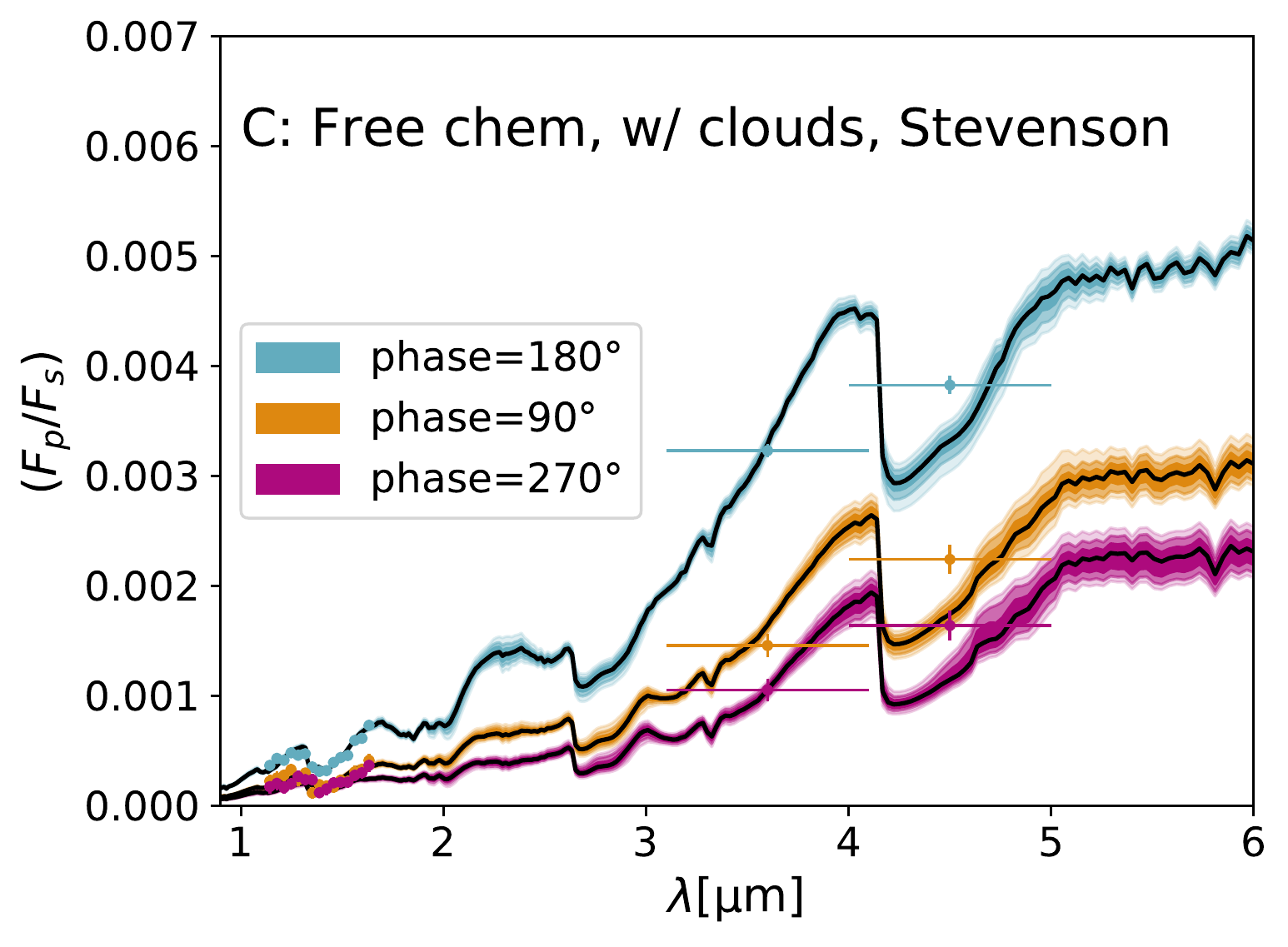}
                        \includegraphics[width=0.4\textwidth]{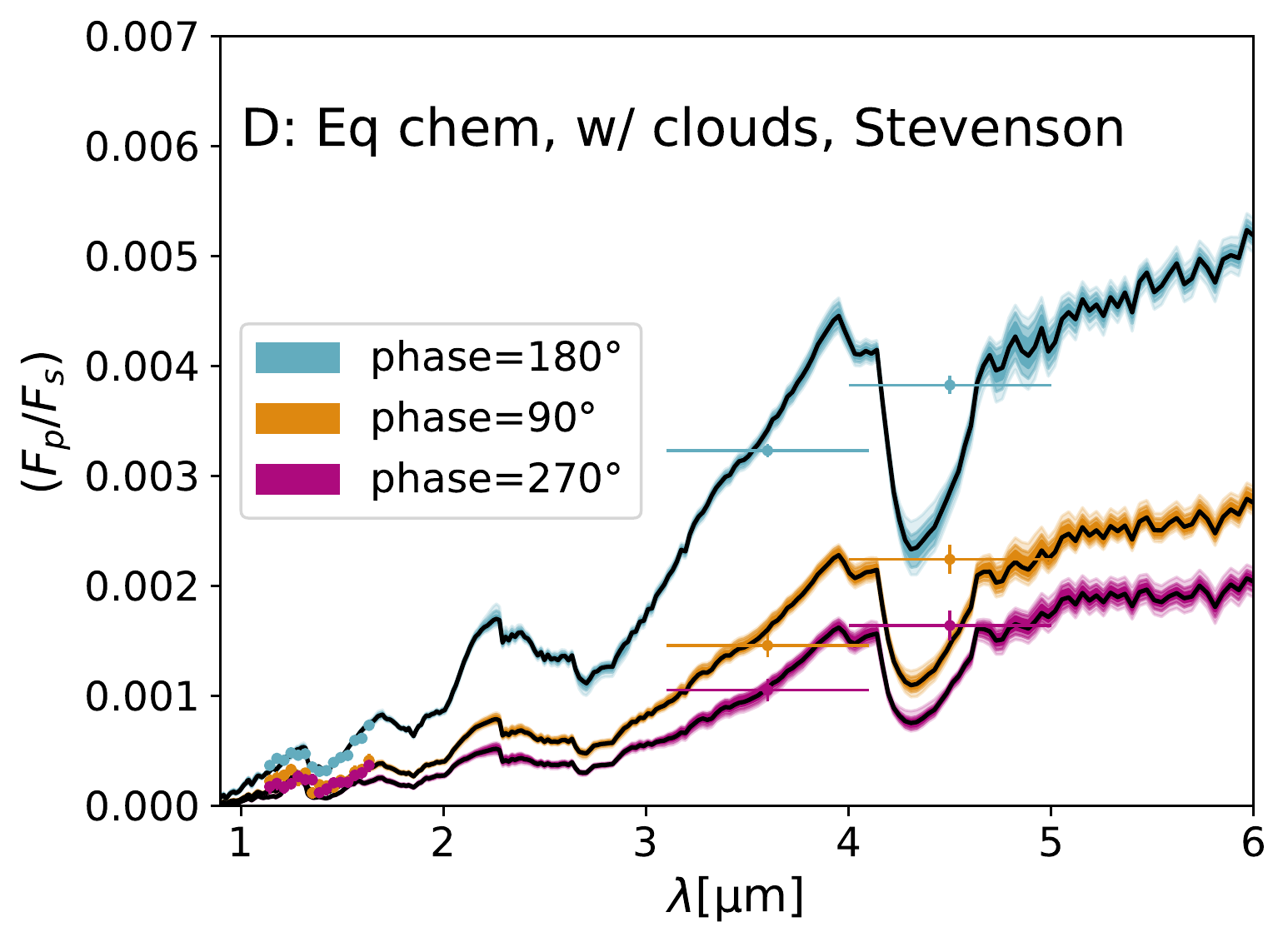}
                        \includegraphics[width=0.4\textwidth]{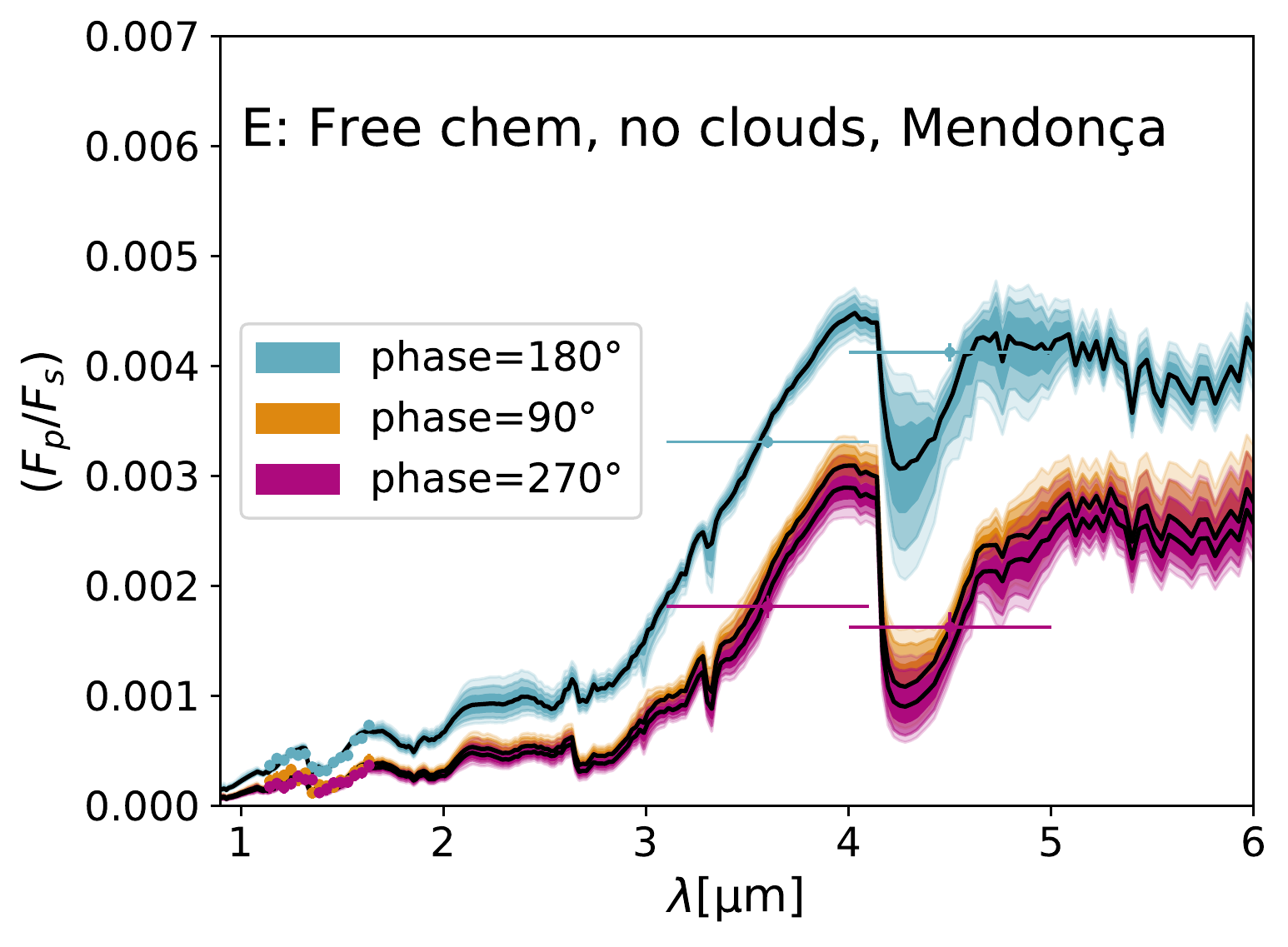}
                        \includegraphics[width=0.4\textwidth]{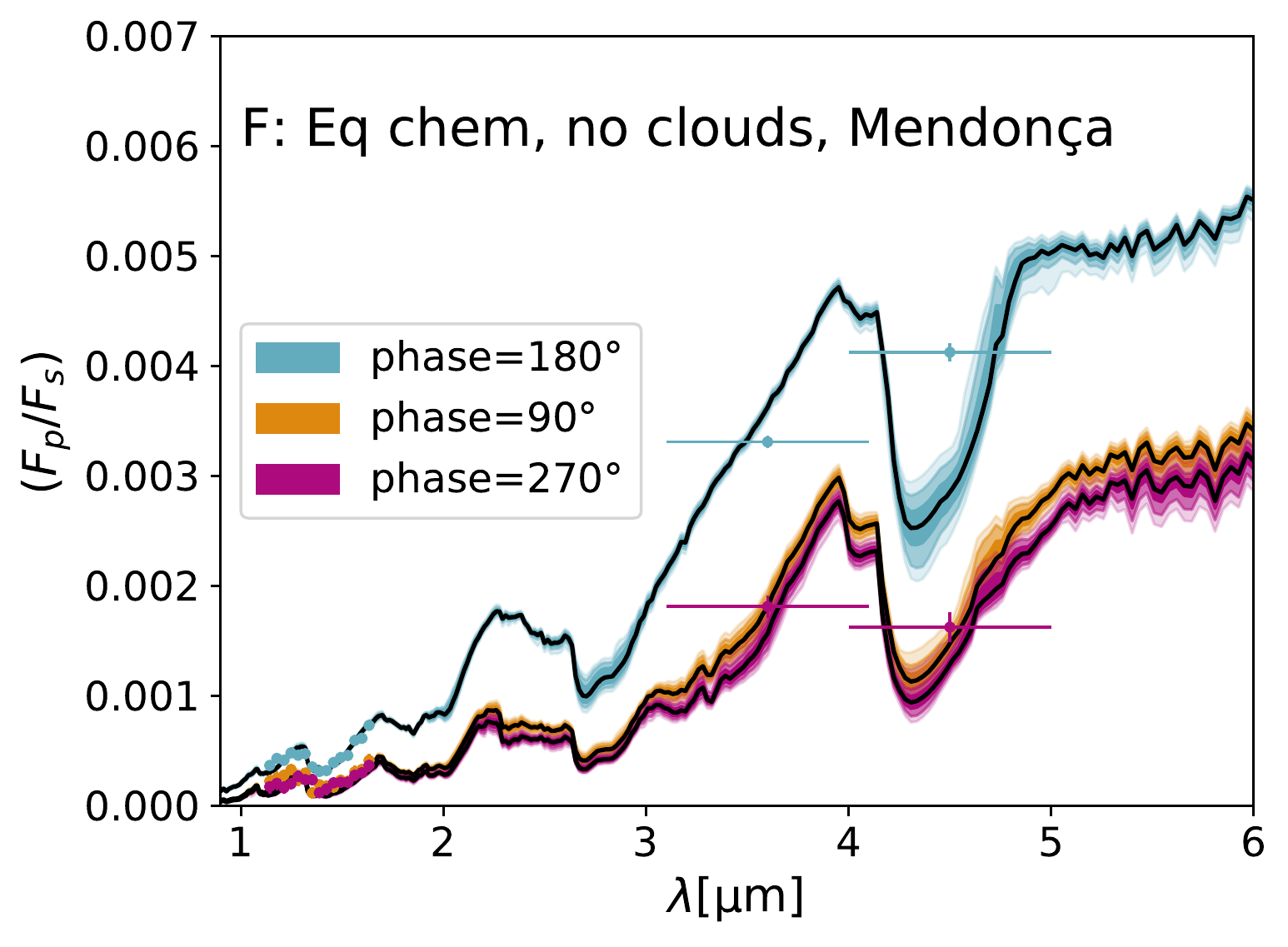}
                        \includegraphics[width=0.4\textwidth]{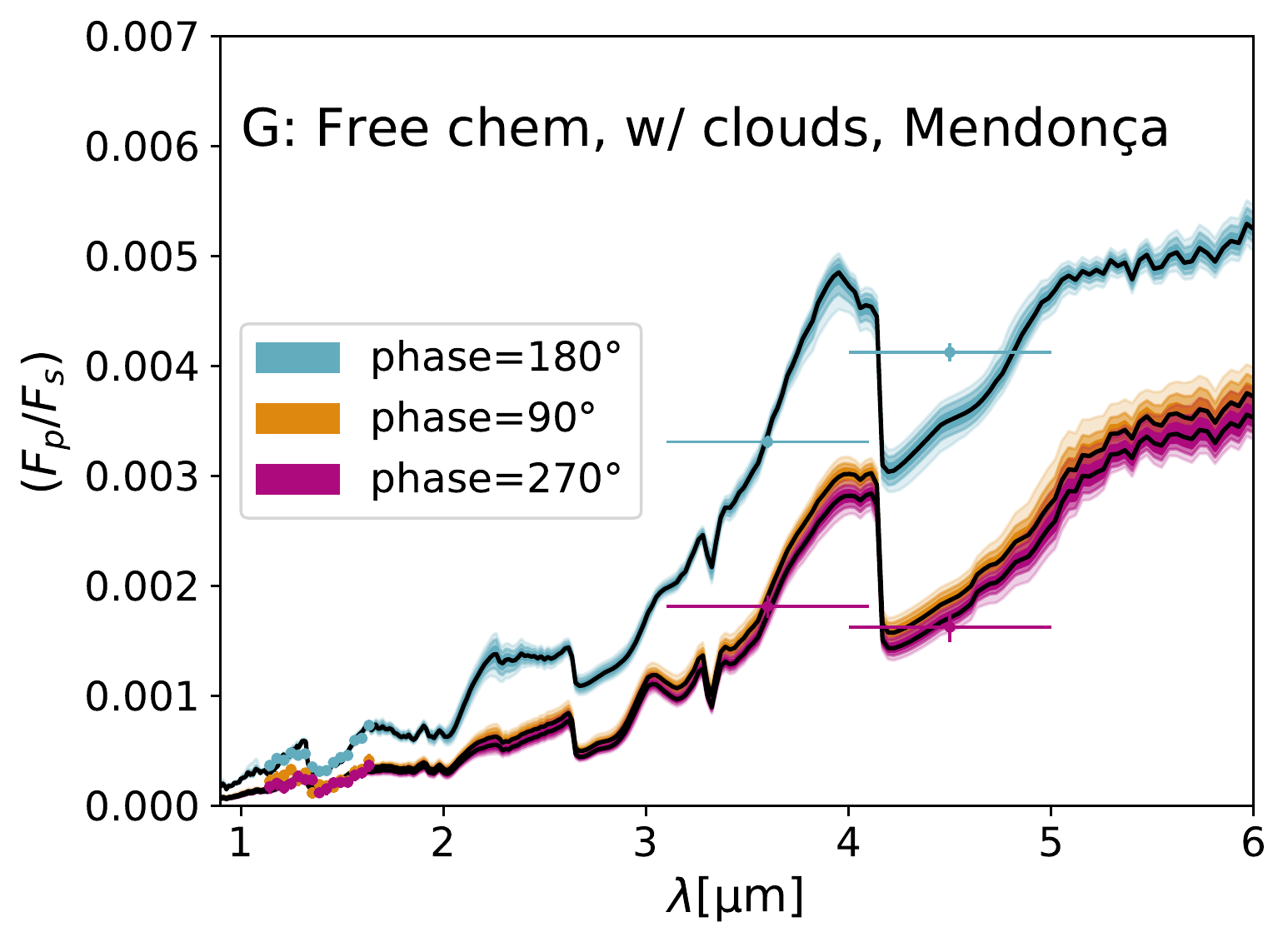}
                        \includegraphics[width=0.4\textwidth]{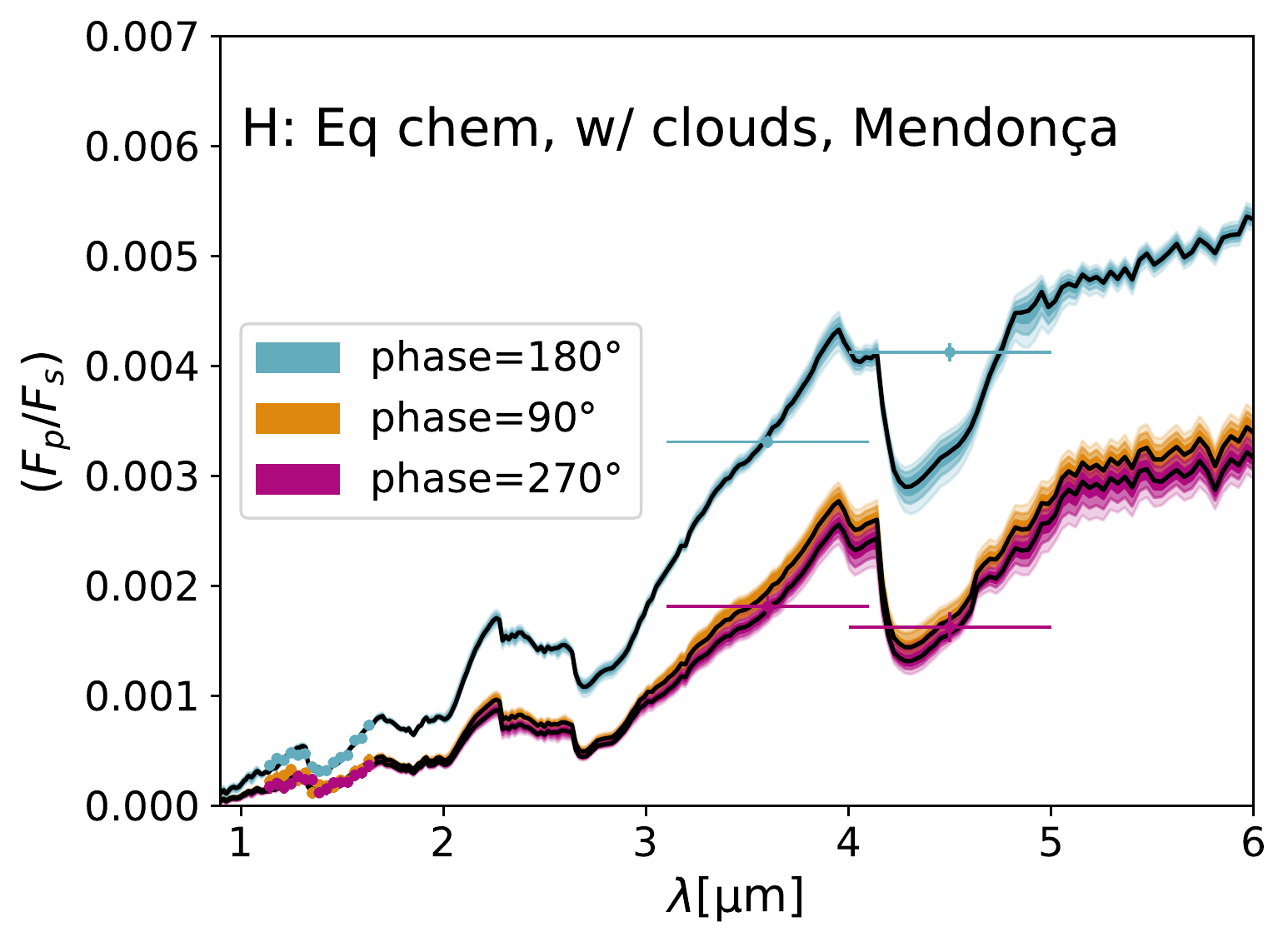}
                        \caption{Retrieved spectra for the 90$\degree$ (evening), 180$\degree$ (dayside) and 270$\degree$ (morning) phases for eight different retrieval setups, as detailed in each panel. The shading represents the 1, 2, and 3 $\sigma$ bounds. Observed data (HST and Spitzer) points are plotted with error bars. Stevenson and Mendon{\c{c}}a refer to the use of Spitzer data analysed by \cite{17StLiBe.wasp43b} and \cite{18MeMaDe.wasp43b}, respectively. }\label{fig:spectra_3phase}
                \end{figure}
                
                \begin{figure}
                        \centering
                        \includegraphics[width=0.4\textwidth]{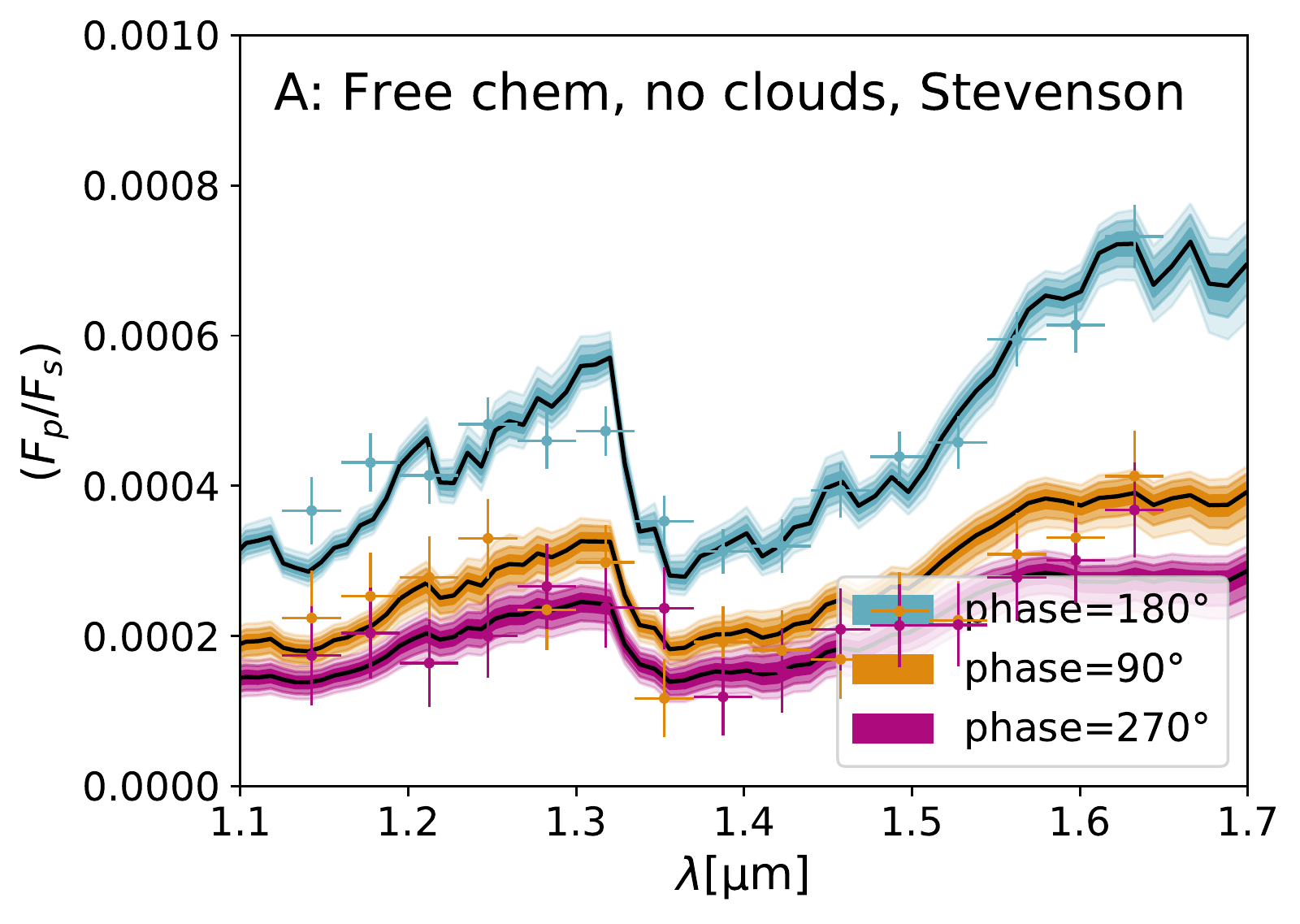}
                        \includegraphics[width=0.4\textwidth]{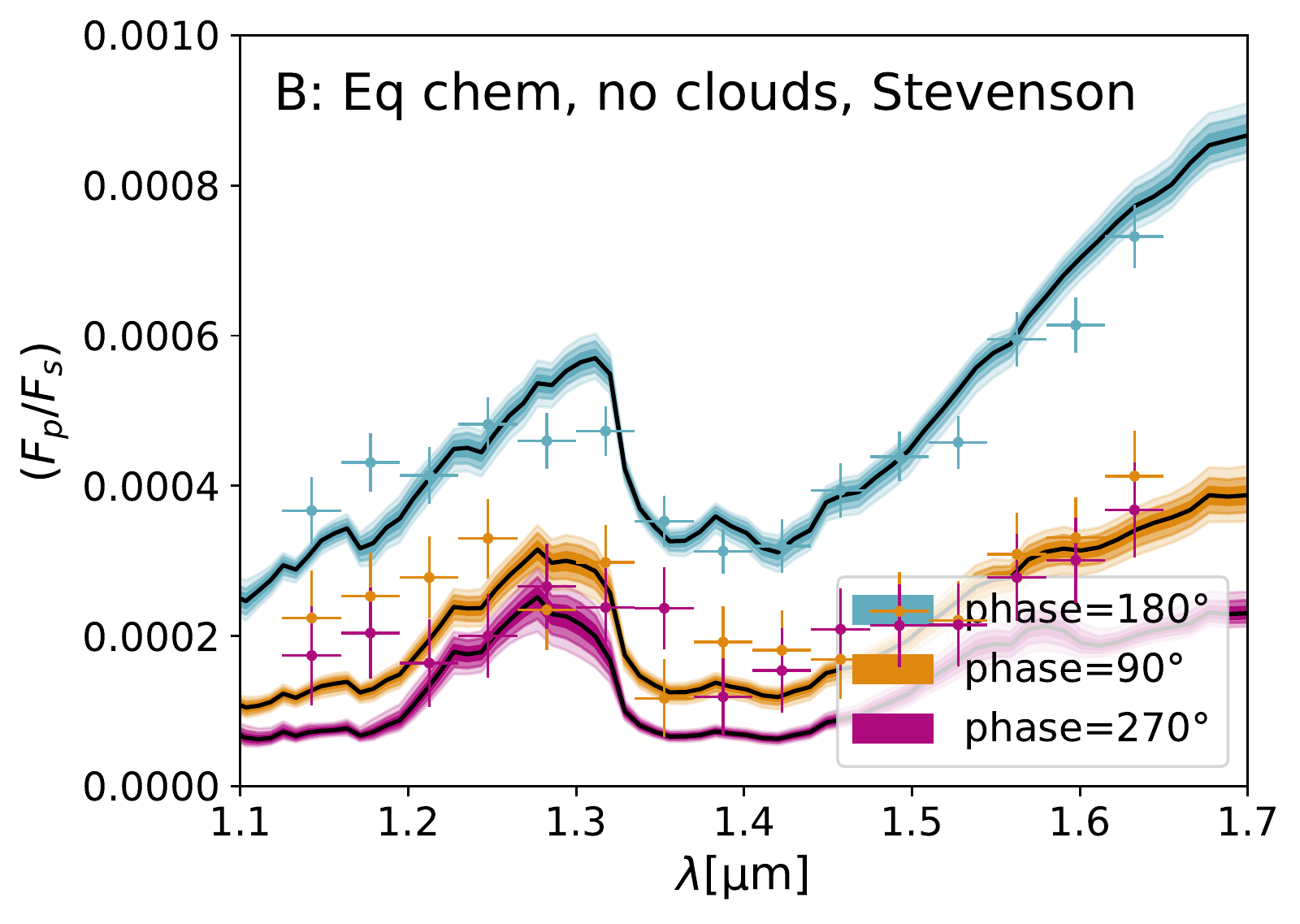}
                        \includegraphics[width=0.4\textwidth]{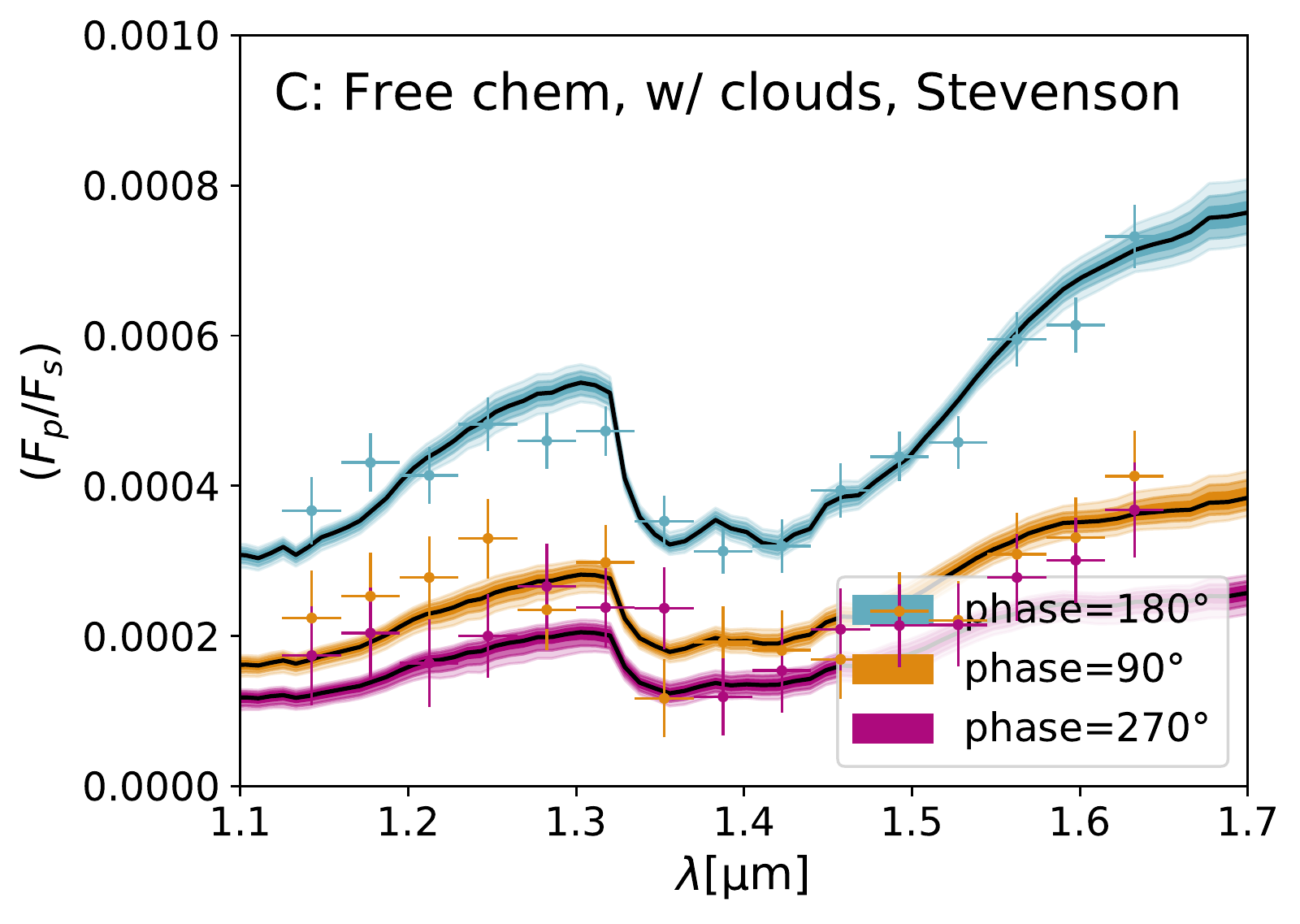}
                        \includegraphics[width=0.4\textwidth]{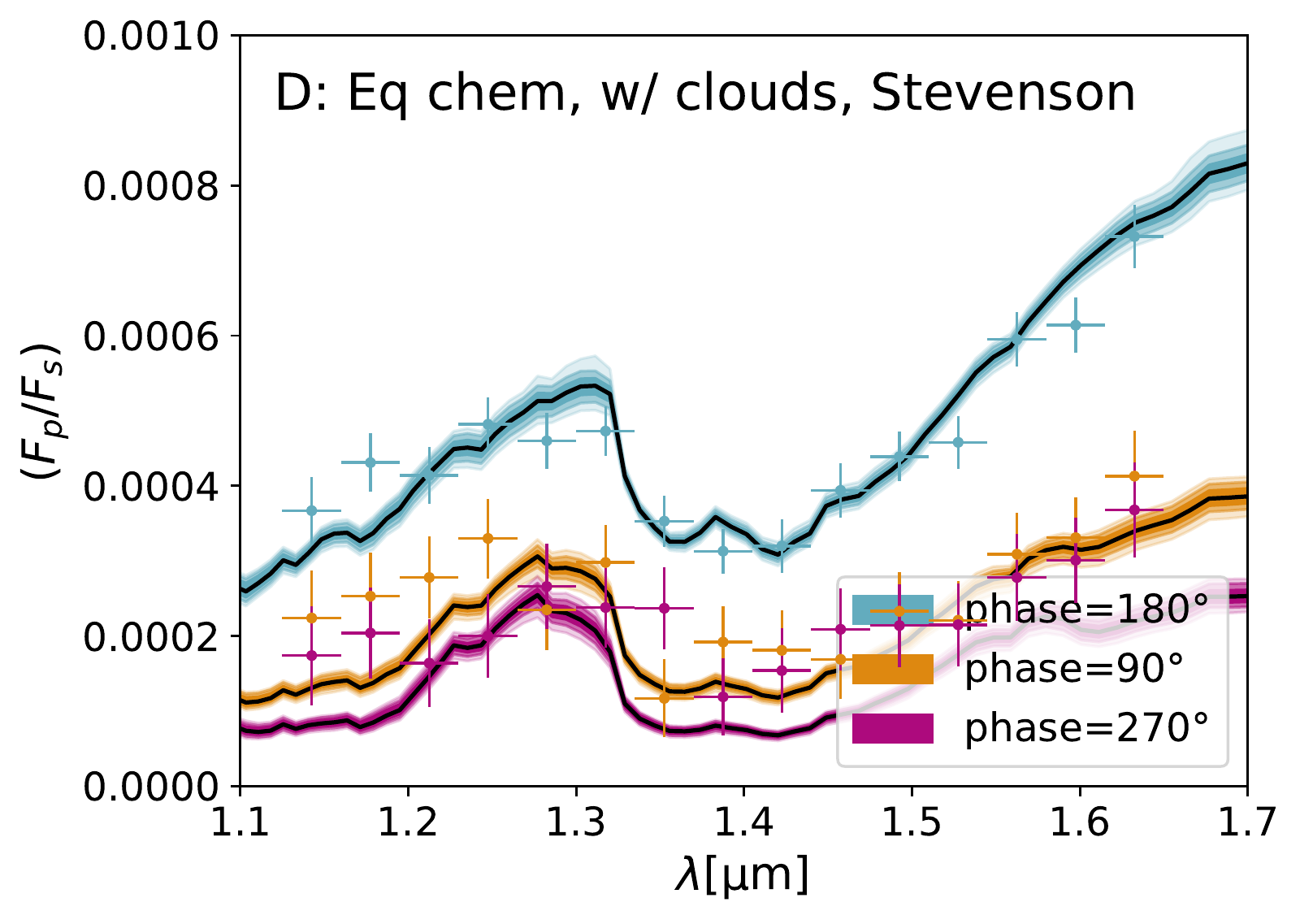}
                        \includegraphics[width=0.4\textwidth]{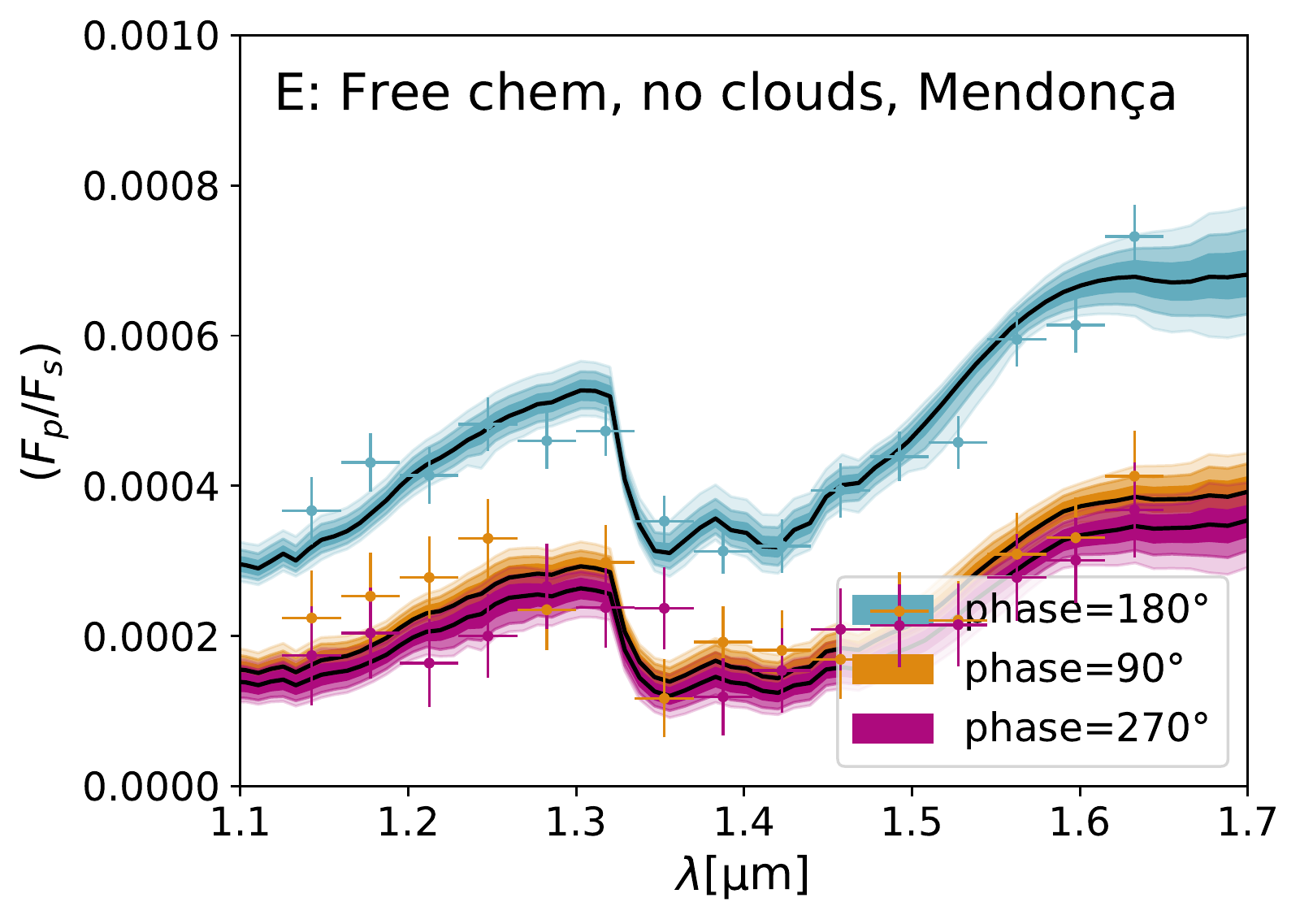}
                        \includegraphics[width=0.4\textwidth]{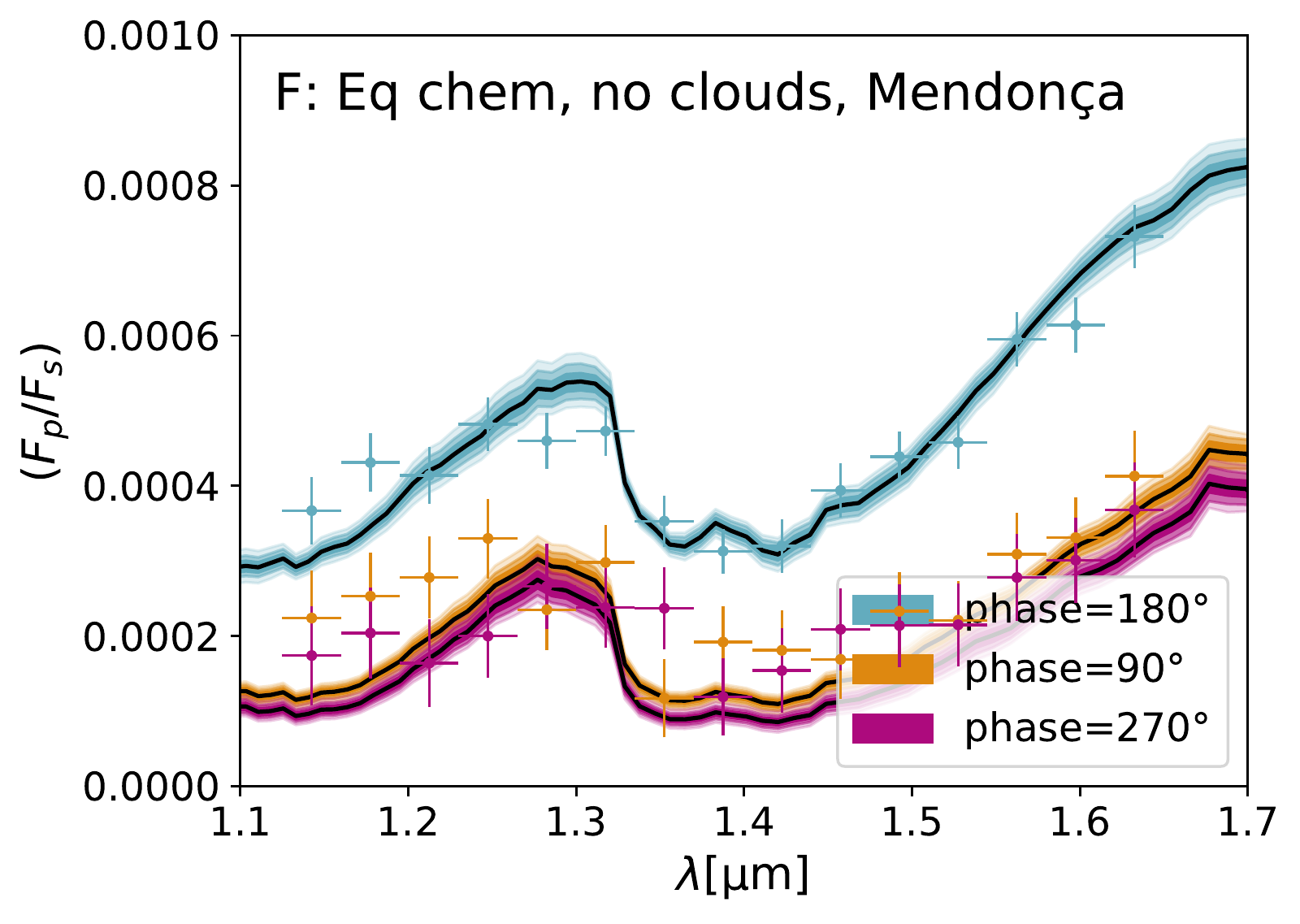}
                        \includegraphics[width=0.4\textwidth]{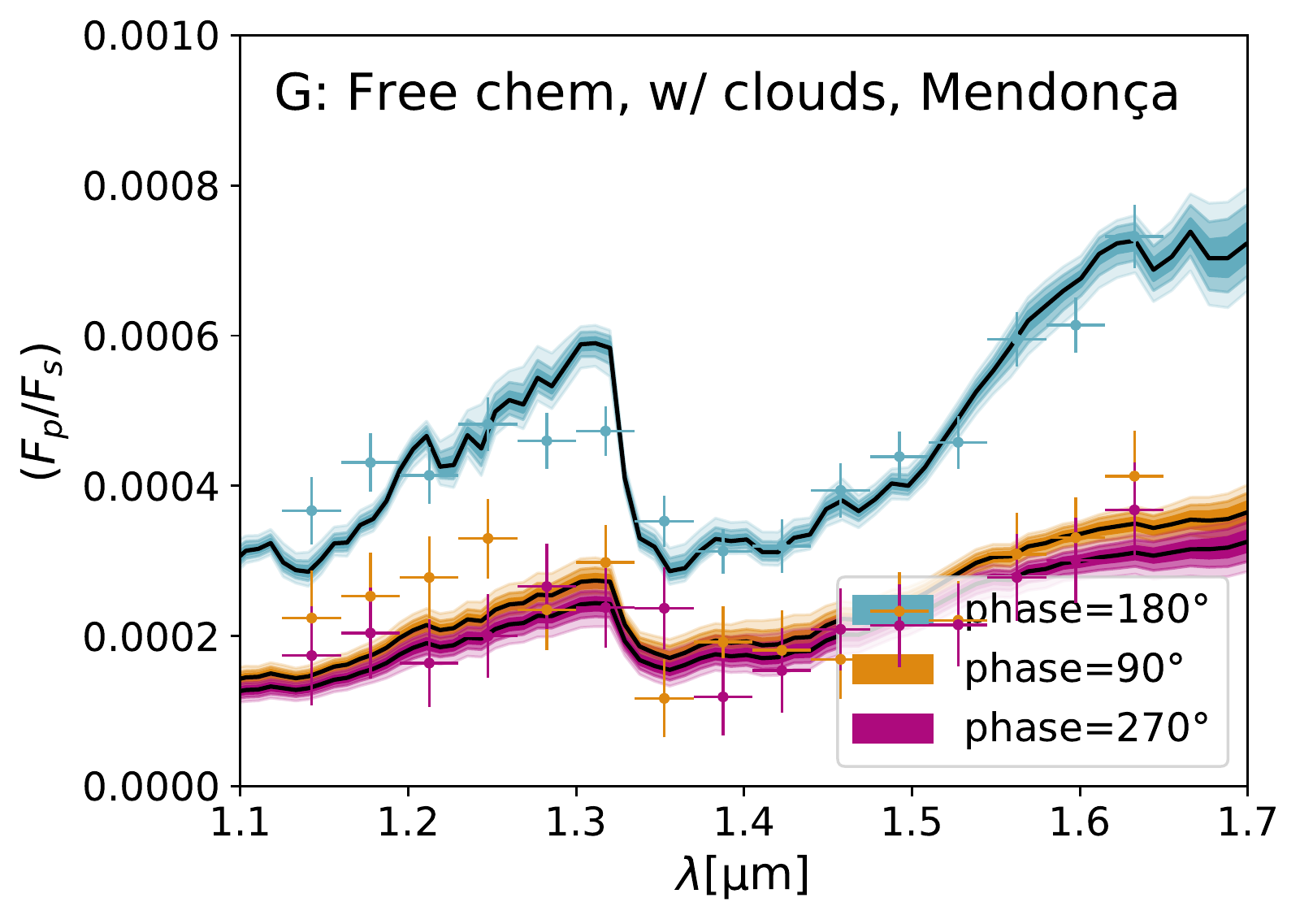}
                        \includegraphics[width=0.4\textwidth]{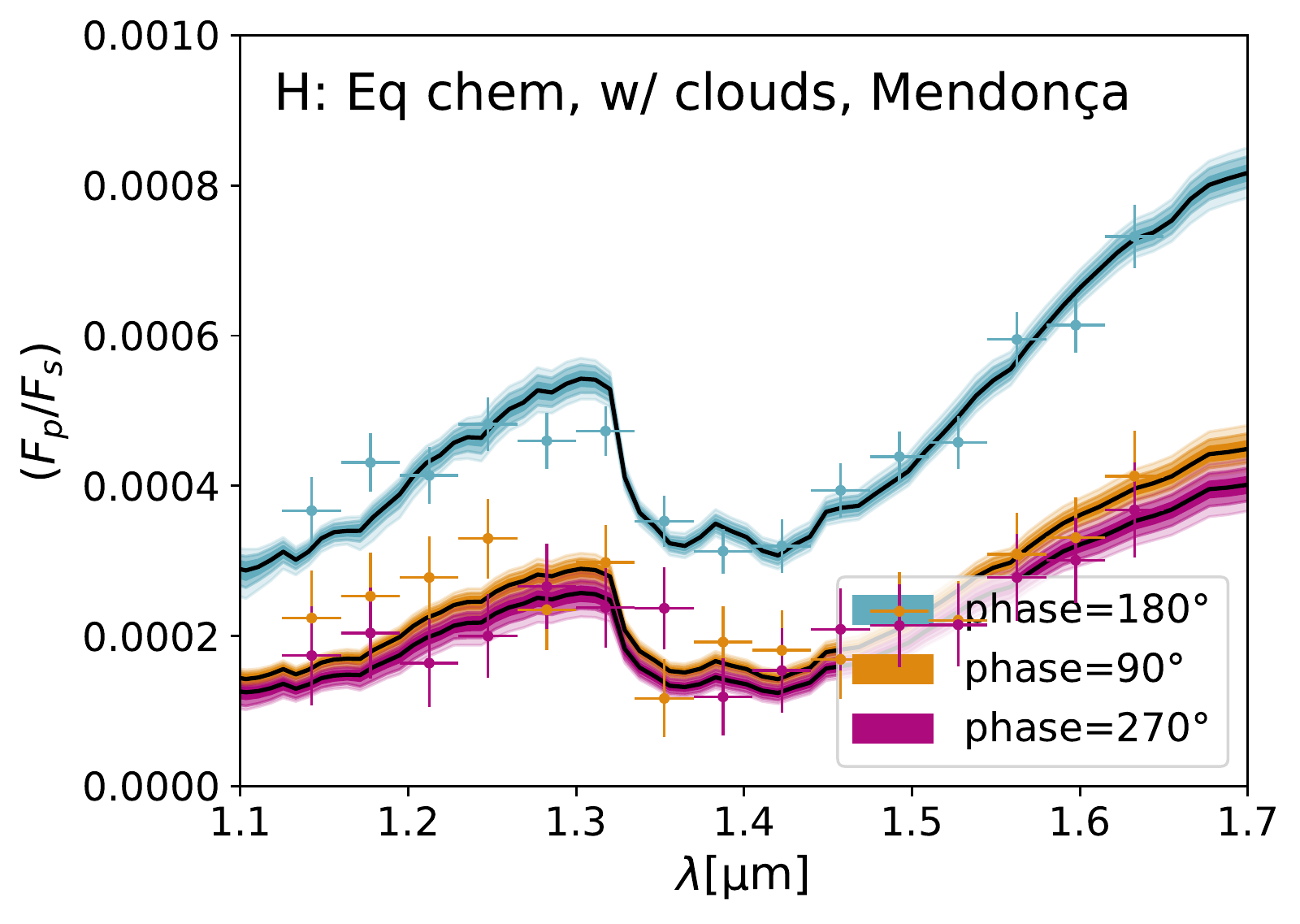}
                        \caption{Retrieved spectra showing the HST/WFC3 data region only for the 90$\degree$ (evening), 180$\degree$ (dayside) and 270$\degree$ (morning) phases, for eight different retrieval setups, as detailed in each panel. The shading represents the 1, 2, and 3 $\sigma$ bounds. Observed data (HST) points are plotted with error bars. Stevenson and Mendon{\c{c}}a refer to the use of Spitzer data analysed by \cite{17StLiBe.wasp43b} and \cite{18MeMaDe.wasp43b}, respectively. }\label{fig:spectra_3phase_zoomed}
                \end{figure}
                
                \begin{figure}
                        \centering
                        \includegraphics[width=0.35\textwidth]{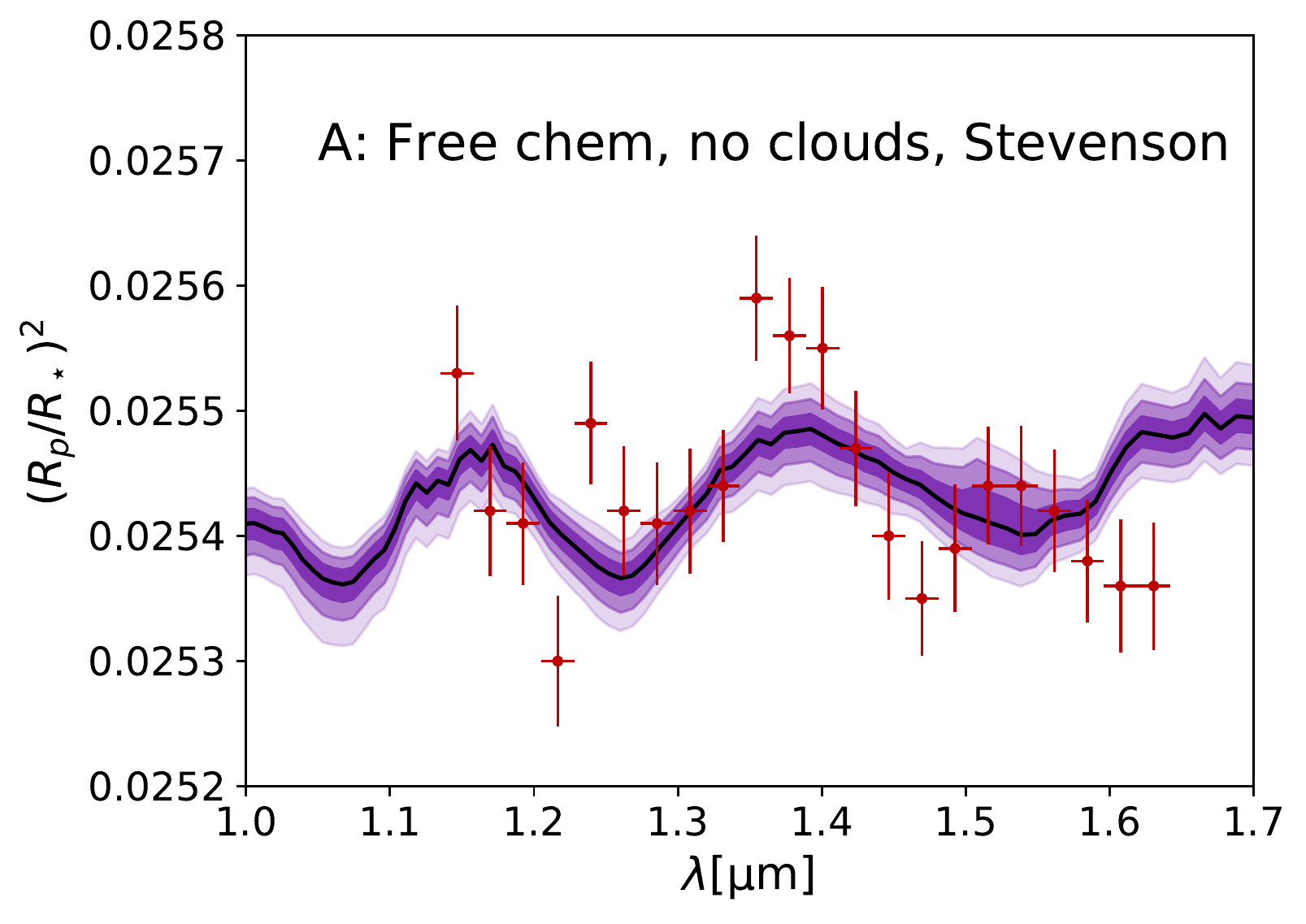}
                        \includegraphics[width=0.35\textwidth]{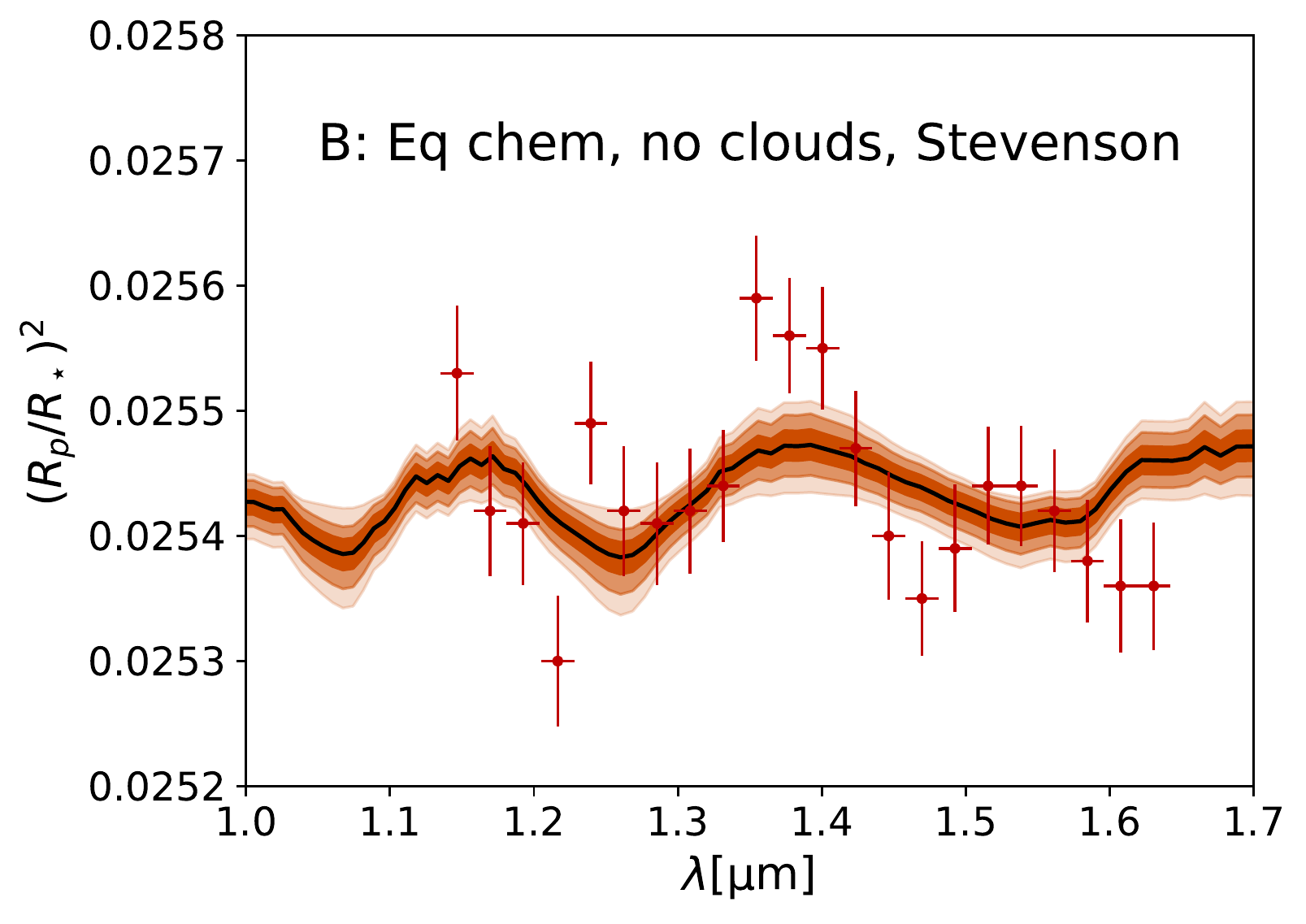}
                        \includegraphics[width=0.35\textwidth]{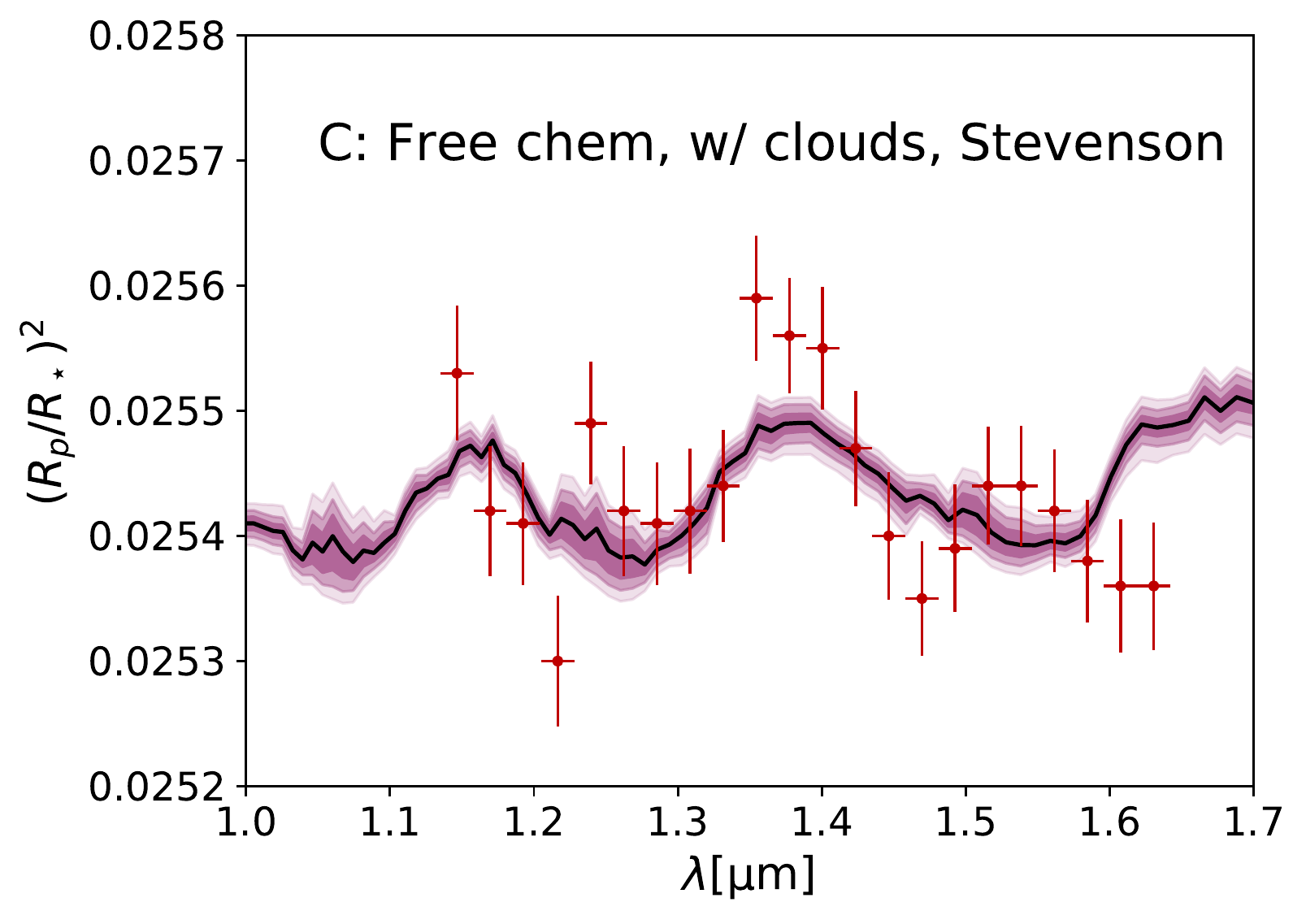}
                        \includegraphics[width=0.35\textwidth]{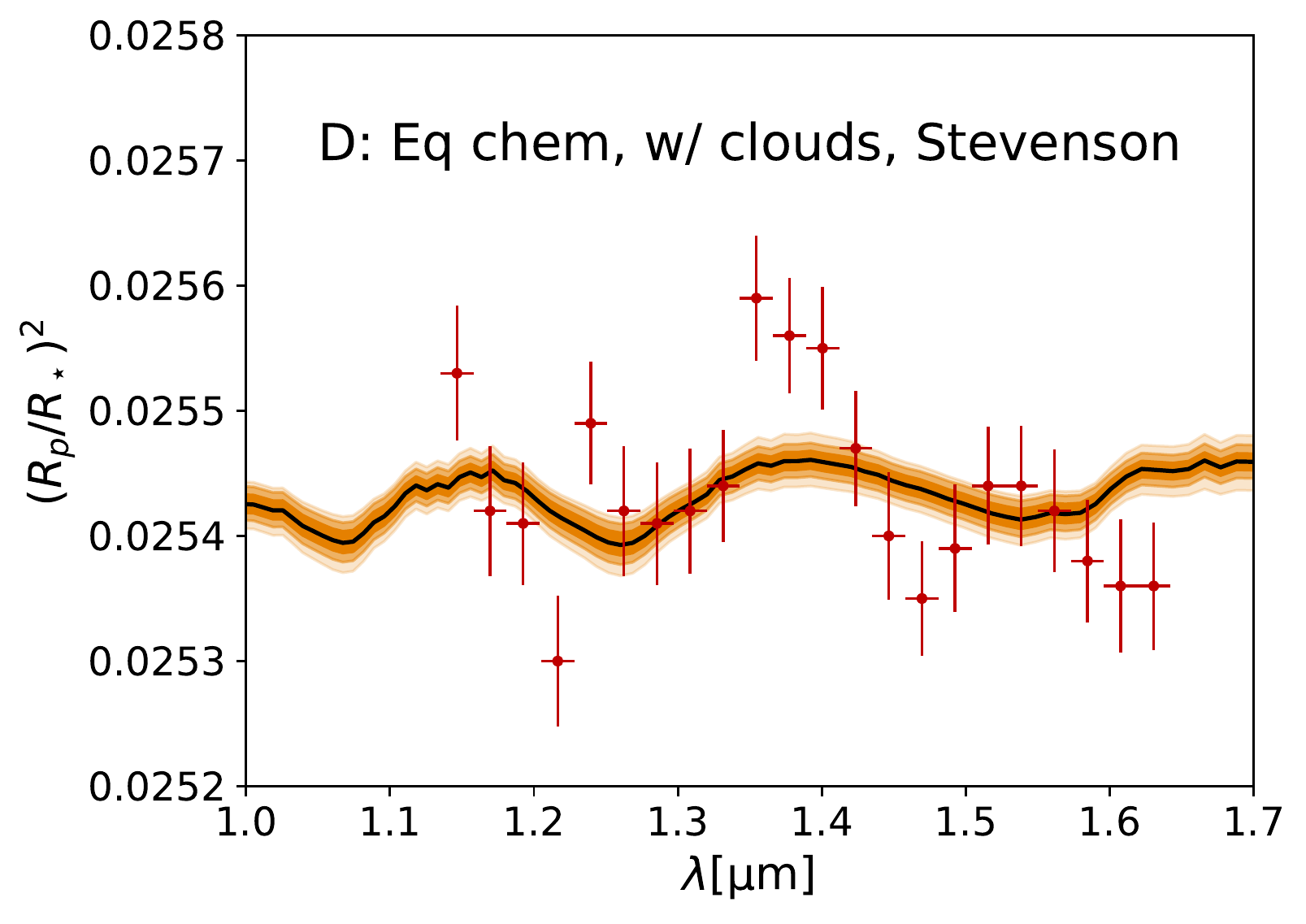}
                        \includegraphics[width=0.35\textwidth]{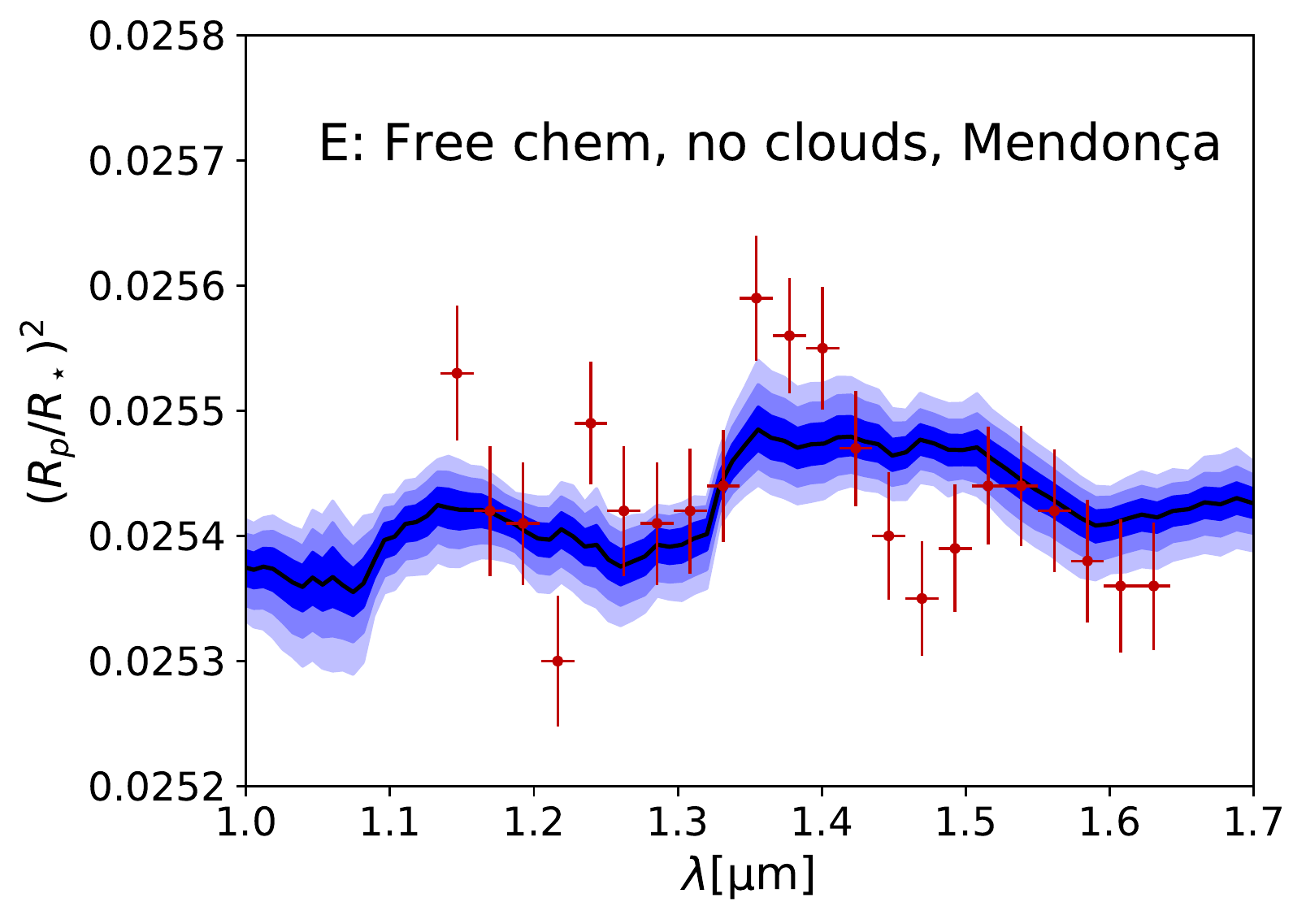}
                        \includegraphics[width=0.35\textwidth]{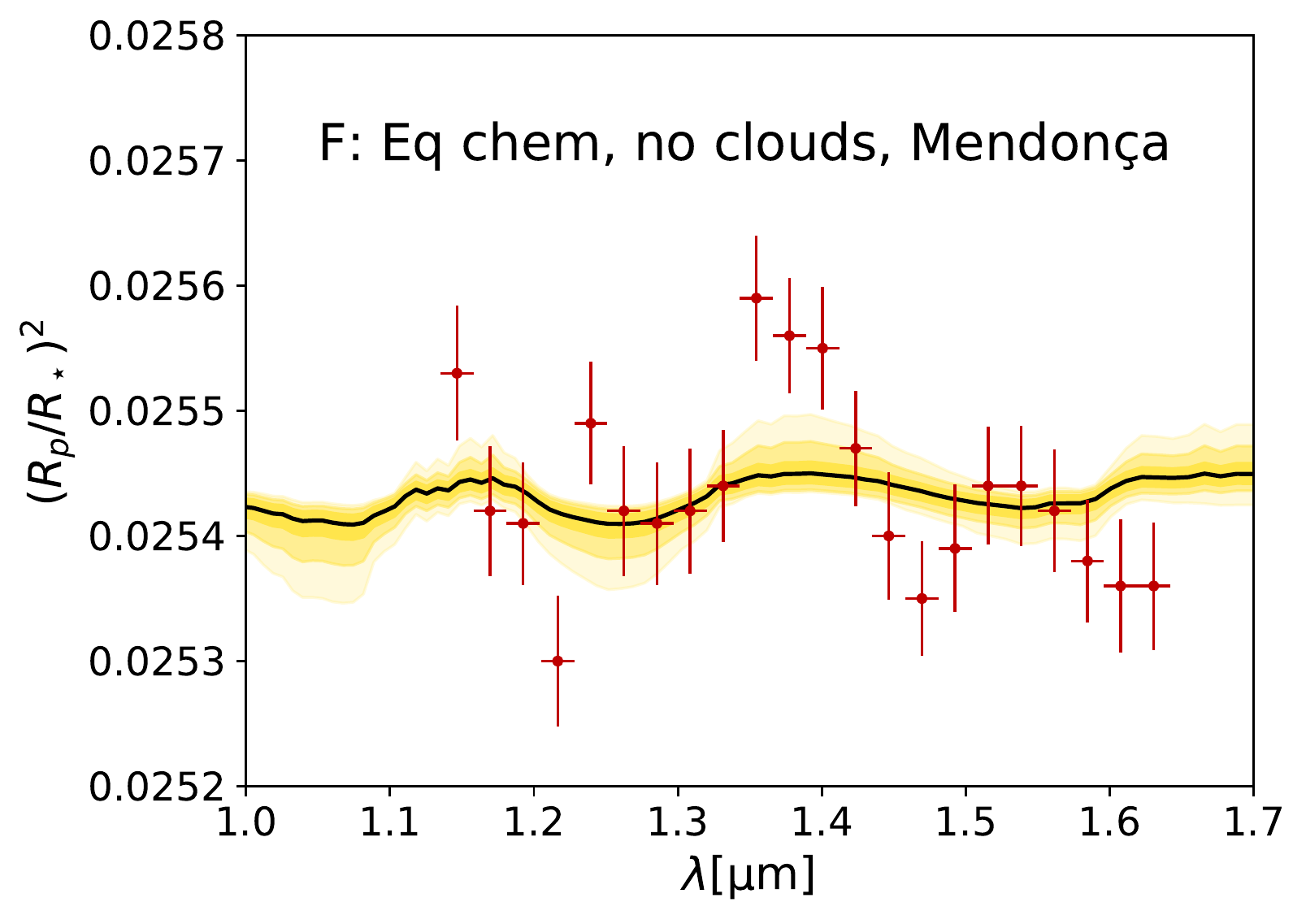}
                        \includegraphics[width=0.35\textwidth]{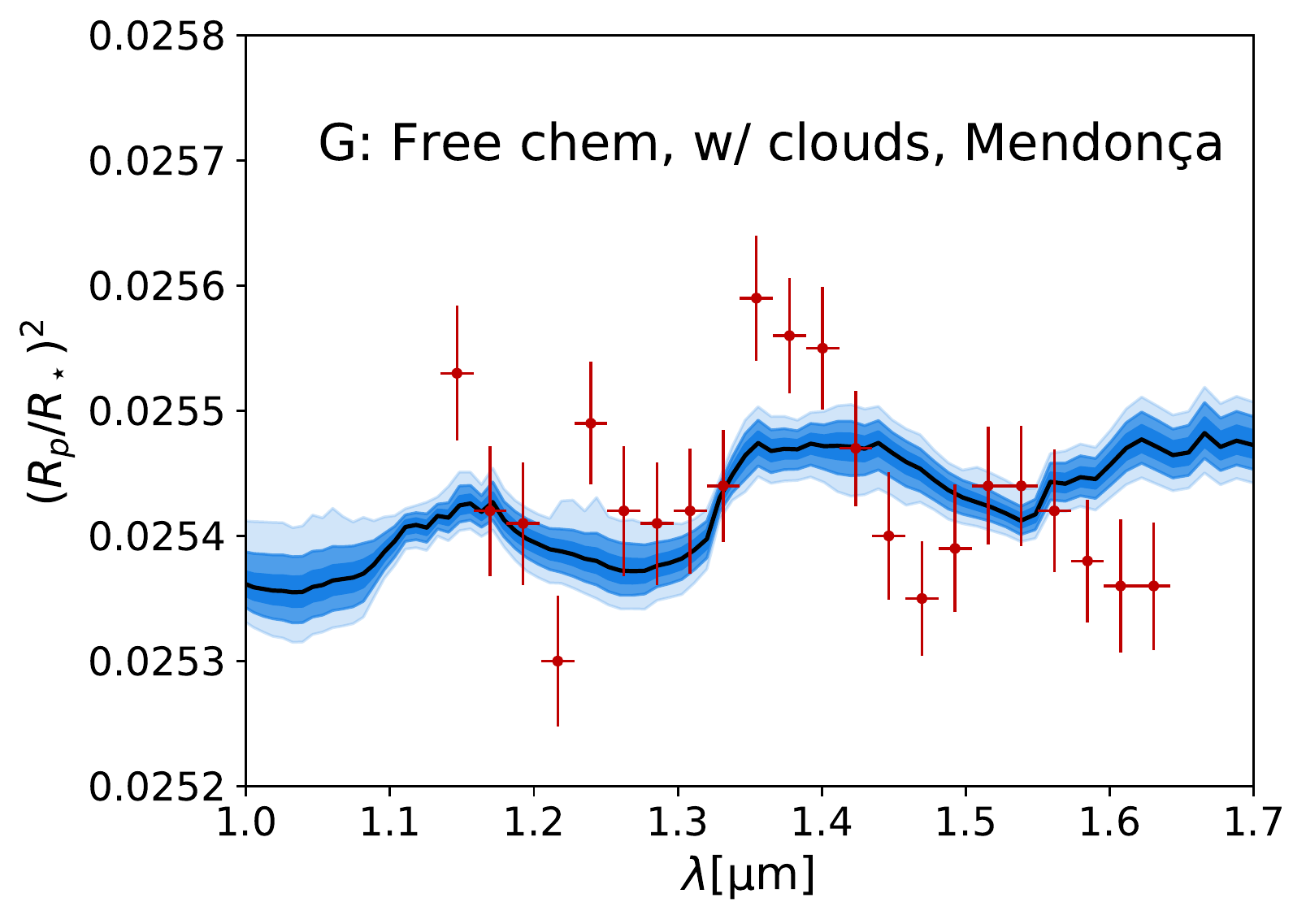}
                        \includegraphics[width=0.35\textwidth]{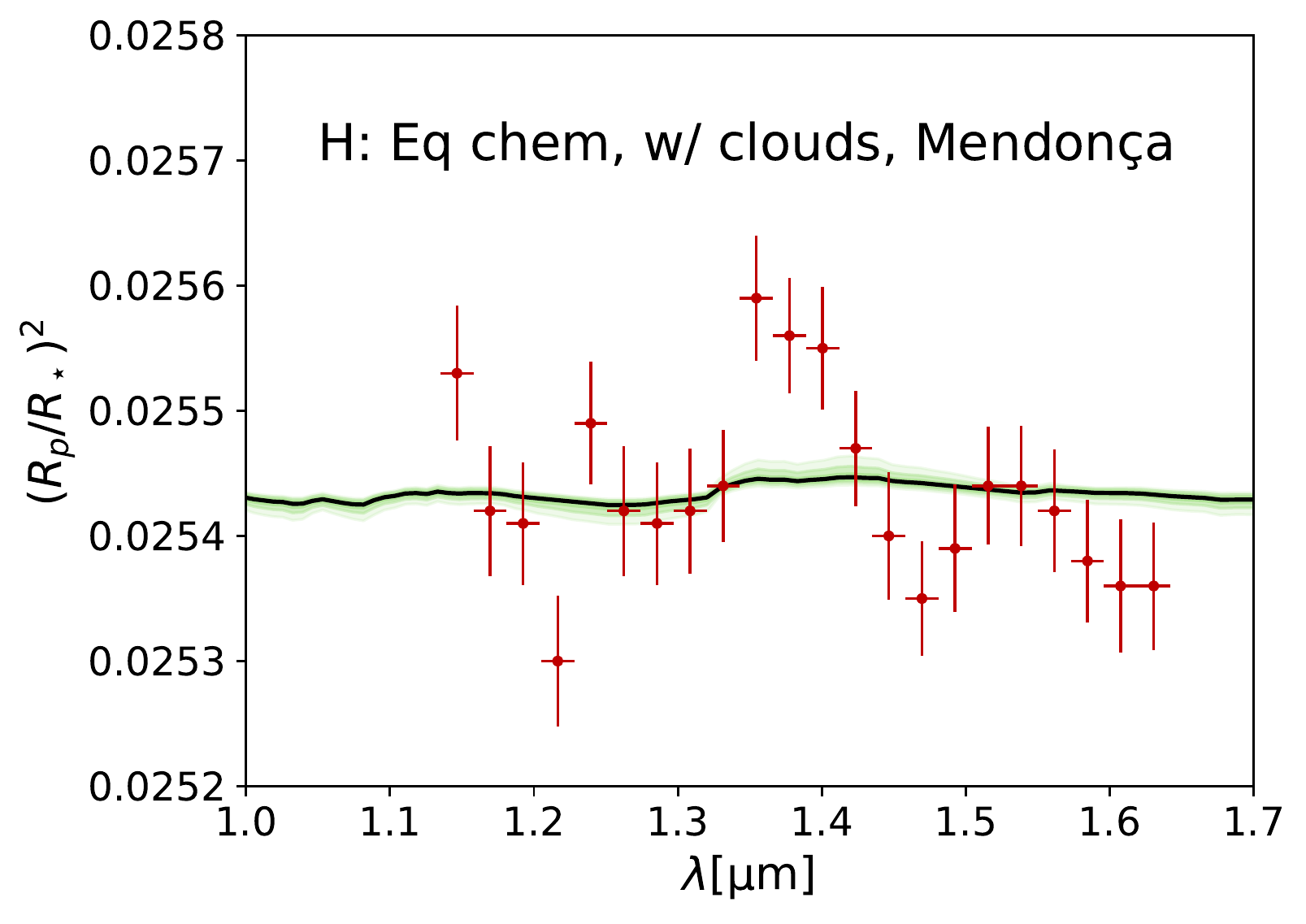}
                        \includegraphics[width=0.35\textwidth]{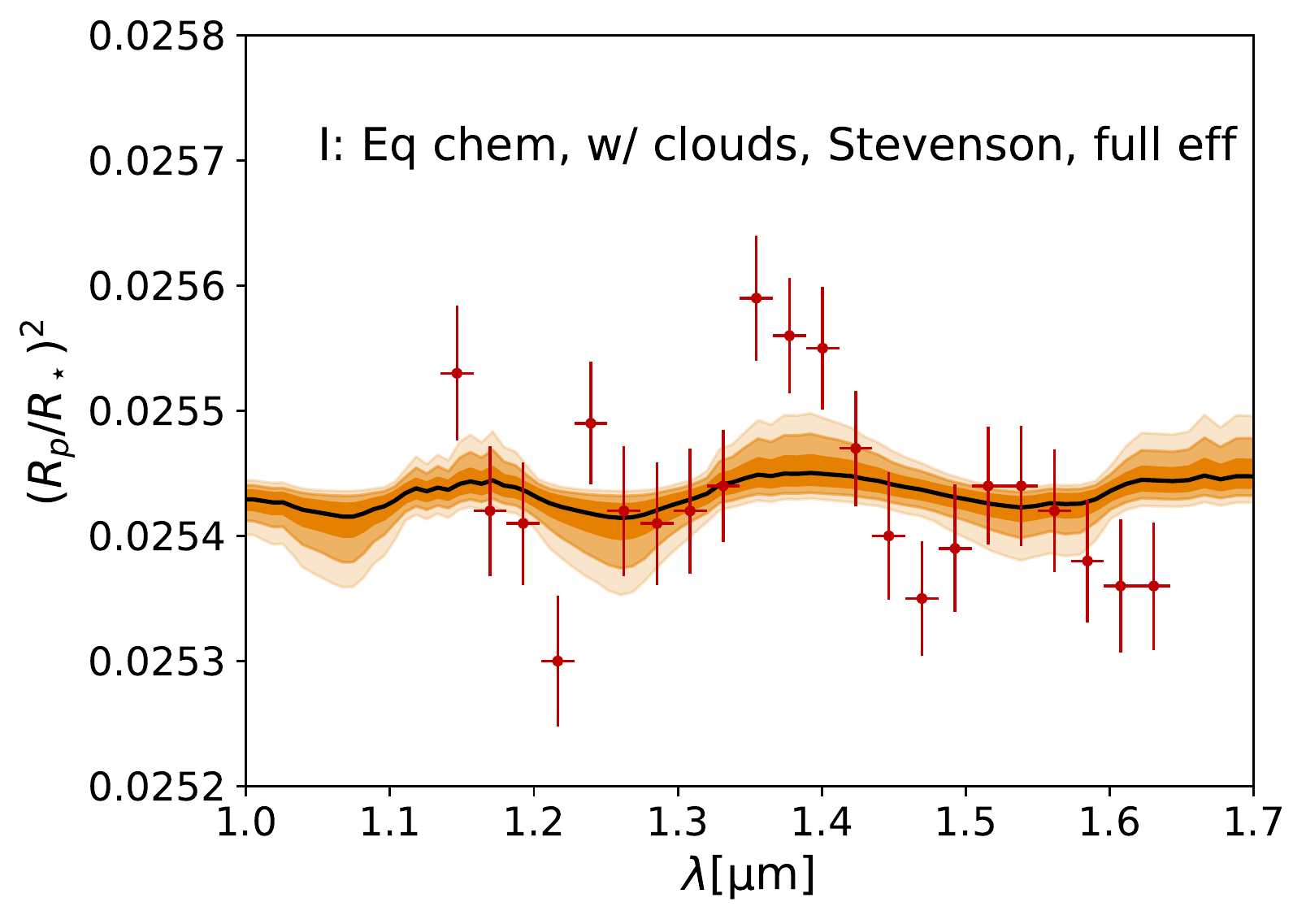}
                        \caption{Retrieved and observed HST/WFC3 transmission spectra for our eight different retrieval setups (plus additional setup I), as detailed in each panel. The shading represents the 1, 2, and 3 $\sigma$ bounds. Observed data (HST/WFC3) points are plotted with error bars. Stevenson and Mendon{\c{c}}a refer to the use of Spitzer data analysed by \cite{17StLiBe.wasp43b} and \cite{18MeMaDe.wasp43b}, respectively. 
                        }\label{fig:spectra_trans}
                \end{figure}

                \begin{figure}
                        \centering
                        \includegraphics[width=0.45\textwidth]{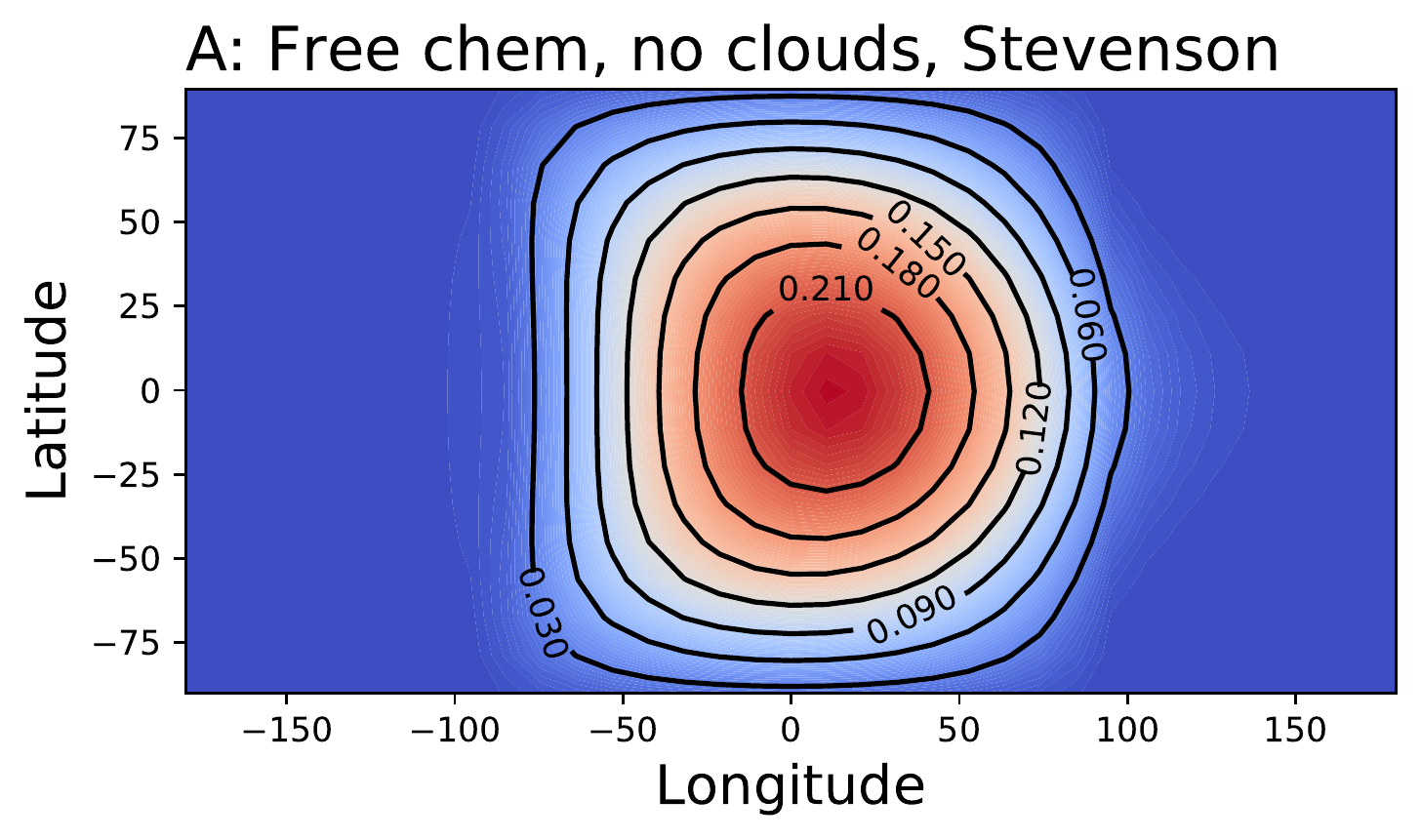}
                        \includegraphics[width=0.45\textwidth]{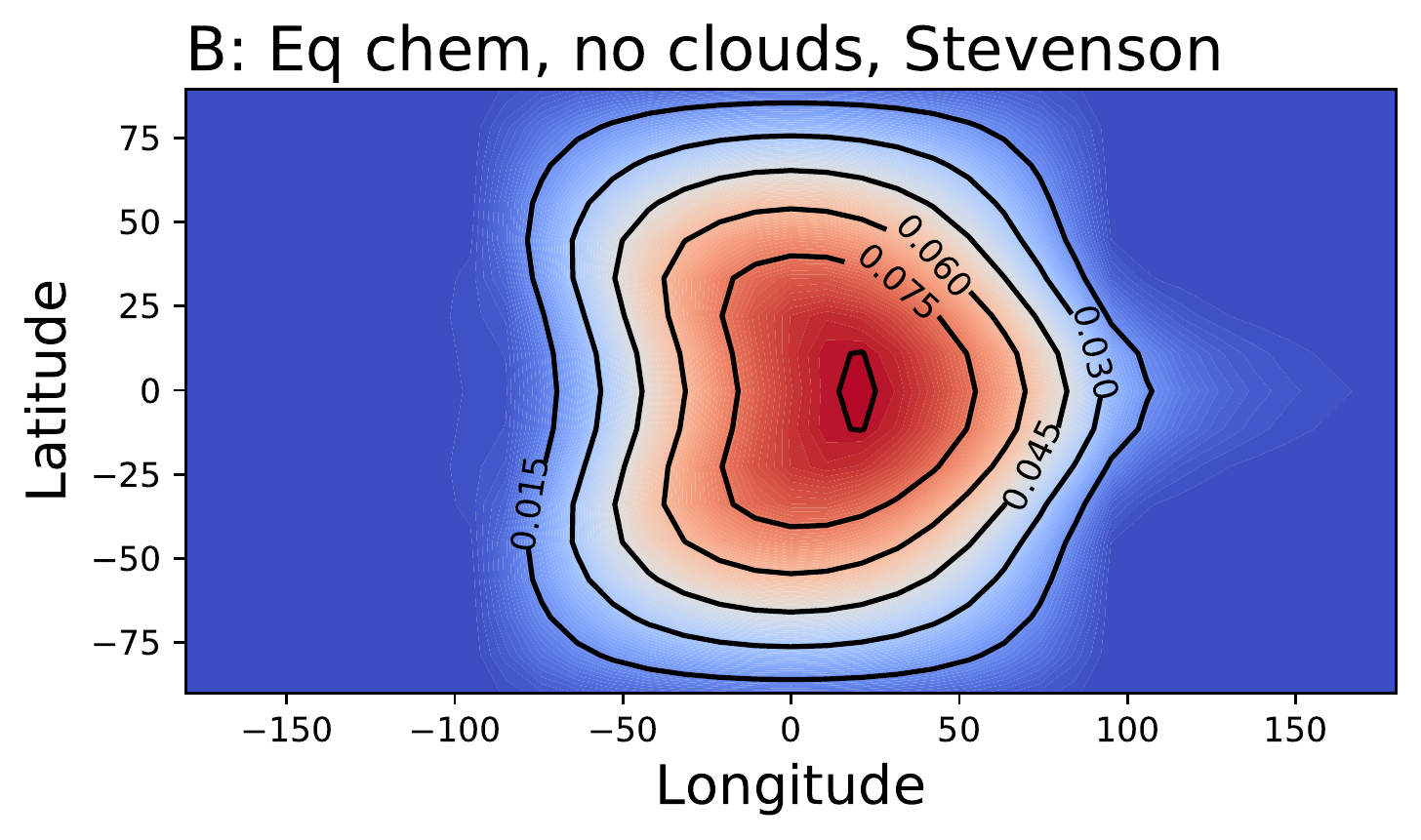}
                        \includegraphics[width=0.45\textwidth]{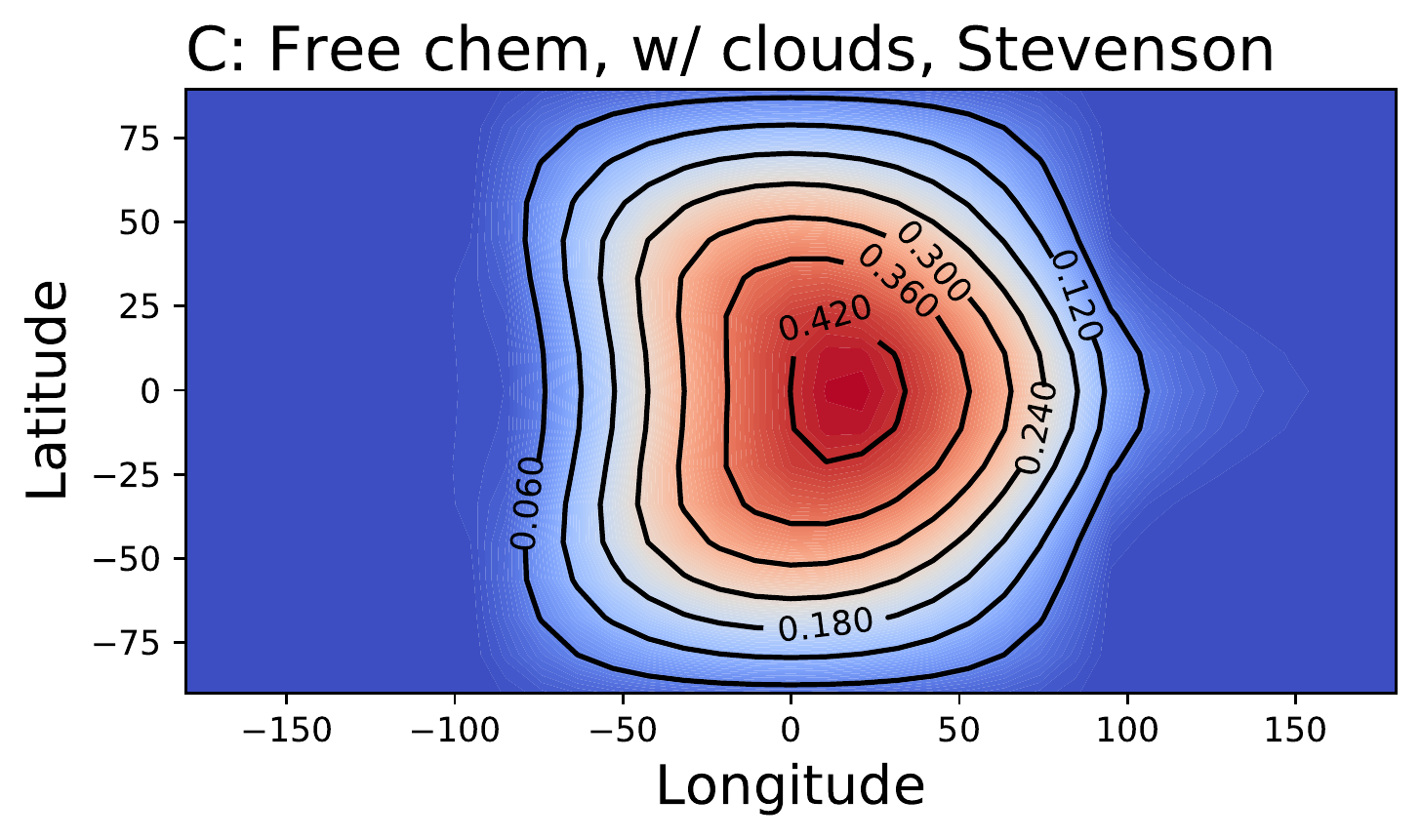}
                        \includegraphics[width=0.45\textwidth]{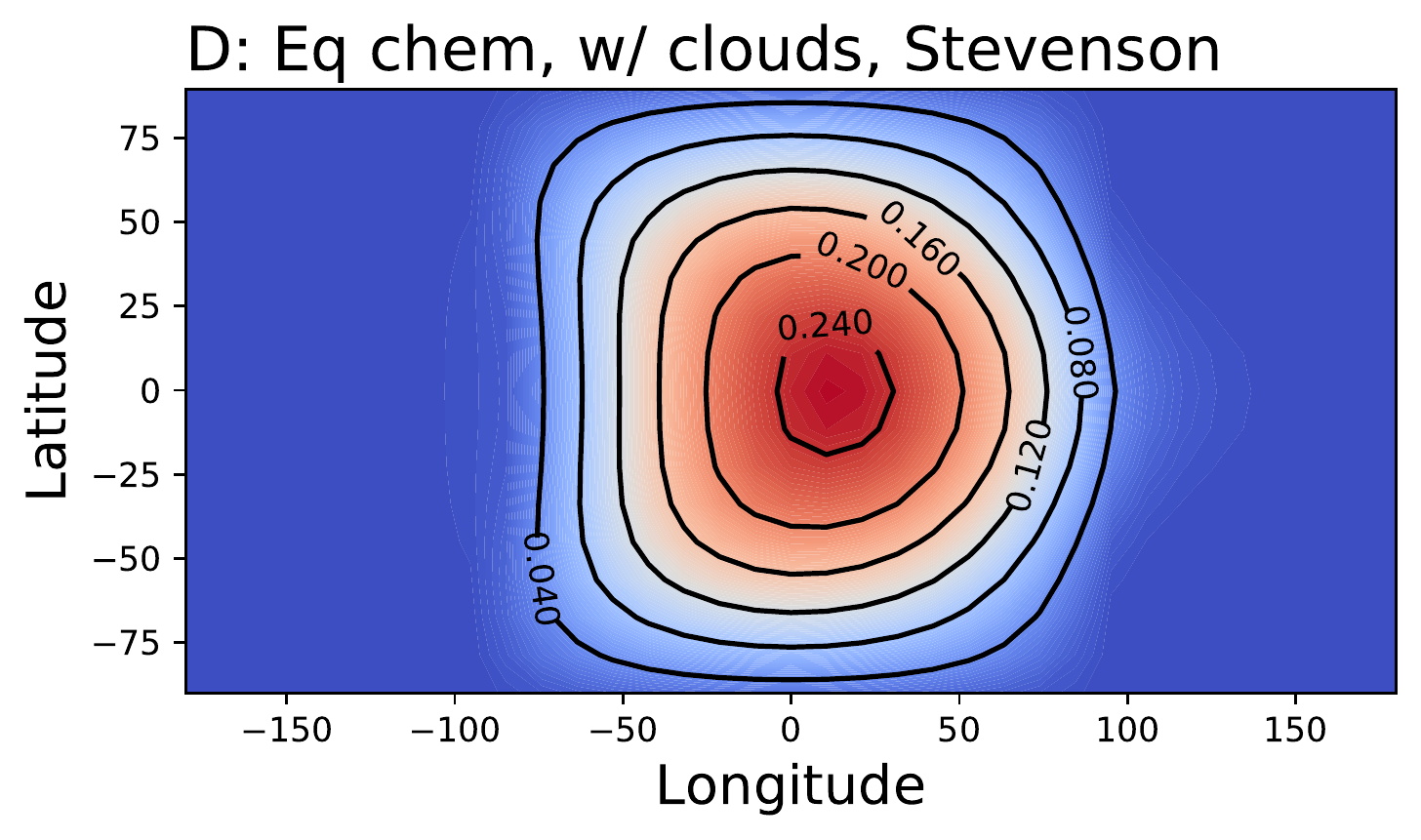}
                        \includegraphics[width=0.45\textwidth]{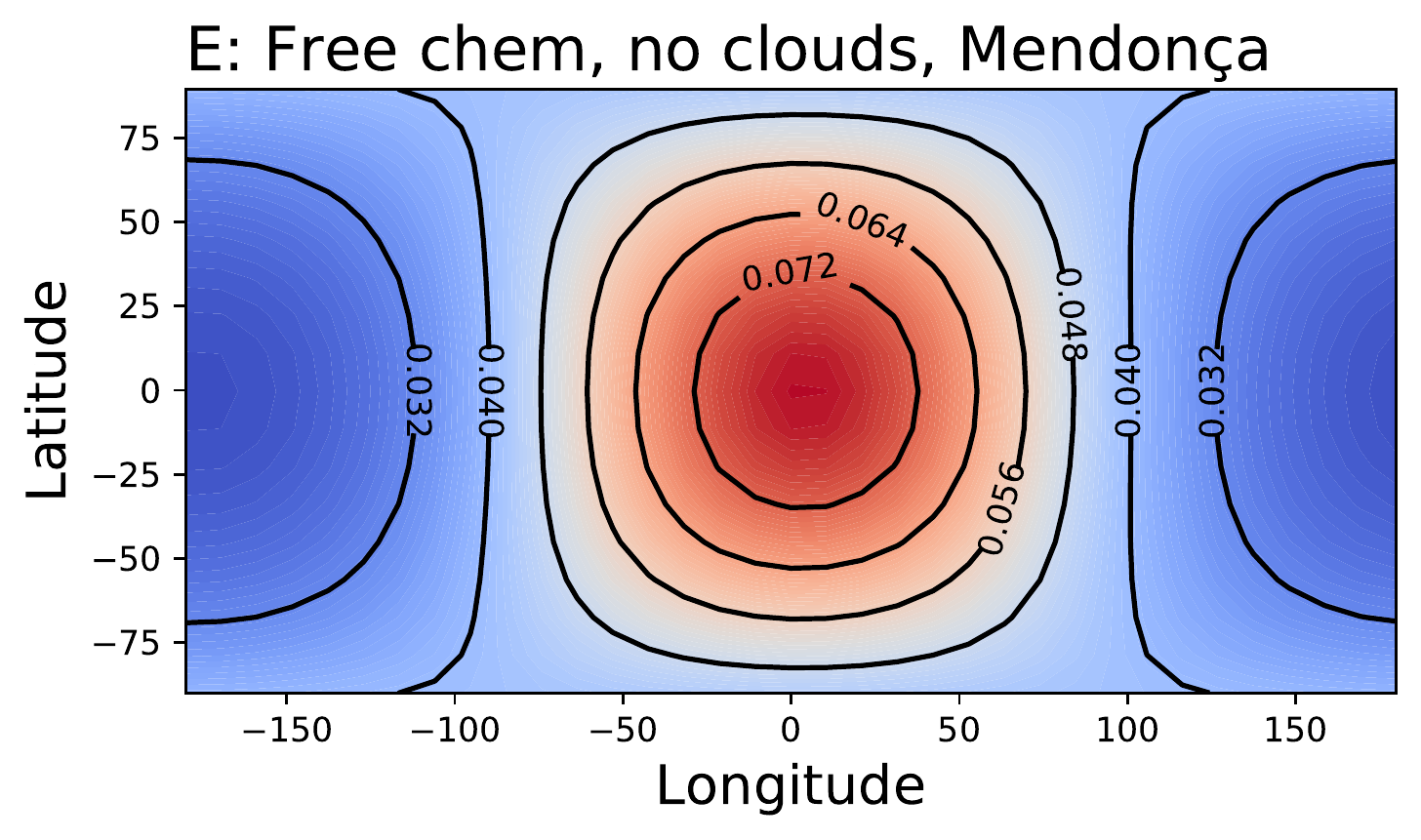}
                        \includegraphics[width=0.45\textwidth]{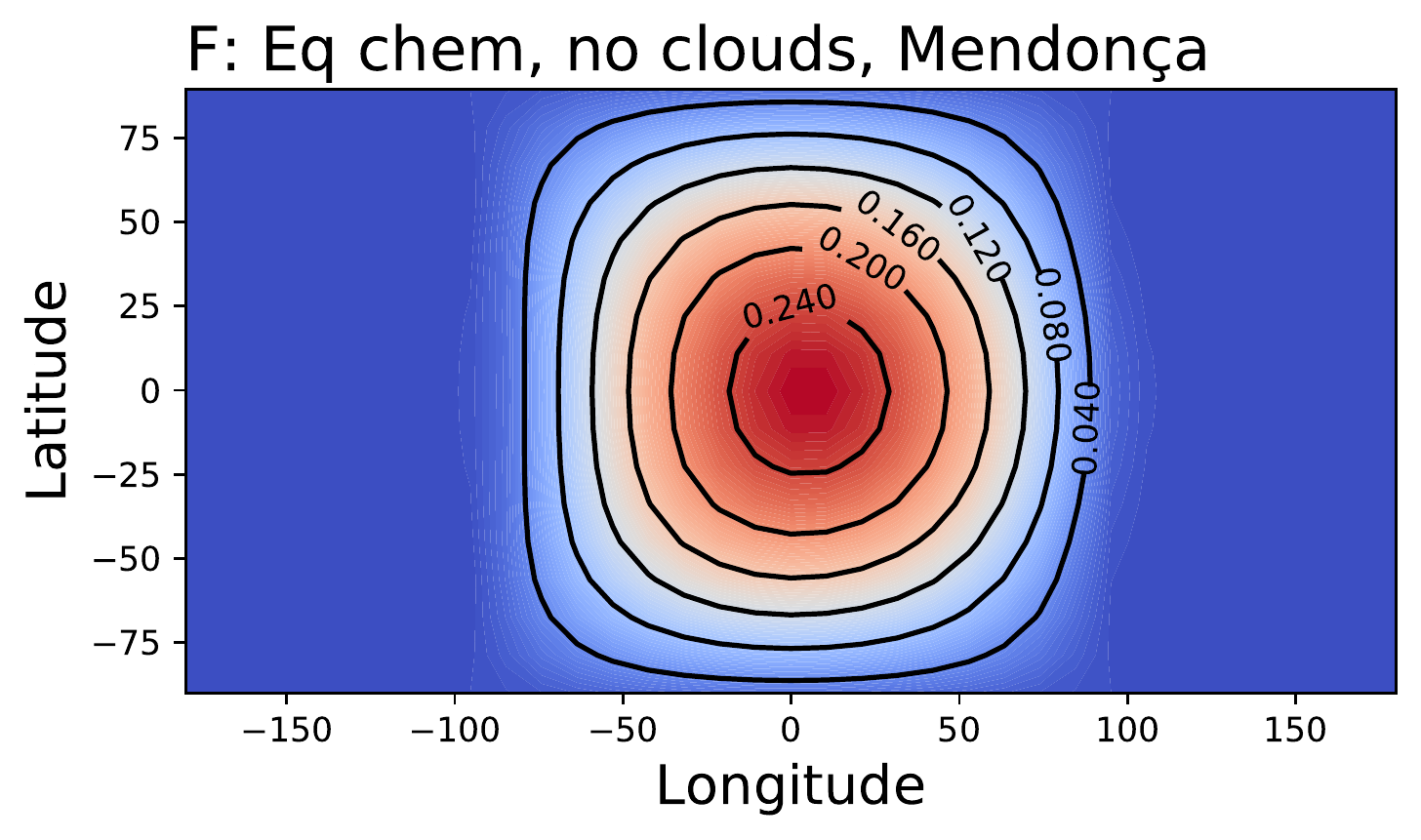}
                        \includegraphics[width=0.45\textwidth]{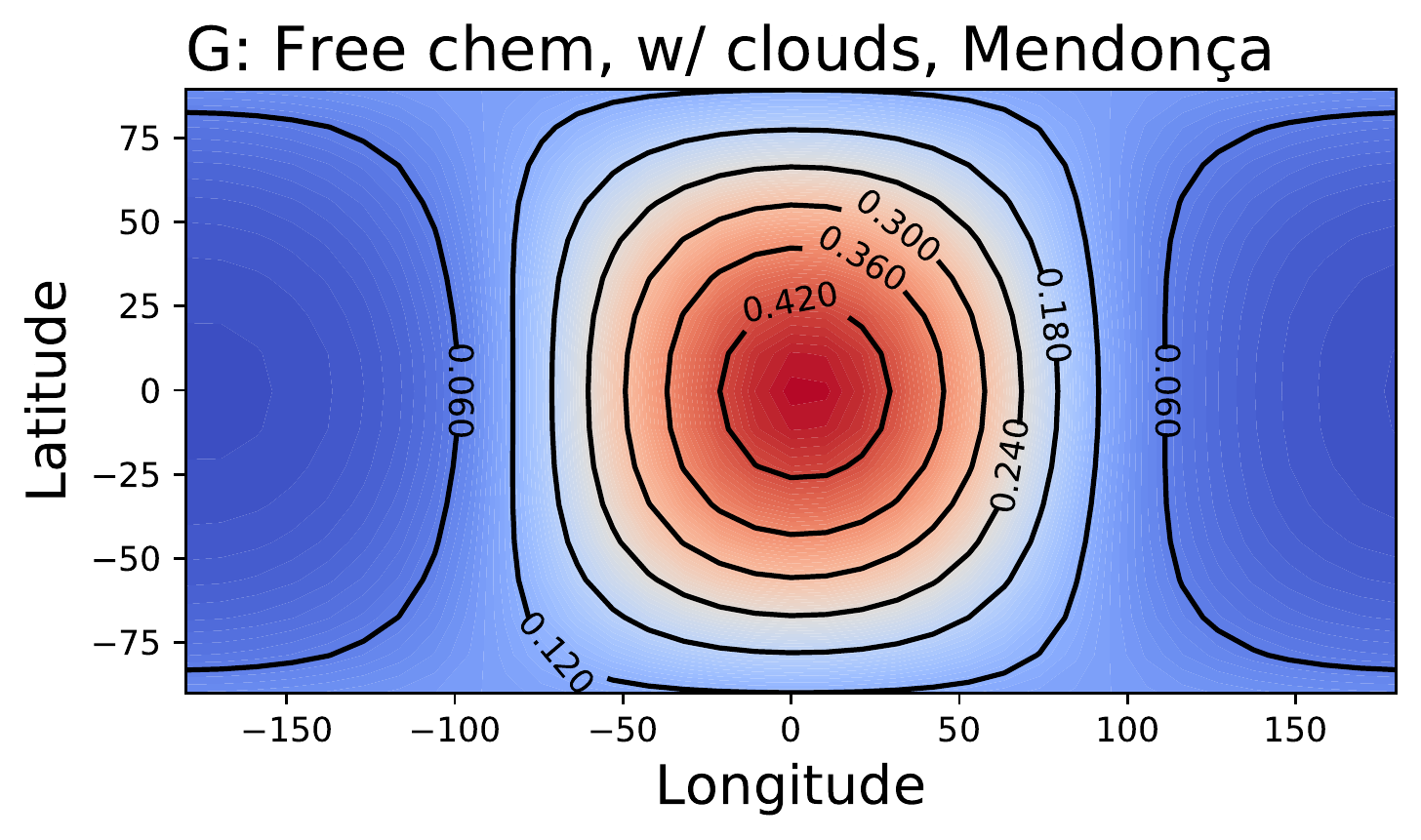}
                        \includegraphics[width=0.45\textwidth]{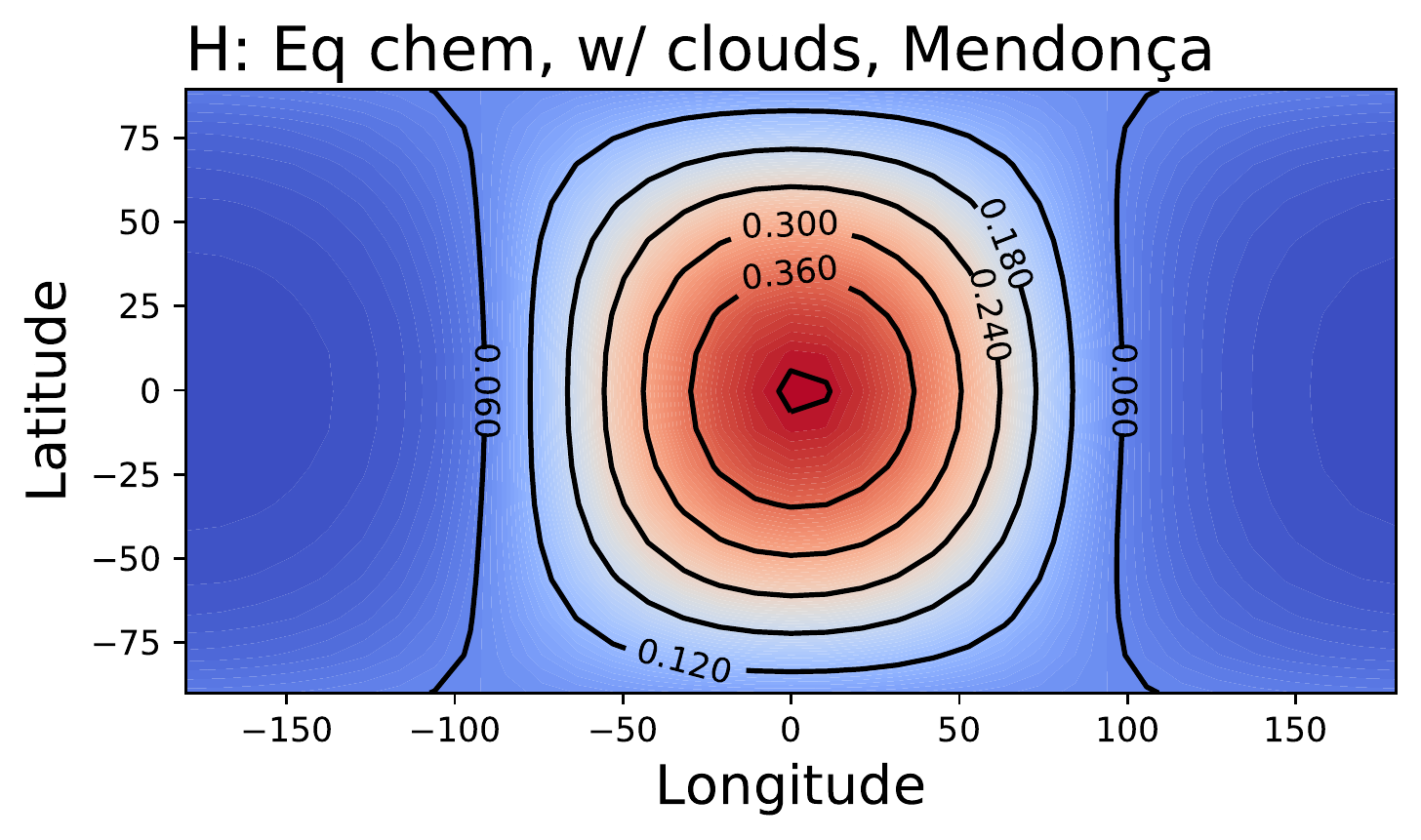}
                        \caption{Contour plots giving retrieved $\beta$ (where $\beta$ specifies how the energy incident from the central star is spread through the atmosphere; see Eq.~\ref{eq:diff1}) as a function of longitude and latitude for the eight different retrieval setups, as detailed above each panel. $\beta$ can be used in Eq~\ref{eq:beta} to compute the temperature at a given pressure layer as a function of longitude and latitude; see Figures~\ref{fig:contour_plots_temp_0.01}~and~\ref{fig:contour_plots_temp_0.12}. Stevenson and Mendon{\c{c}}a refer to the use of Spitzer data analysed by \cite{17StLiBe.wasp43b} and \cite{18MeMaDe.wasp43b}, respectively.}\label{fig:contour_plots}
                \end{figure}
                
                \begin{figure}
                        \centering
                        \includegraphics[width=0.45\textwidth]{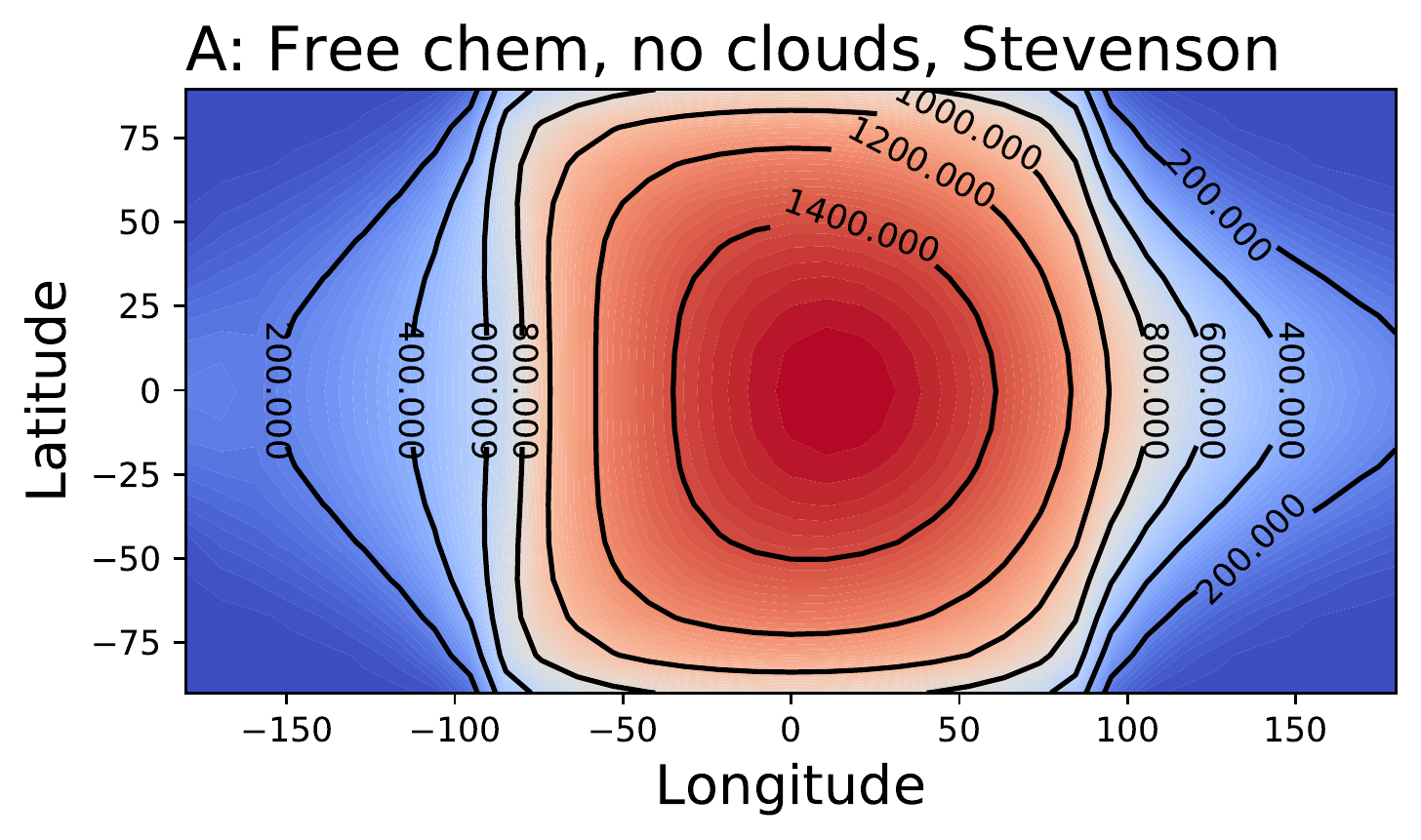}
                        \includegraphics[width=0.45\textwidth]{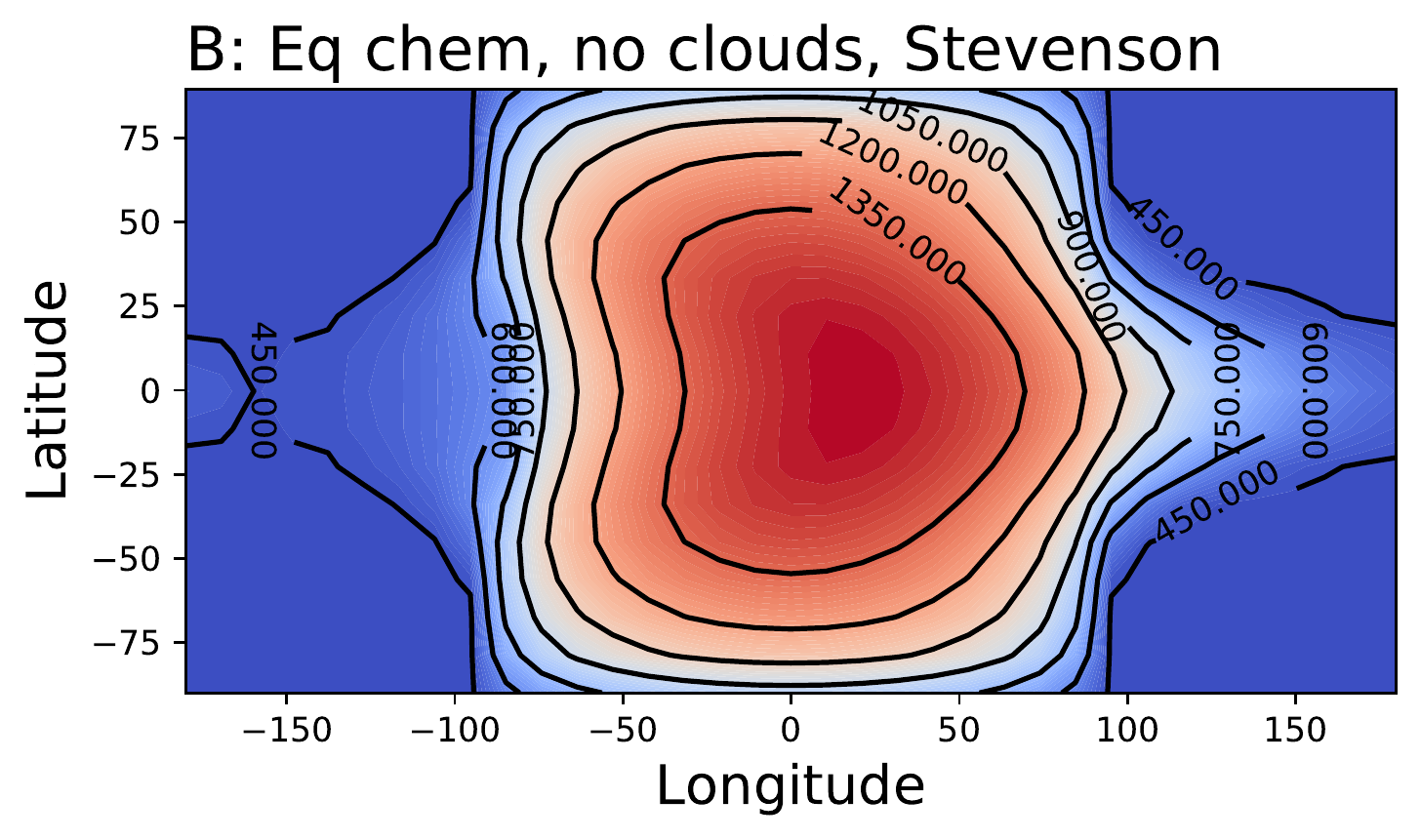}
                        \includegraphics[width=0.45\textwidth]{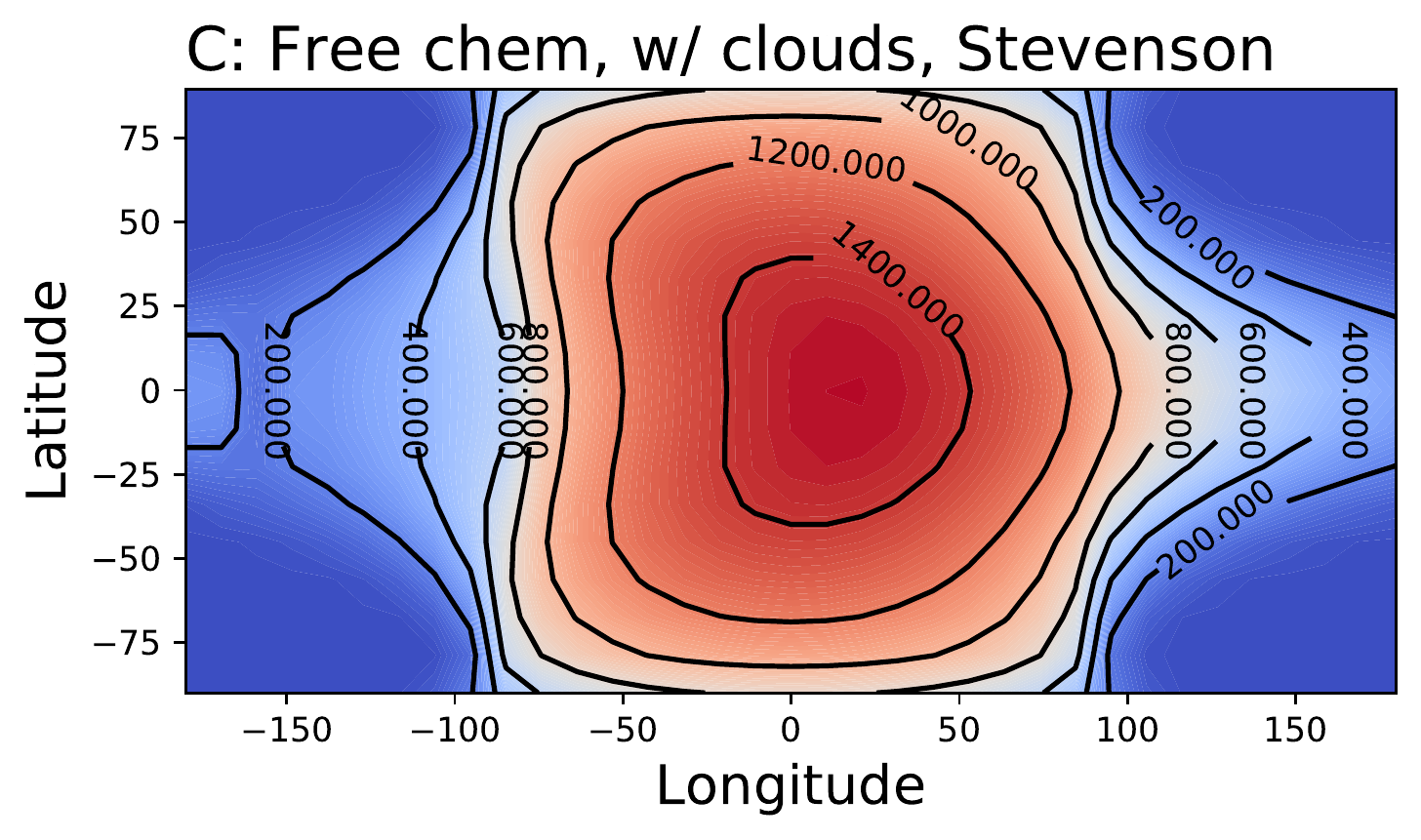}
                        \includegraphics[width=0.45\textwidth]{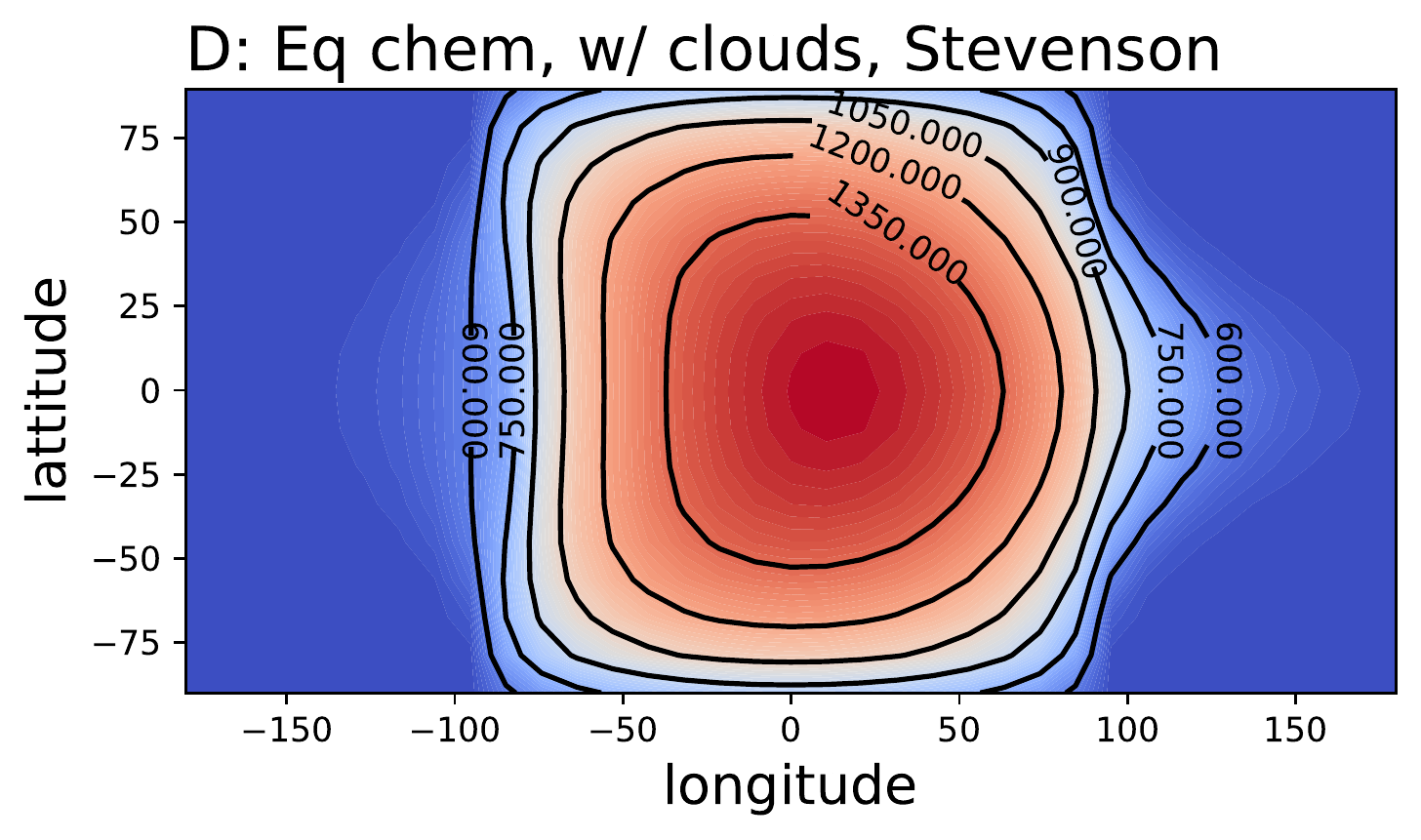}
                        \includegraphics[width=0.45\textwidth]{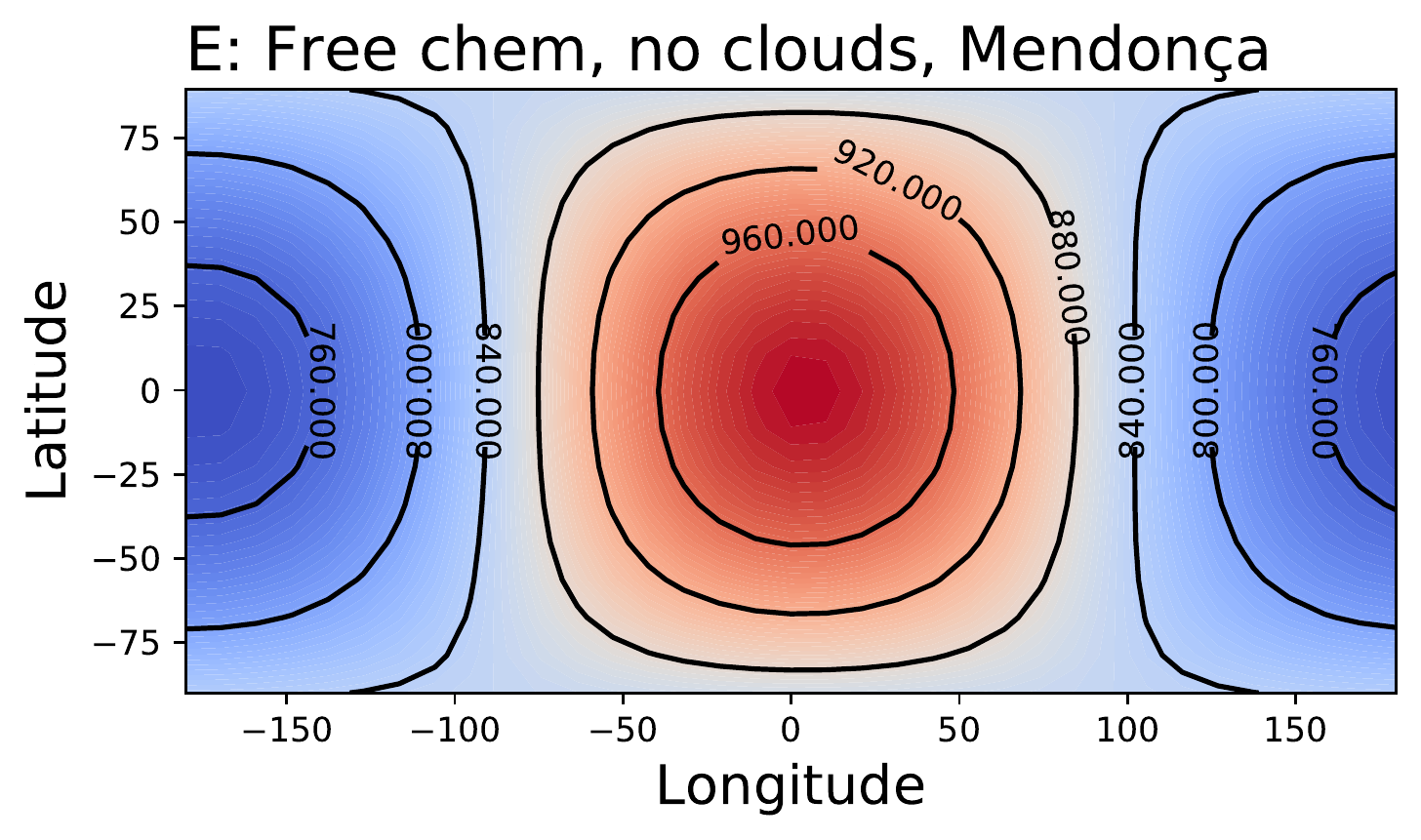}
                        \includegraphics[width=0.45\textwidth]{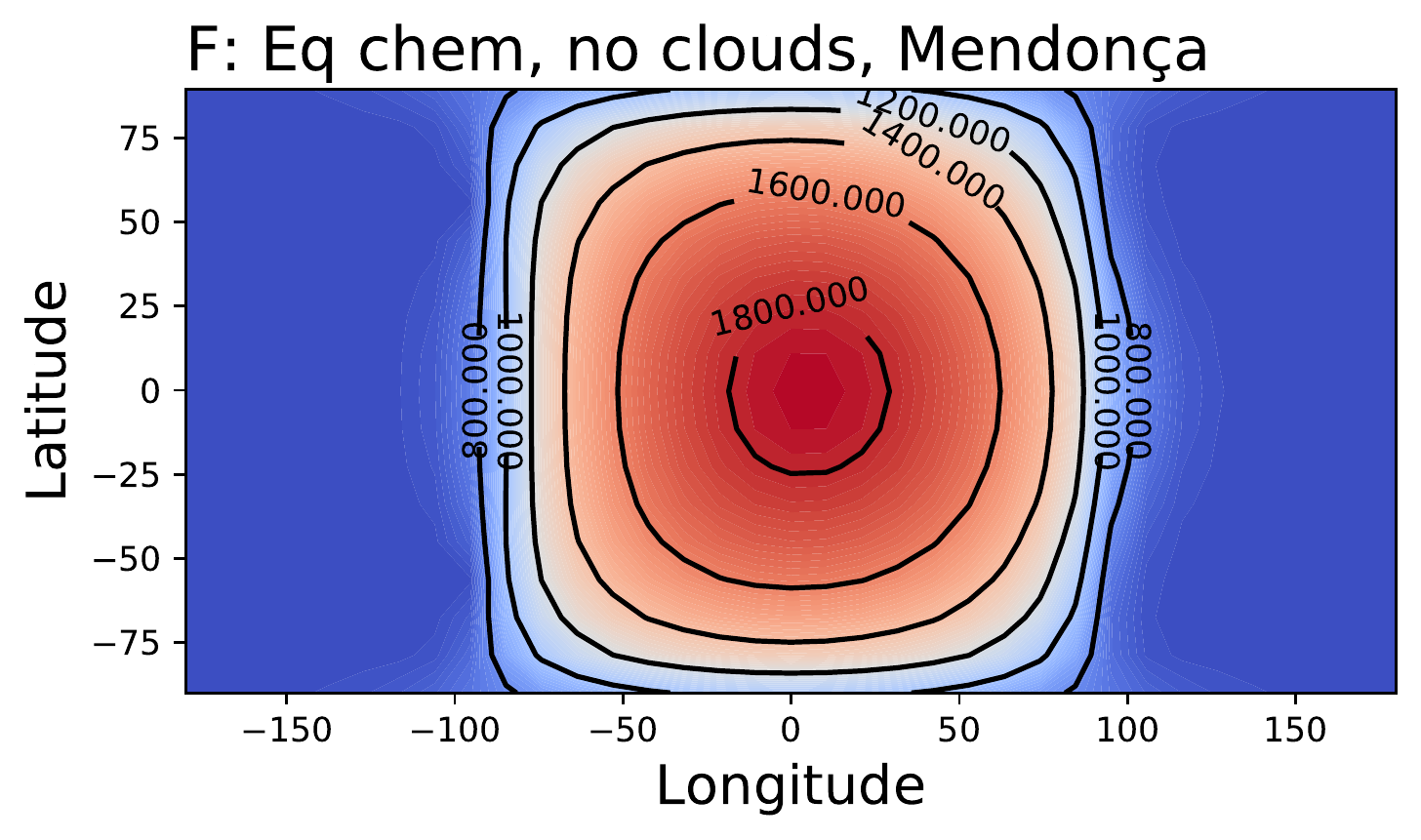}
                        \includegraphics[width=0.45\textwidth]{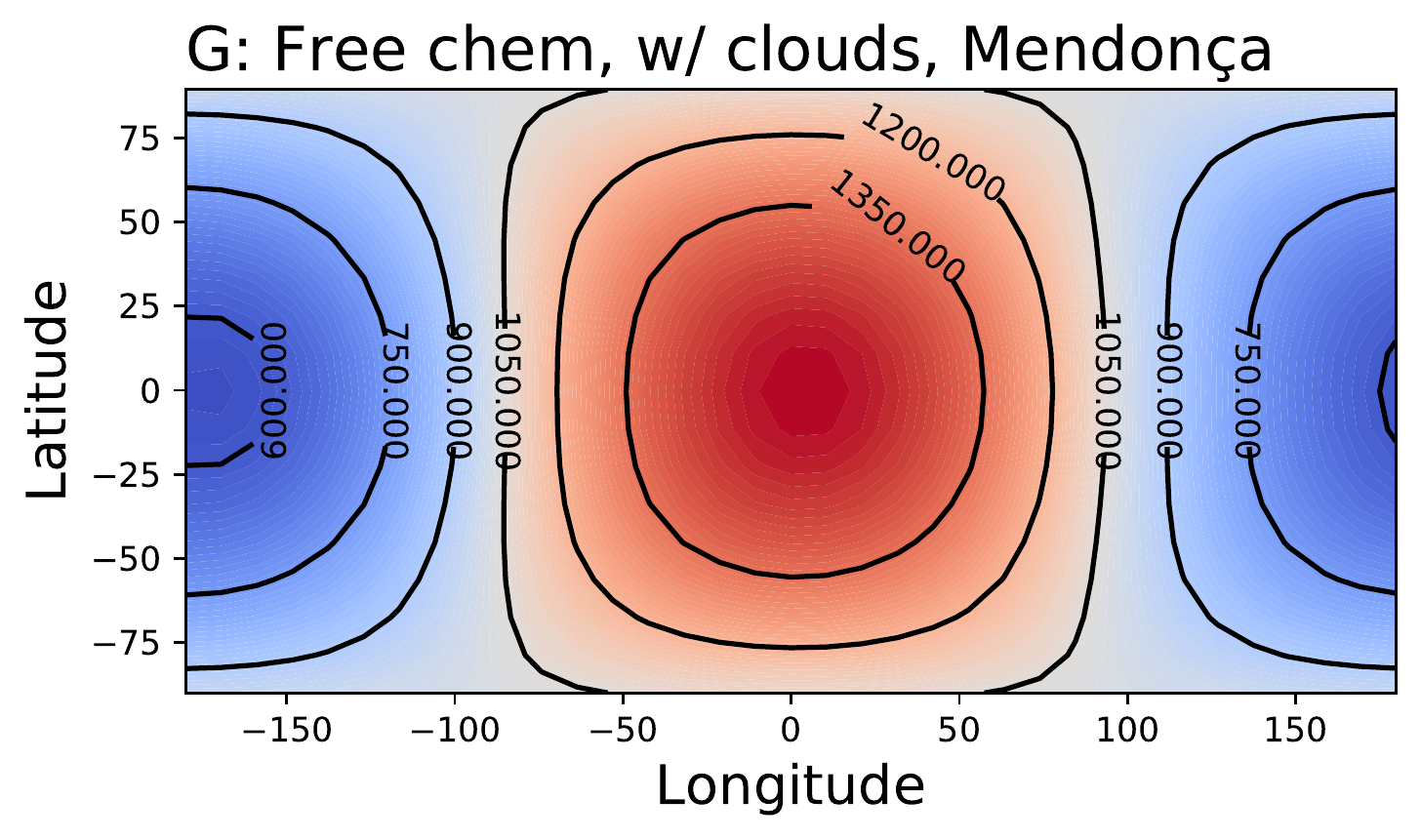}
                        \includegraphics[width=0.45\textwidth]{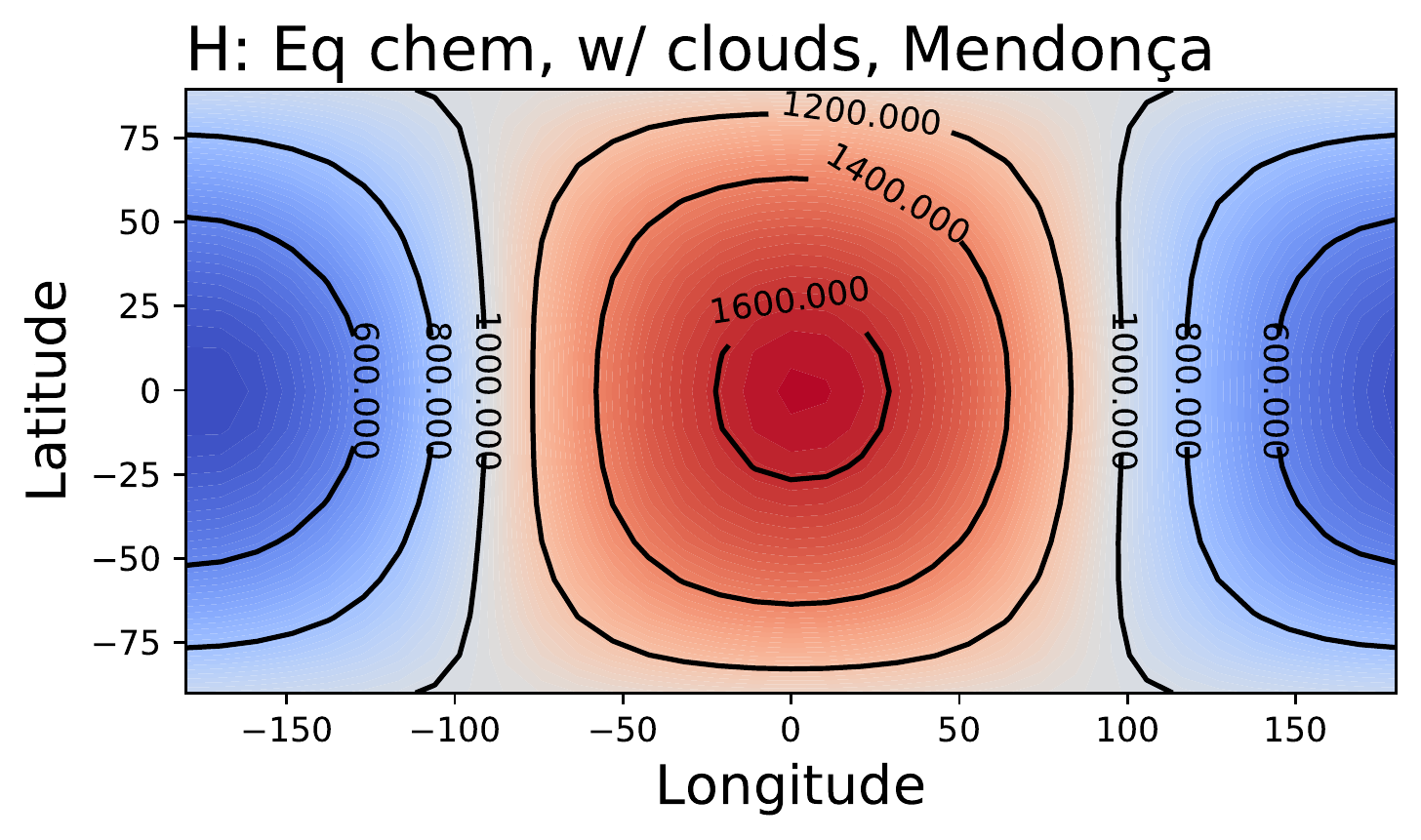}
                        \caption{Contour plots giving retrieved temperature at an altitude of 0.01~bar (see Eq.~\ref{eq:beta}) as a function of longitude and latitude for the eight different retrieval setups, as detailed above each panel. Stevenson and Mendon{\c{c}}a refer to the use of Spitzer data analysed by \cite{17StLiBe.wasp43b} and \cite{18MeMaDe.wasp43b}, respectively.}\label{fig:contour_plots_temp_0.01}
                \end{figure}

                \begin{figure}
                        \centering
                        \includegraphics[width=0.45\textwidth]{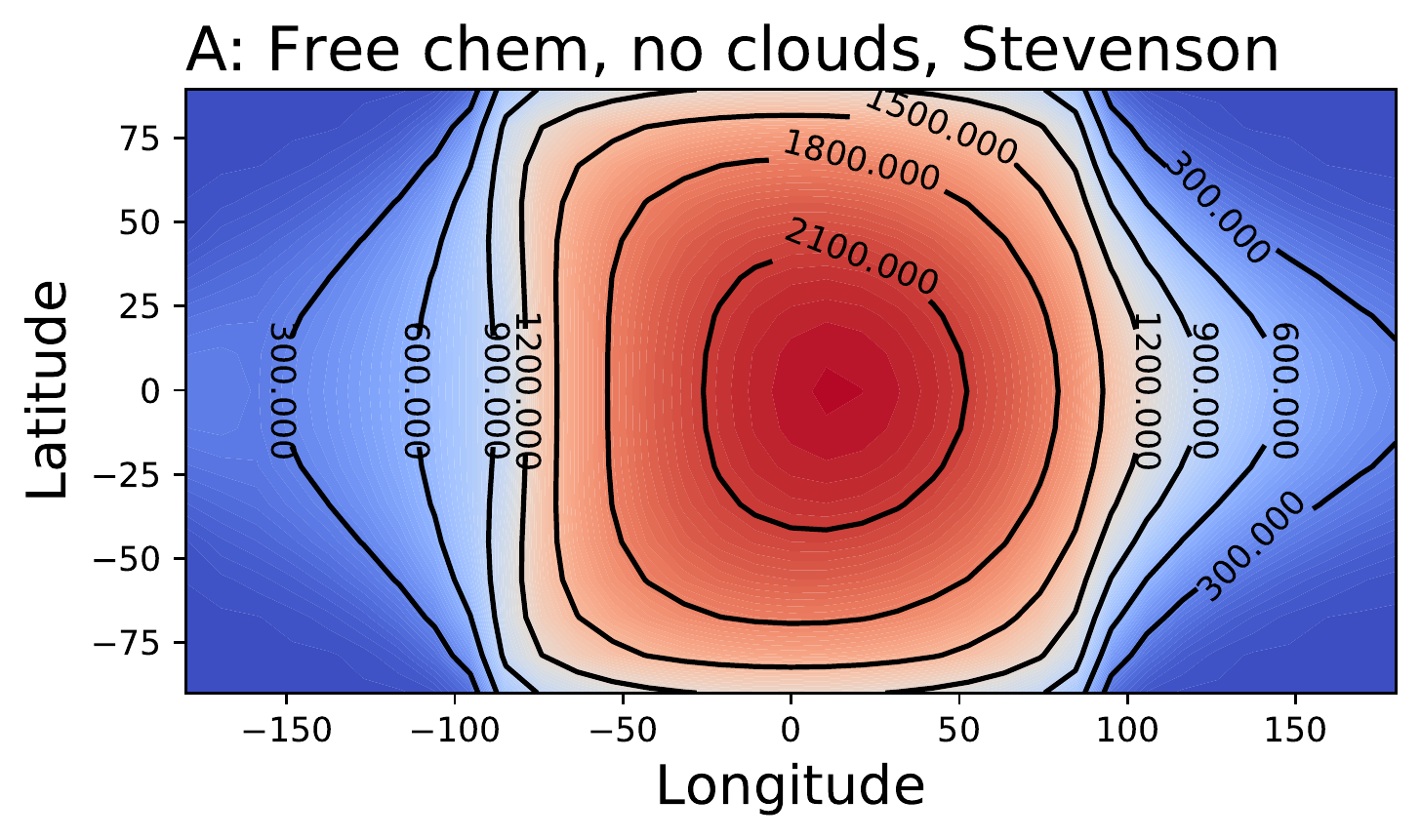}
                        \includegraphics[width=0.45\textwidth]{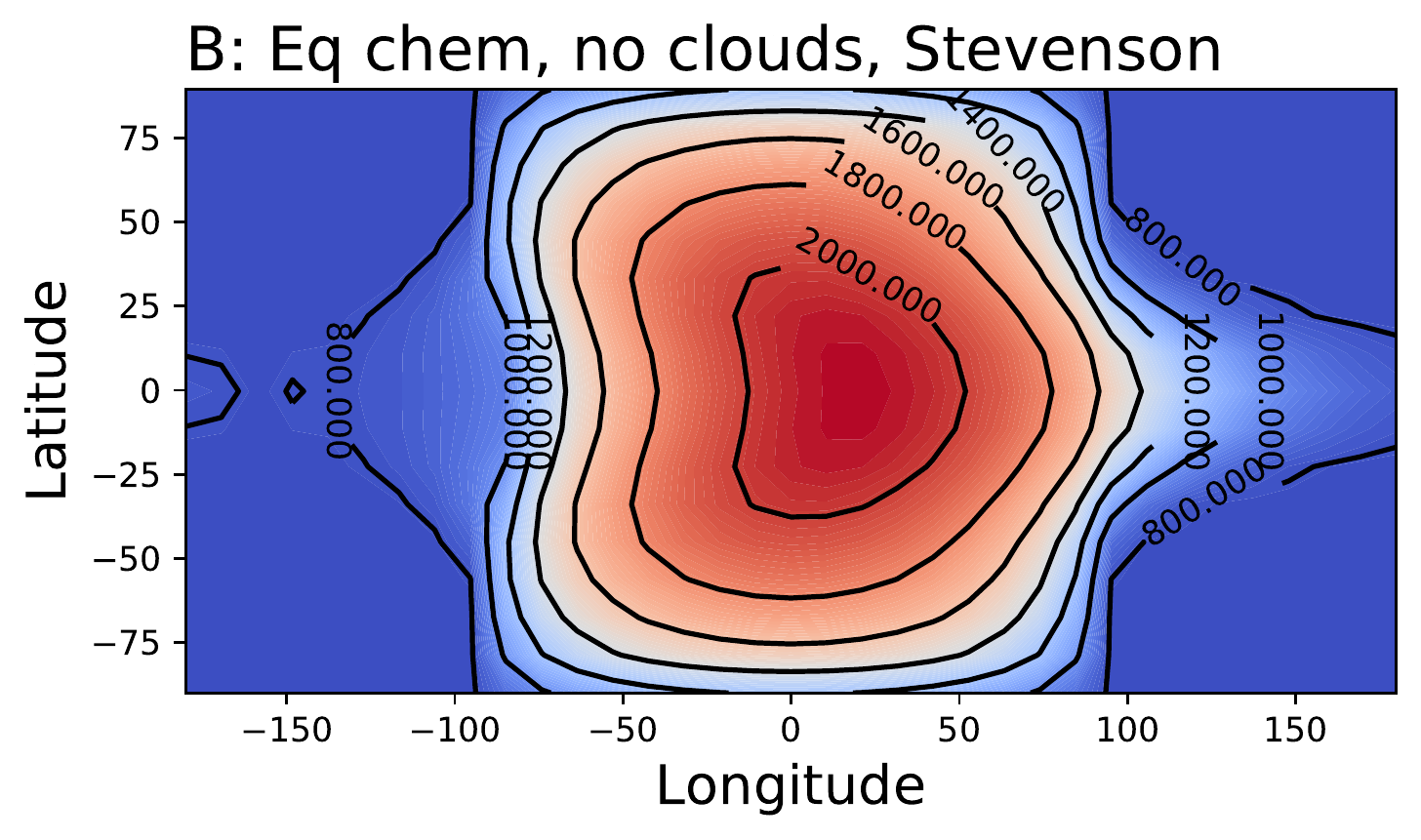}
                        \includegraphics[width=0.45\textwidth]{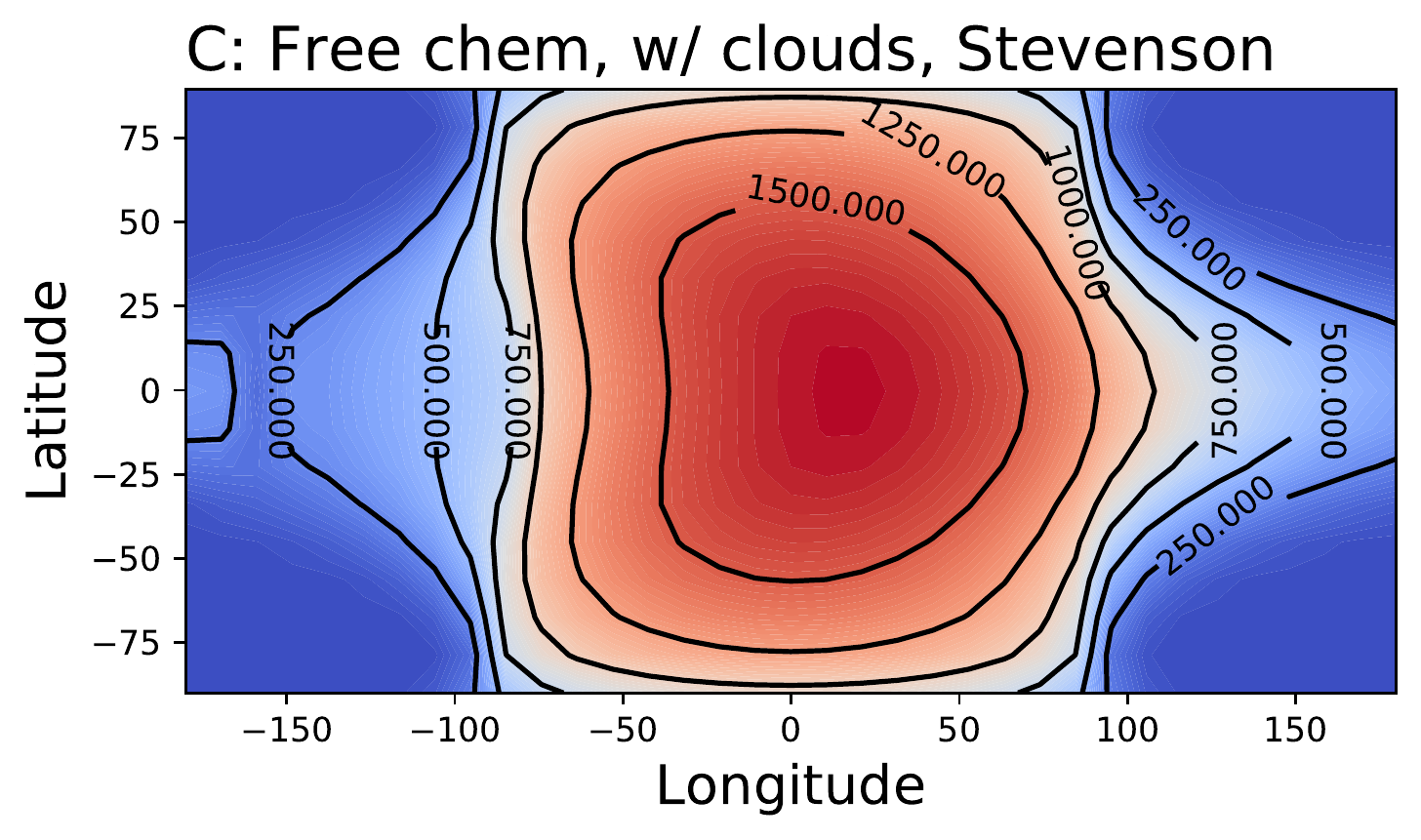}
                        \includegraphics[width=0.45\textwidth]{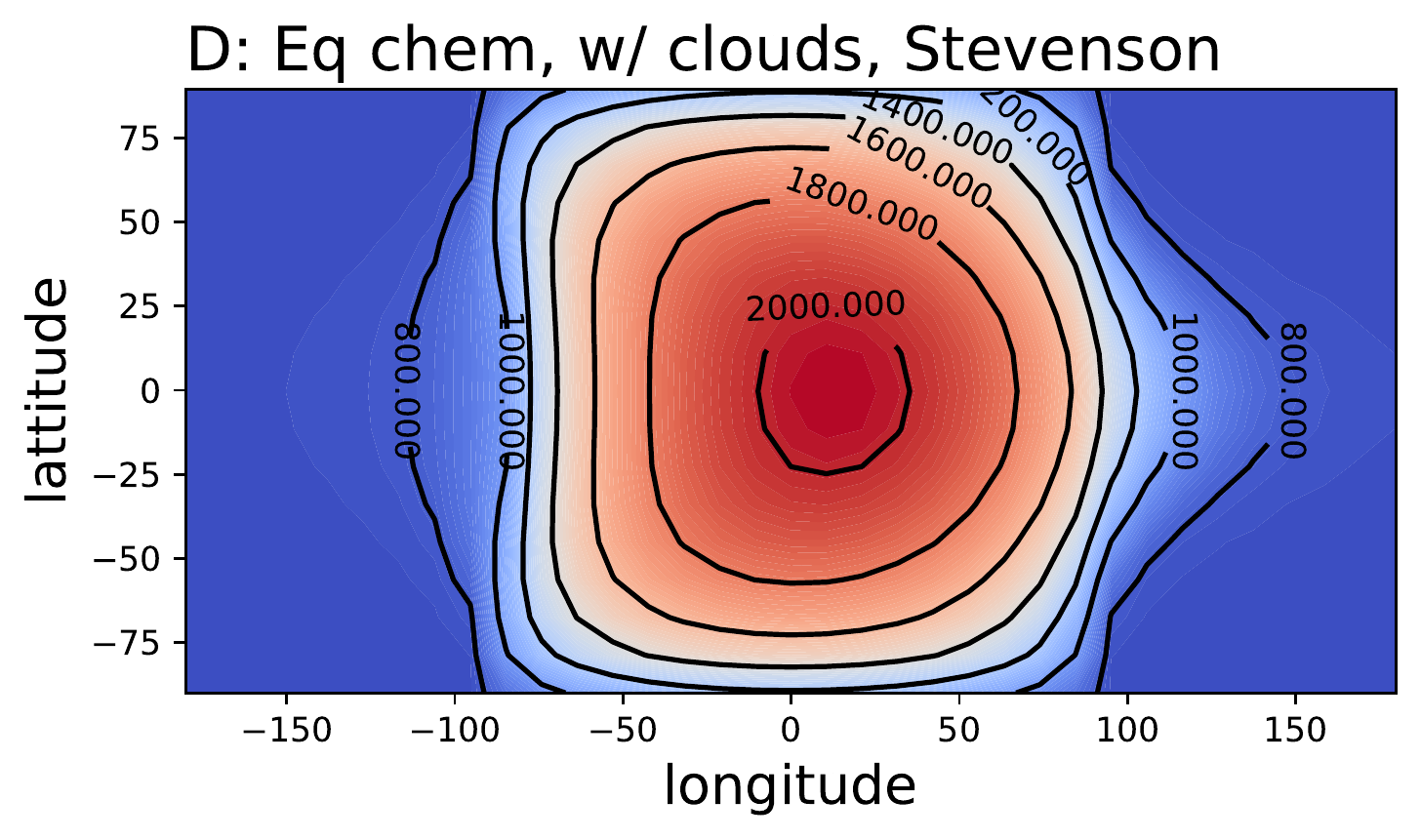}
                        \includegraphics[width=0.45\textwidth]{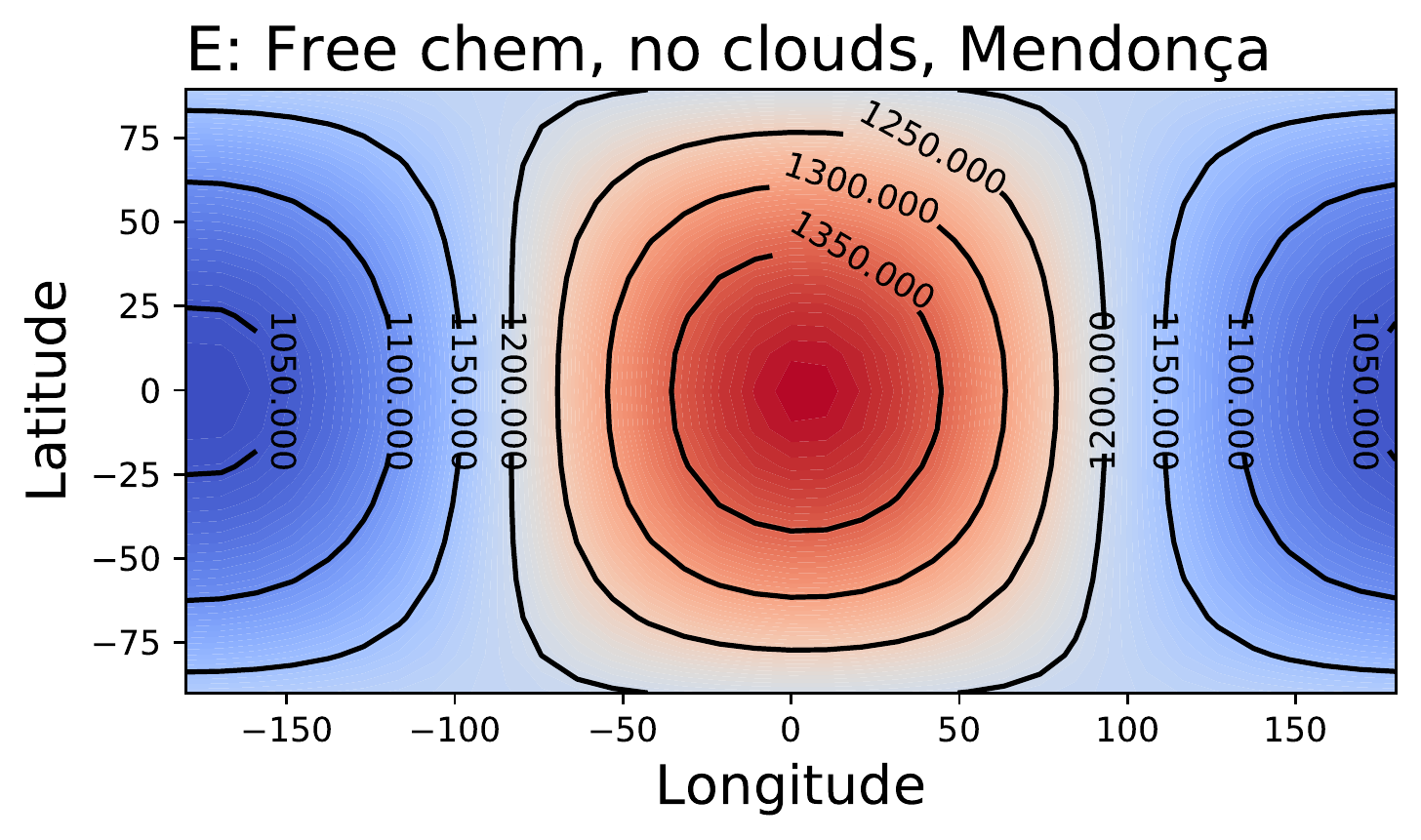}
                        \includegraphics[width=0.45\textwidth]{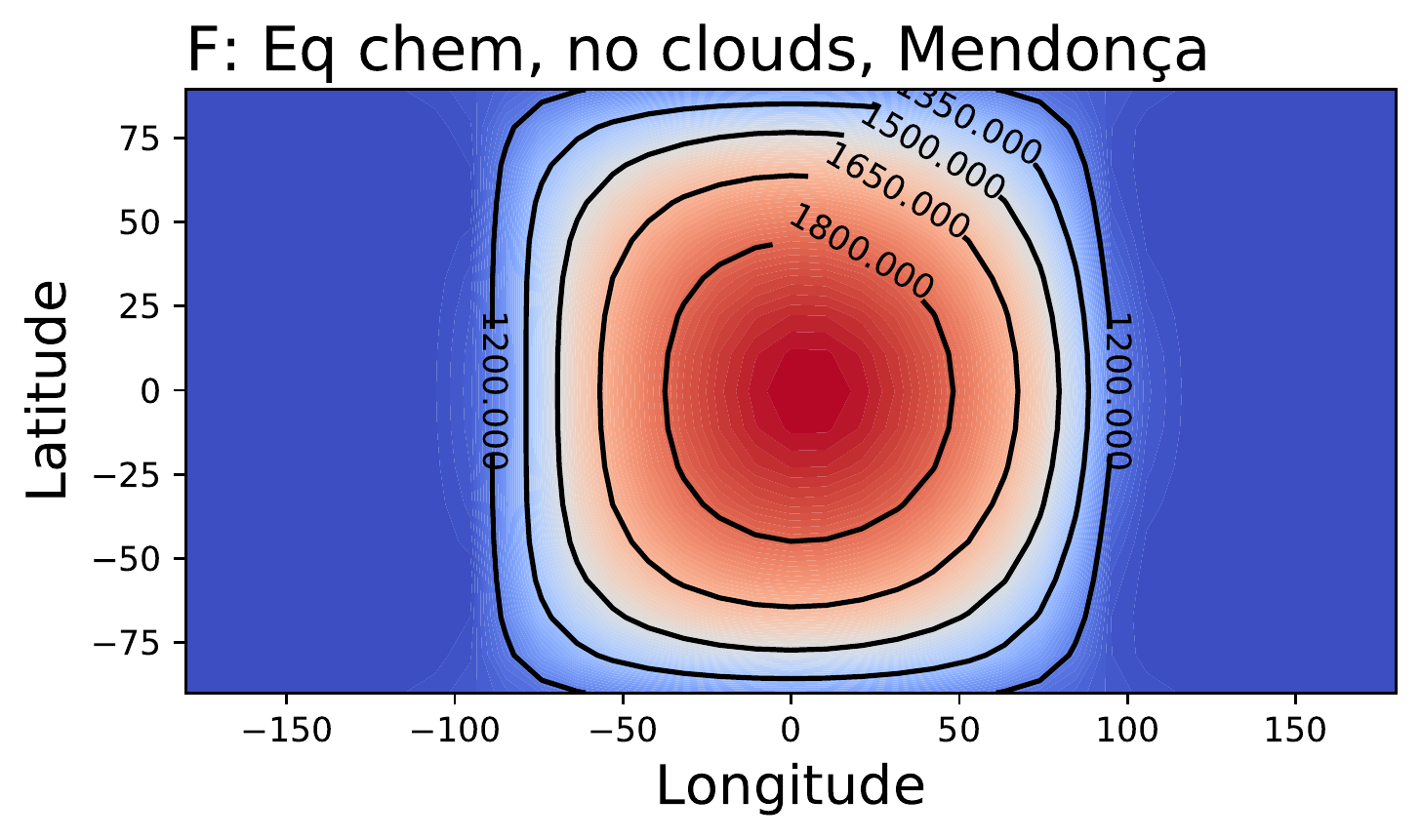}
                        \includegraphics[width=0.45\textwidth]{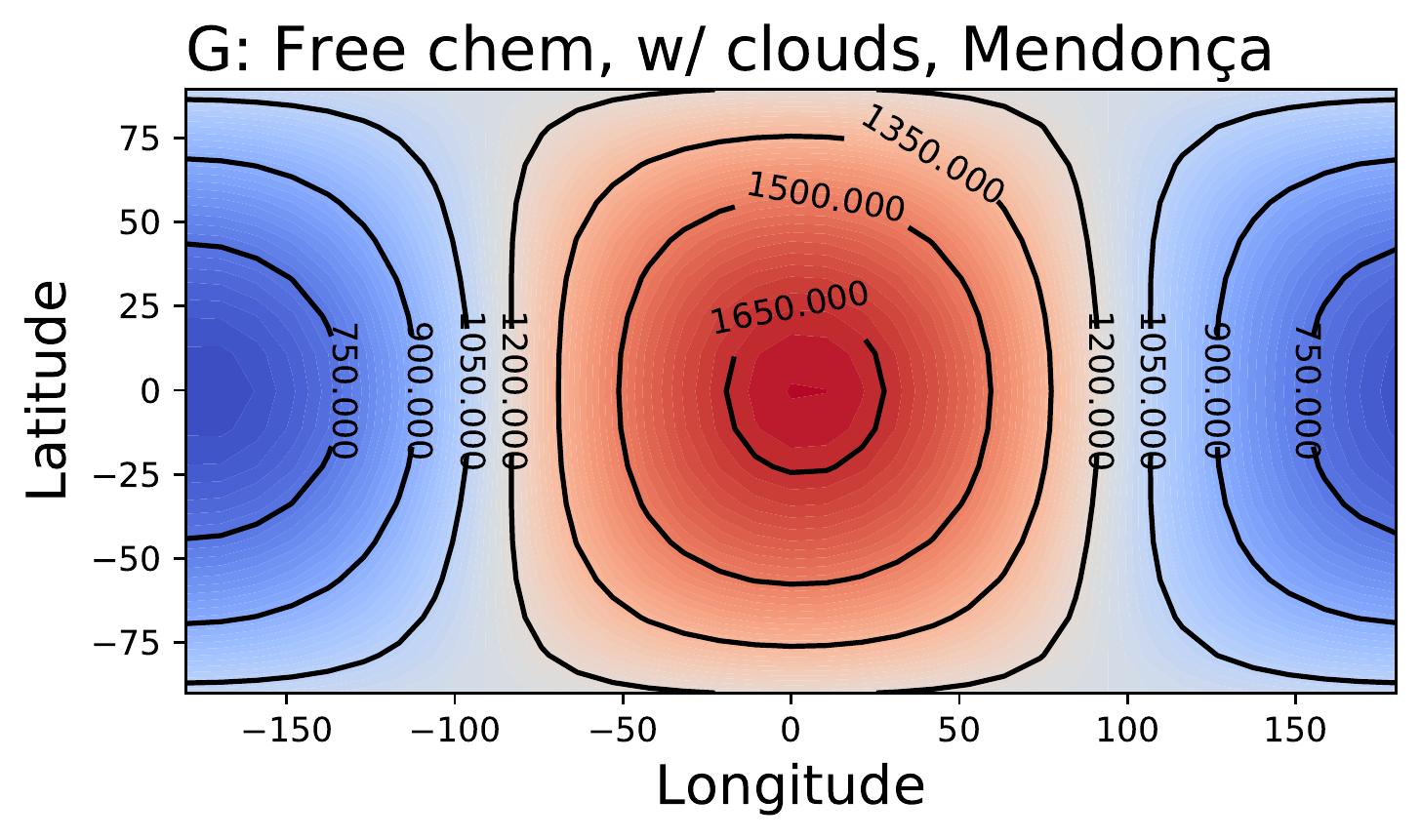}
                        \includegraphics[width=0.45\textwidth]{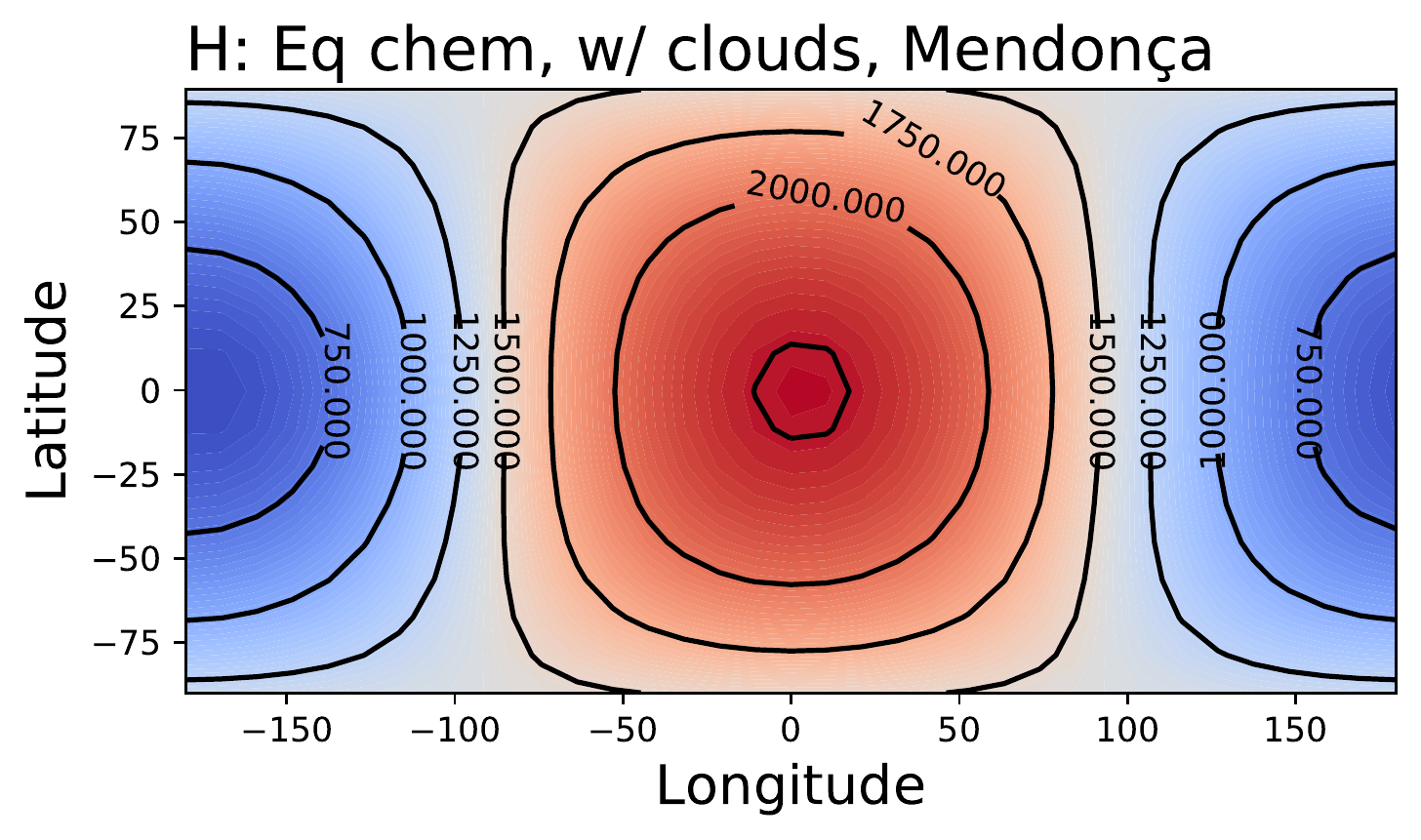}
                        \caption{Same as Fig~\ref{fig:contour_plots_temp_0.01}, but for an altitude level of 0.12~bar}\label{fig:contour_plots_temp_0.12}
                        
                \end{figure}

                \begin{figure}
                        \centering
                        \includegraphics[width=0.4\textwidth]{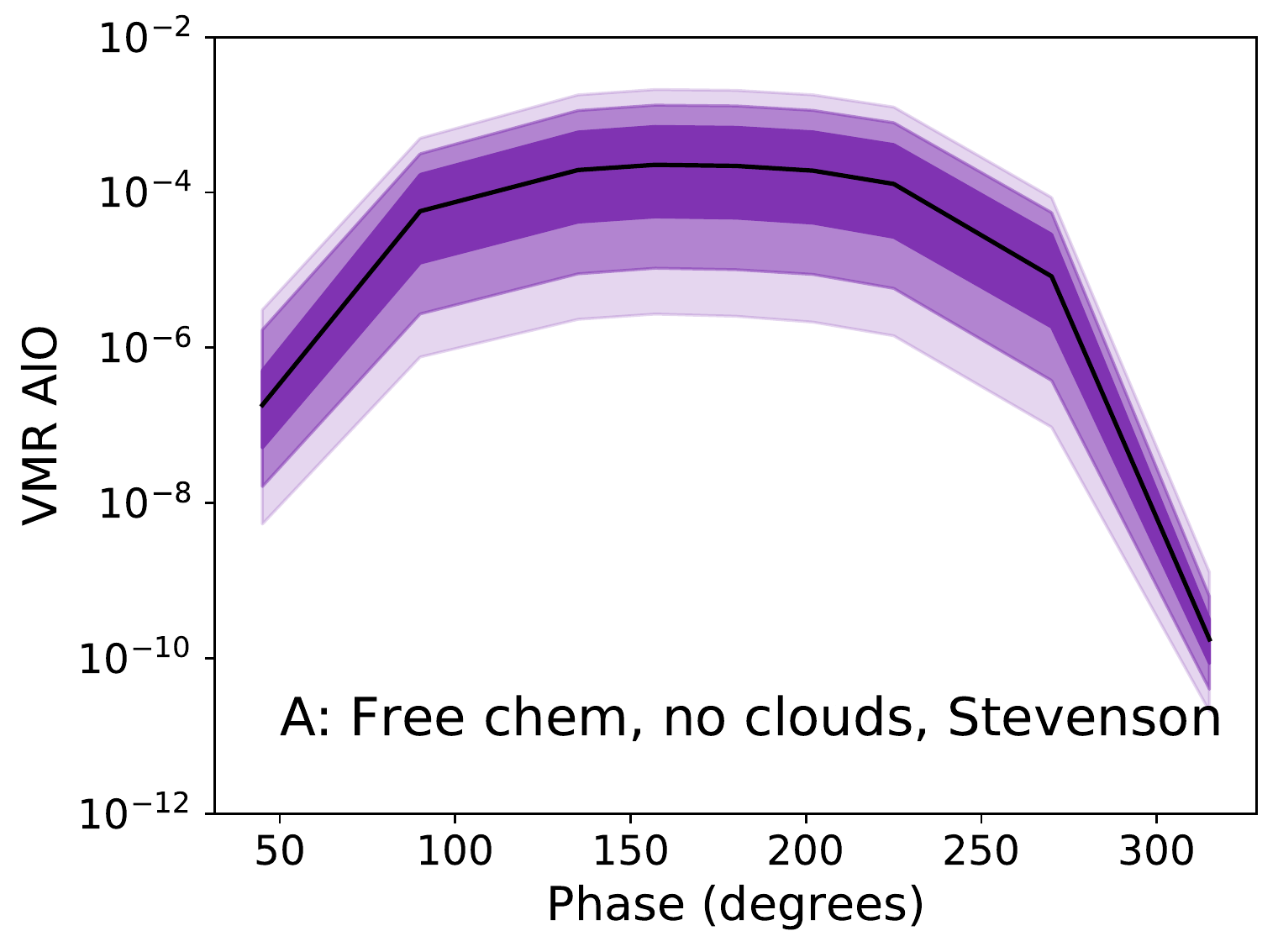}
                        \includegraphics[width=0.4\textwidth]{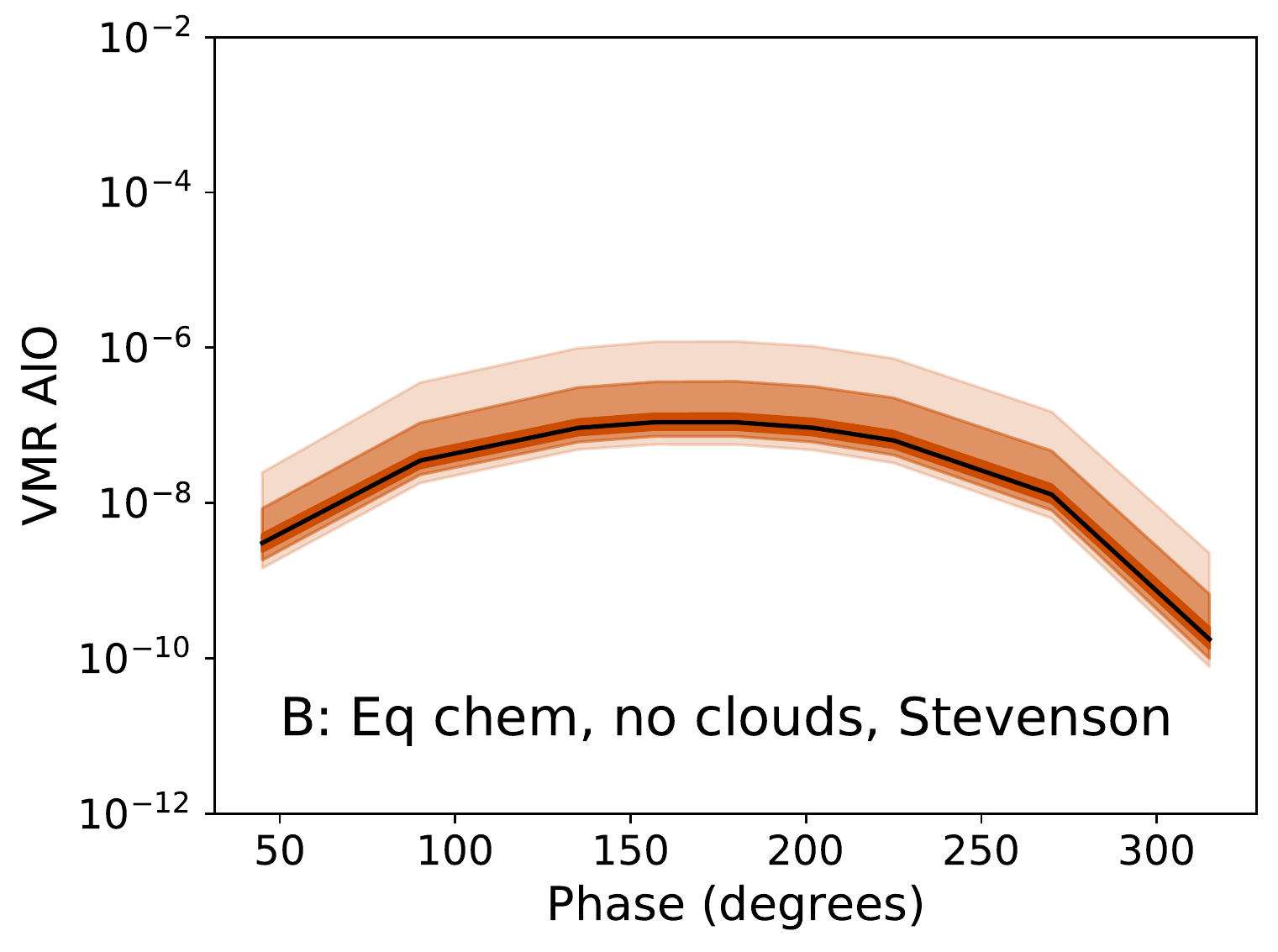}
                        \includegraphics[width=0.4\textwidth]{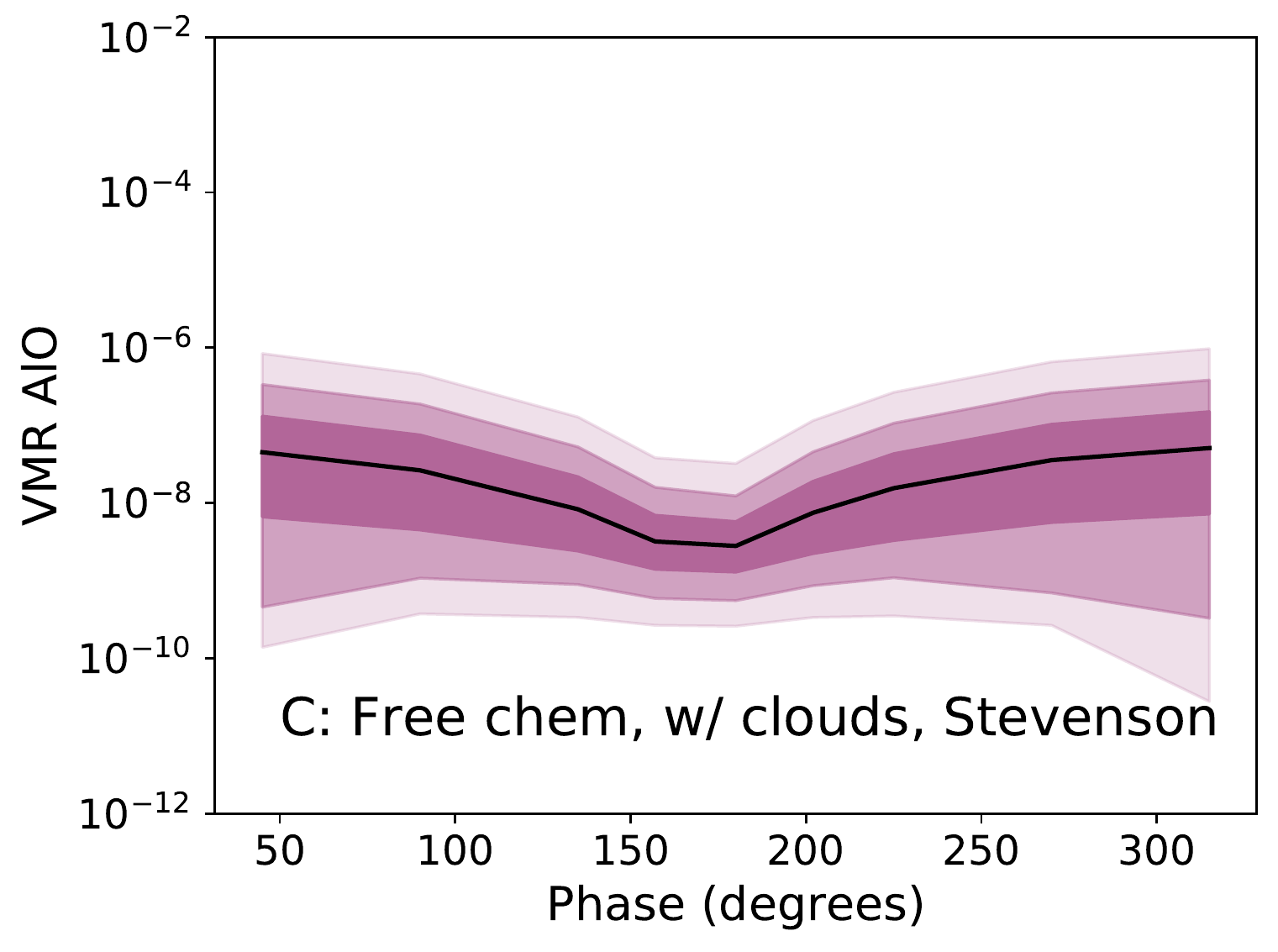}
                        \includegraphics[width=0.4\textwidth]{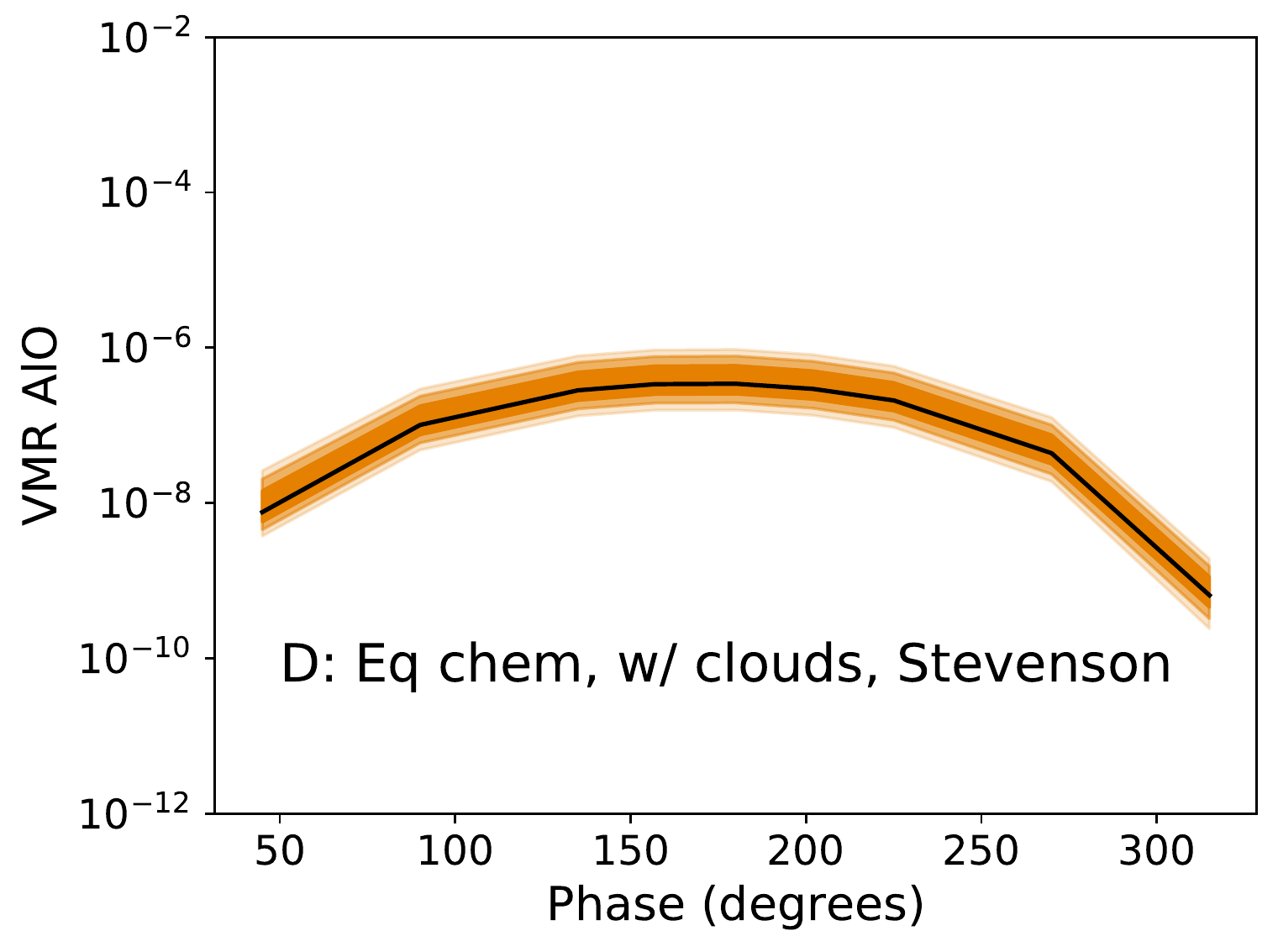}
                        \includegraphics[width=0.4\textwidth]{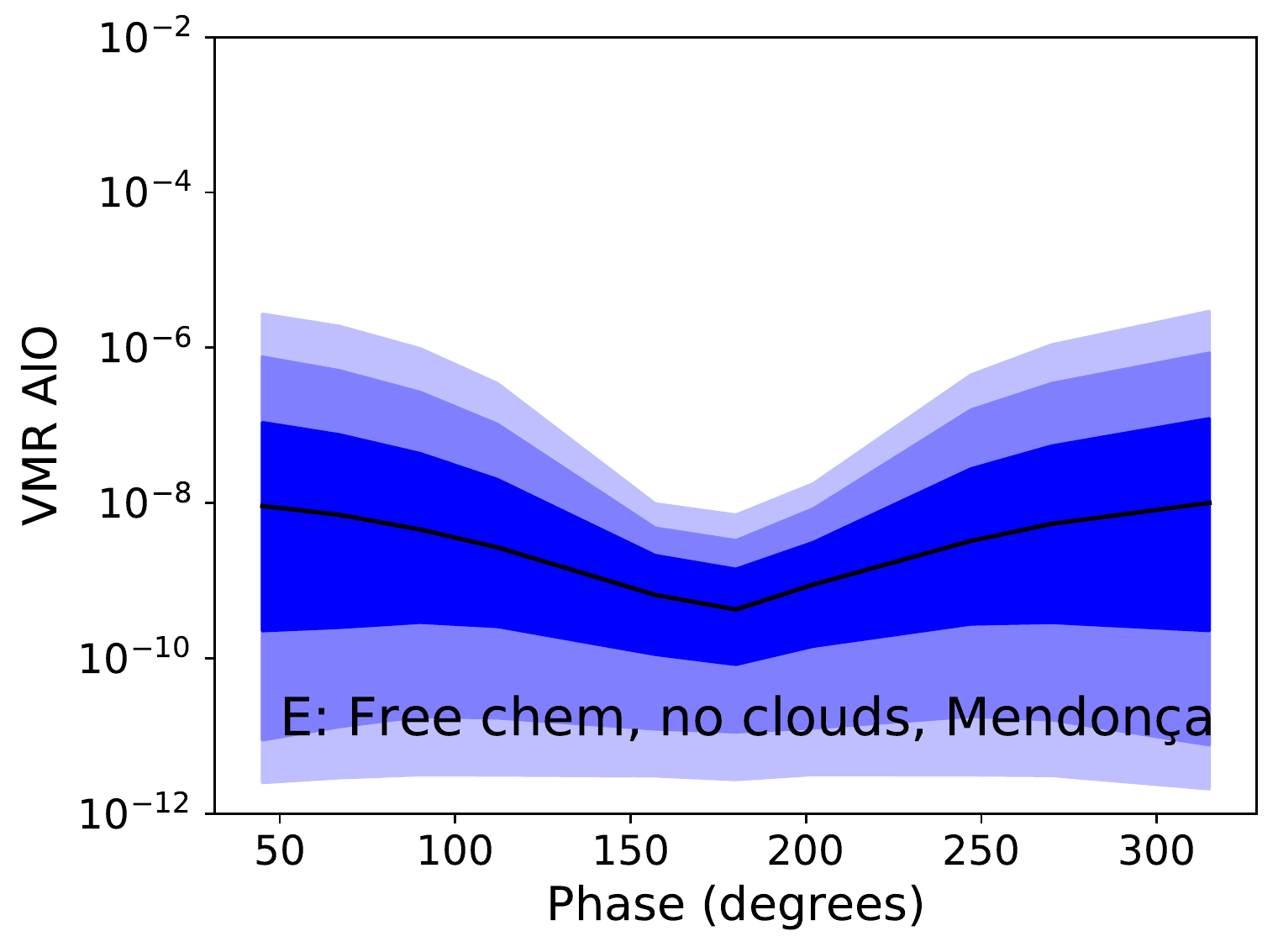}
                        \includegraphics[width=0.4\textwidth]{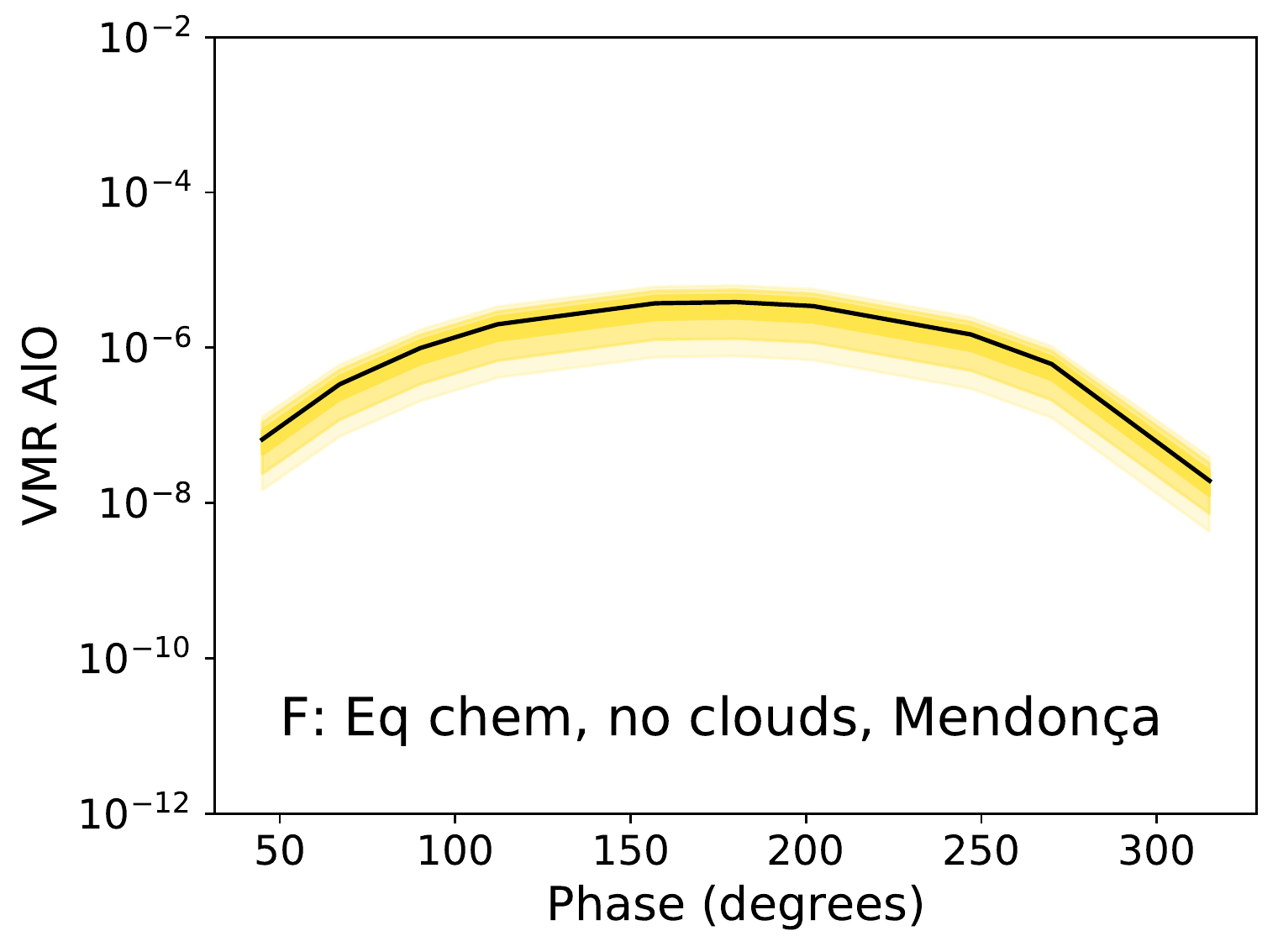}
                        \includegraphics[width=0.4\textwidth]{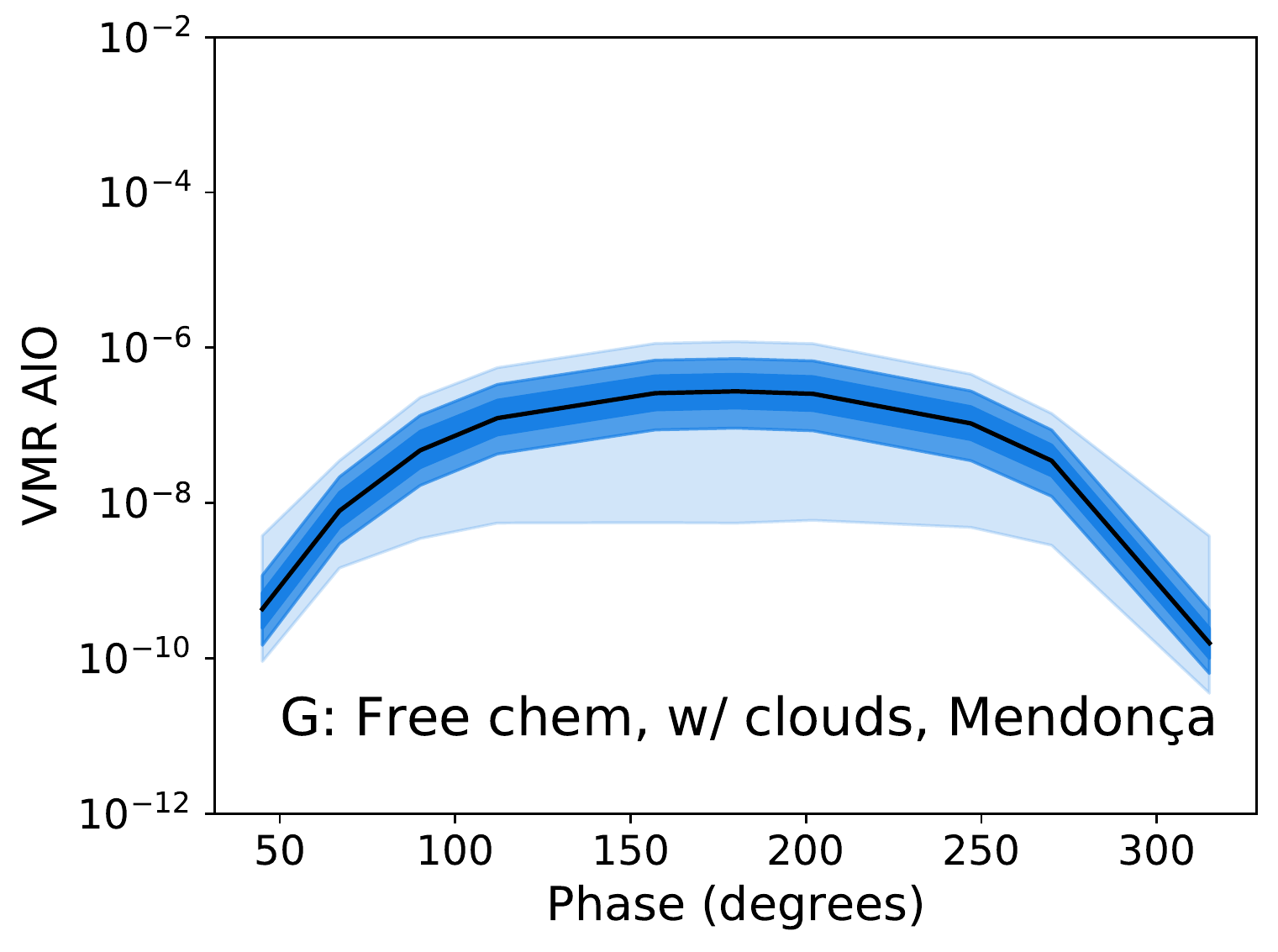}
                        \includegraphics[width=0.4\textwidth]{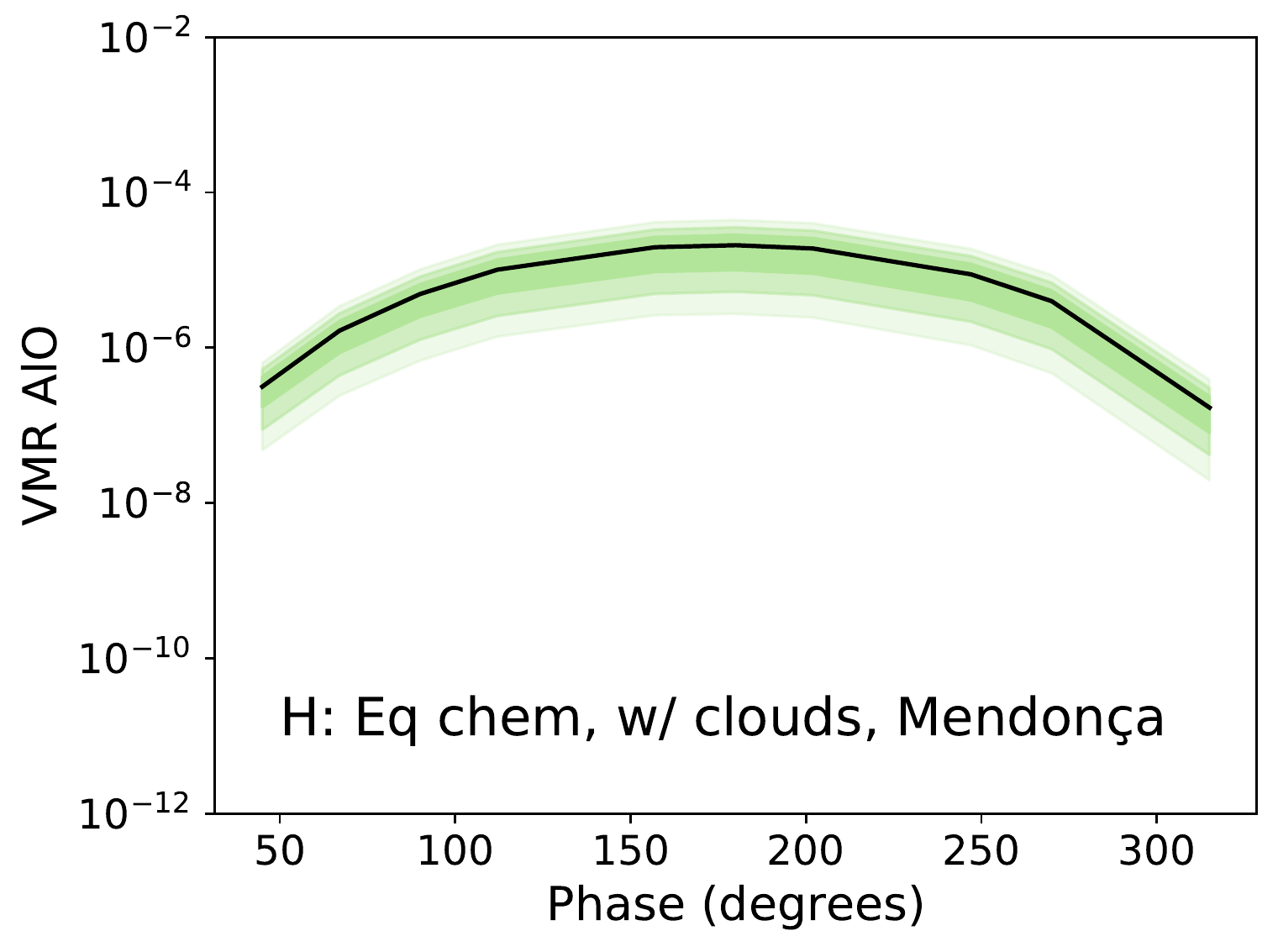}
                        \caption{Variation of retrieved AlO abundance (VMR) as a function of phase ($\degree$) for the eight different retrieval setups, as detailed in each panel. The VMRs are averaged over the visible part of the disk, centred at each of the given phases, and taken at a pressure-layer cut of 0.18~bar at the equator. Stevenson and Mendon{\c{c}}a refer to the use of Spitzer data analysed by \cite{17StLiBe.wasp43b} and \cite{18MeMaDe.wasp43b}, respectively.}\label{fig:AlO_COratio}
                \end{figure}

                \begin{figure}
                        \centering
                        \includegraphics[width=0.4\textwidth]{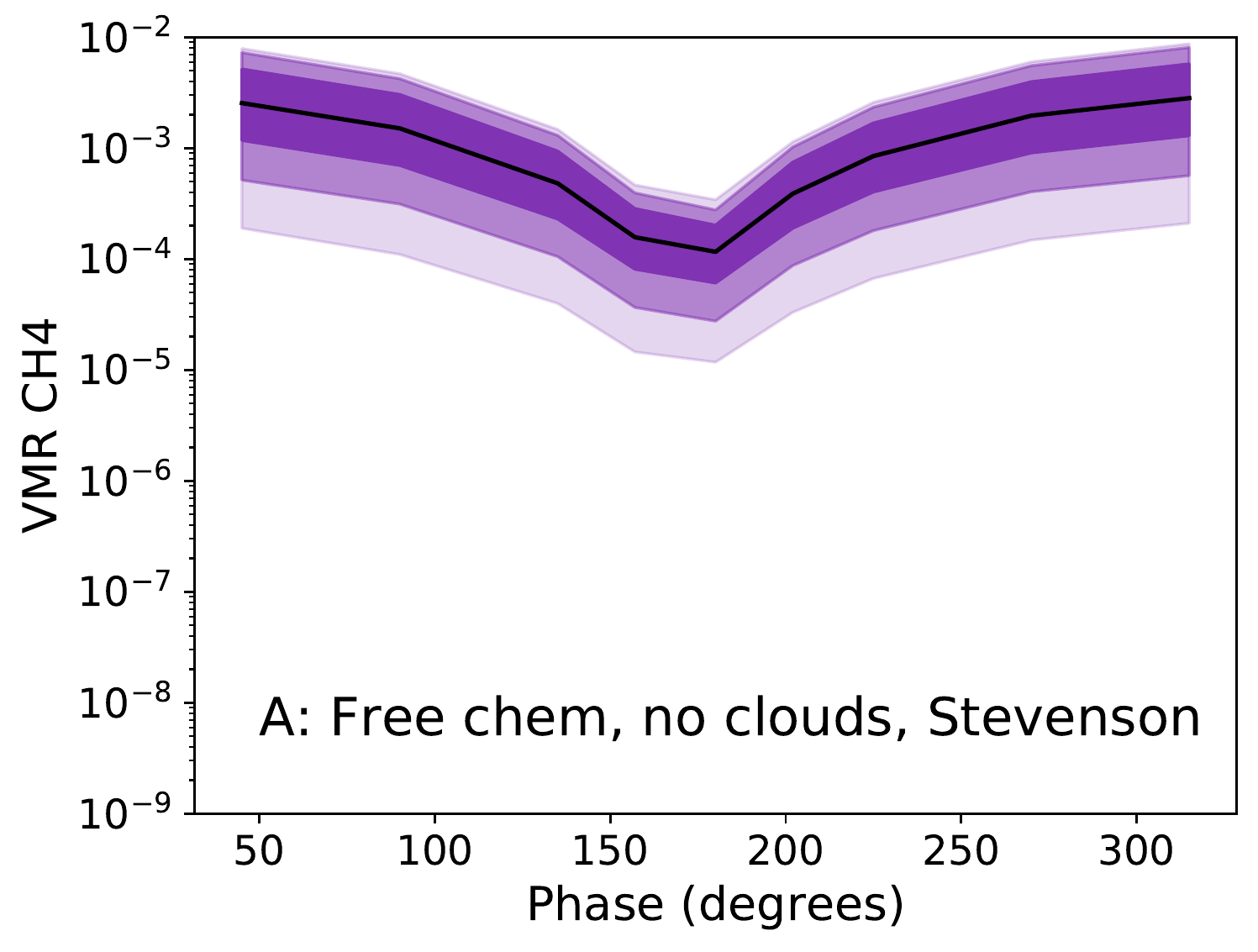}
                        \includegraphics[width=0.4\textwidth]{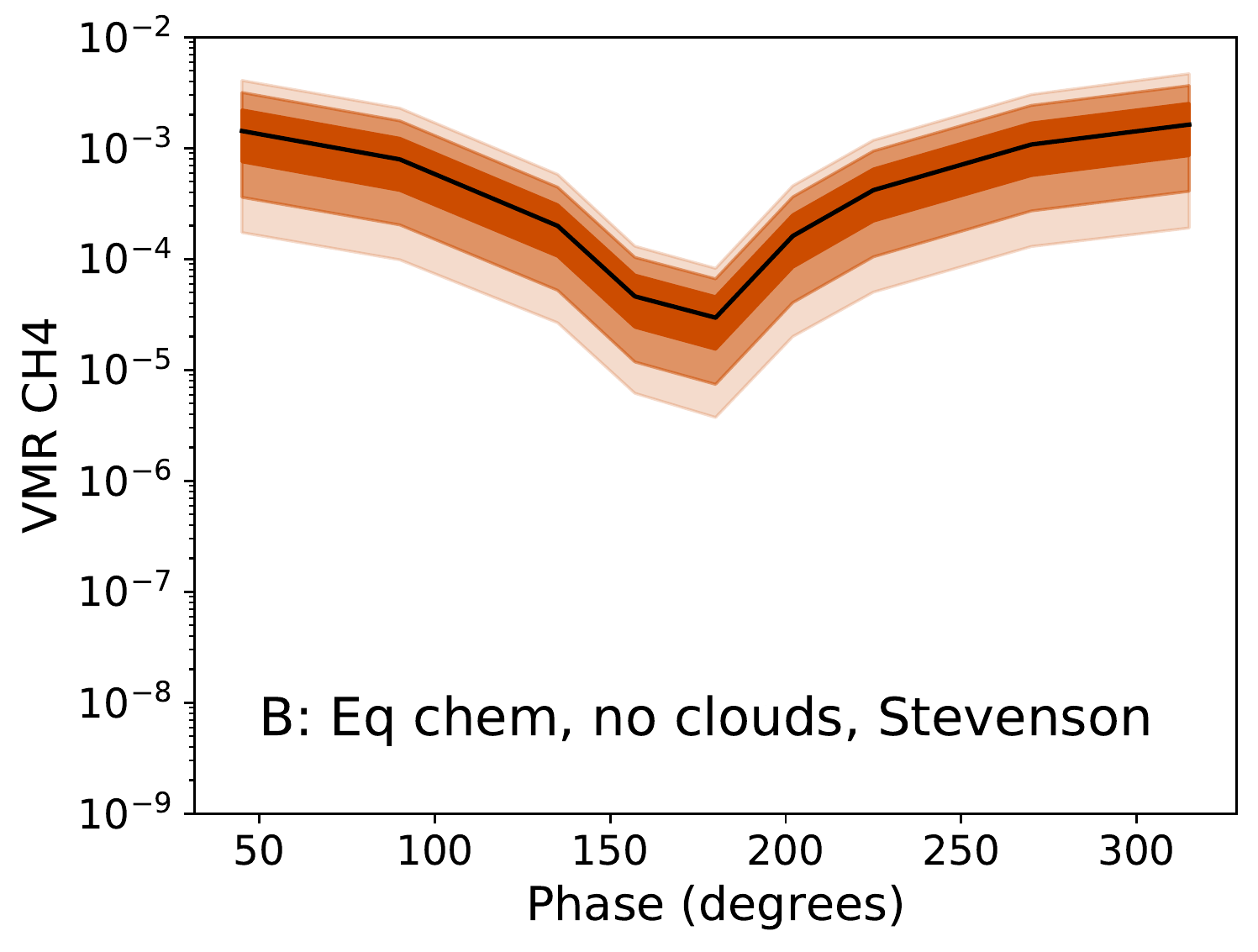}
                        \includegraphics[width=0.4\textwidth]{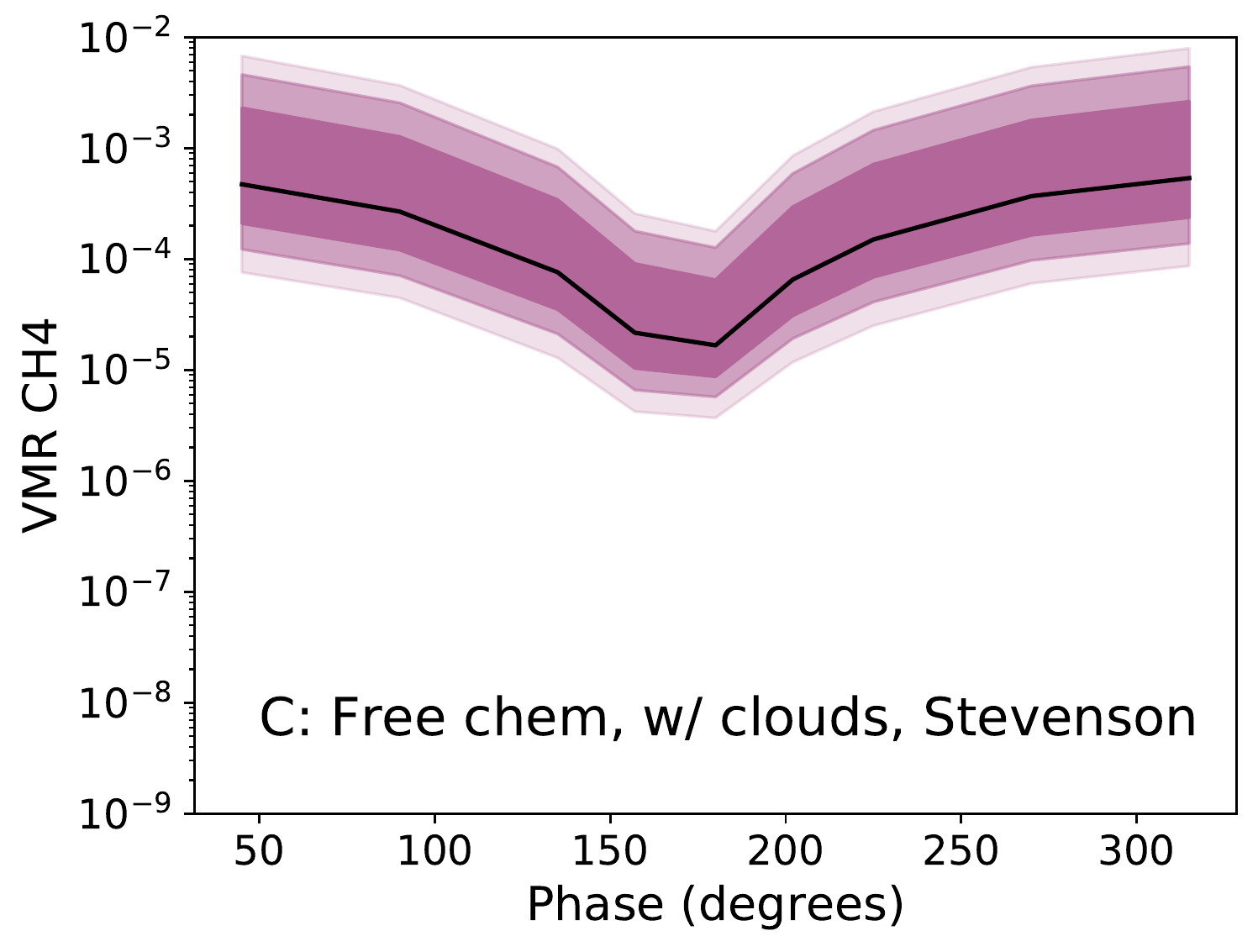}
                        \includegraphics[width=0.4\textwidth]{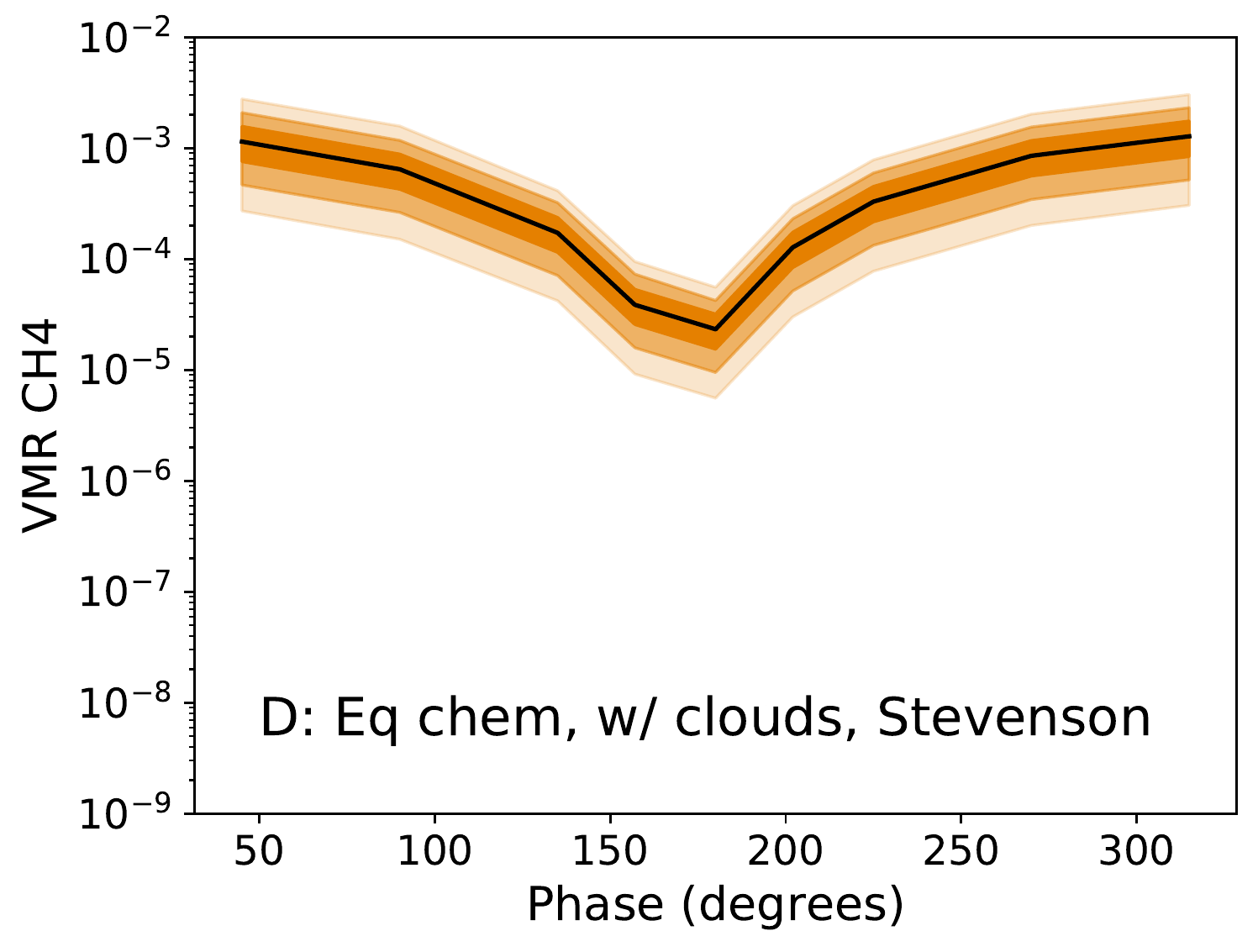}
                        \includegraphics[width=0.4\textwidth]{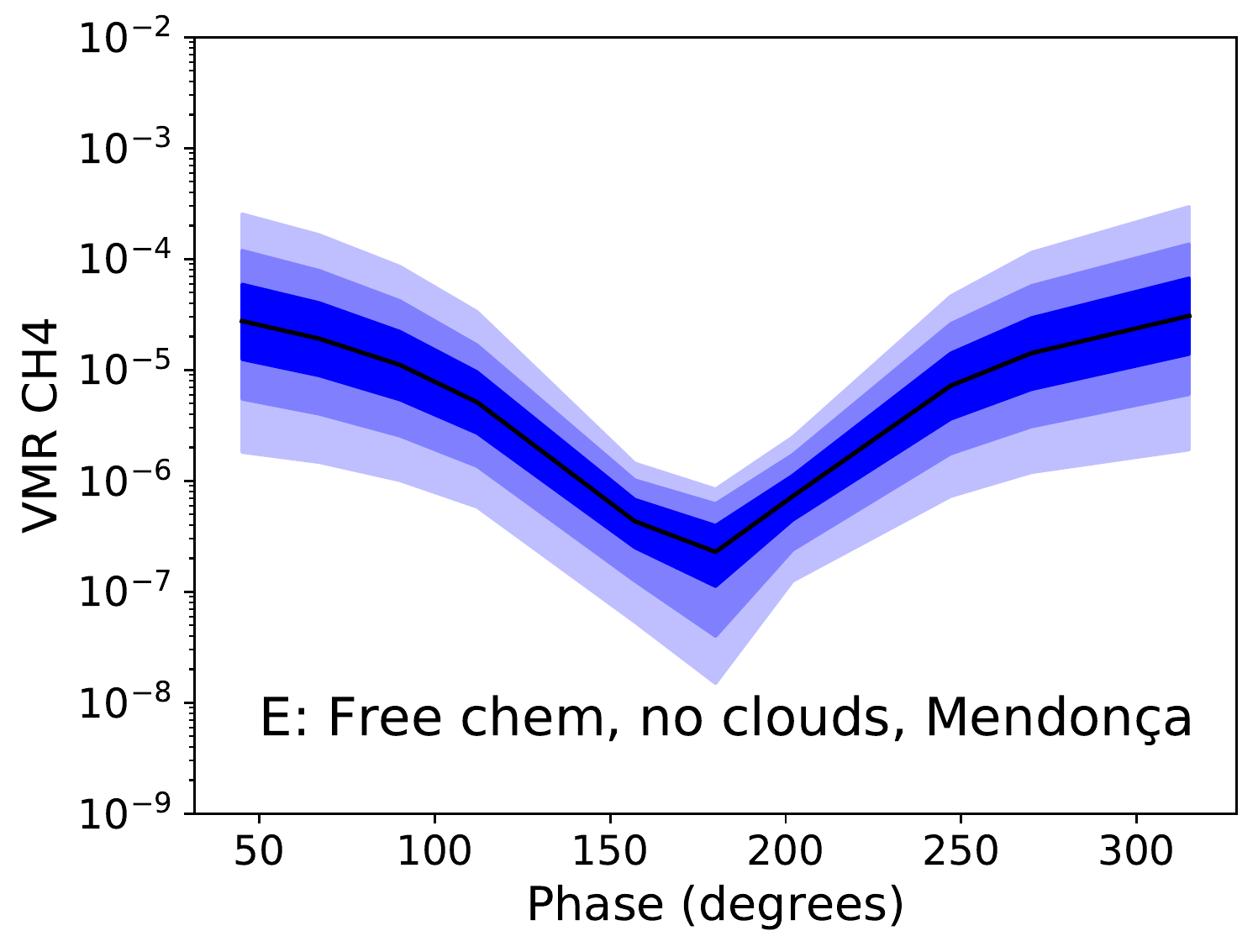}
                        \includegraphics[width=0.4\textwidth]{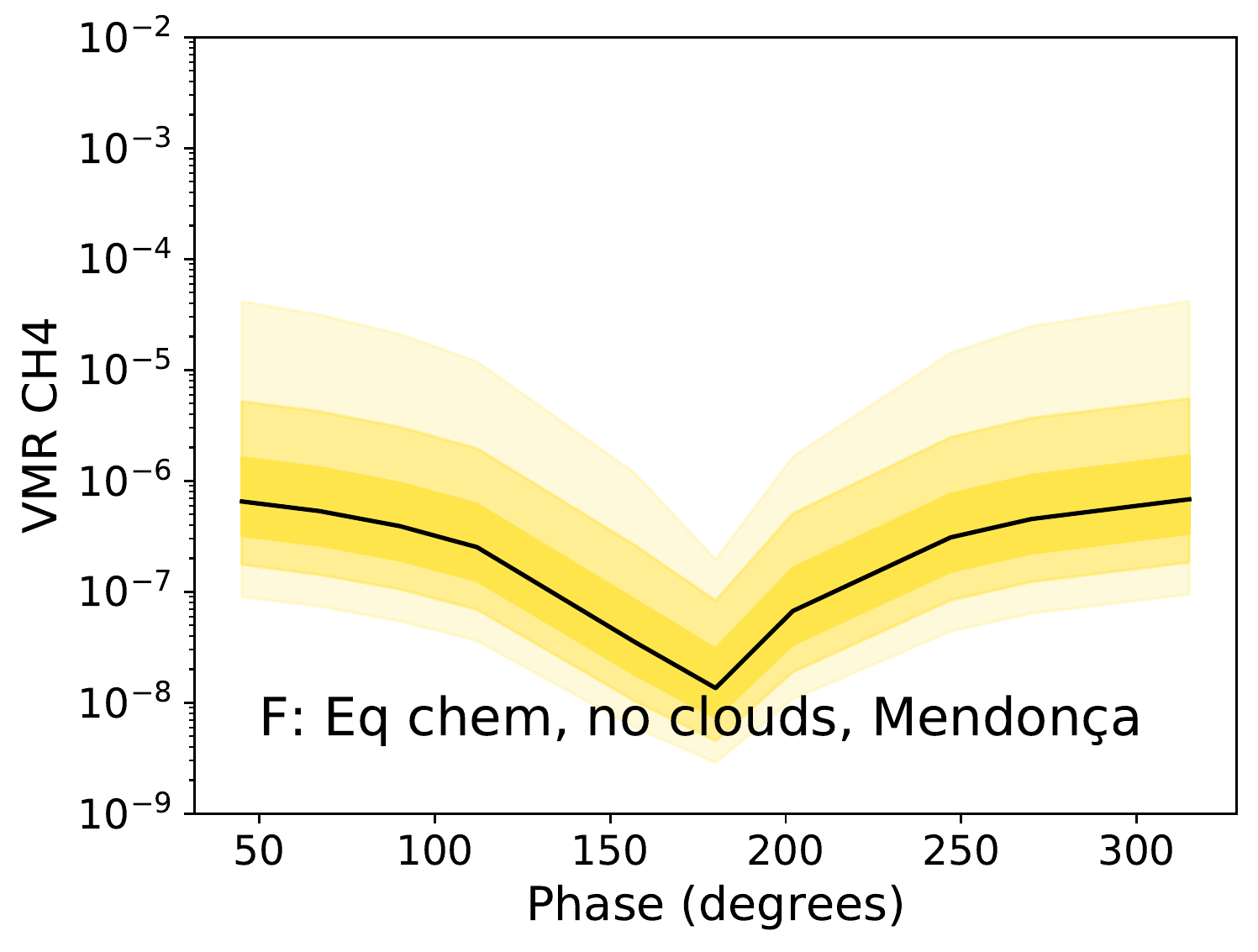}
                        \includegraphics[width=0.4\textwidth]{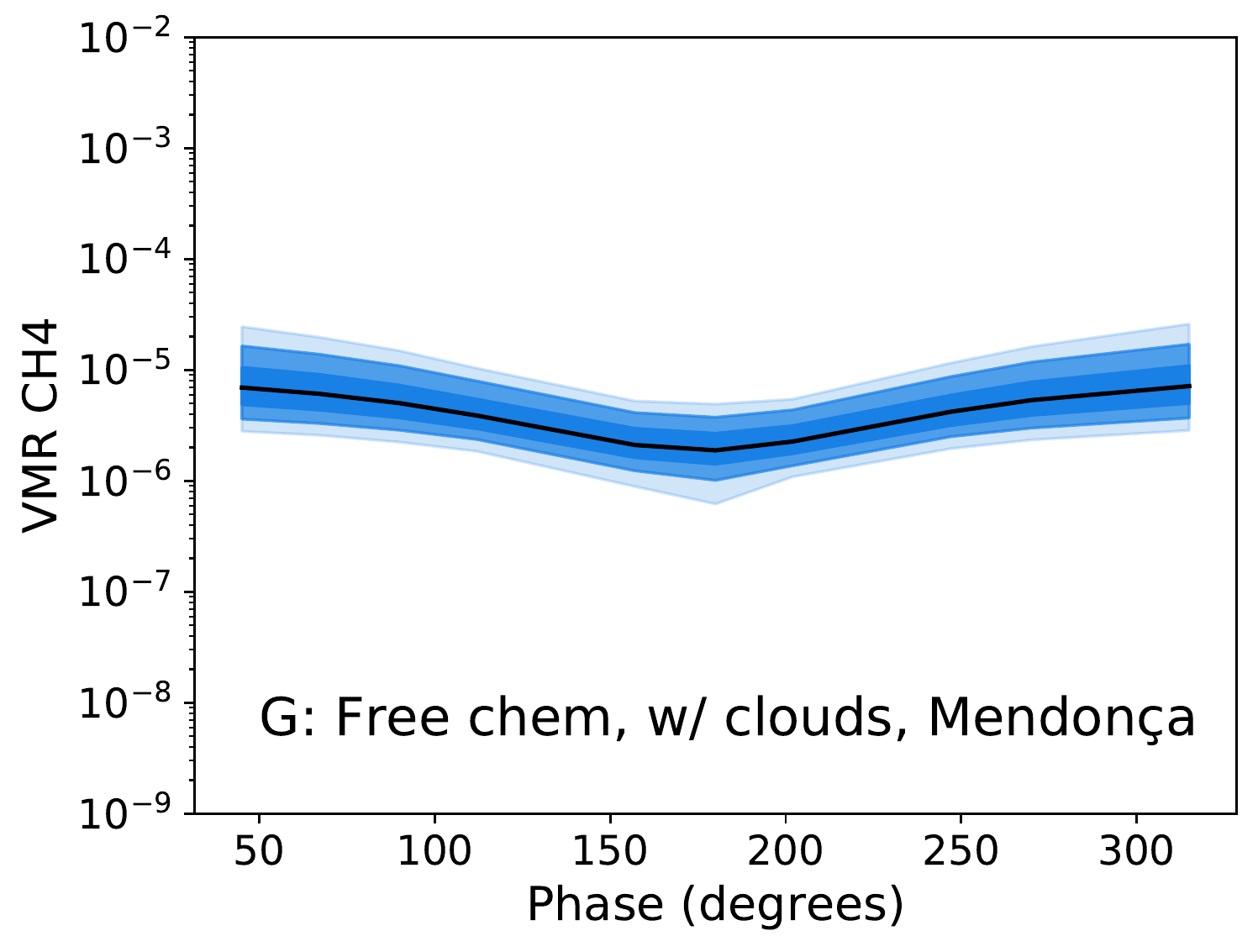}
                        \includegraphics[width=0.4\textwidth]{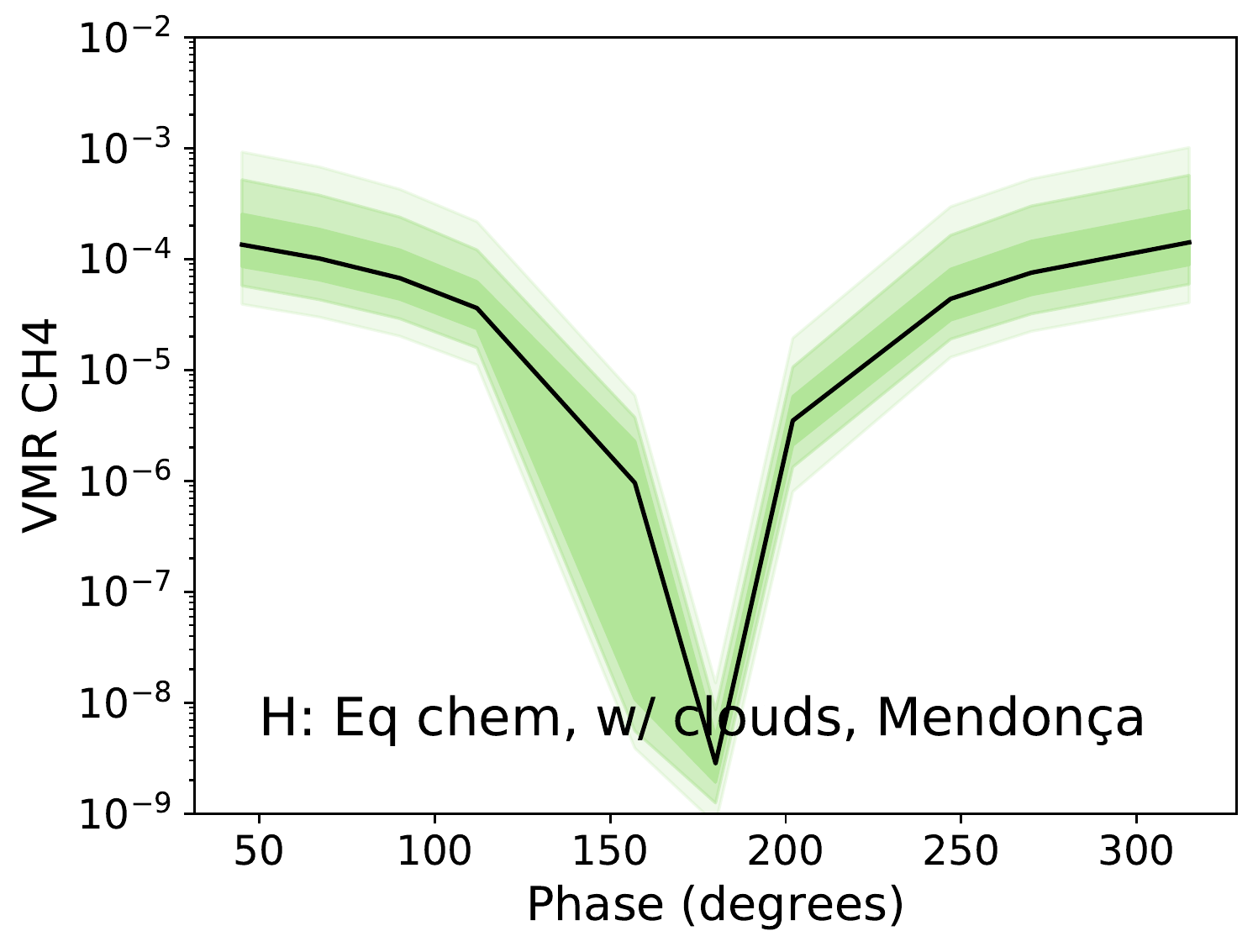}
                        \caption{Same as Fig~\ref{fig:AlO_COratio}, but for CH$_4$.}\label{fig:CH4_COratio}
                \end{figure}
                
                \begin{figure}
                        \centering
                        \includegraphics[width=0.4\textwidth]{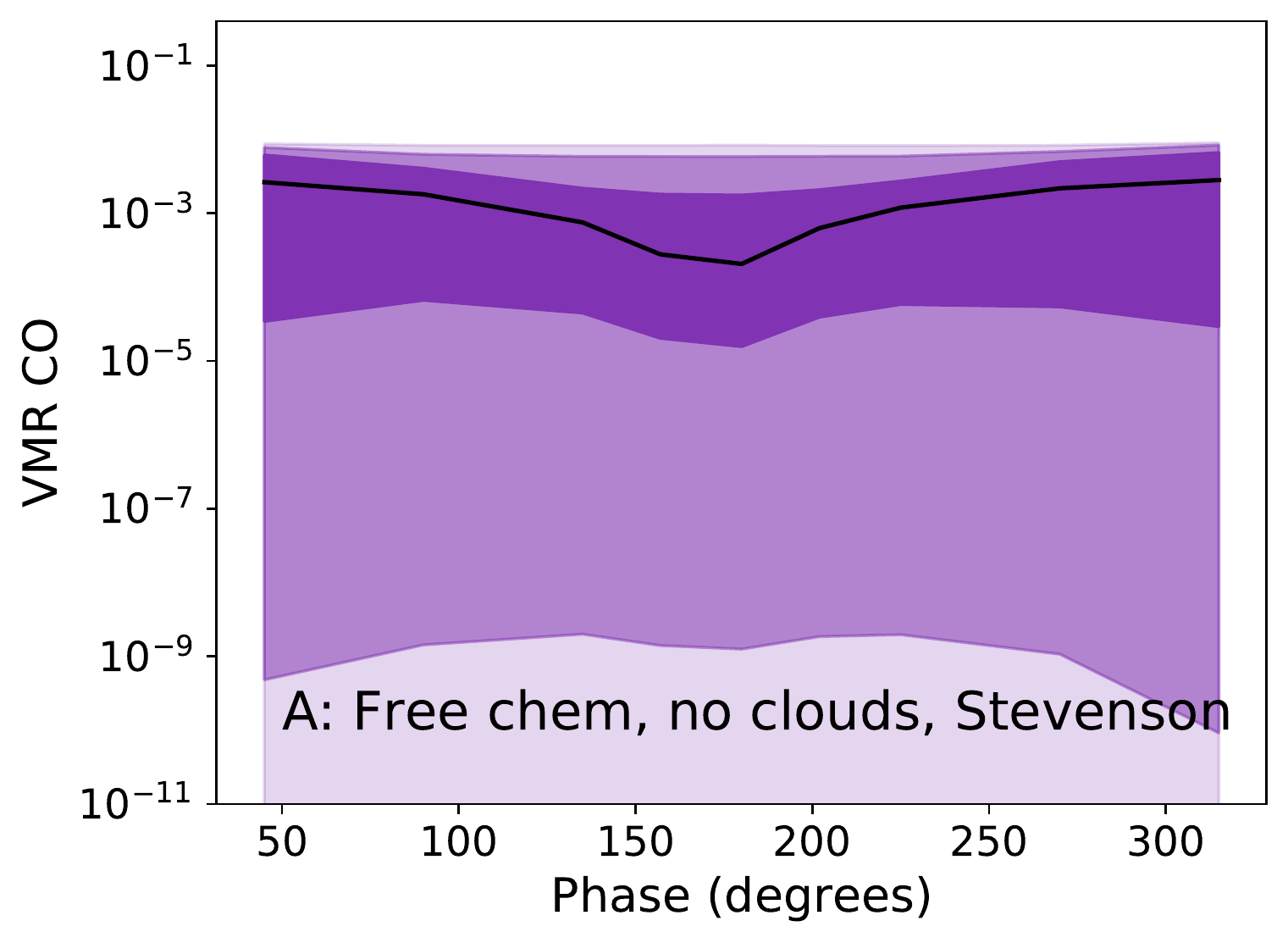}
                        \includegraphics[width=0.4\textwidth]{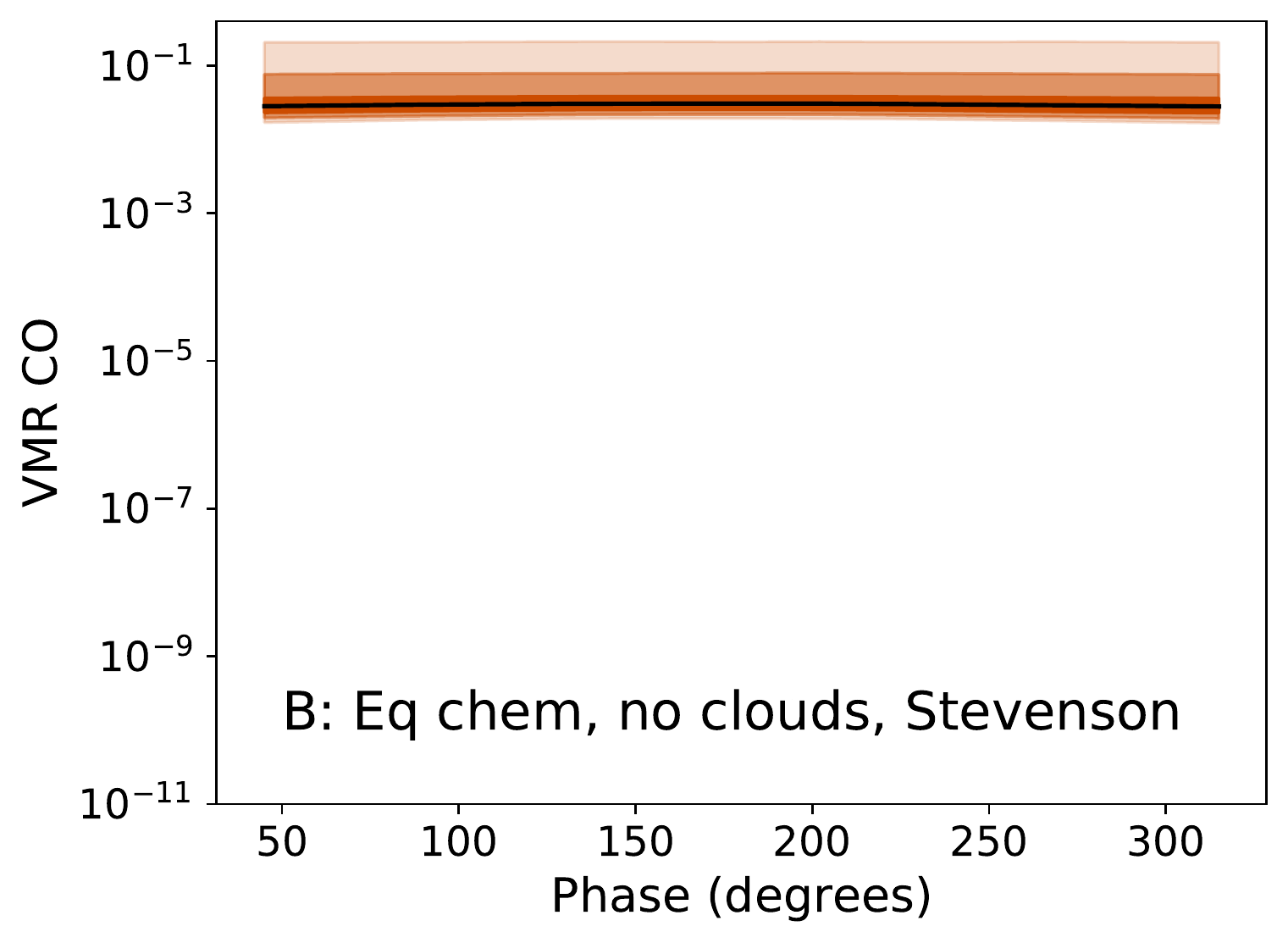}
                        \includegraphics[width=0.4\textwidth]{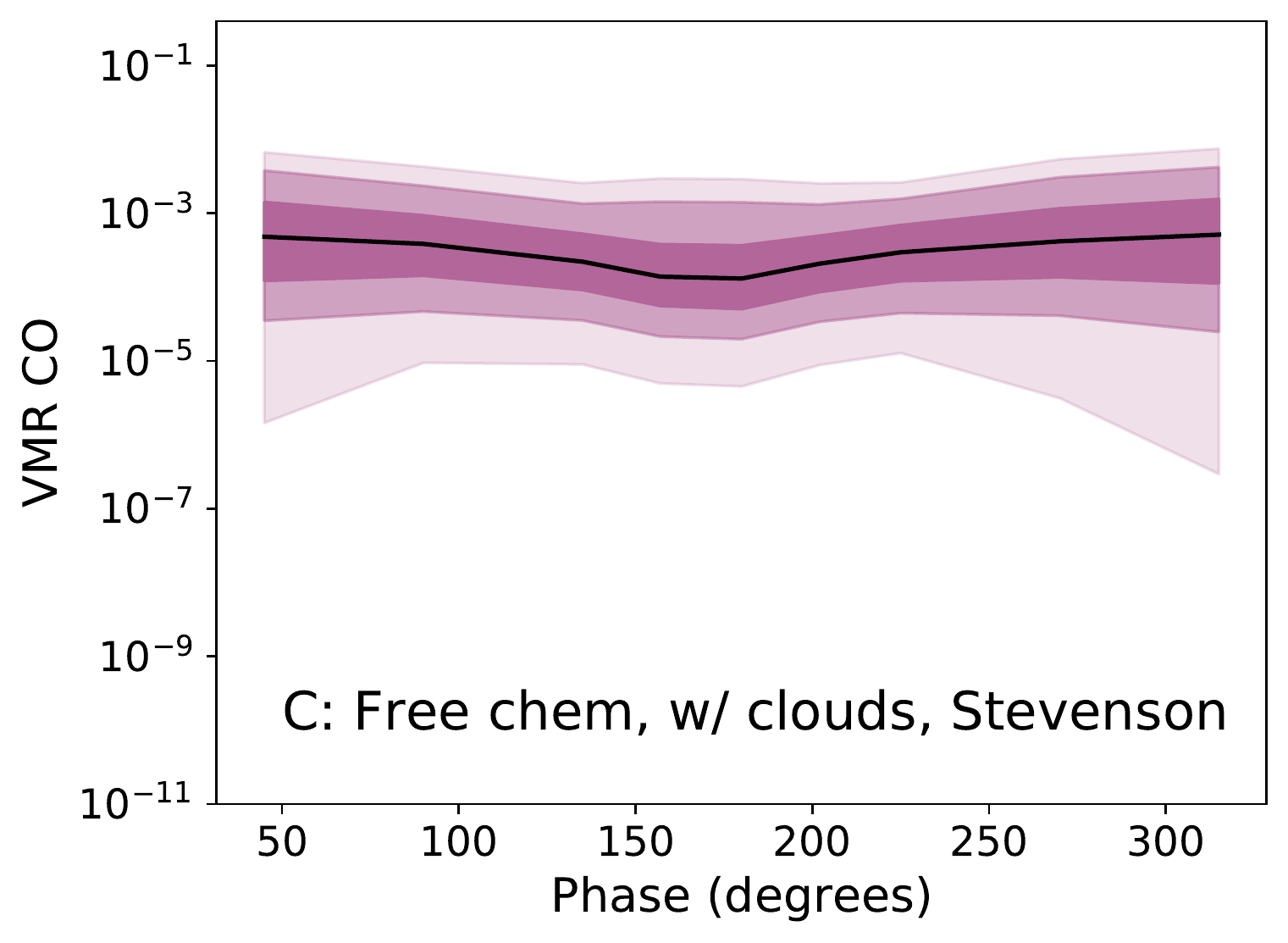}
                        \includegraphics[width=0.4\textwidth]{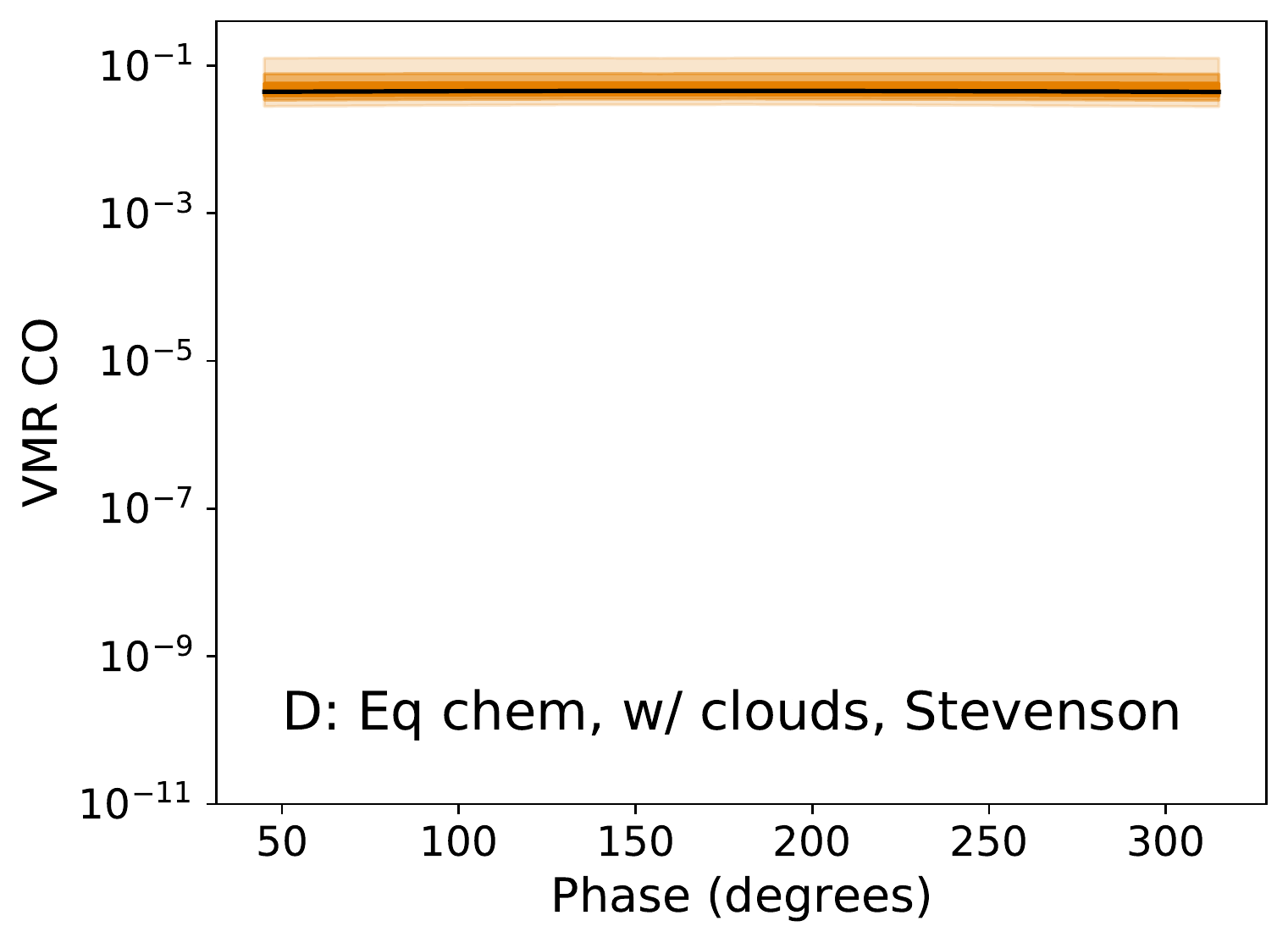}
                        \includegraphics[width=0.4\textwidth]{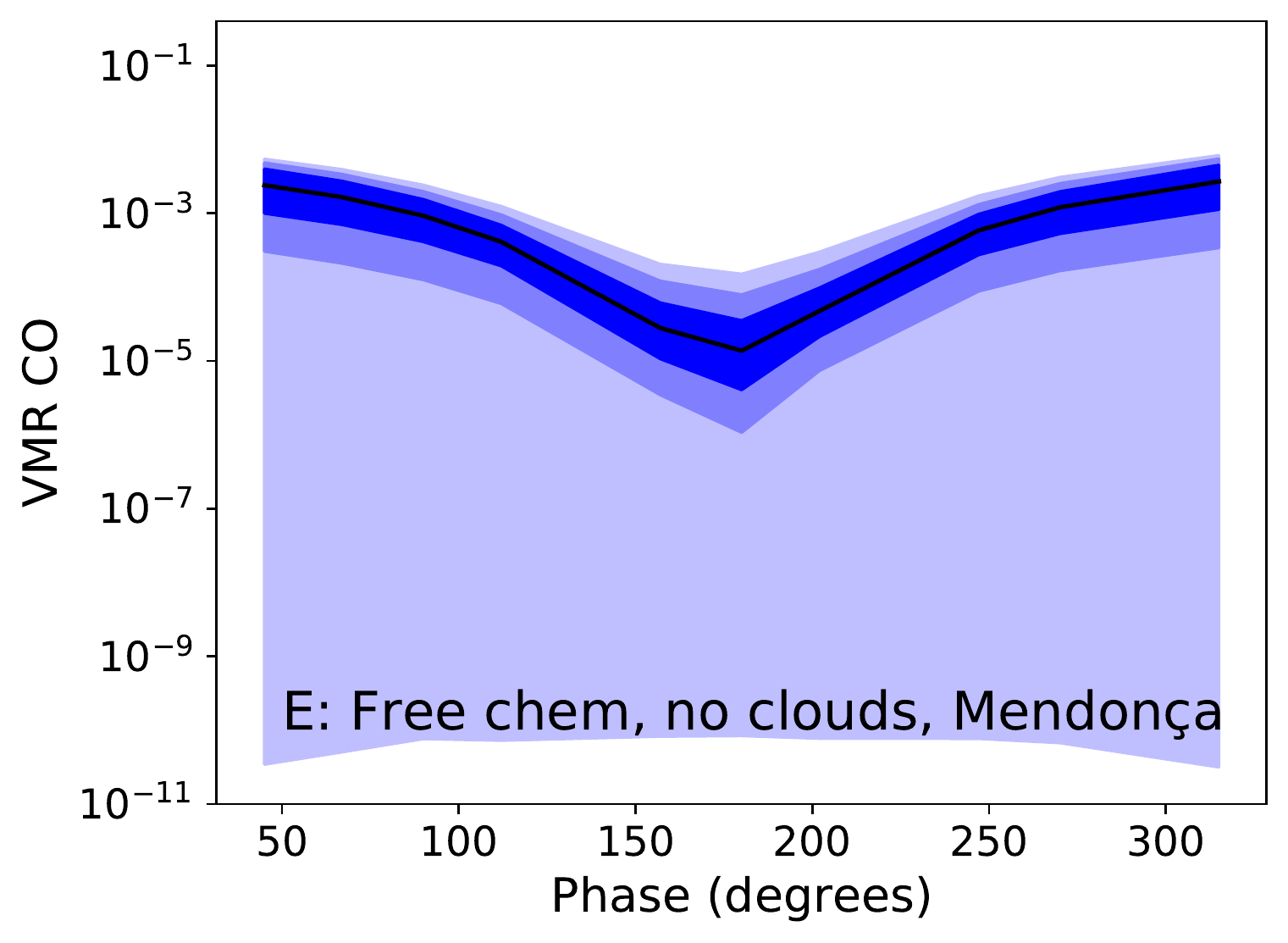}
                        \includegraphics[width=0.4\textwidth]{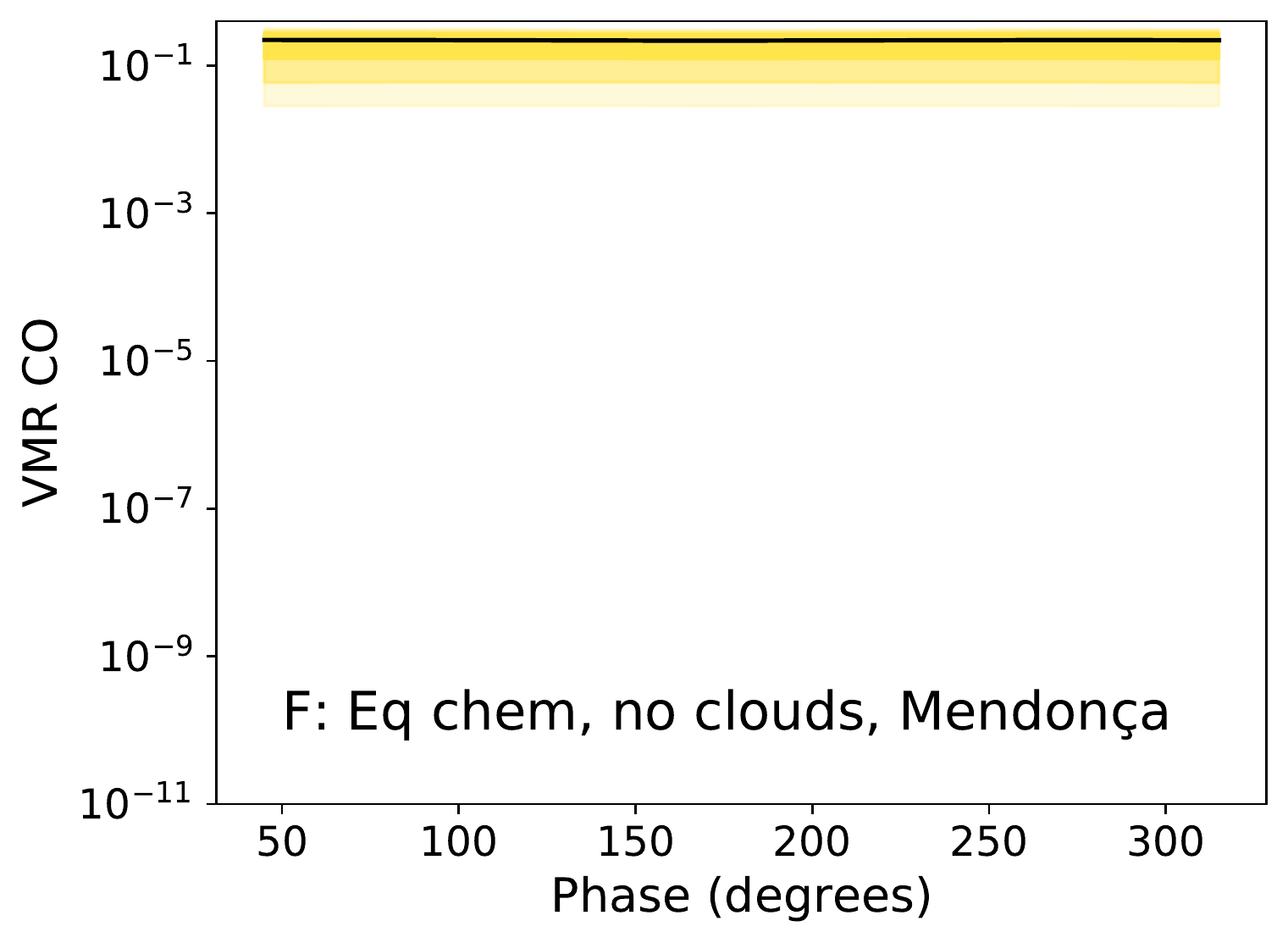}
                        \includegraphics[width=0.4\textwidth]{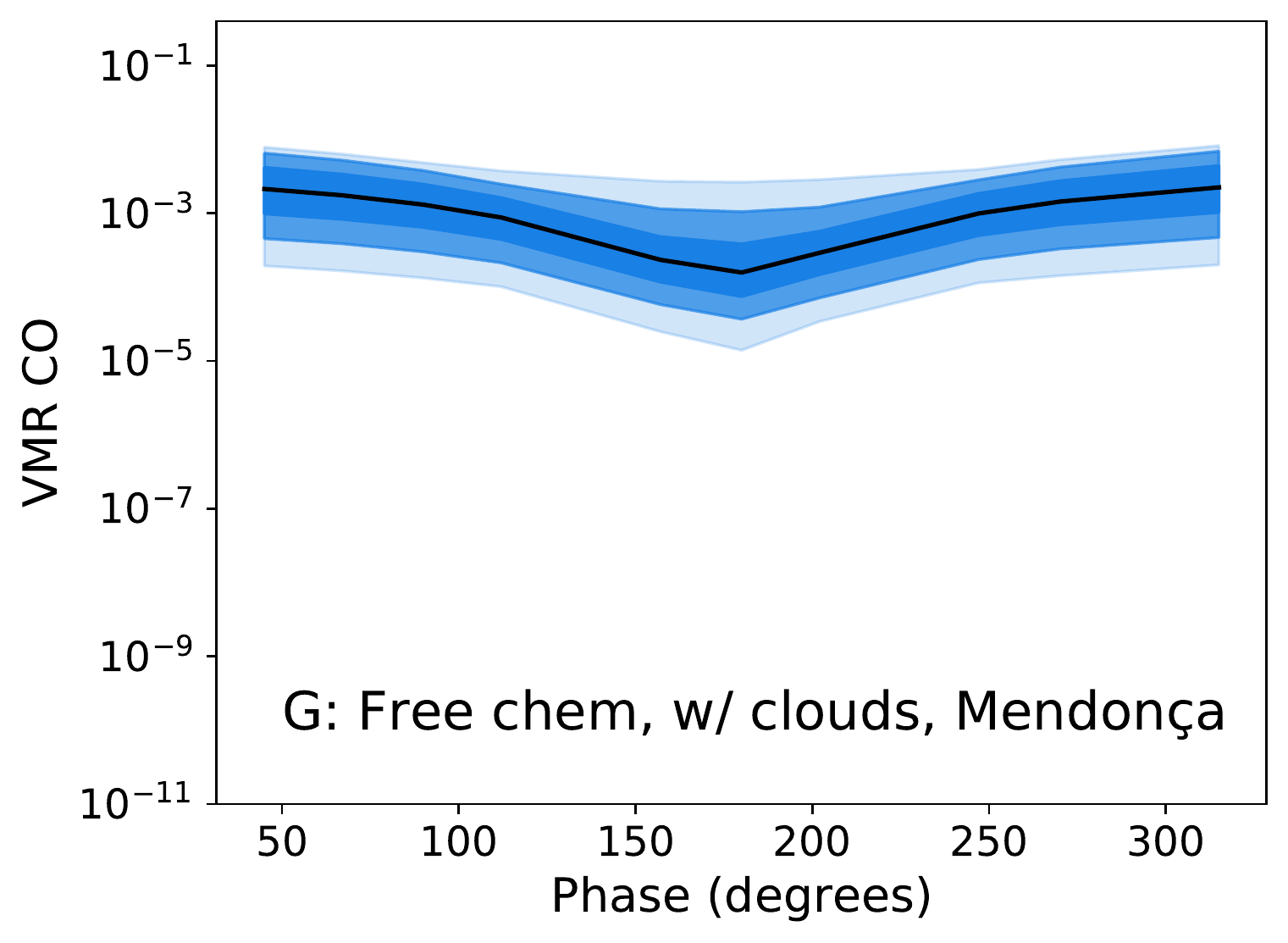}
                        \includegraphics[width=0.4\textwidth]{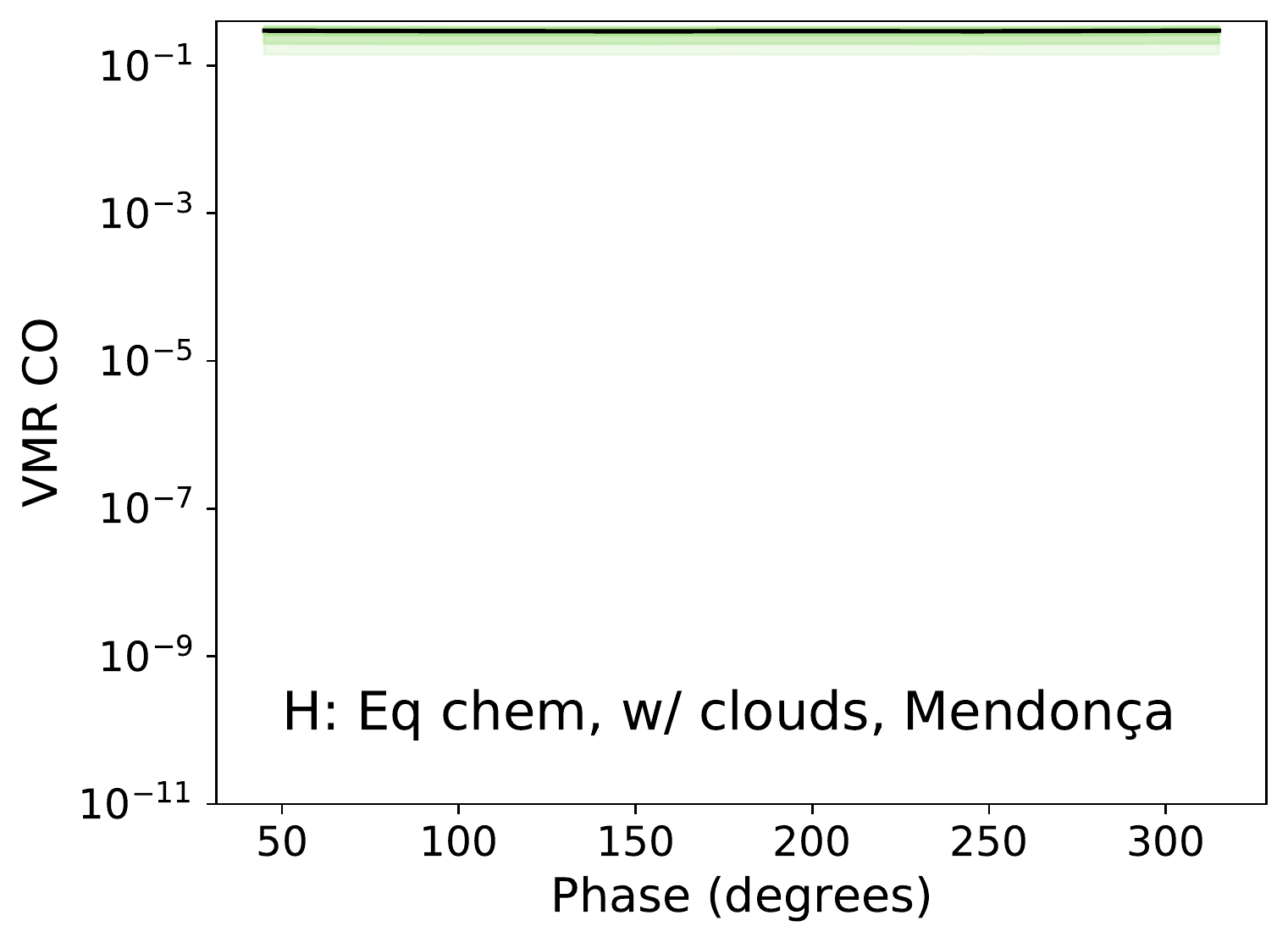}
                        \caption{Same as Fig~\ref{fig:AlO_COratio}, but for CO}\label{fig:CO_COratio}
                \end{figure}
                
                \begin{figure}
                        \centering
                        \includegraphics[width=0.4\textwidth]{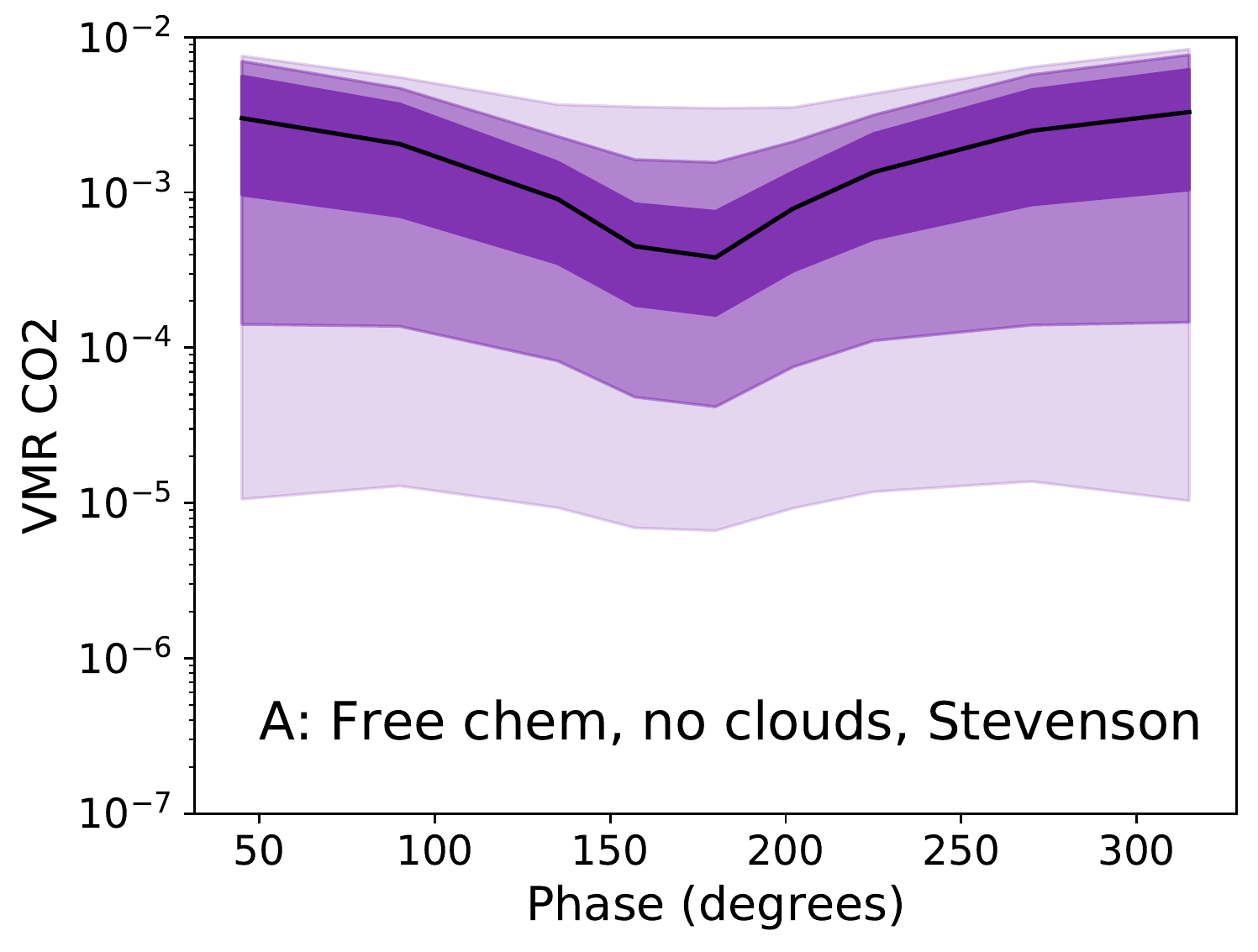}
                        \includegraphics[width=0.4\textwidth]{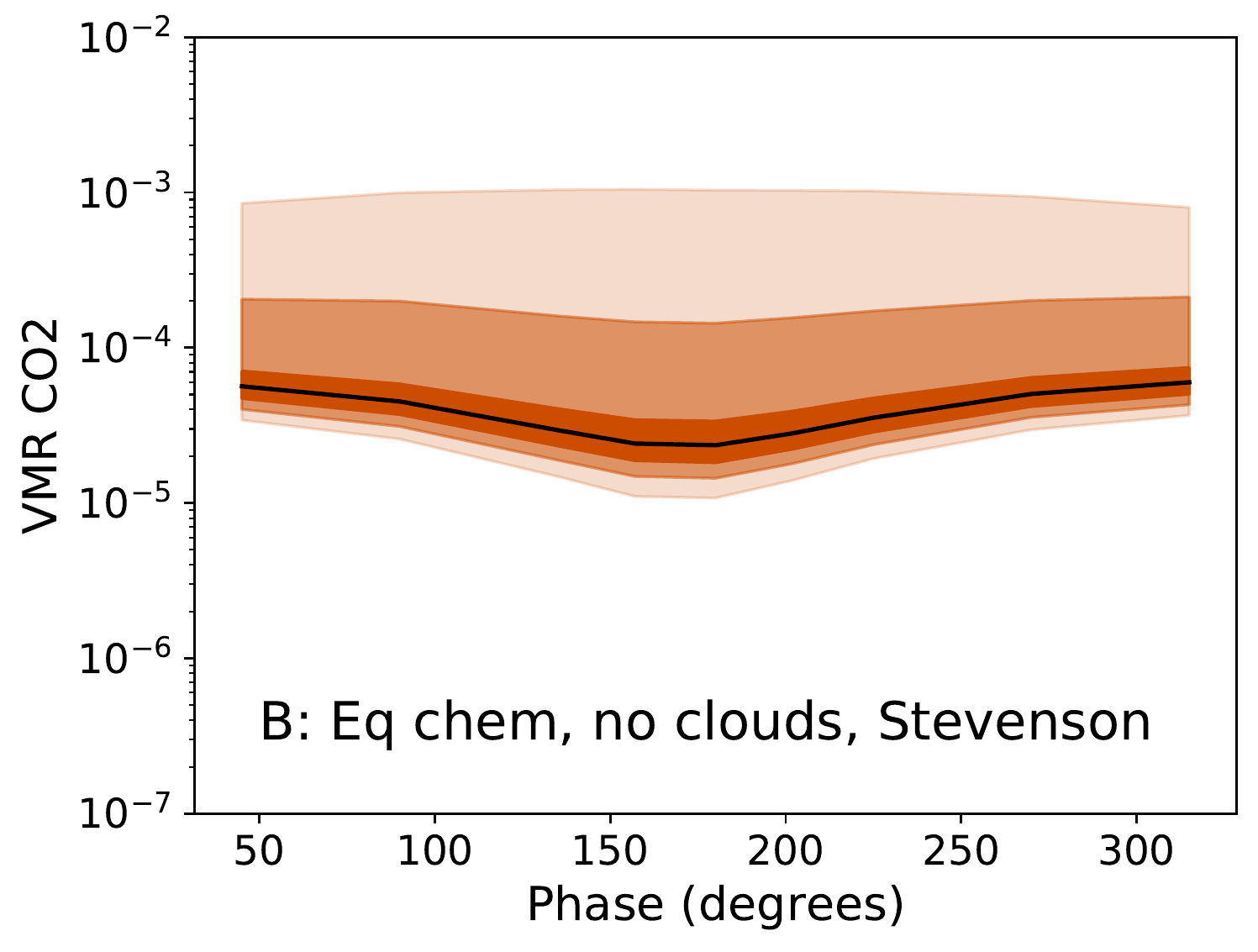}
                        \includegraphics[width=0.4\textwidth]{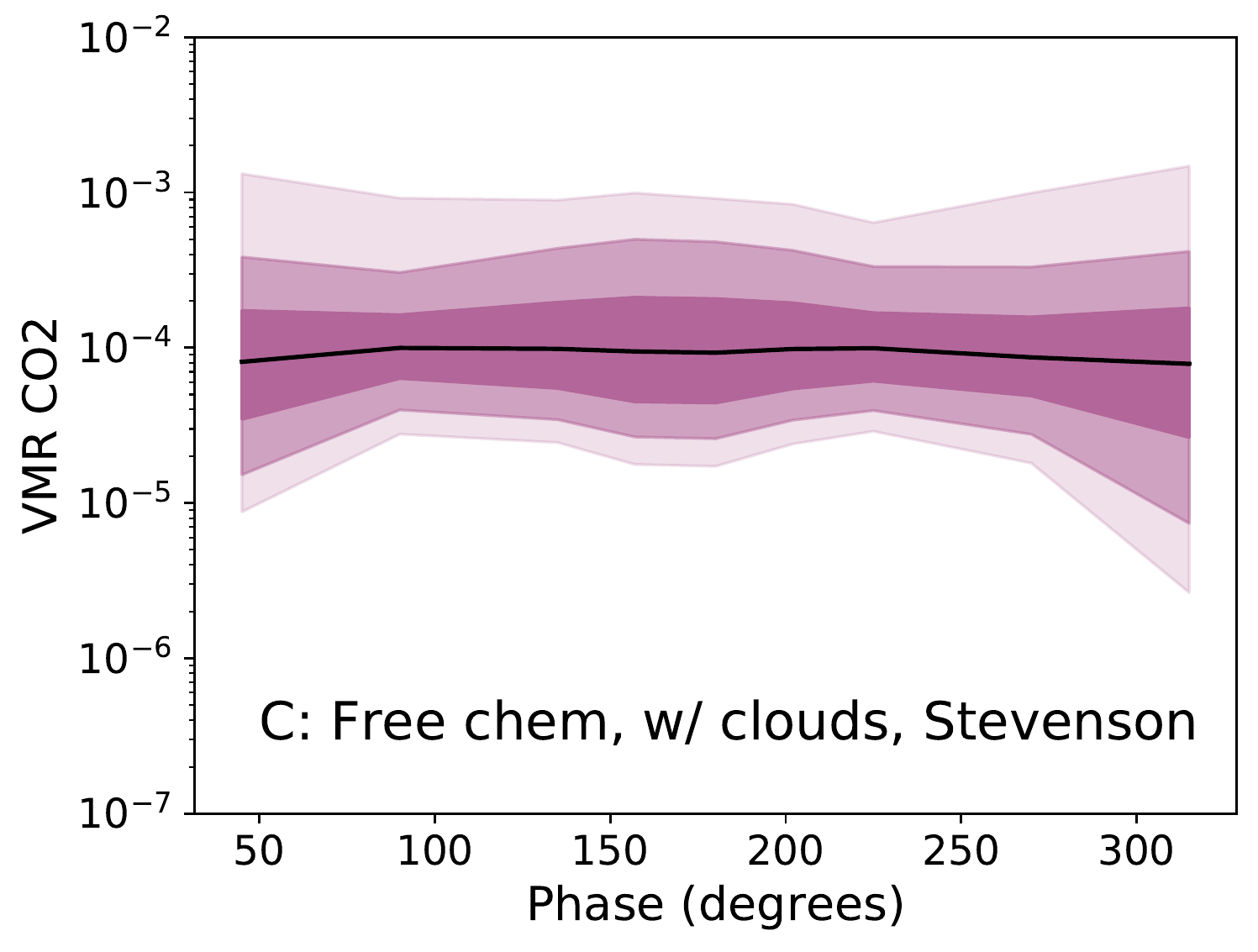}
                        \includegraphics[width=0.4\textwidth]{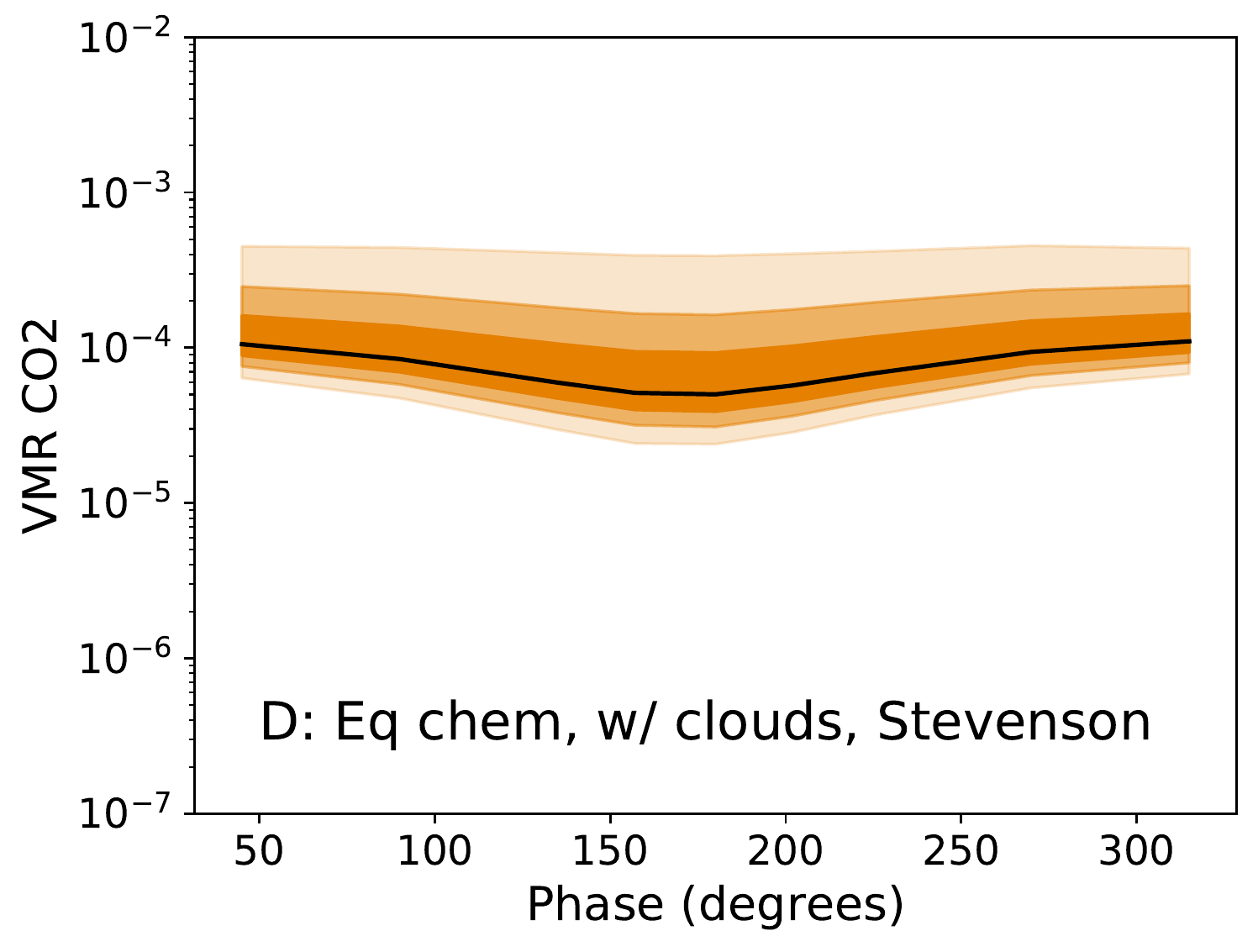}
                        \includegraphics[width=0.4\textwidth]{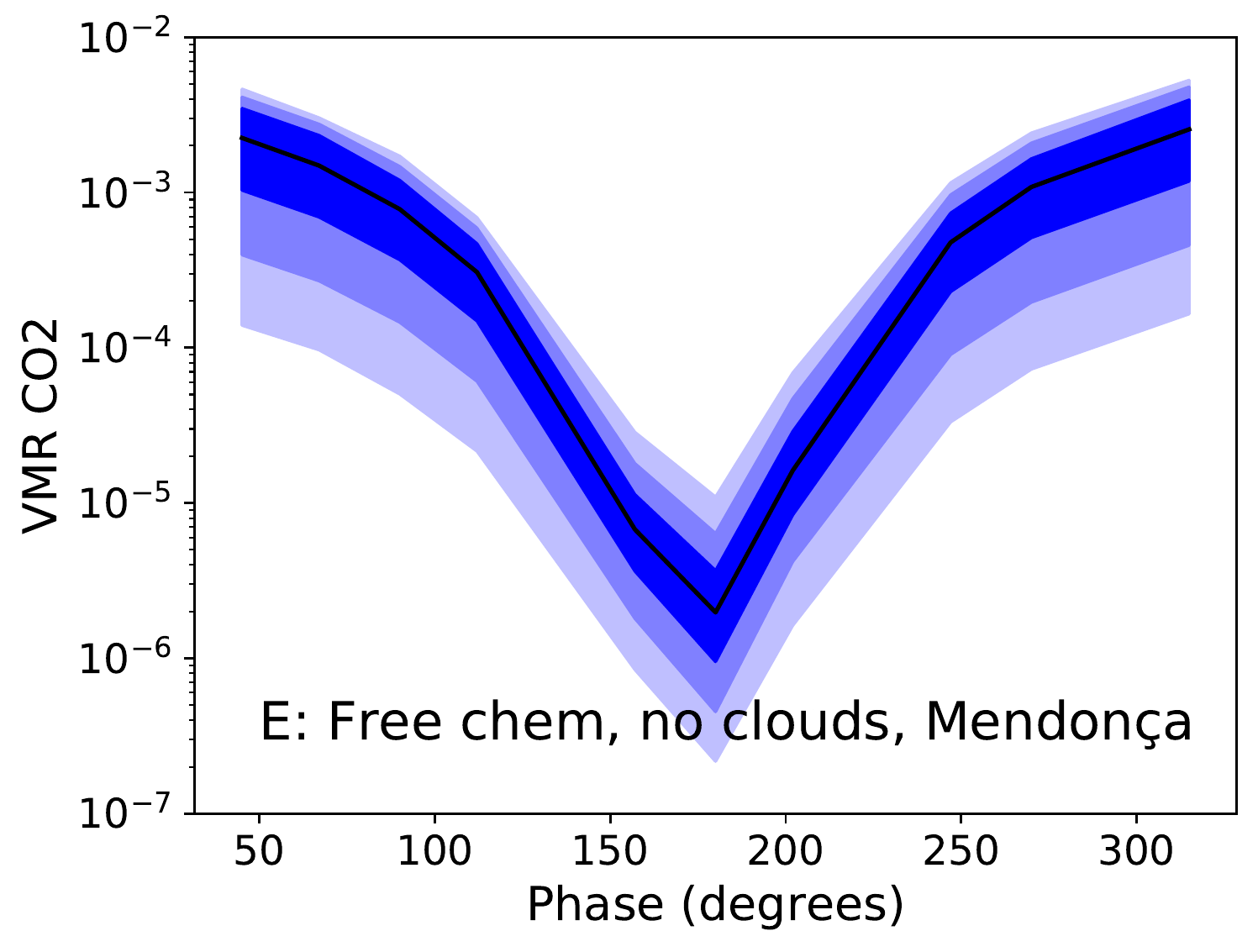}
                        \includegraphics[width=0.4\textwidth]{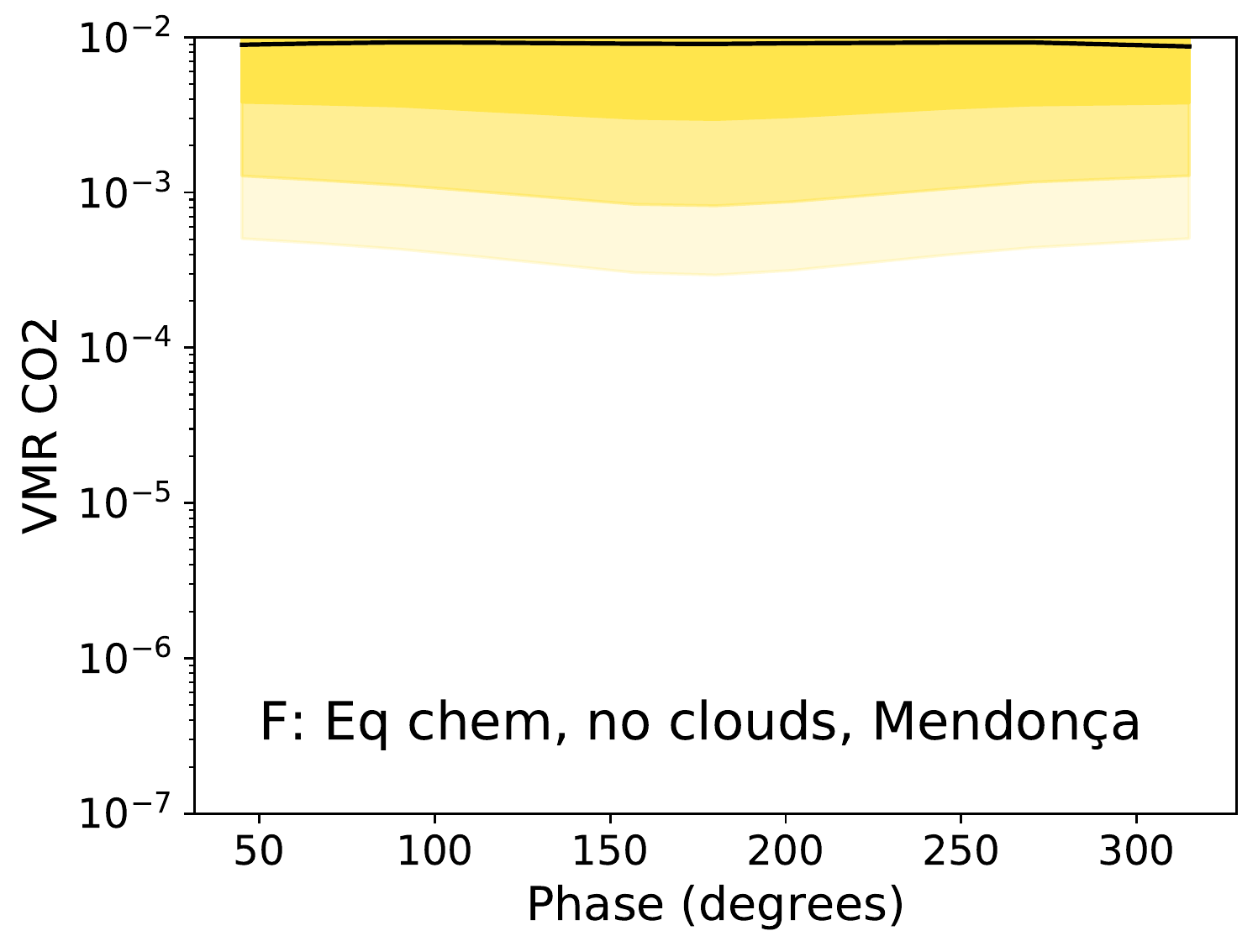}
                        \includegraphics[width=0.4\textwidth]{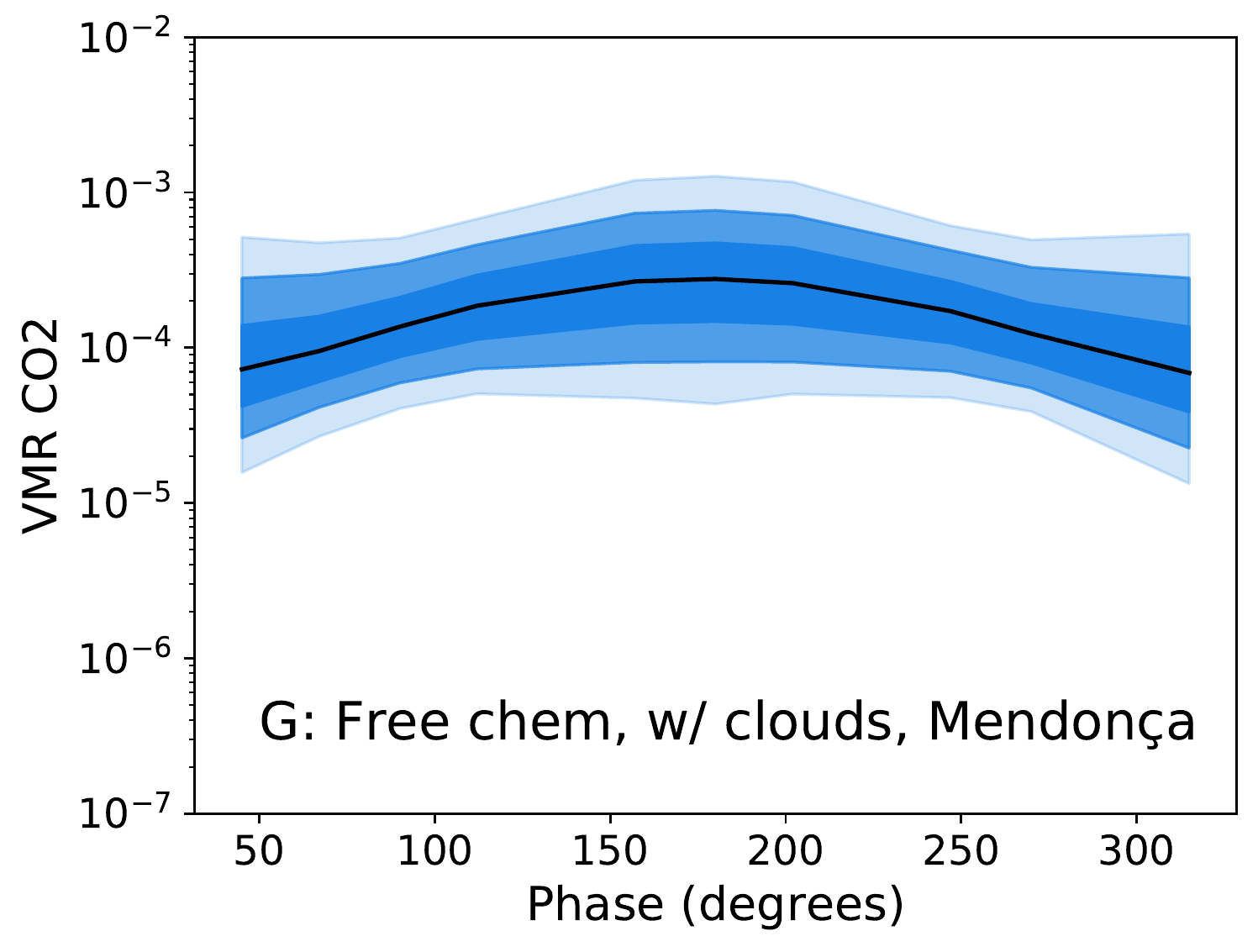}
                        \includegraphics[width=0.4\textwidth]{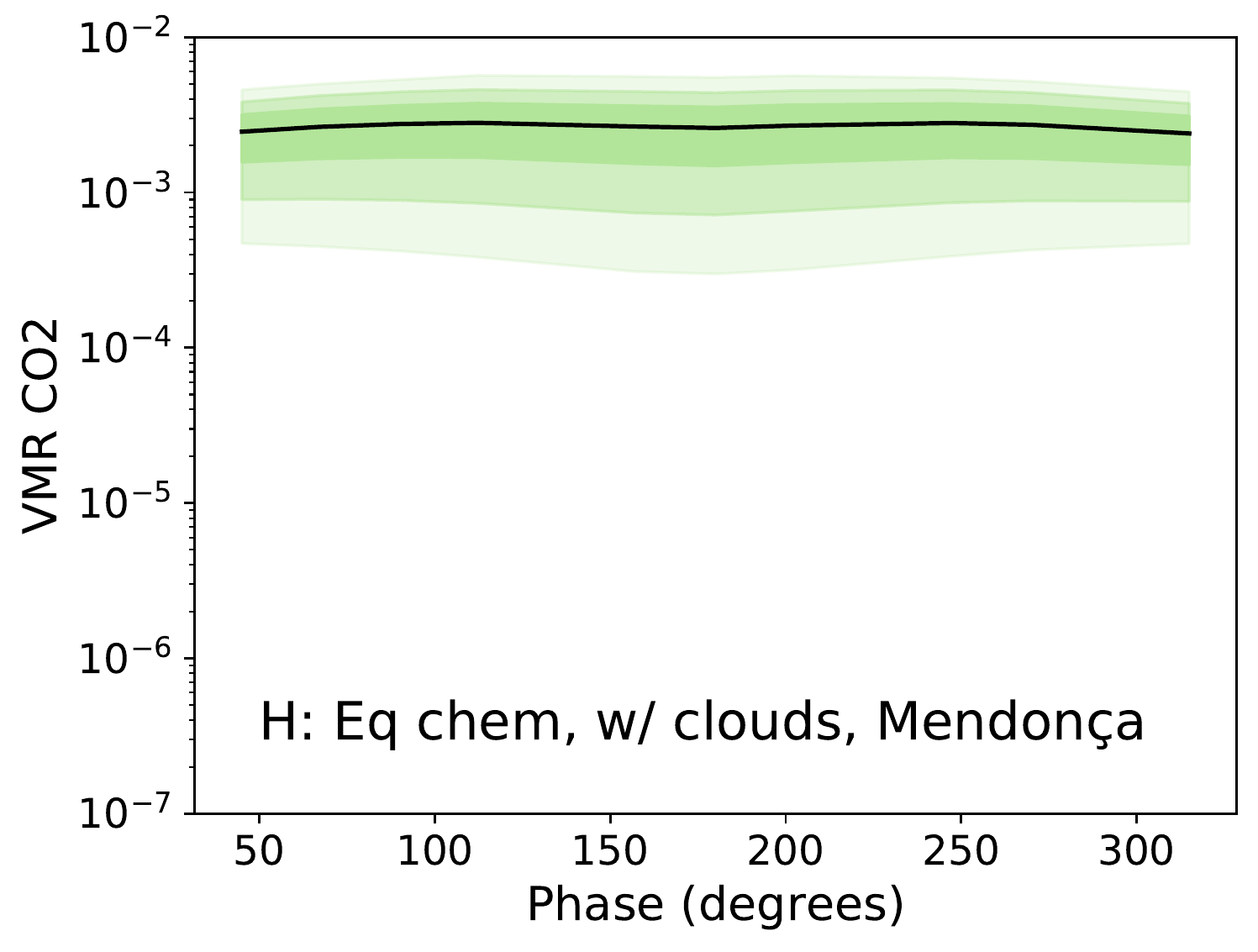}
                        \caption{Same as Fig~\ref{fig:AlO_COratio}, but for CO$_2$}\label{fig:CO2_COratio}
                \end{figure}
                
                \begin{figure}
                        \centering
                        \includegraphics[width=0.4\textwidth]{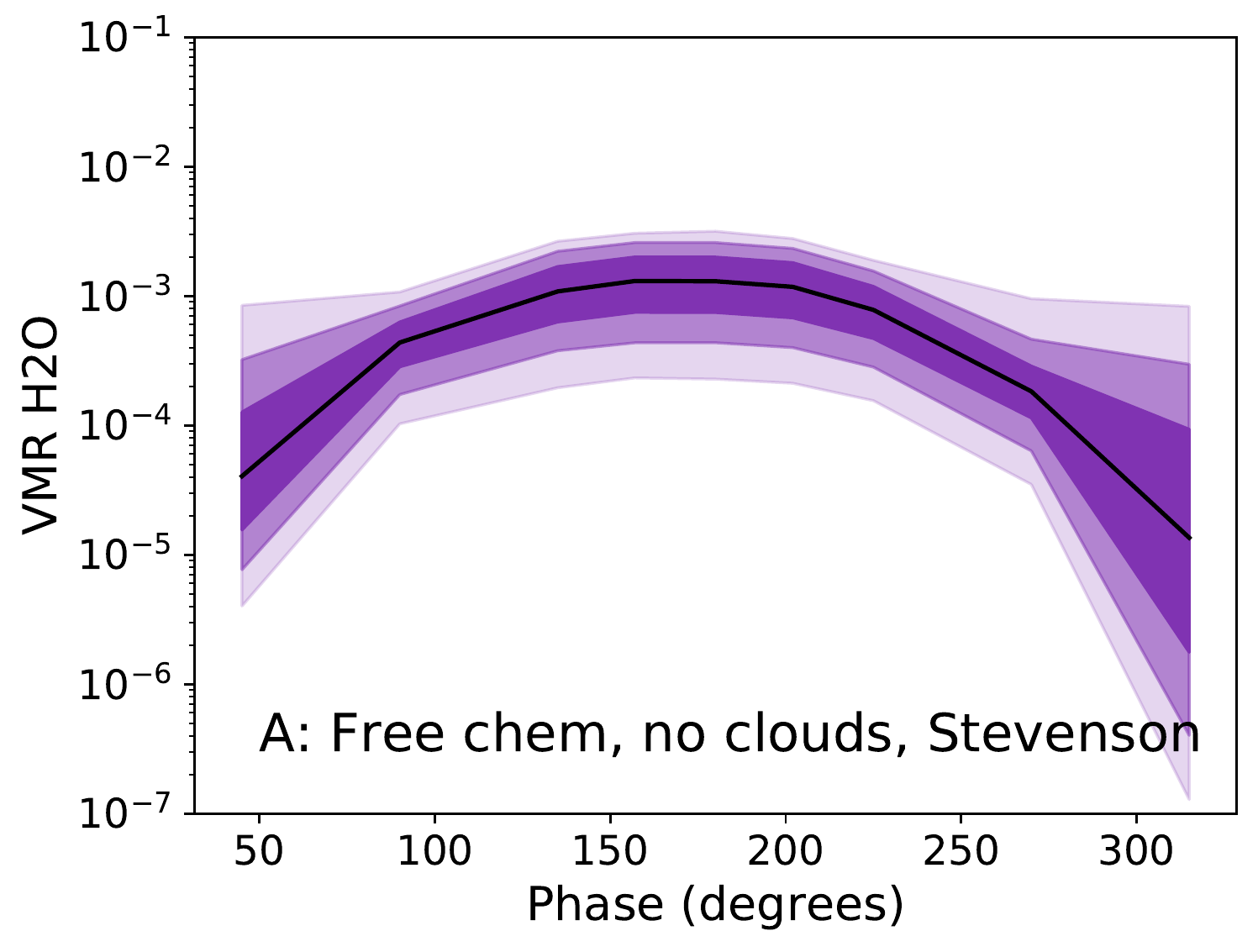}
                        \includegraphics[width=0.4\textwidth]{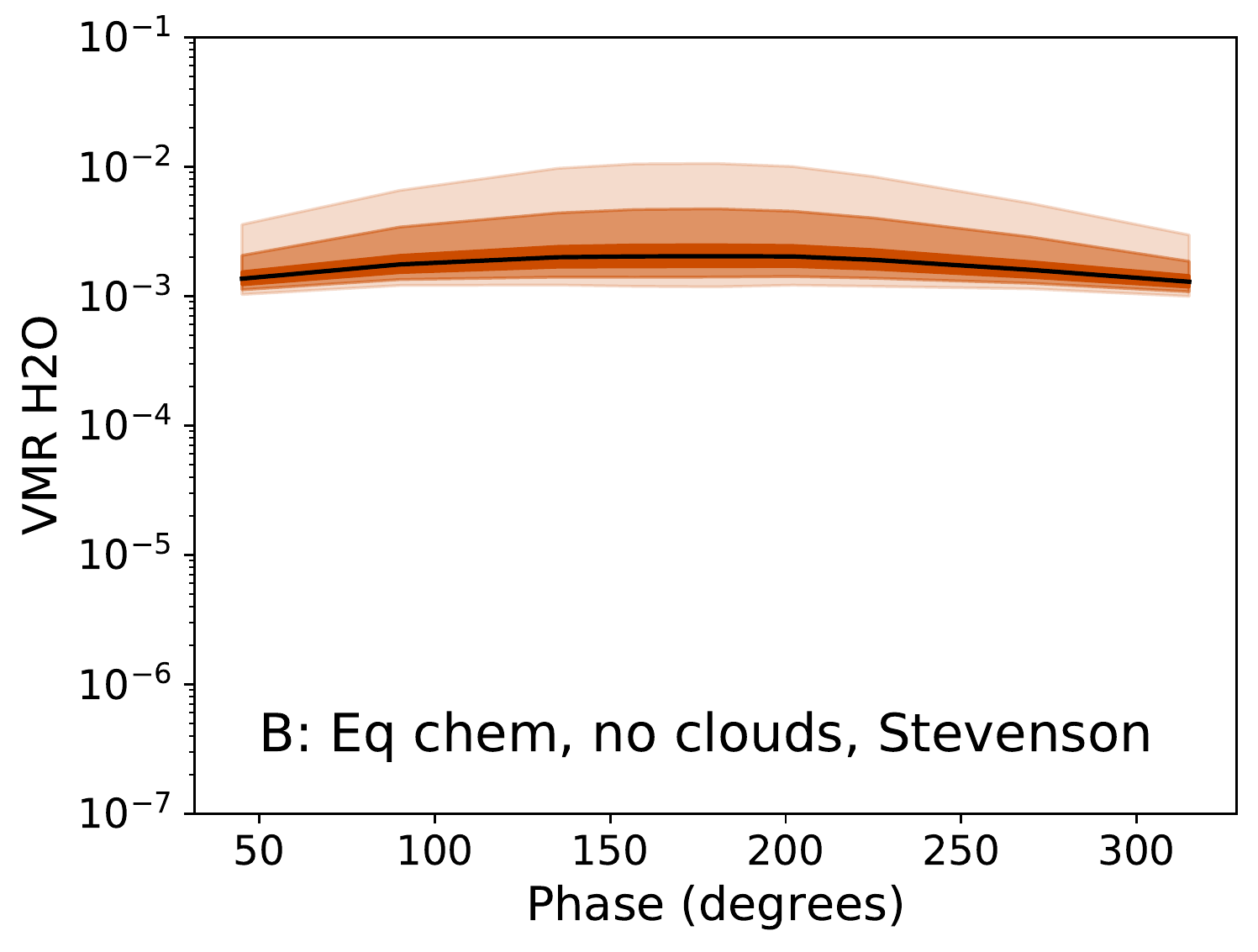}
                        \includegraphics[width=0.4\textwidth]{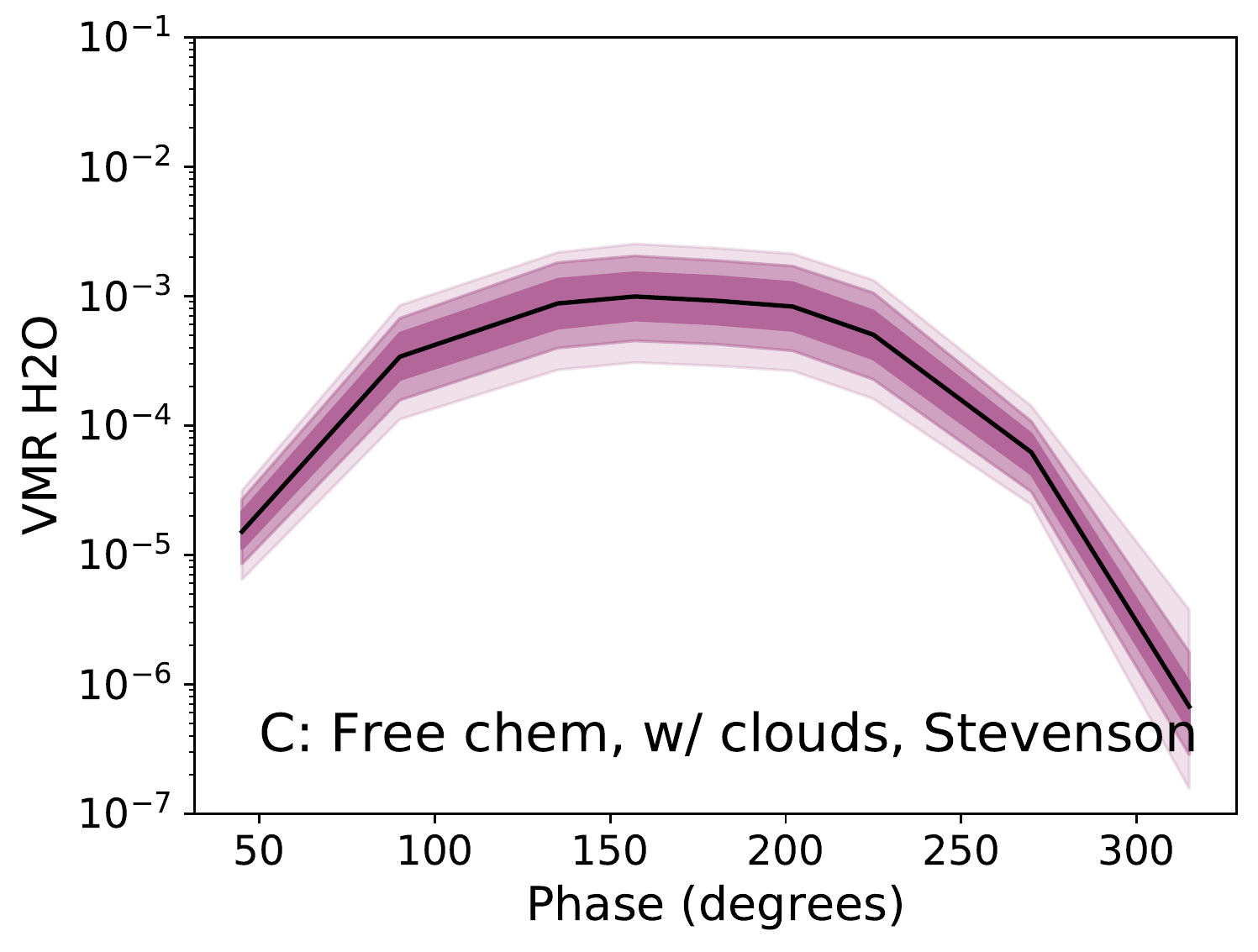}
                        \includegraphics[width=0.4\textwidth]{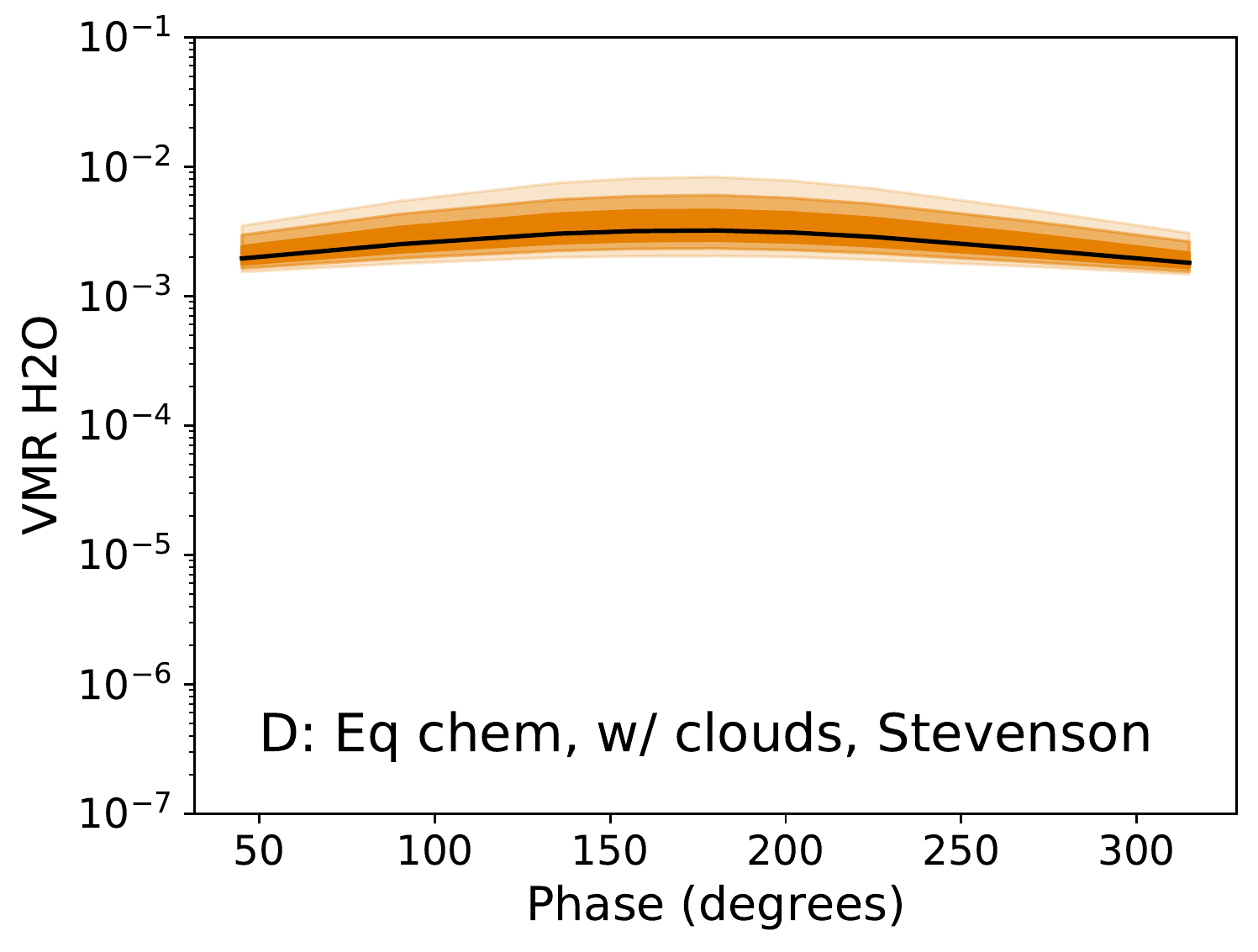}
                        \includegraphics[width=0.4\textwidth]{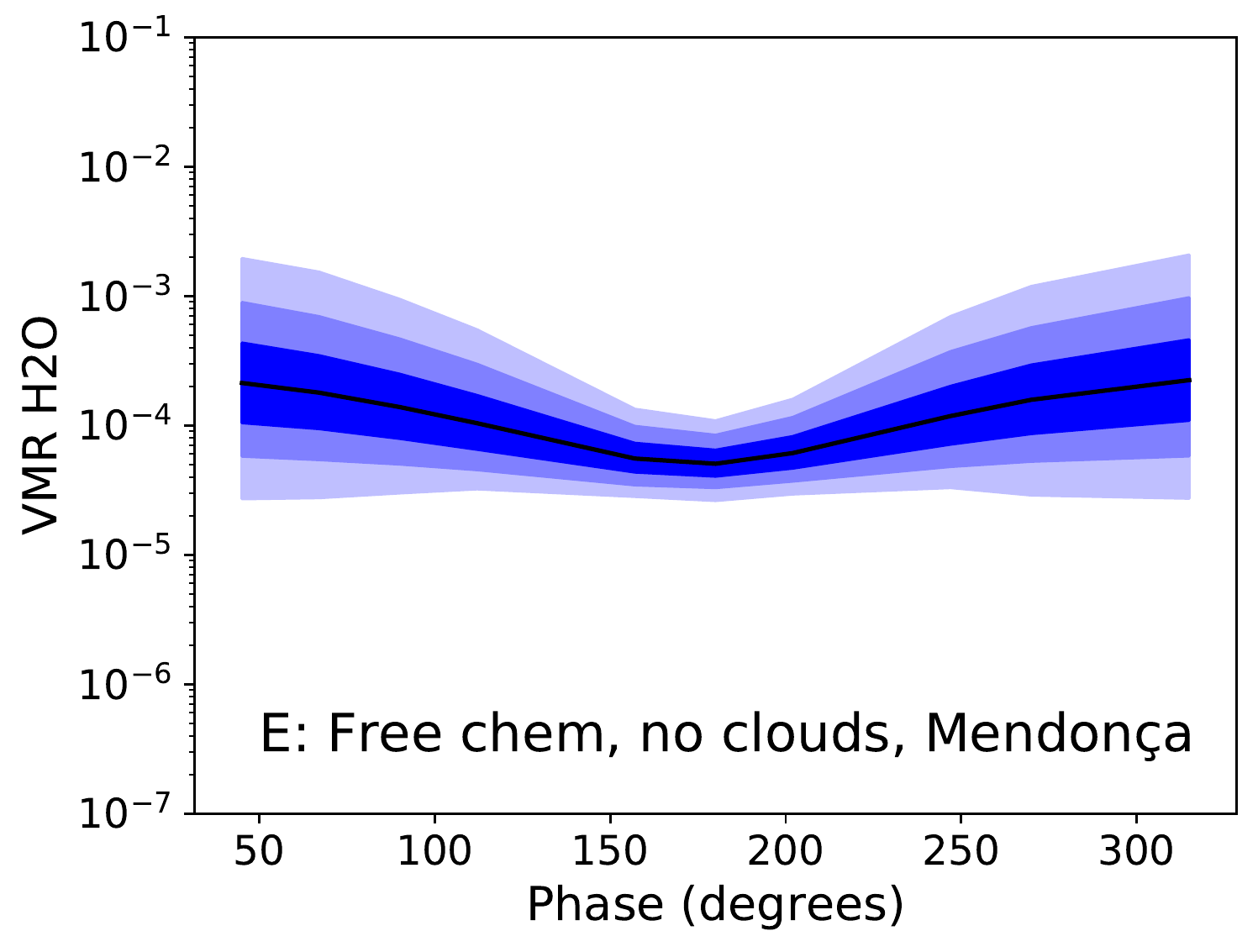}
                        \includegraphics[width=0.4\textwidth]{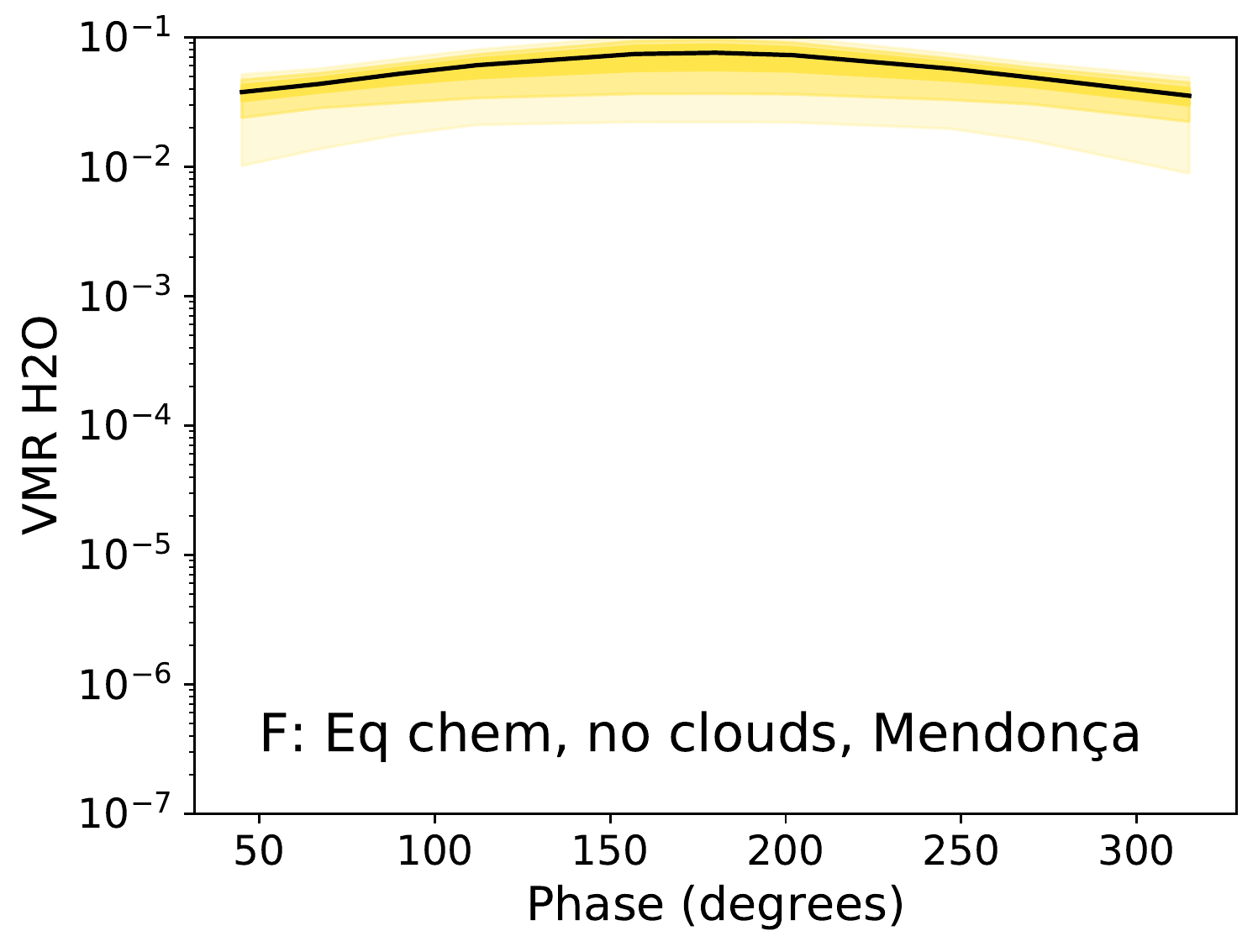}
                        \includegraphics[width=0.4\textwidth]{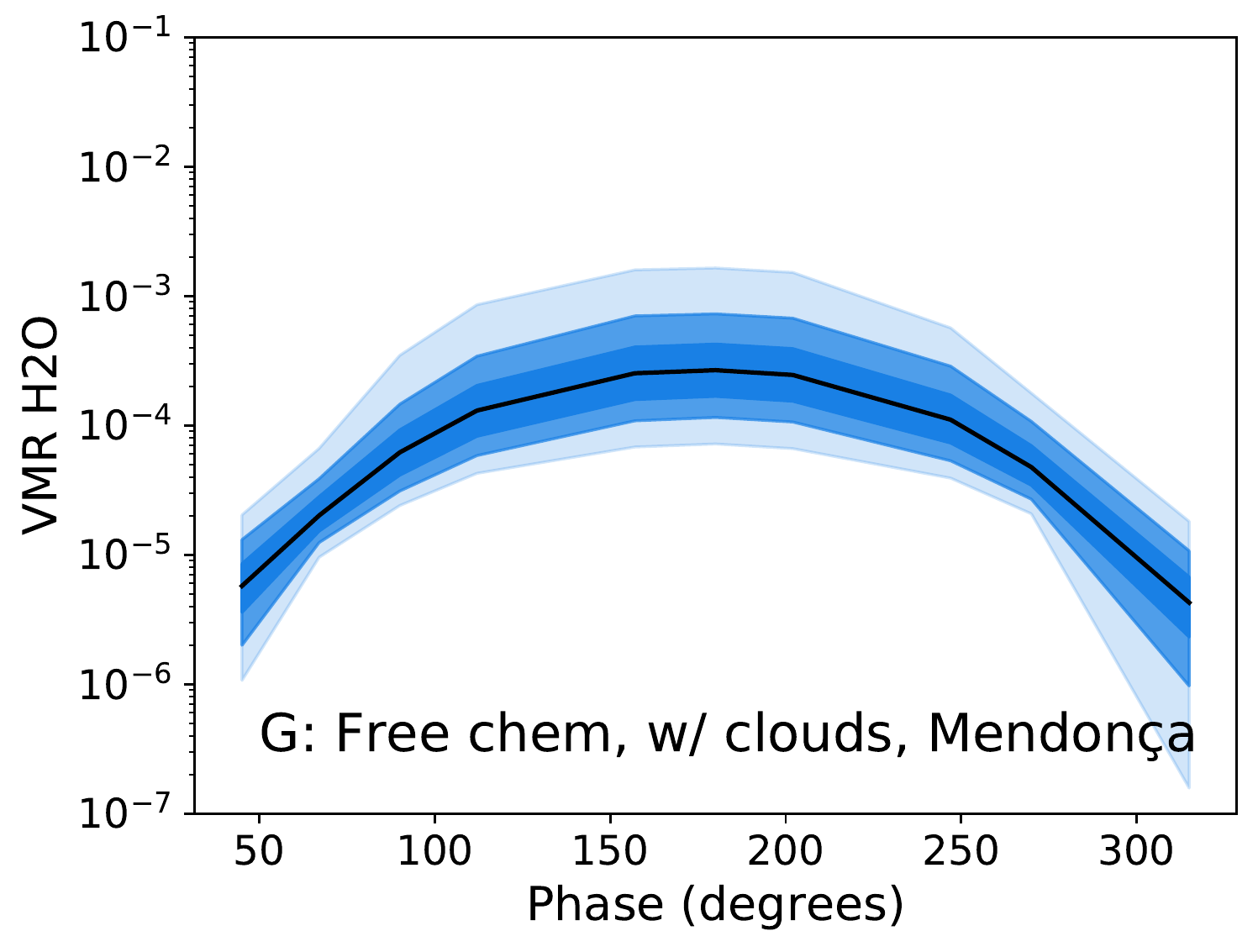}
                        \includegraphics[width=0.4\textwidth]{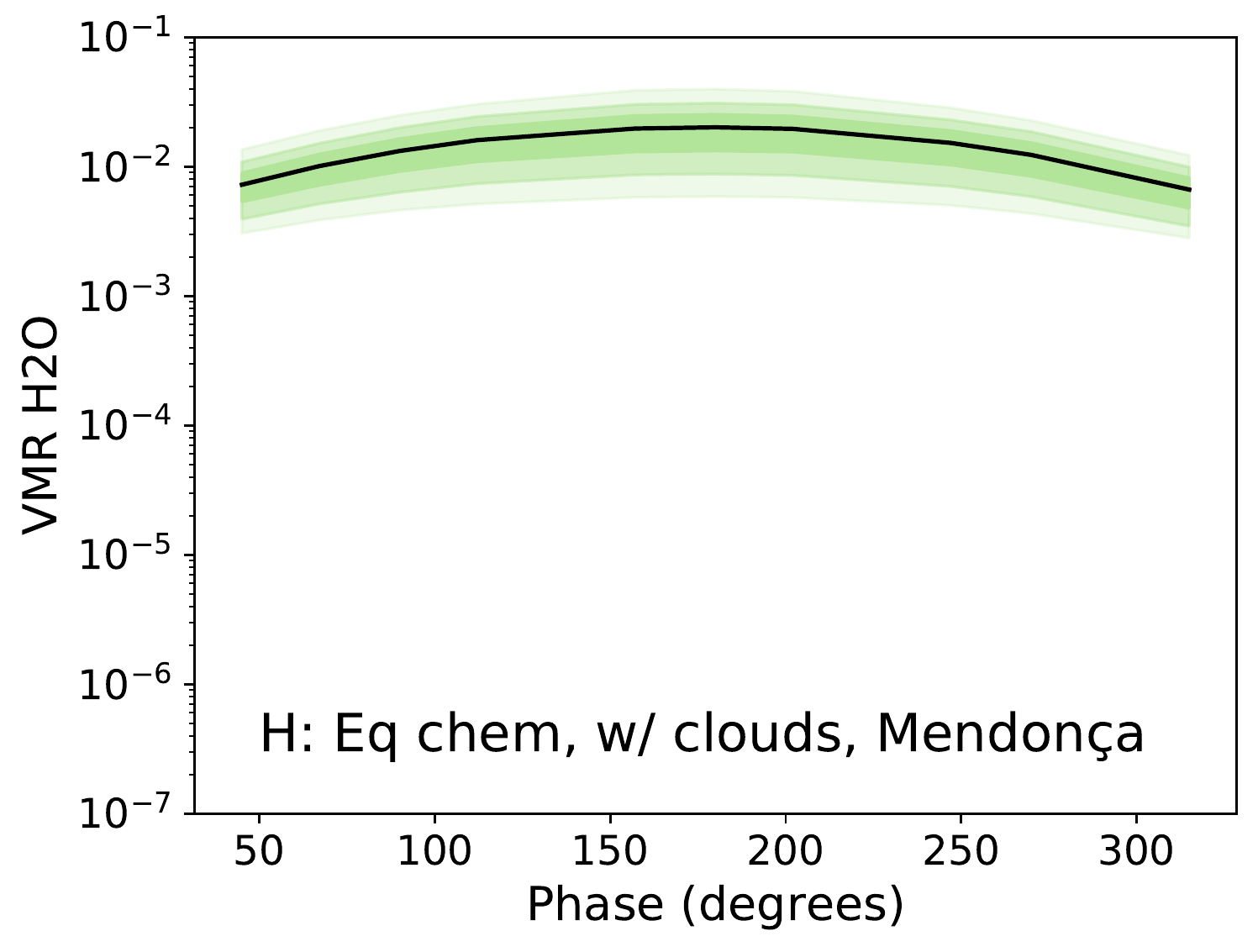}
                        \caption{Same as Fig~\ref{fig:AlO_COratio}, but for H$_2$O}\label{fig:H2O_COratio}
                \end{figure}

                \begin{figure}
                        \centering
                        \includegraphics[width=0.4\textwidth]{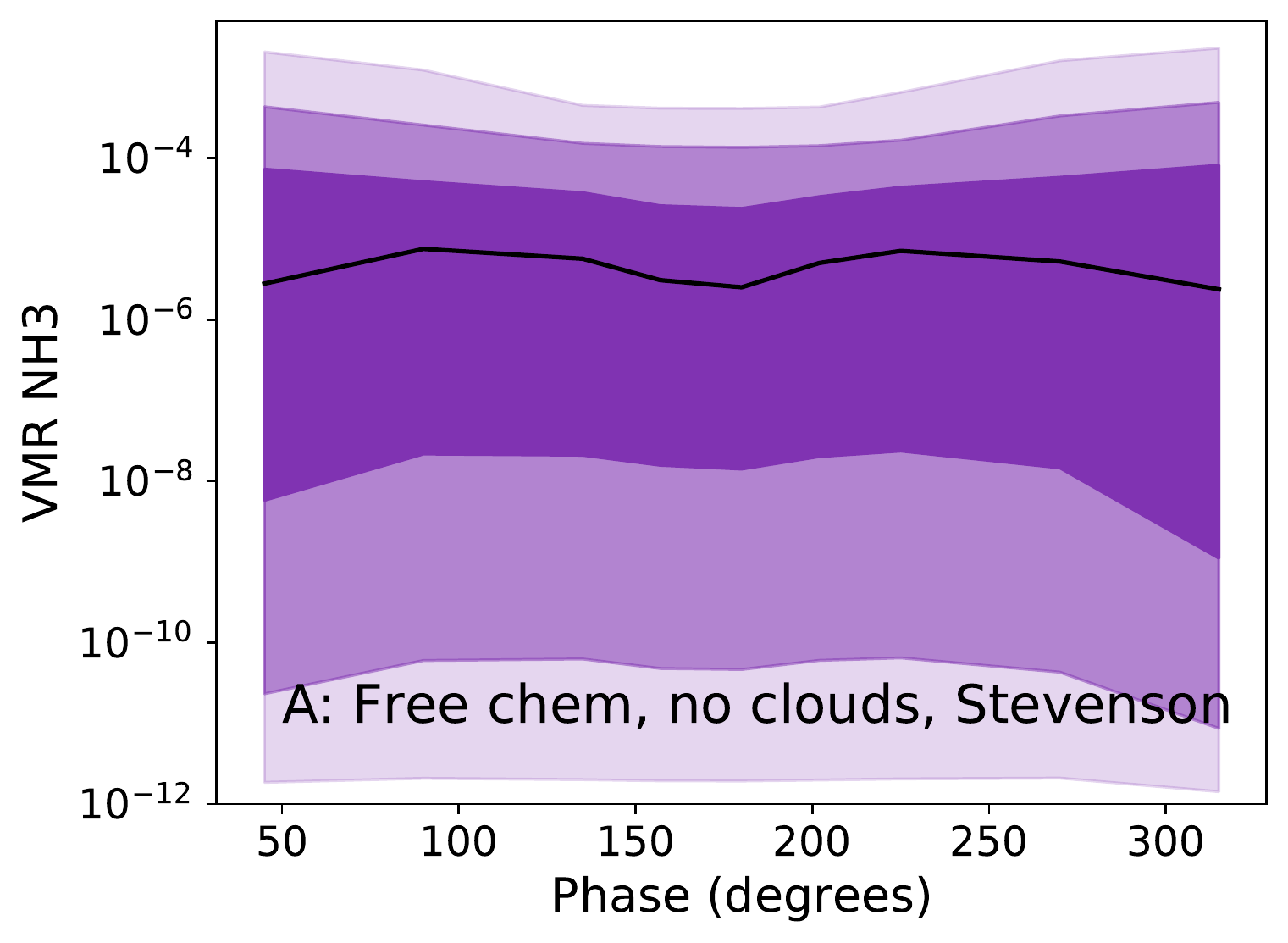}
                        \includegraphics[width=0.4\textwidth]{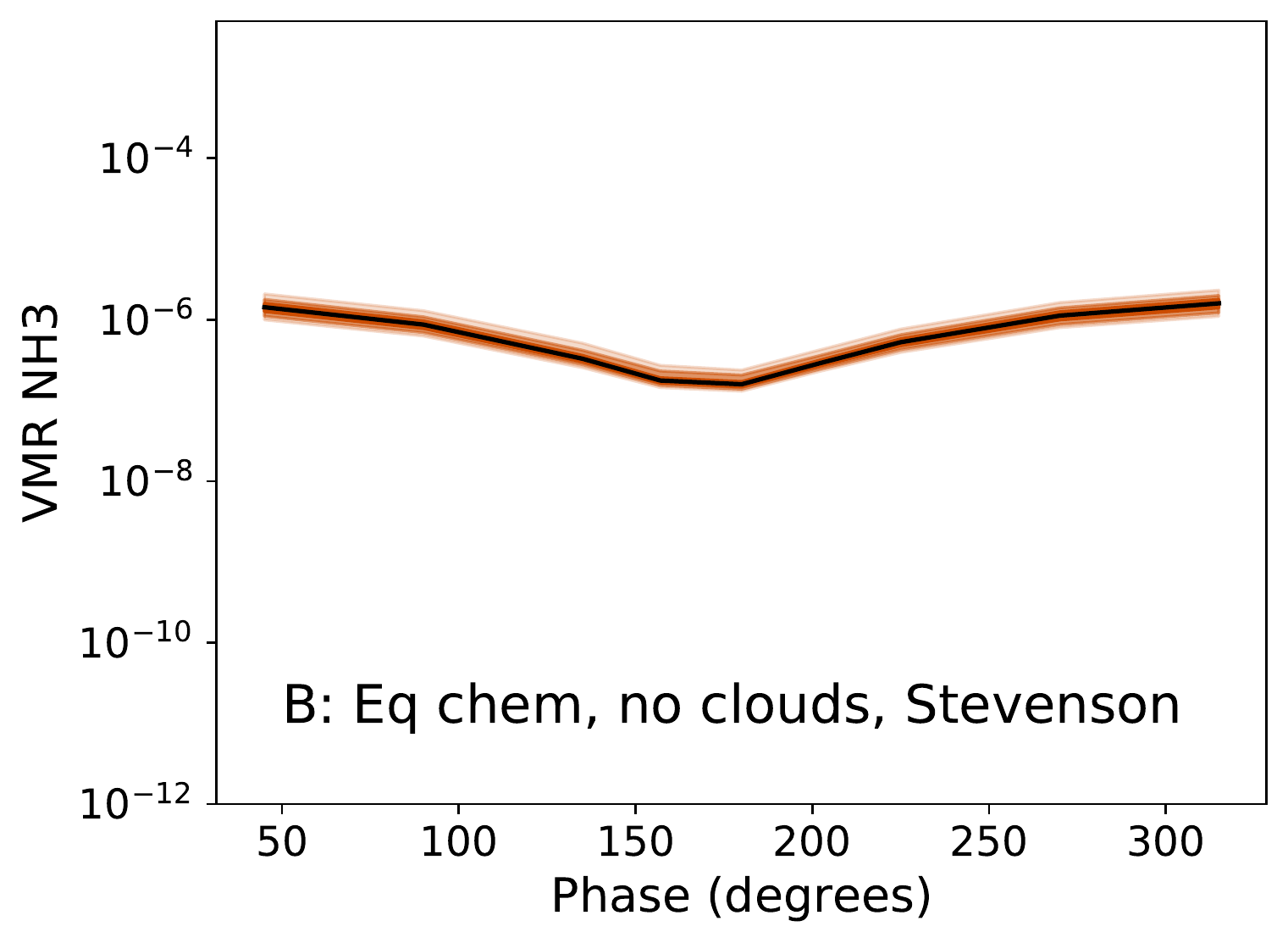}
                        \includegraphics[width=0.4\textwidth]{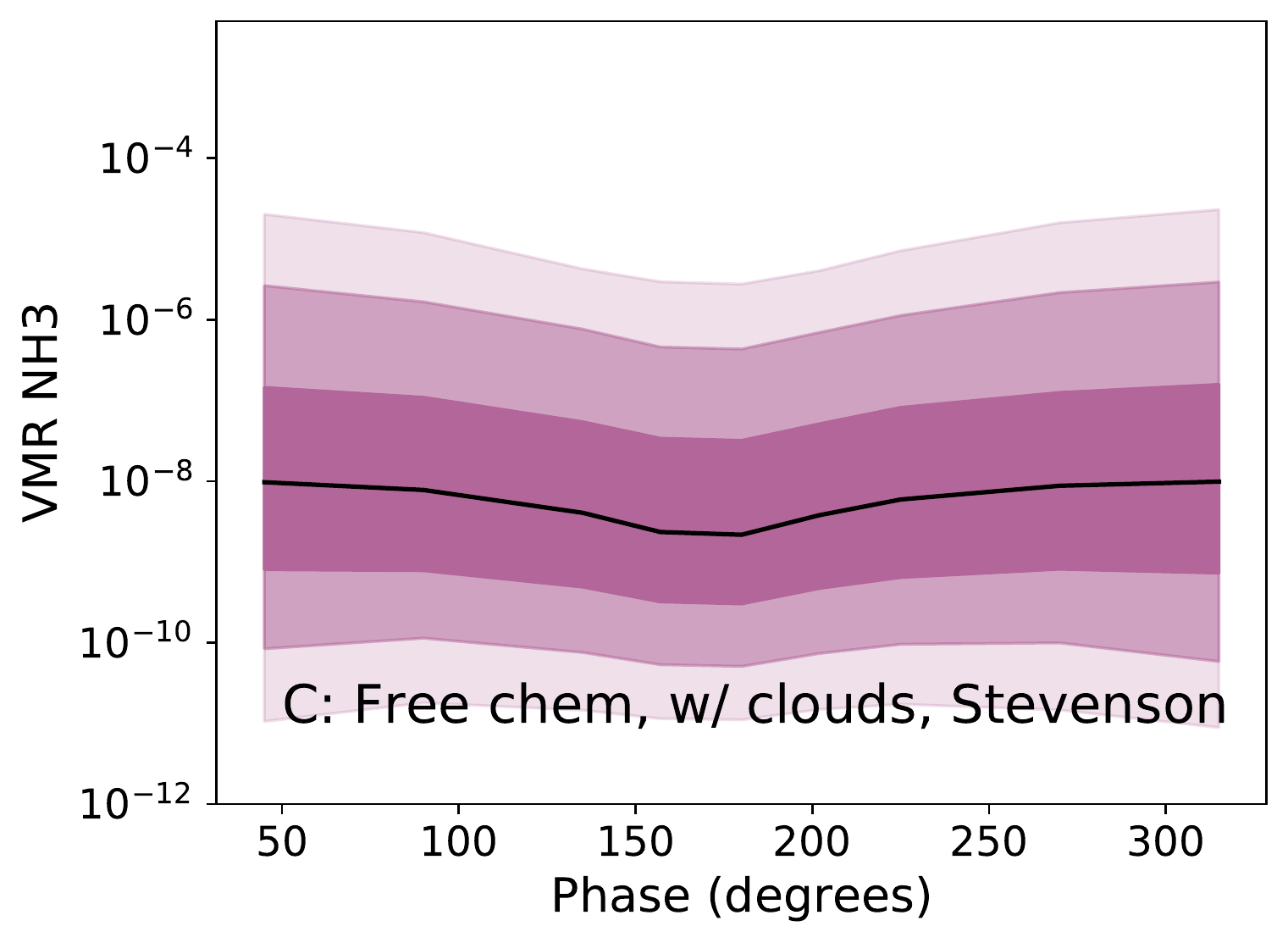}
                        \includegraphics[width=0.4\textwidth]{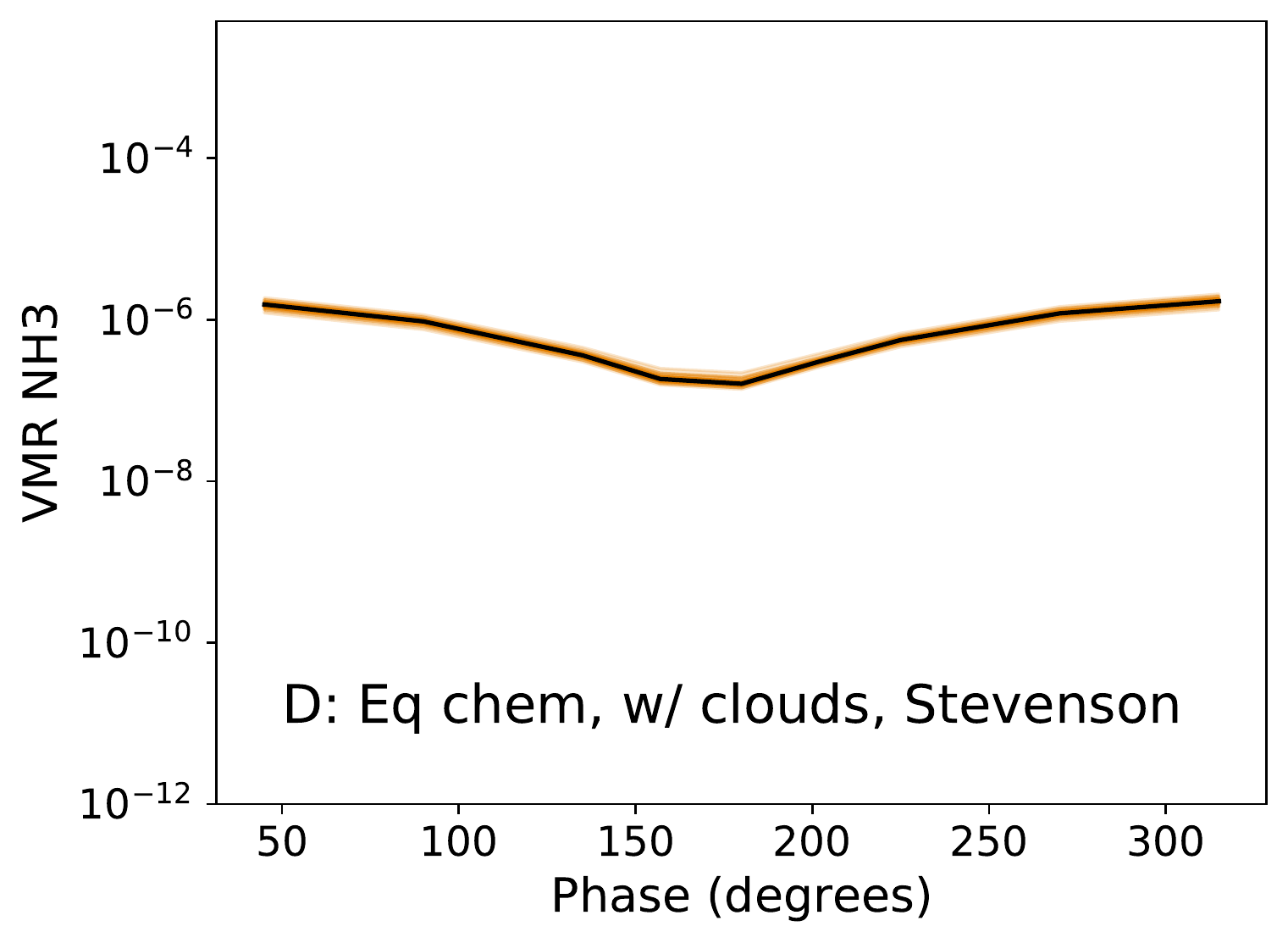}
                        \includegraphics[width=0.4\textwidth]{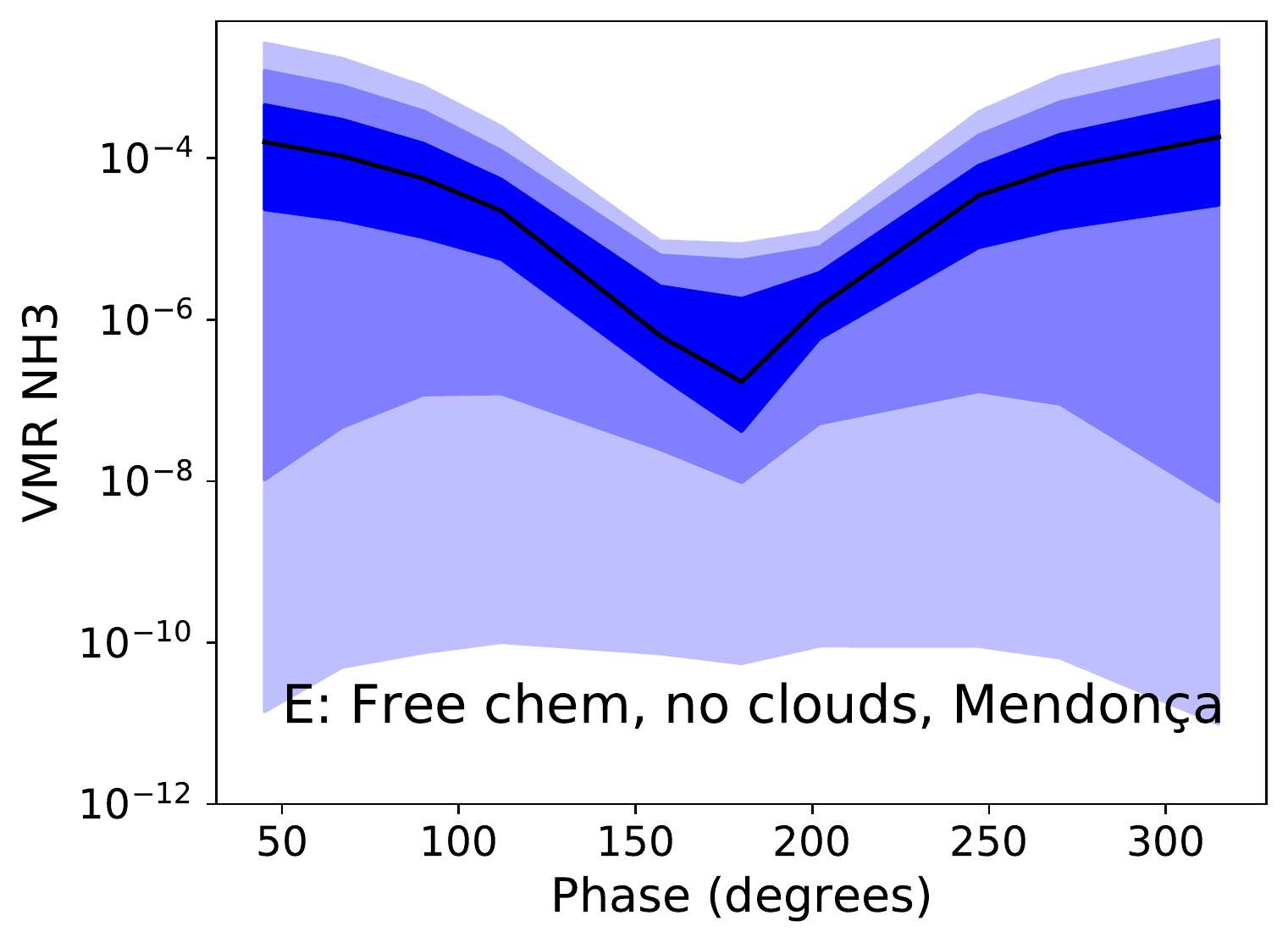}
                        \includegraphics[width=0.4\textwidth]{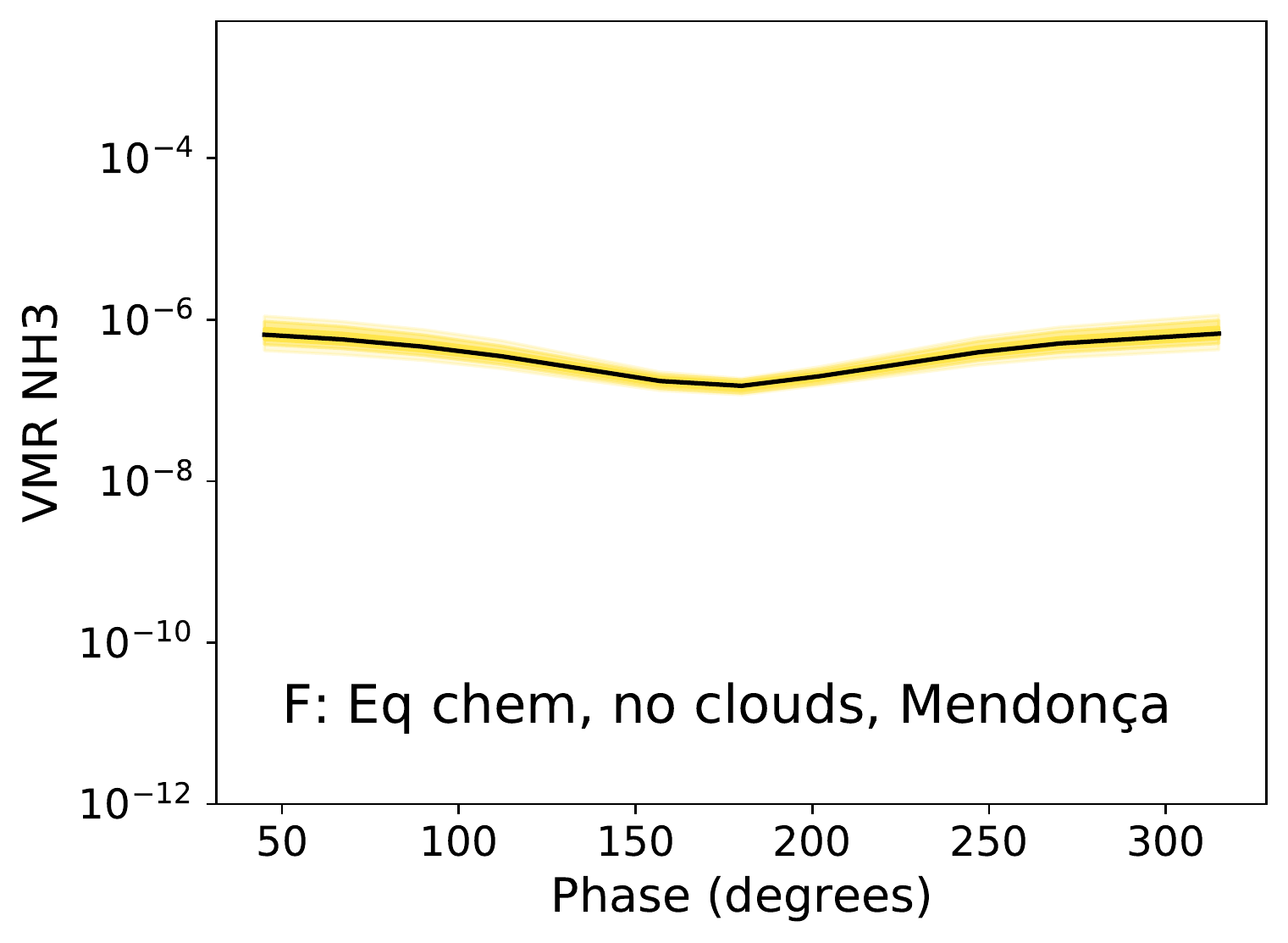}
                        \includegraphics[width=0.4\textwidth]{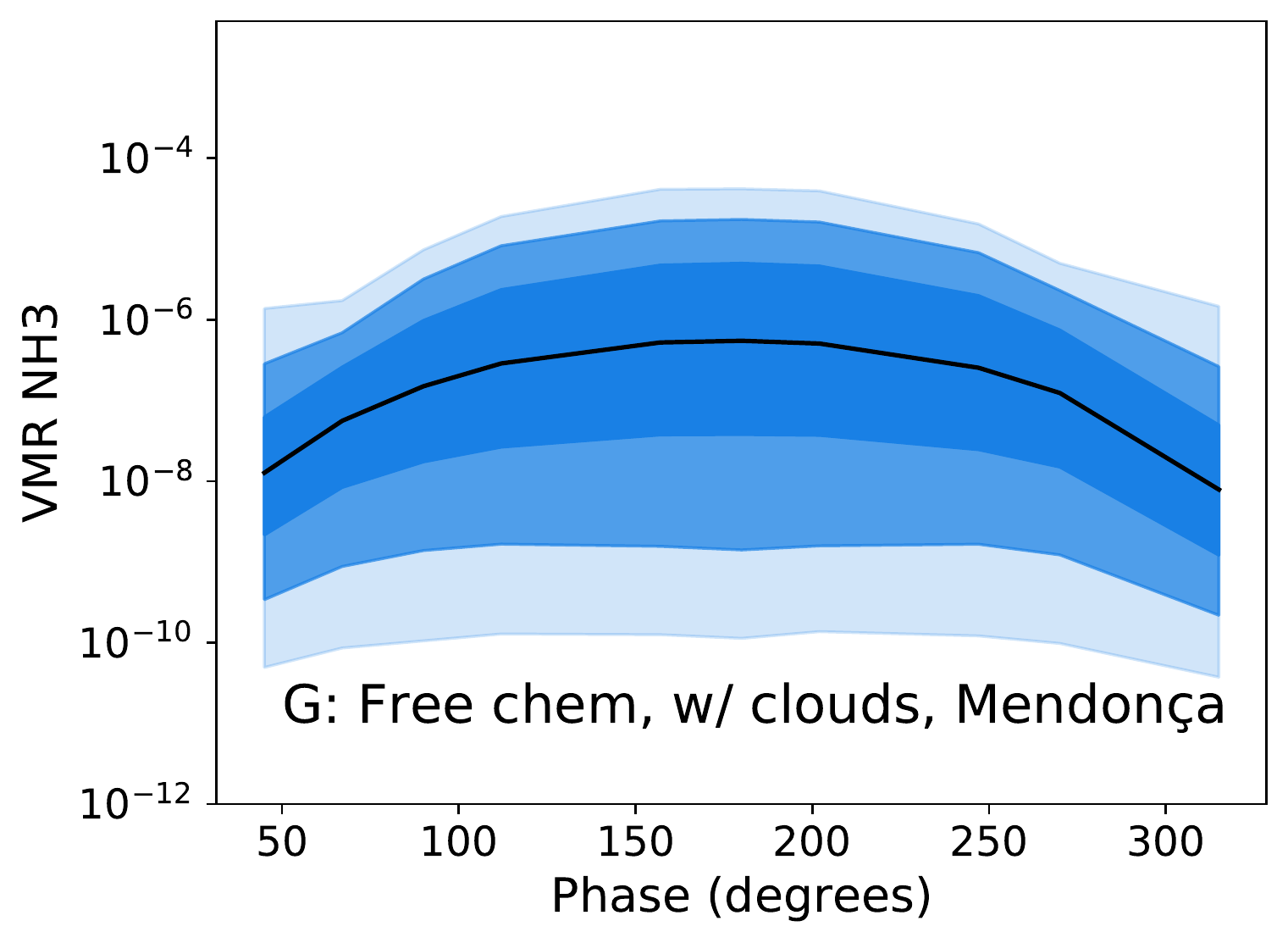}
                        \includegraphics[width=0.4\textwidth]{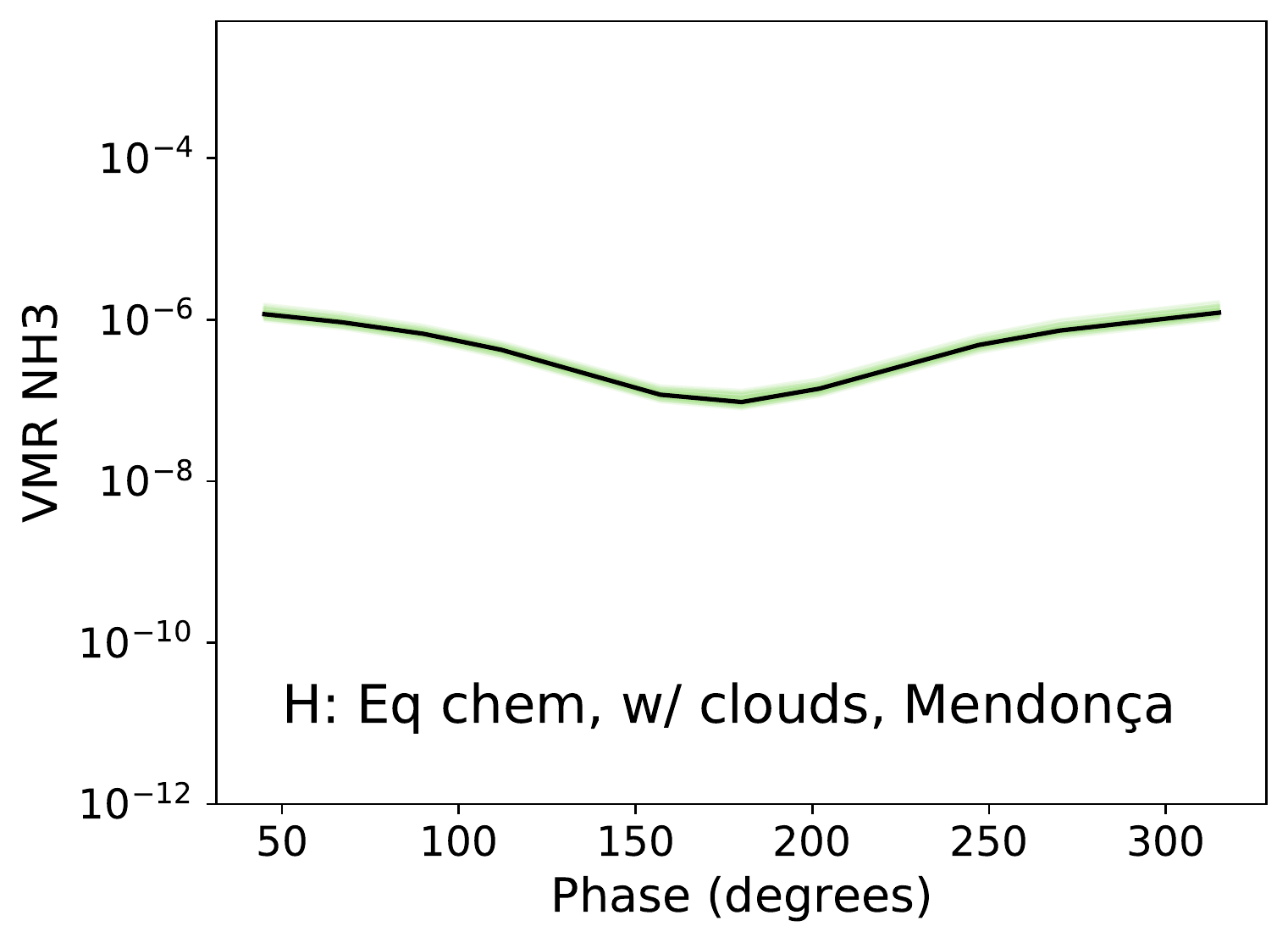}
                        \caption{Same as Fig~\ref{fig:AlO_COratio}, but for NH$_3$}\label{fig:NH3_COratio}
                \end{figure}

                \begin{figure}
                        \centering
                        \includegraphics[width=0.35\textwidth]{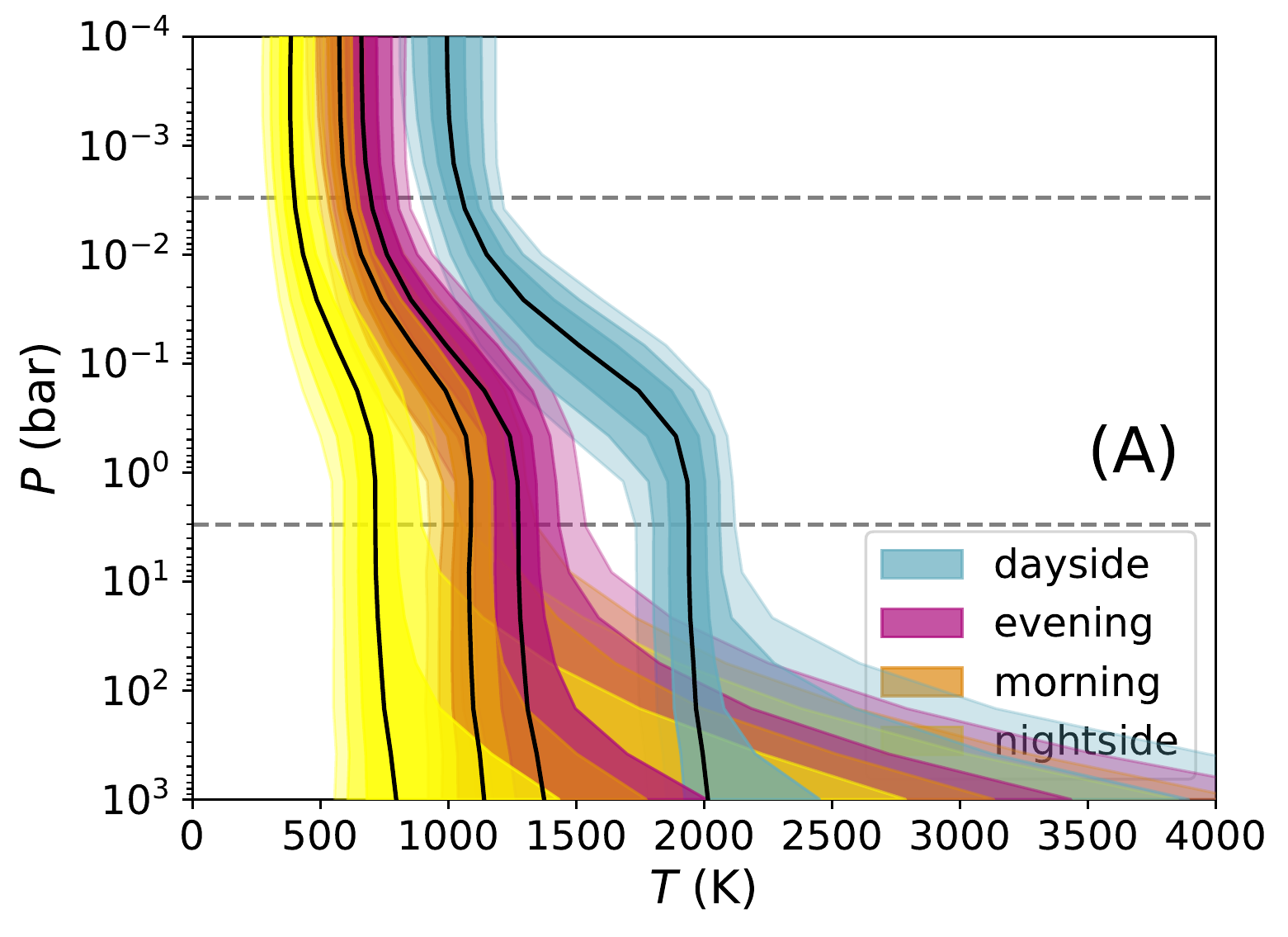}
                        \includegraphics[width=0.35\textwidth]{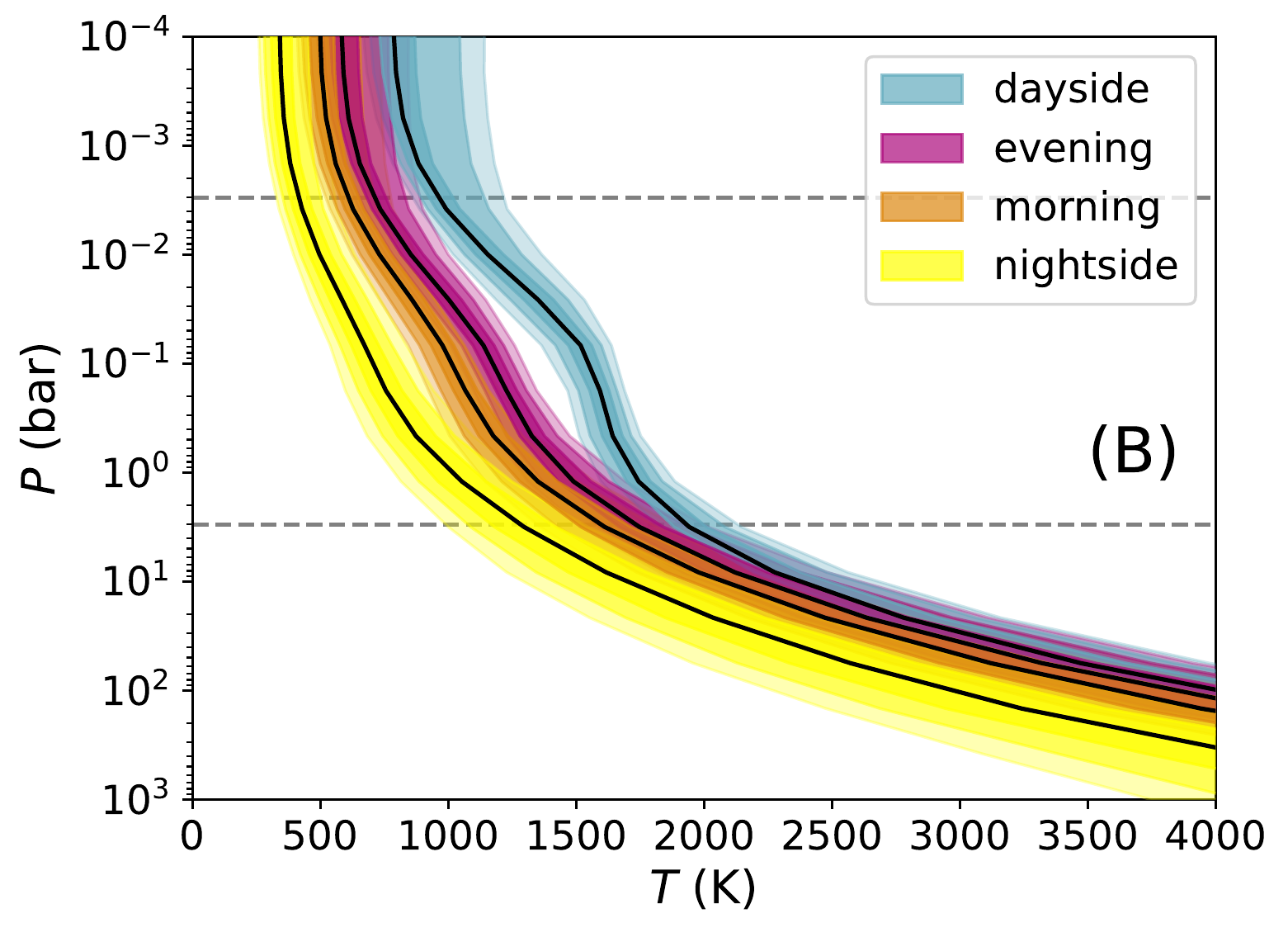}
                        \includegraphics[width=0.35\textwidth]{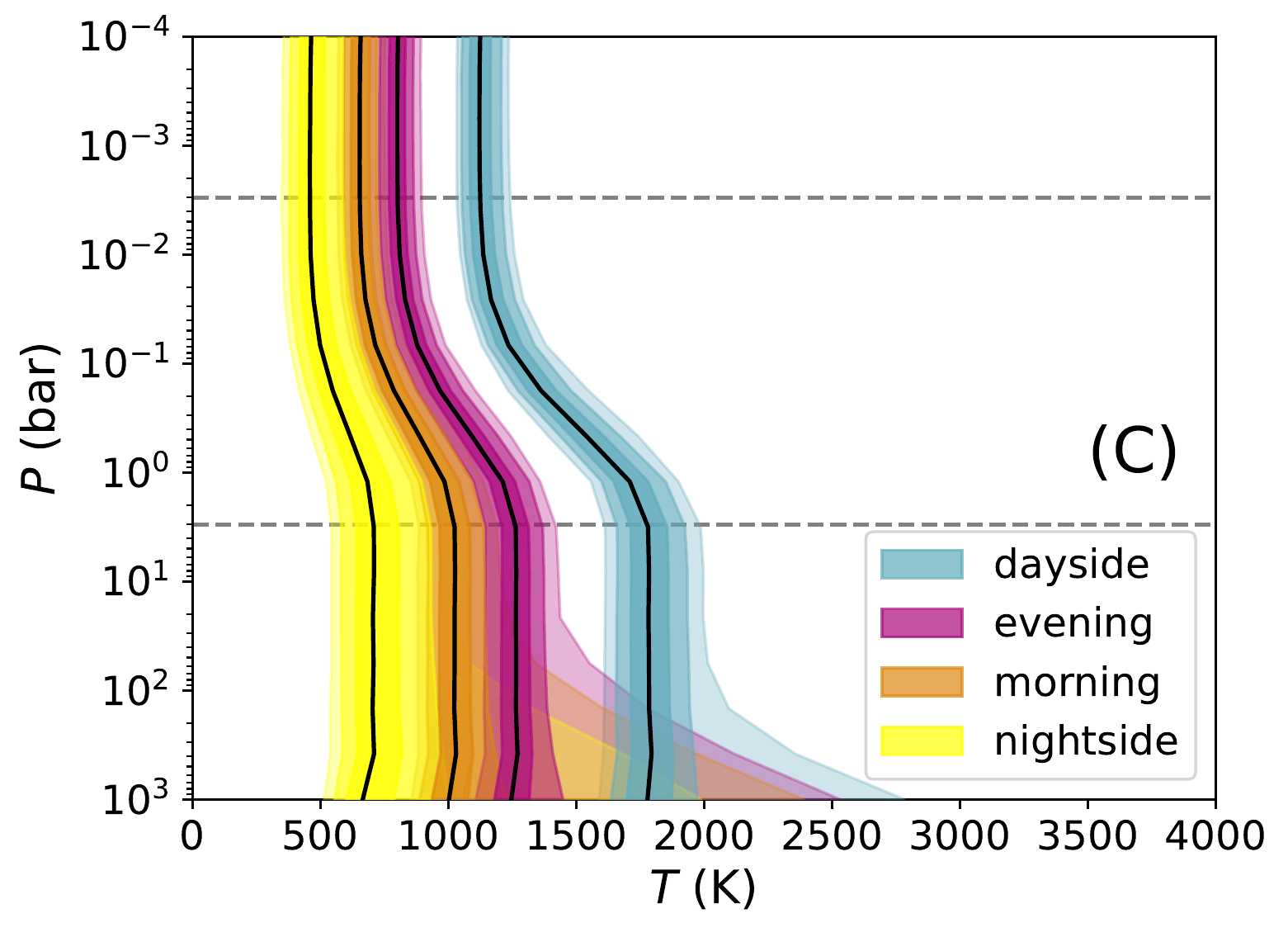}
                        \includegraphics[width=0.35\textwidth]{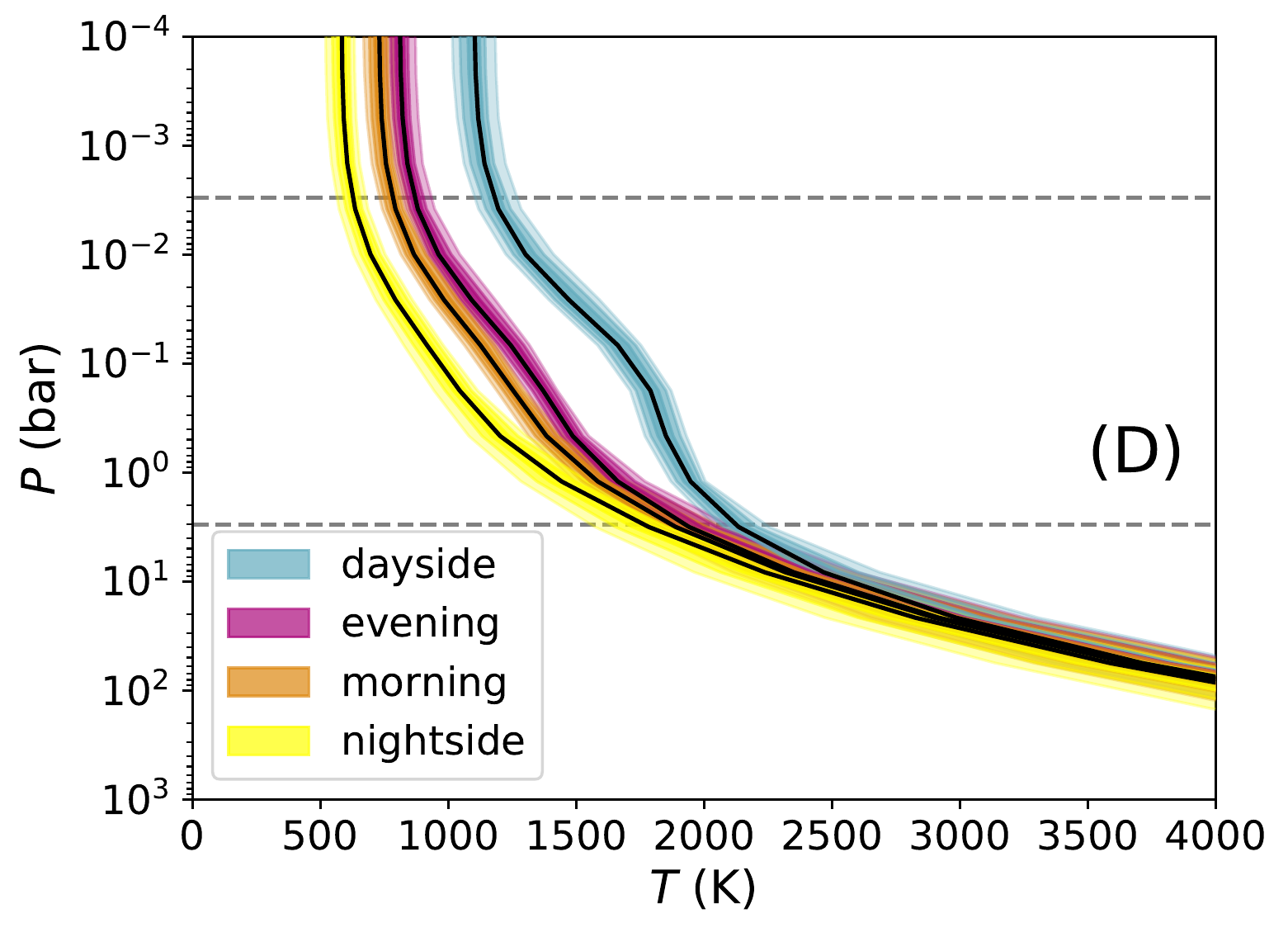}
                        \includegraphics[width=0.35\textwidth]{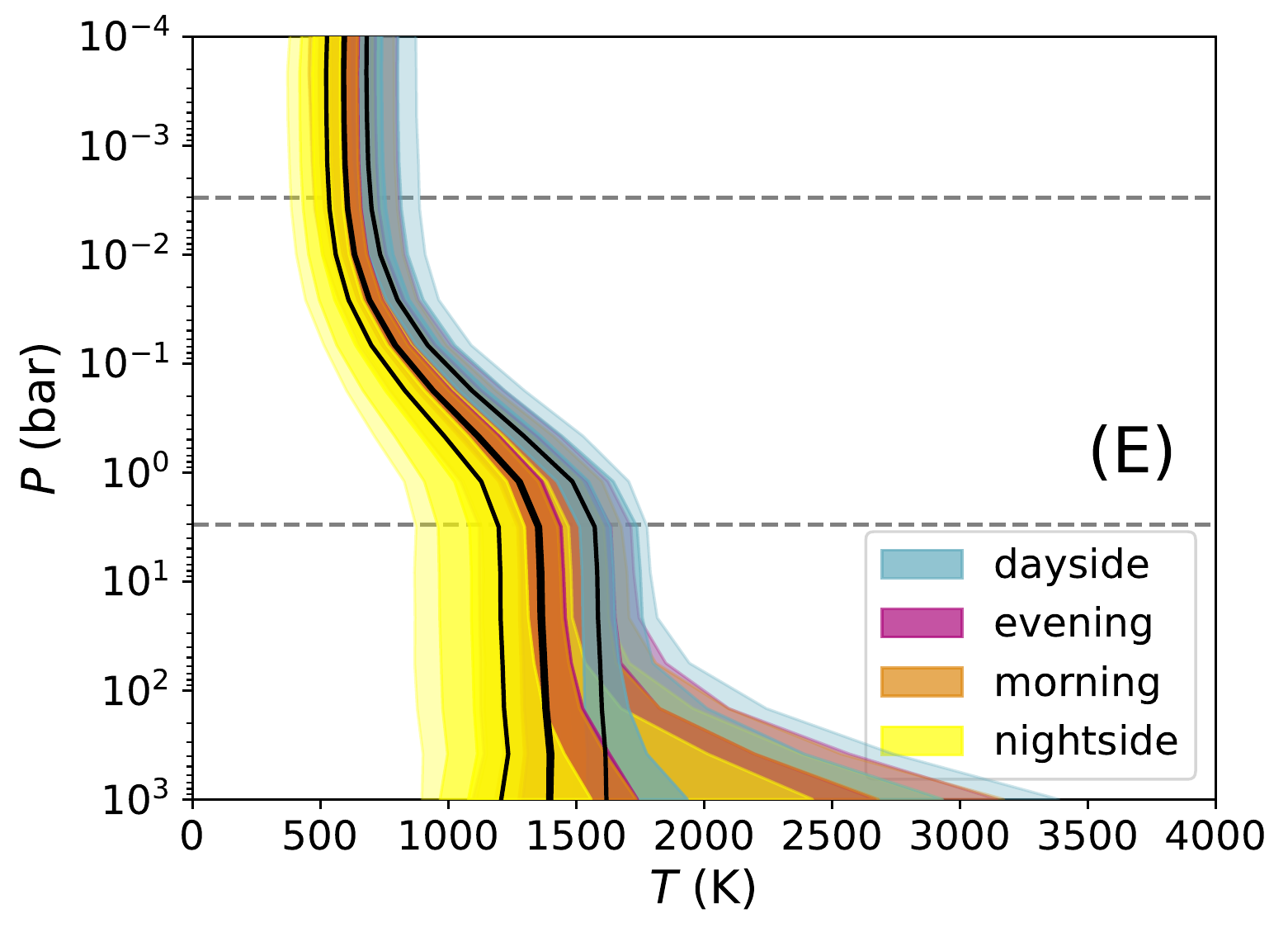}
                        \includegraphics[width=0.35\textwidth]{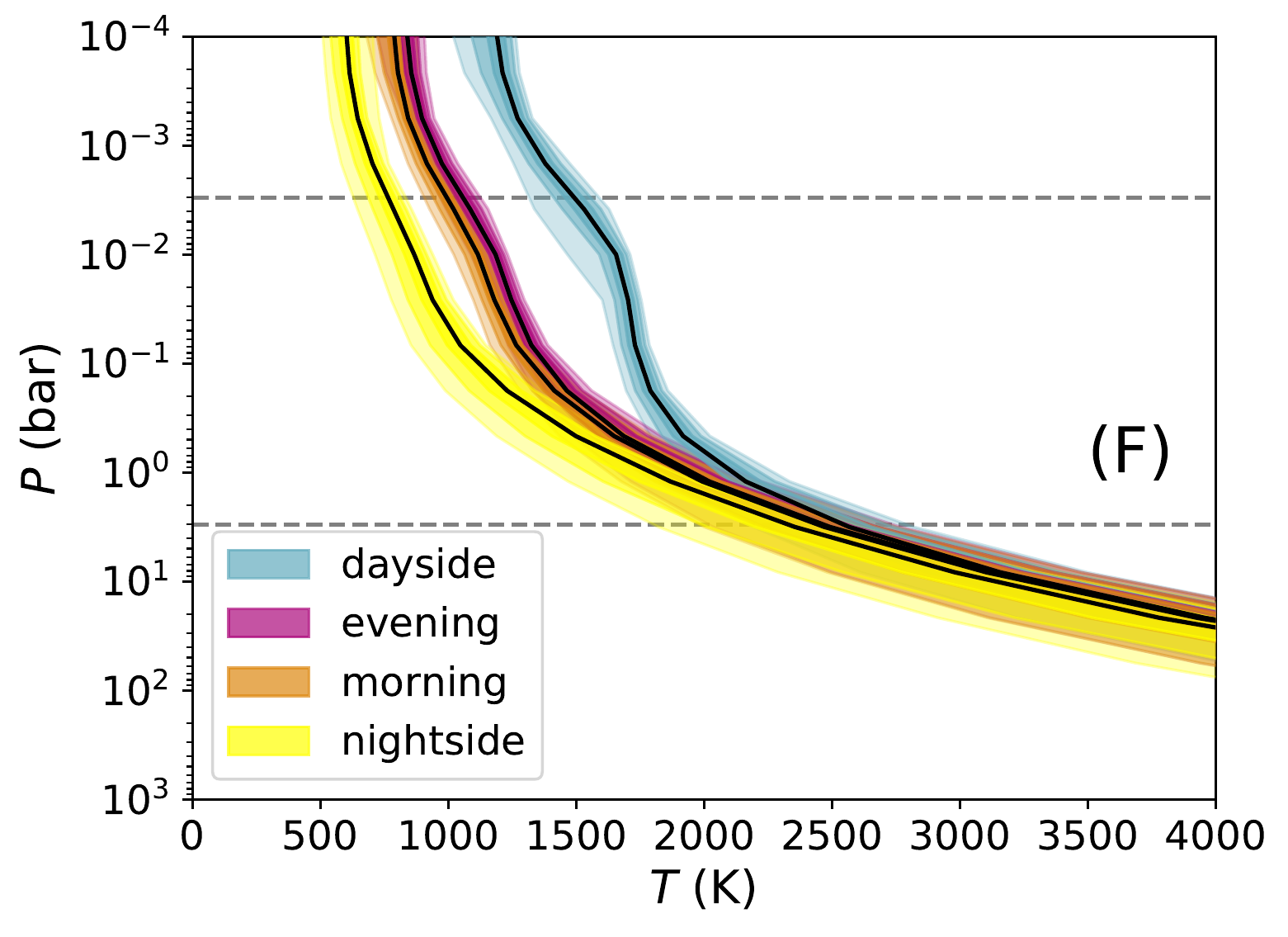}
                        \includegraphics[width=0.35\textwidth]{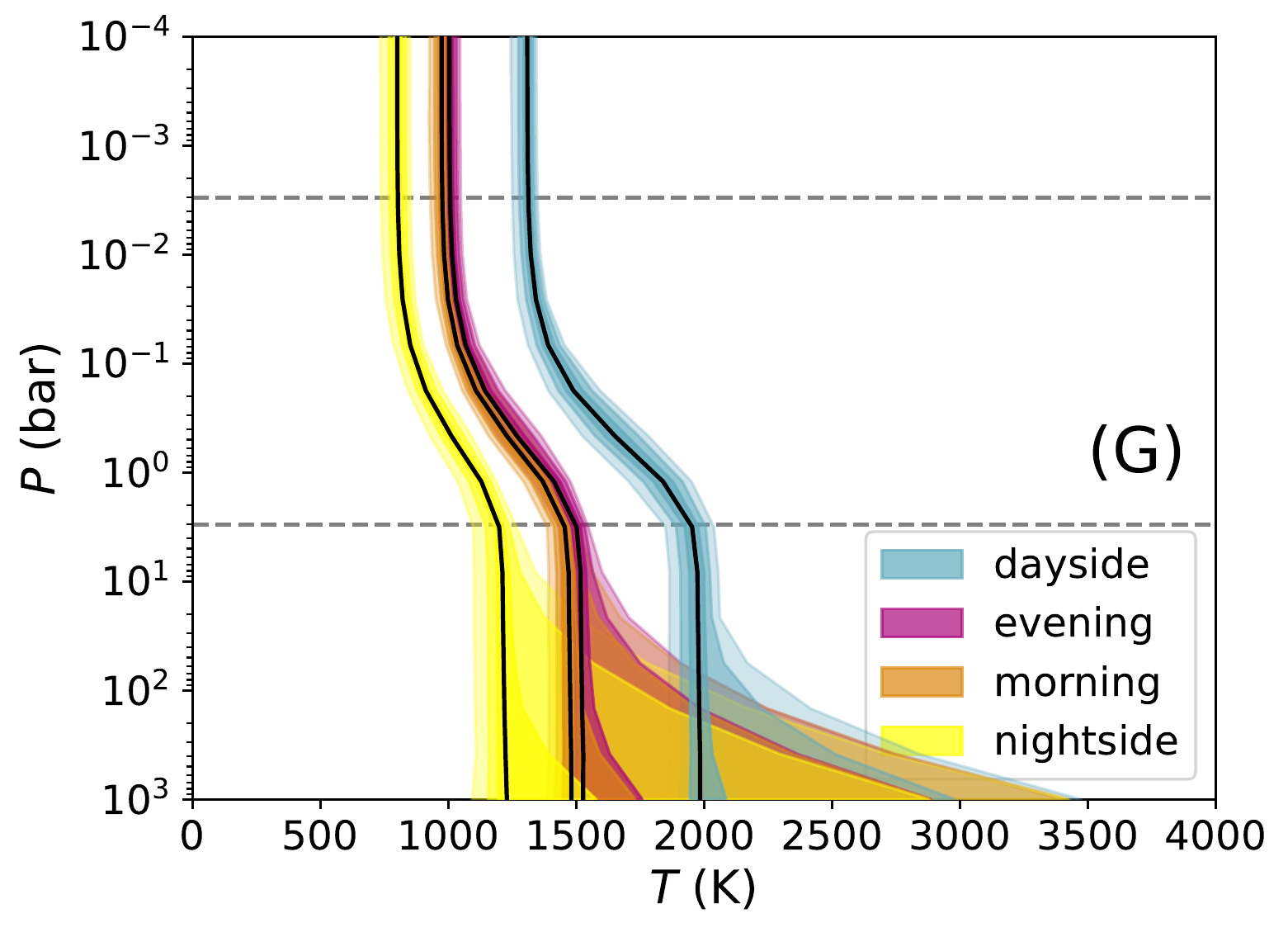}
                        \includegraphics[width=0.35\textwidth]{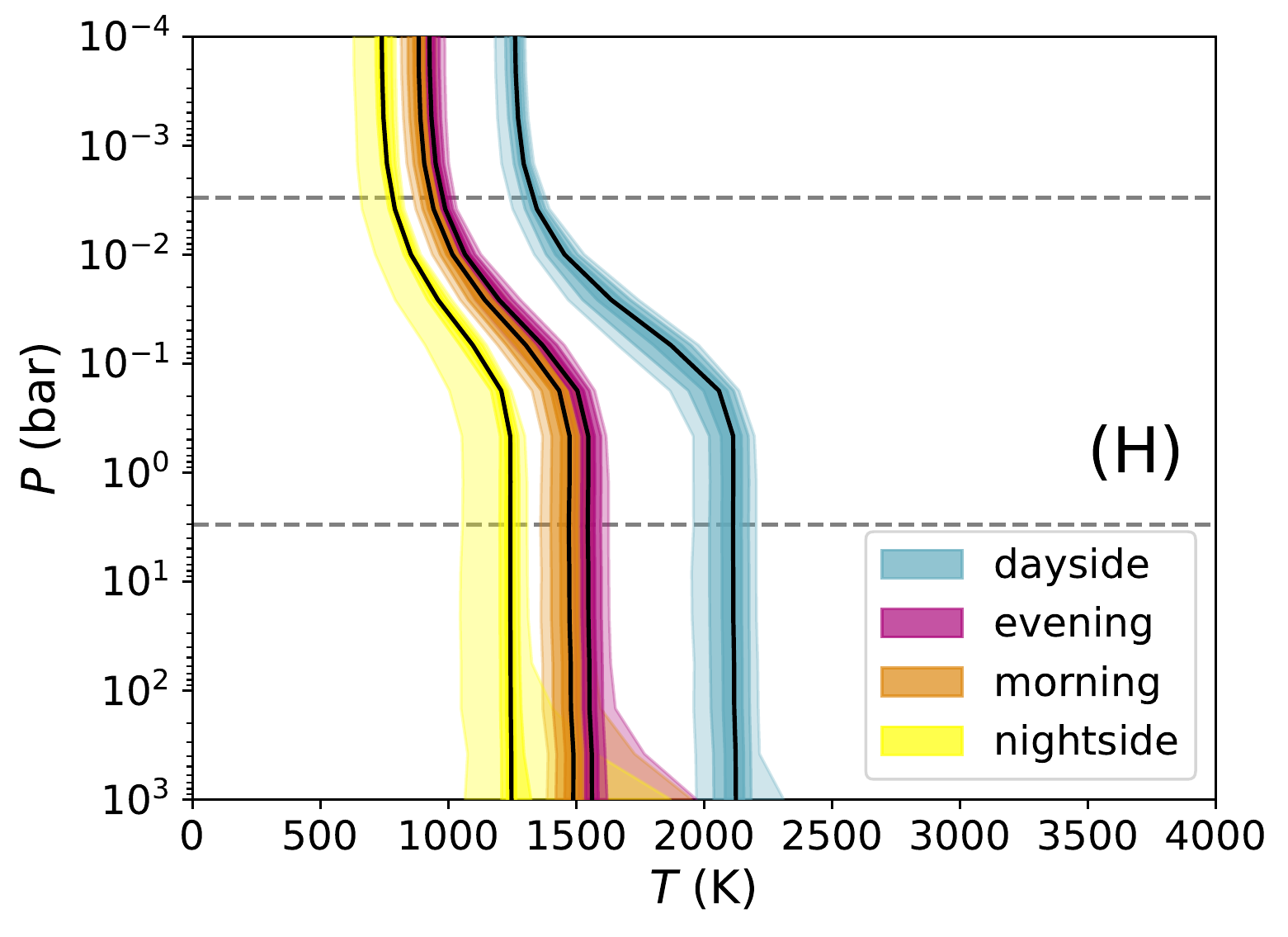}
                        \includegraphics[width=0.35\textwidth]{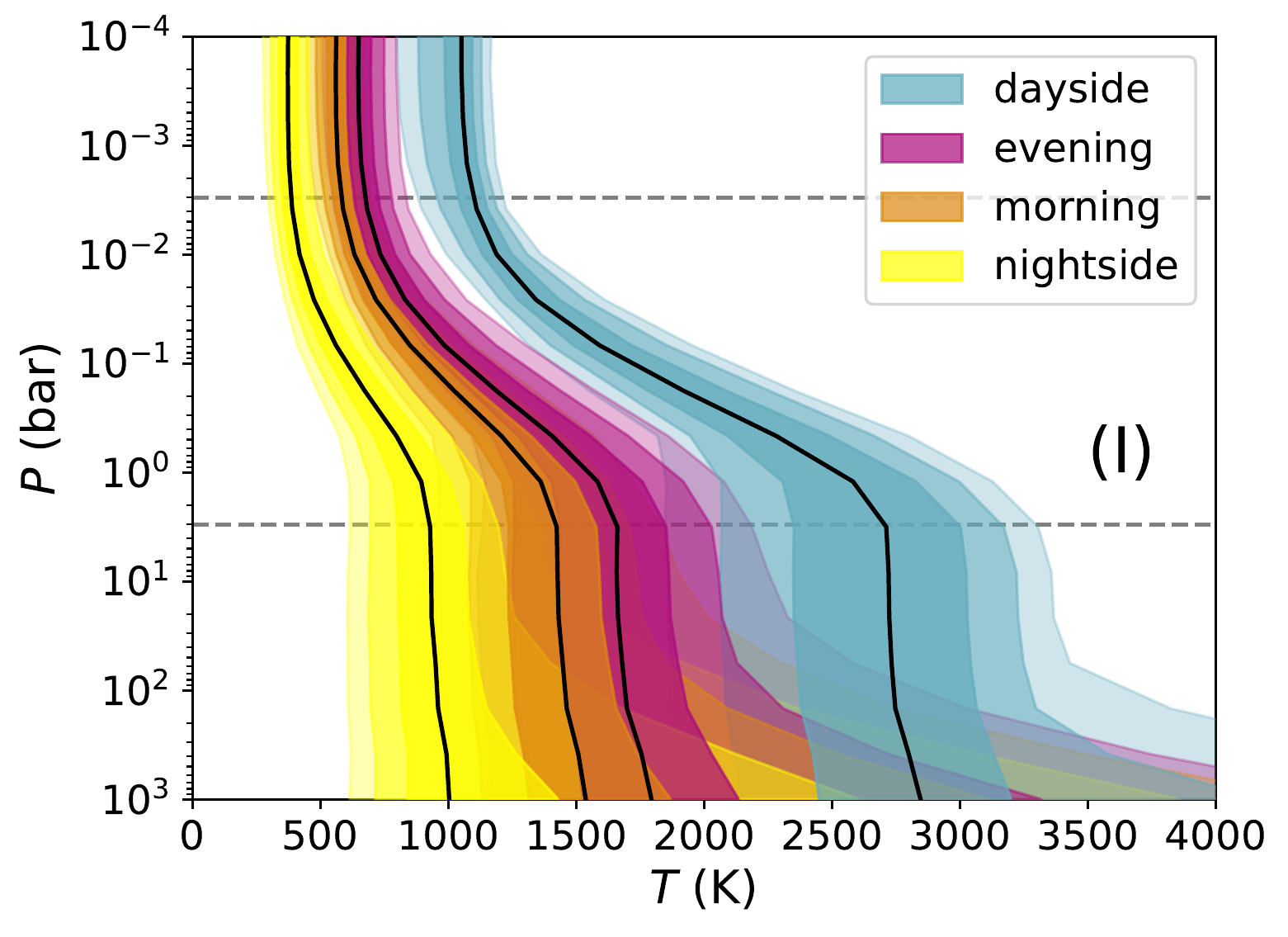}
                        \caption{Retrieved pressure-temperature profiles for the 90$\degree$ (evening), 180$\degree$ (dayside), 270$\degree$ (morning), and 315$\degree$ (nightside) phases for eight different retrieval setups, as labelled in each panel. 
                                We note that the pressure-temperature profiles are artificially too well constrained in many of these figures, particularly in the lower part of the atmosphere. 
                        }\label{fig:TP_4phase}
                \end{figure}

                \begin{figure}
                        \centering
                        \includegraphics[width=0.3\textwidth]{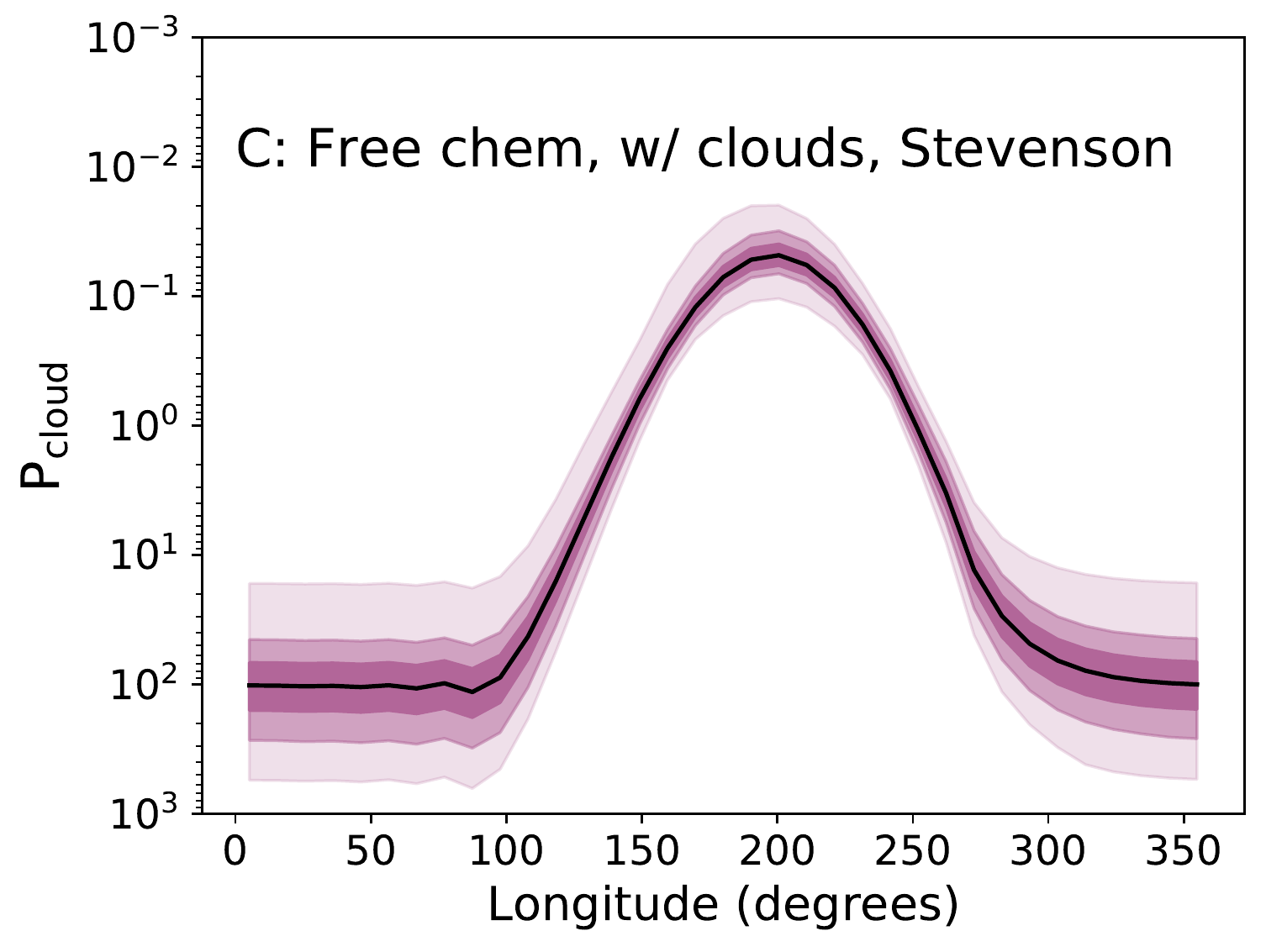}
                        \includegraphics[width=0.3\textwidth]{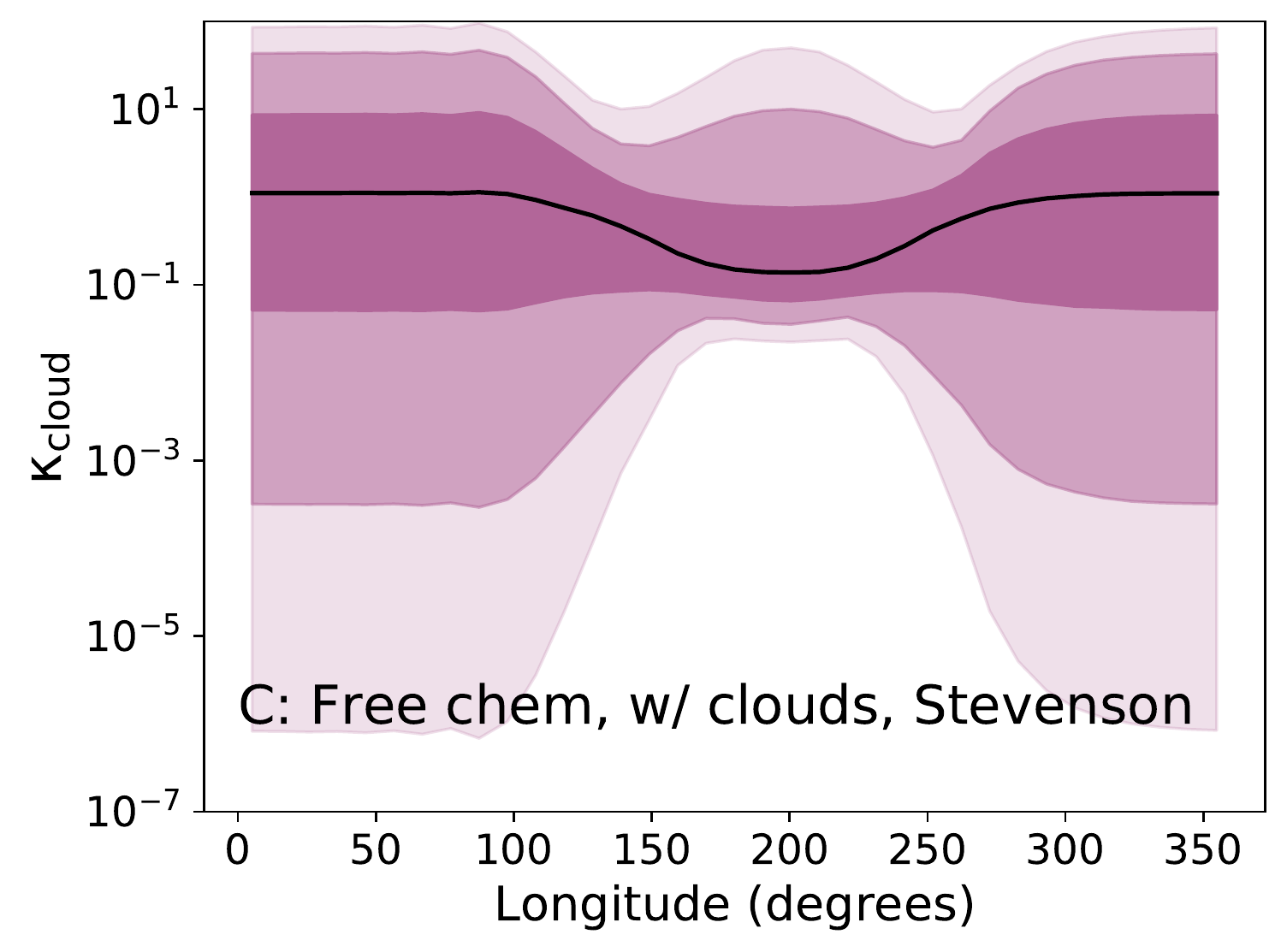}
                        \includegraphics[width=0.3\textwidth]{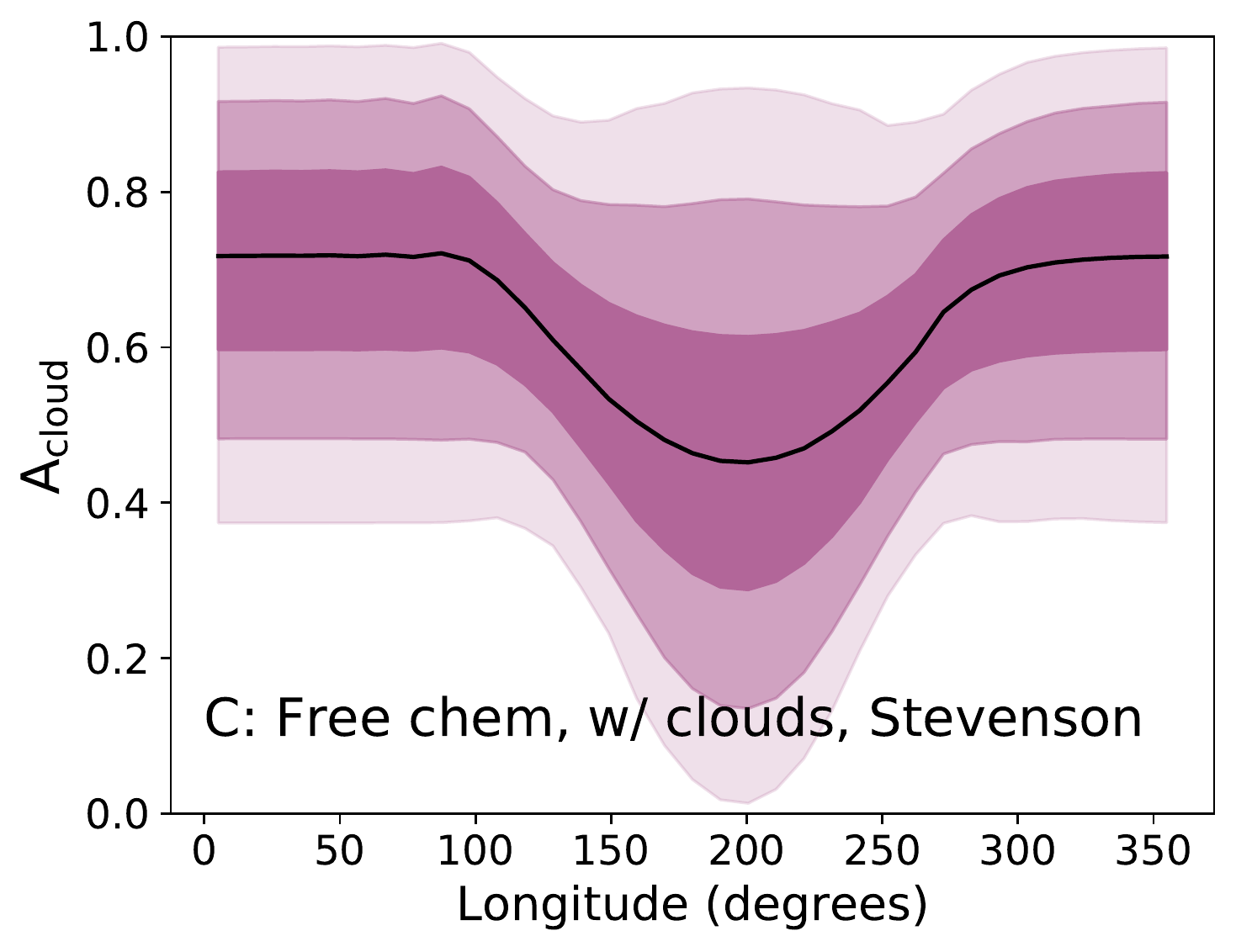}
                        \includegraphics[width=0.3\textwidth]{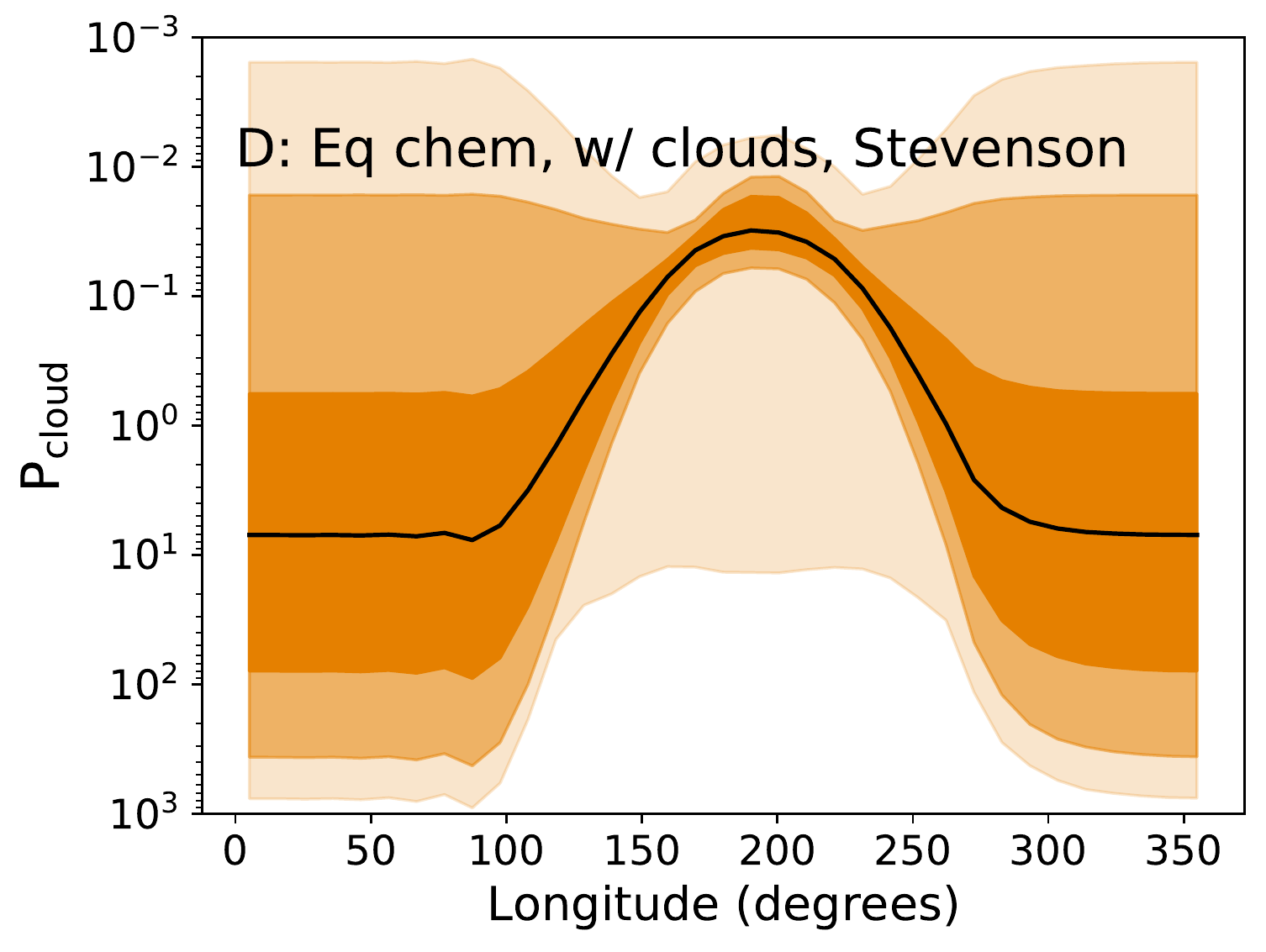}
                        \includegraphics[width=0.3\textwidth]{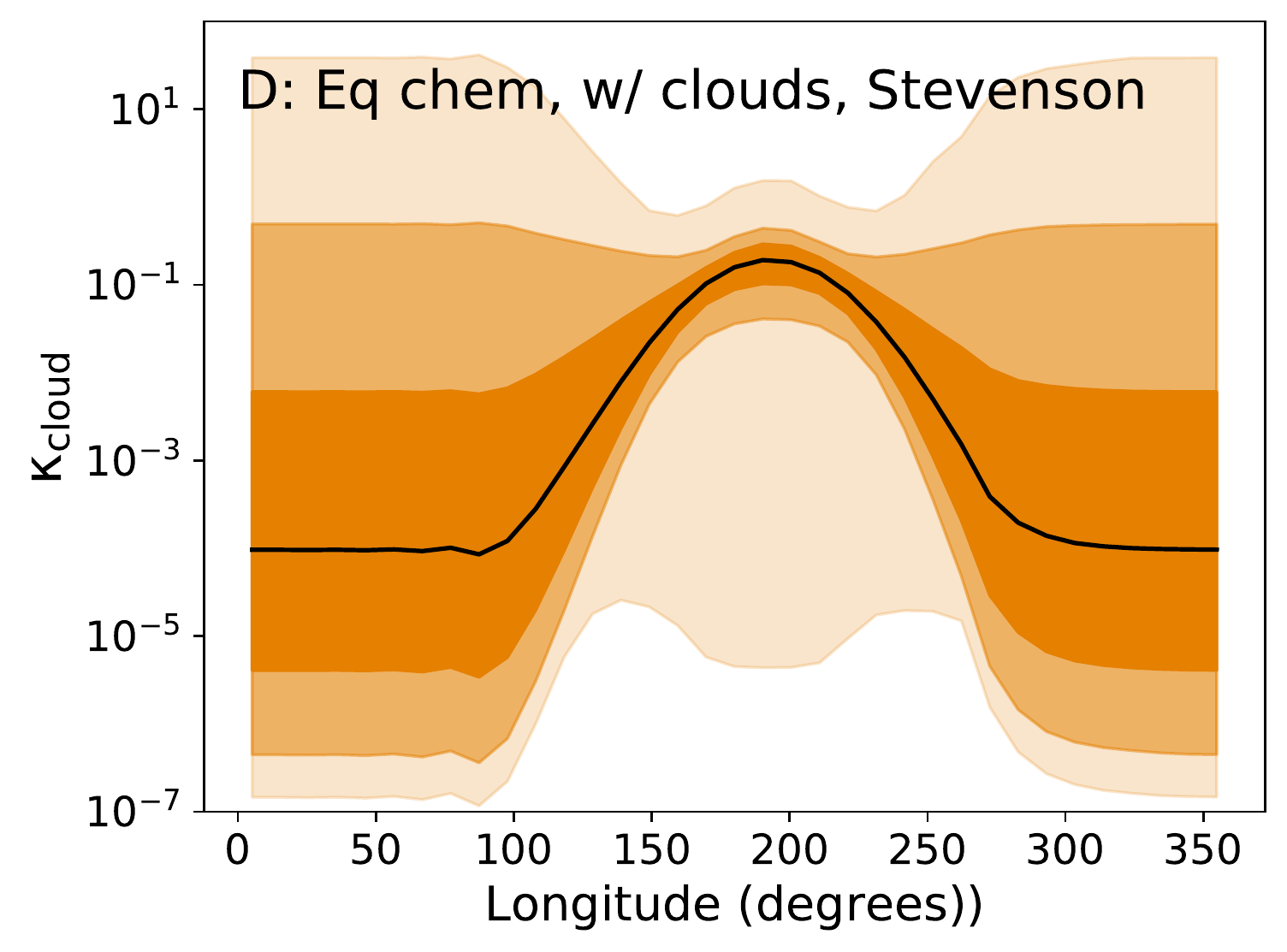}
                        \includegraphics[width=0.3\textwidth]{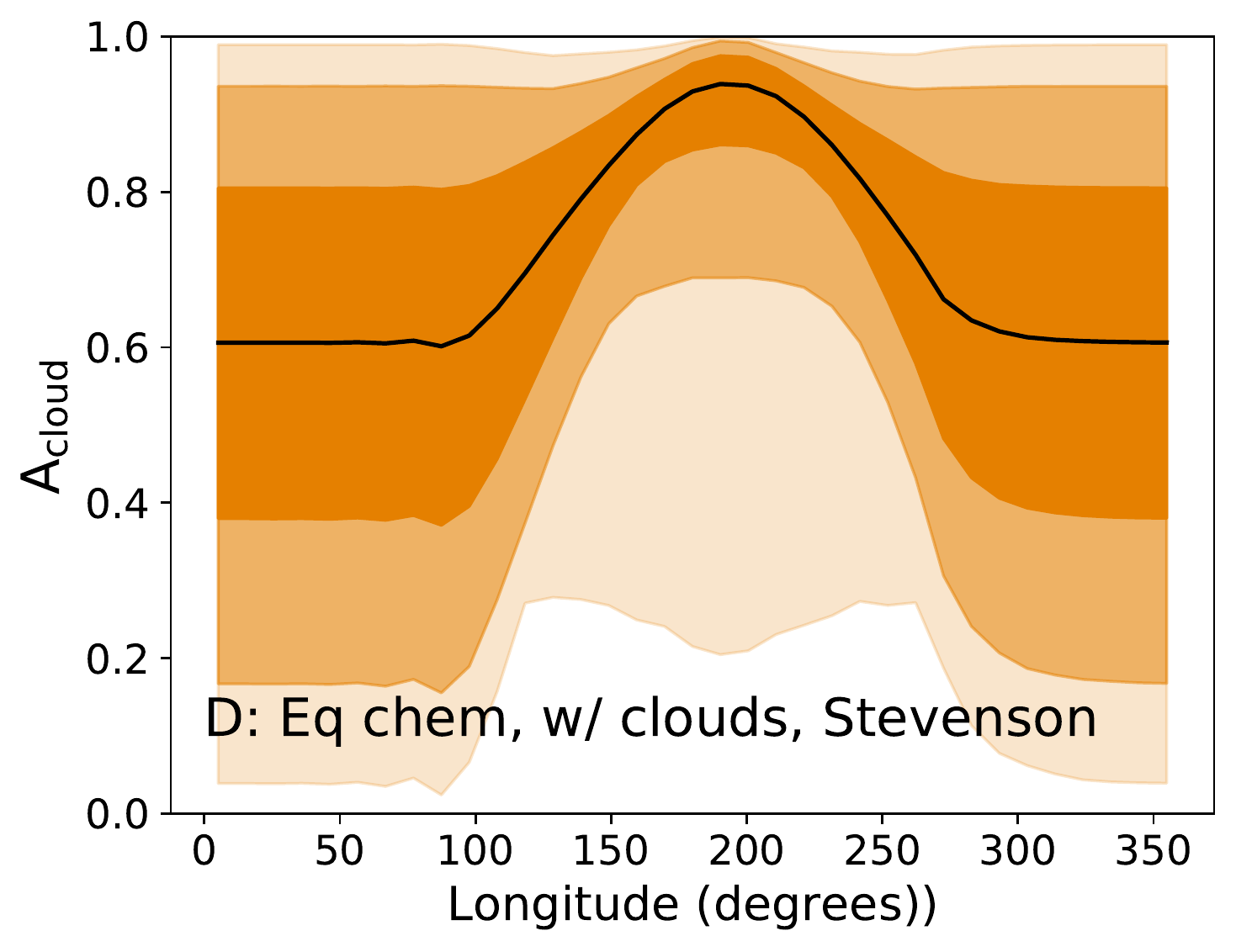}
                        \includegraphics[width=0.3\textwidth]{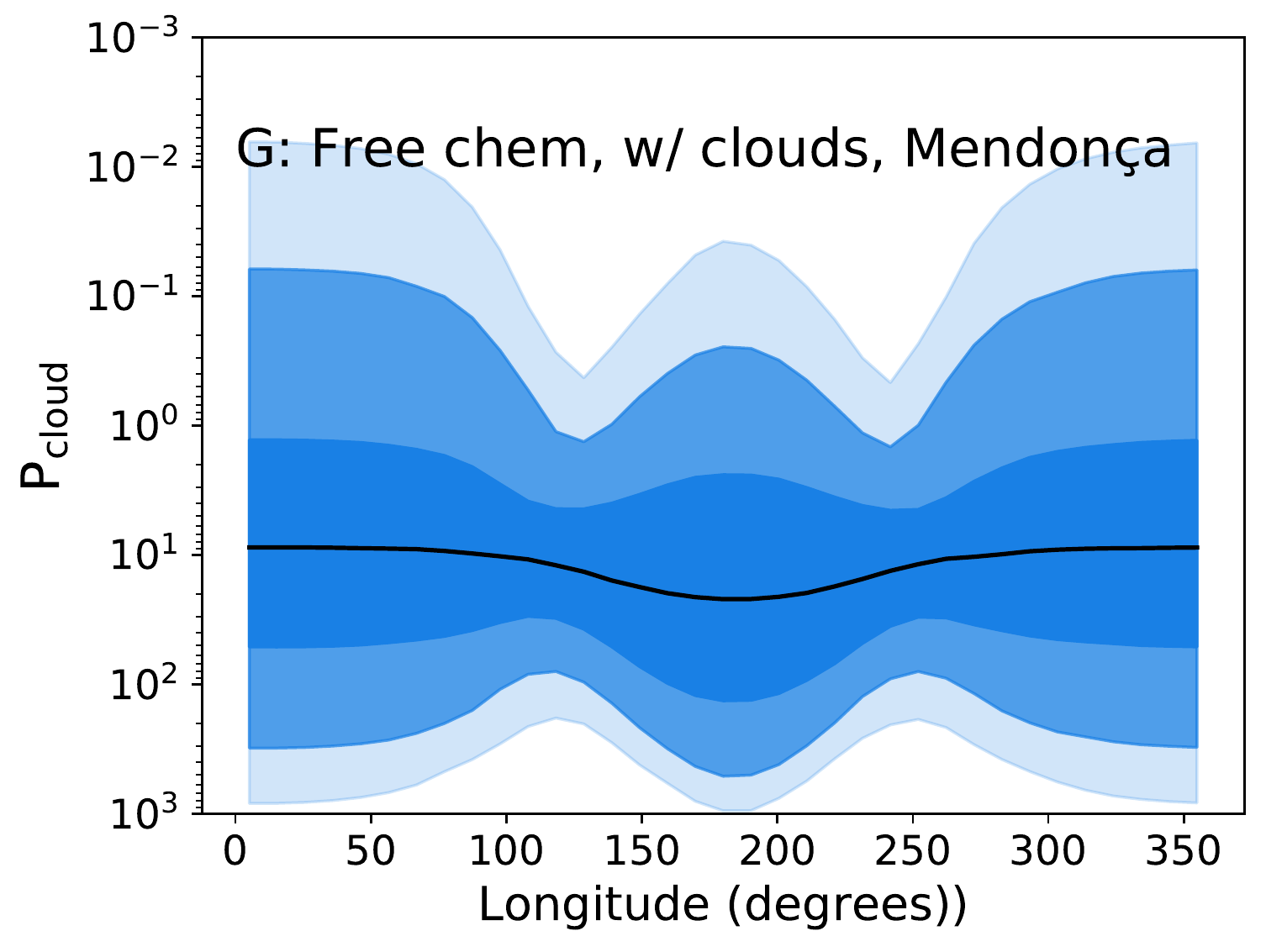}
                        \includegraphics[width=0.3\textwidth]{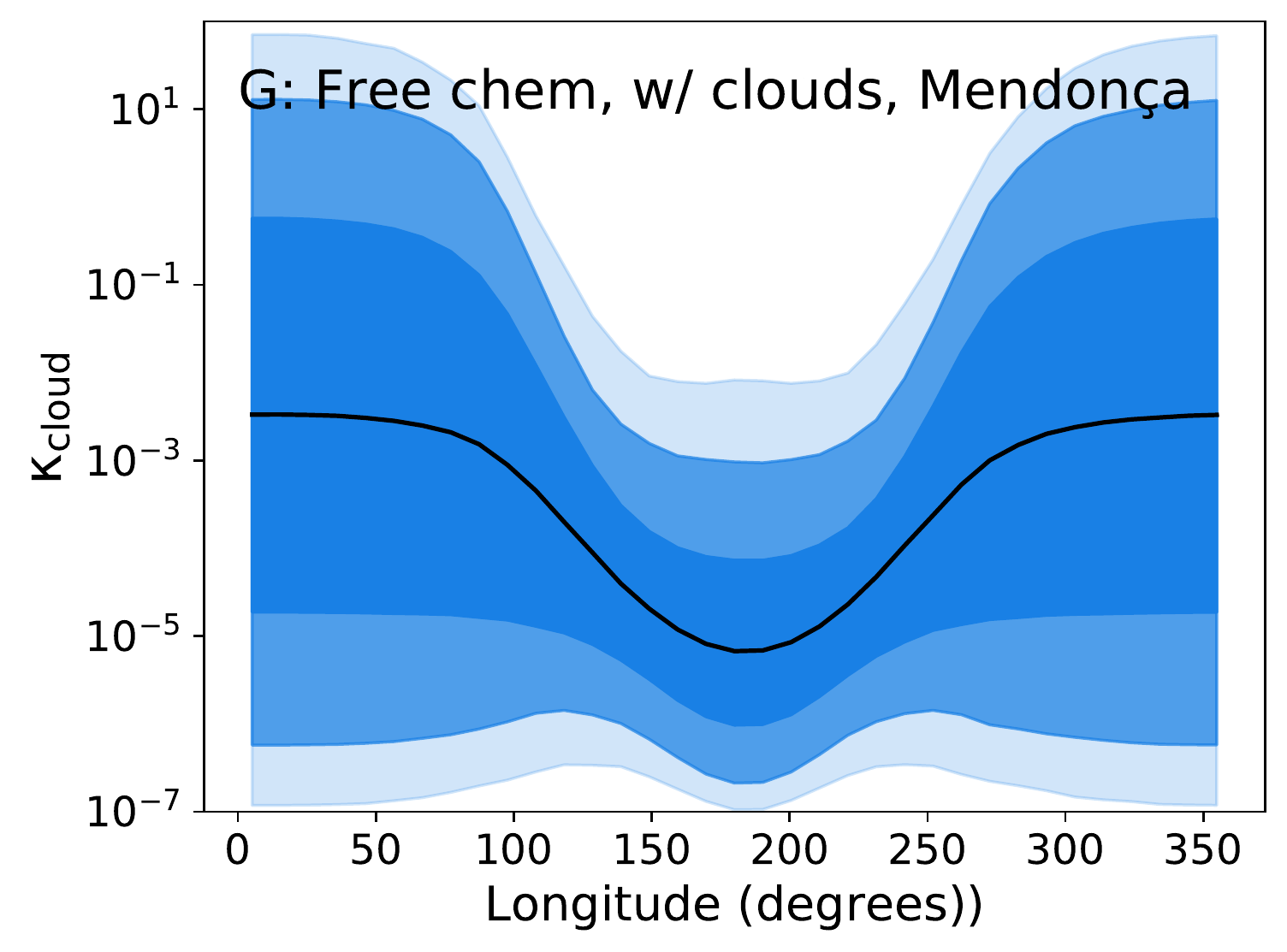}
                        \includegraphics[width=0.3\textwidth]{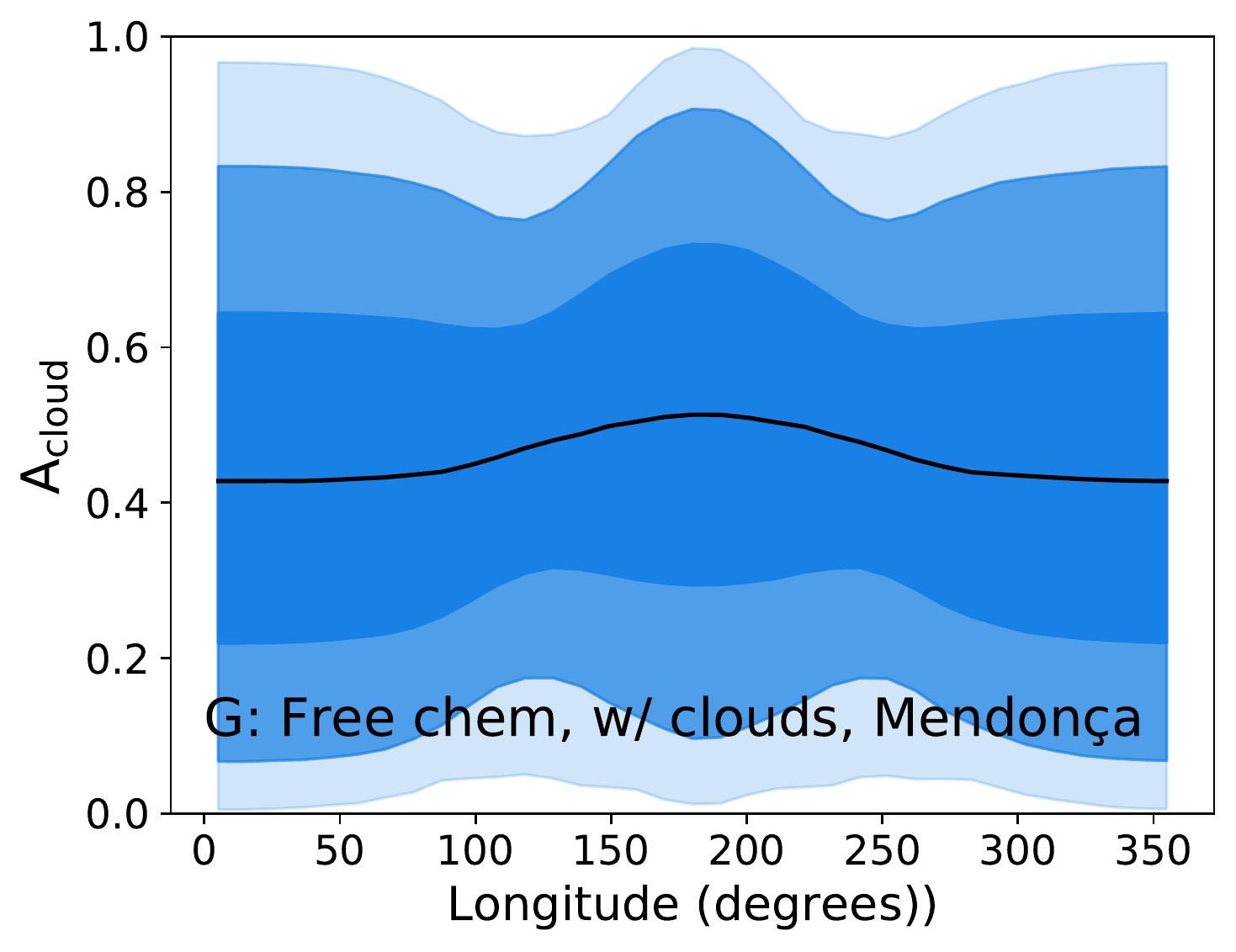}
                        \includegraphics[width=0.3\textwidth]{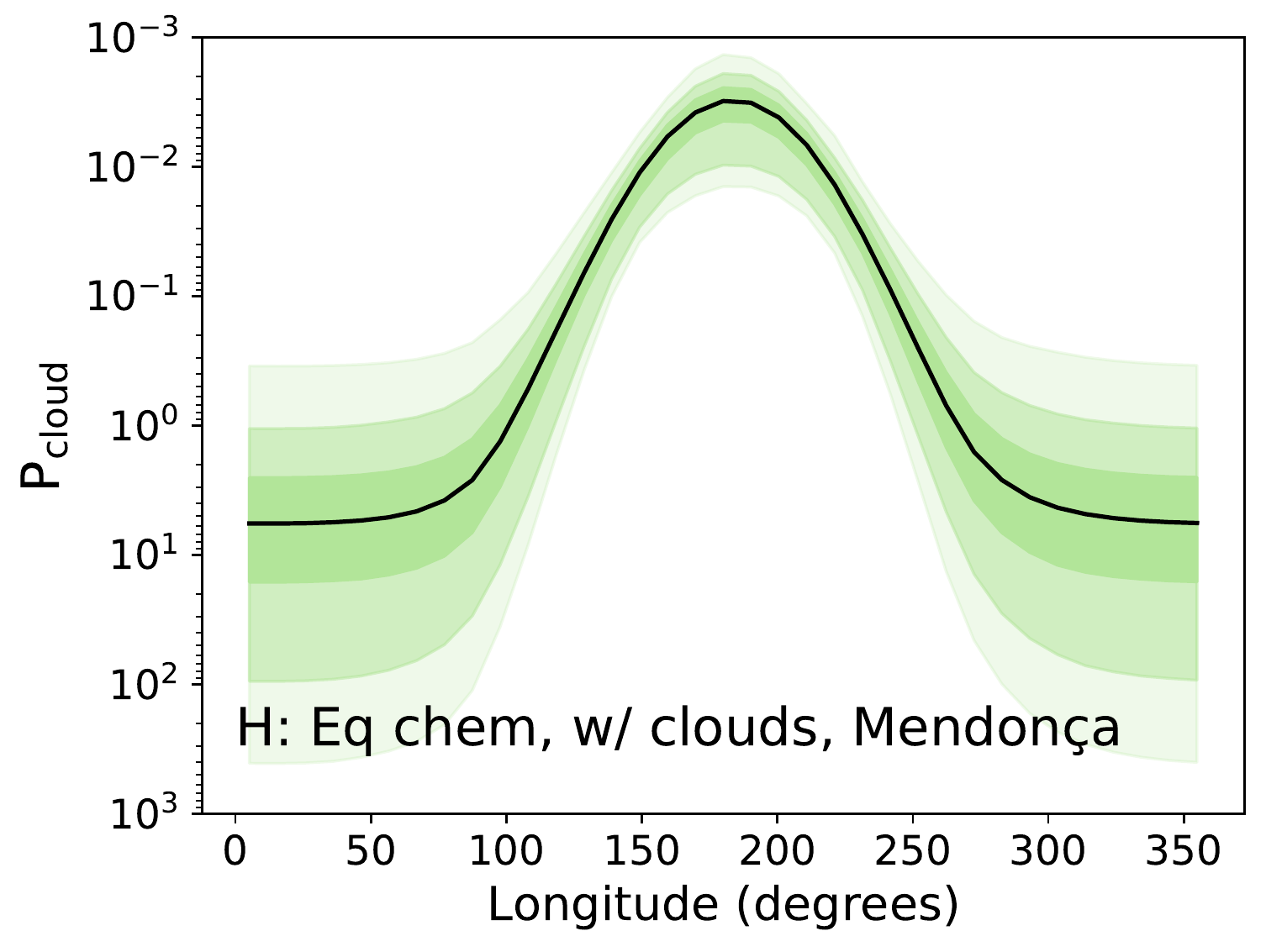}
                        \includegraphics[width=0.3\textwidth]{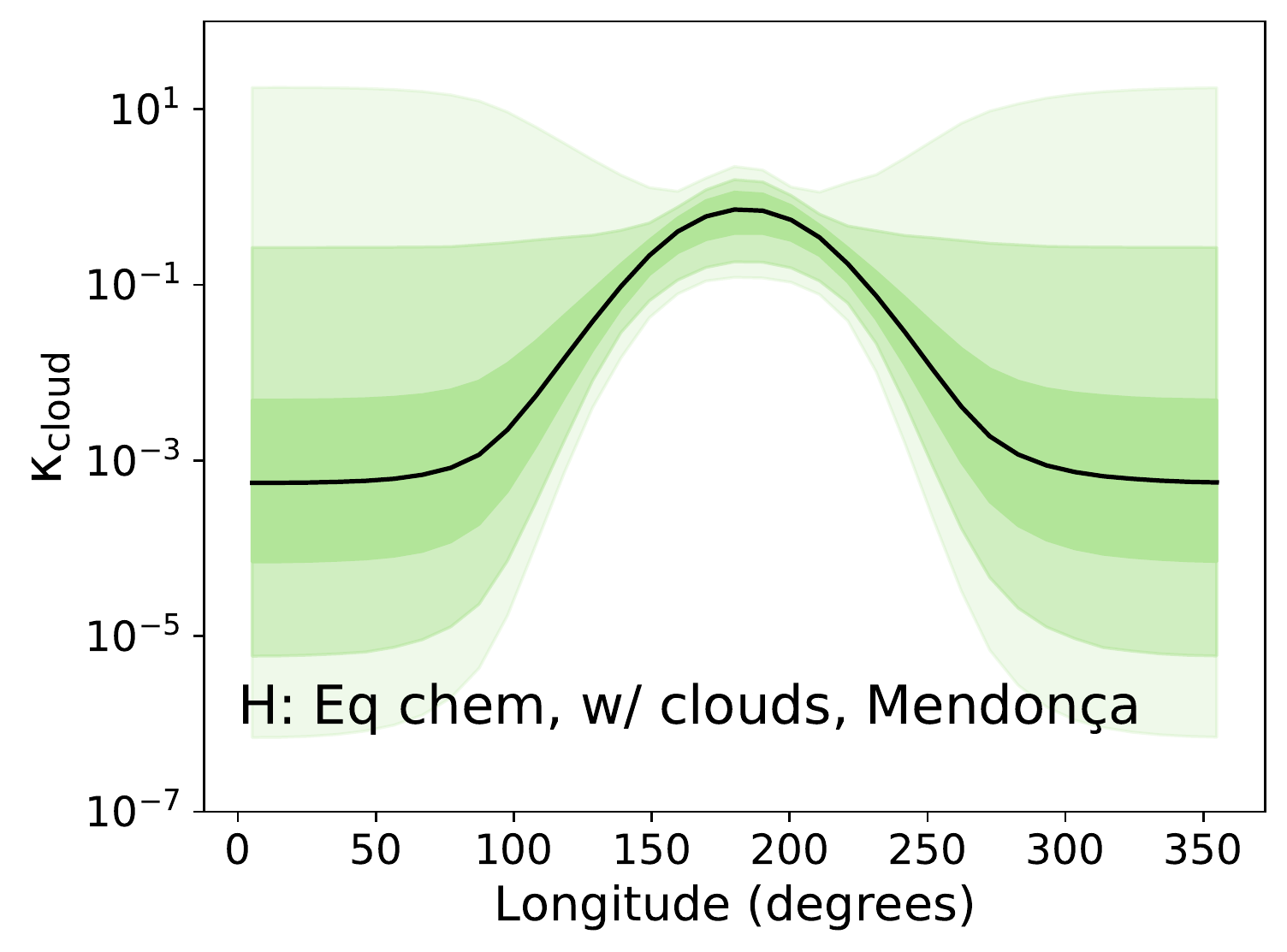}
                        \includegraphics[width=0.3\textwidth]{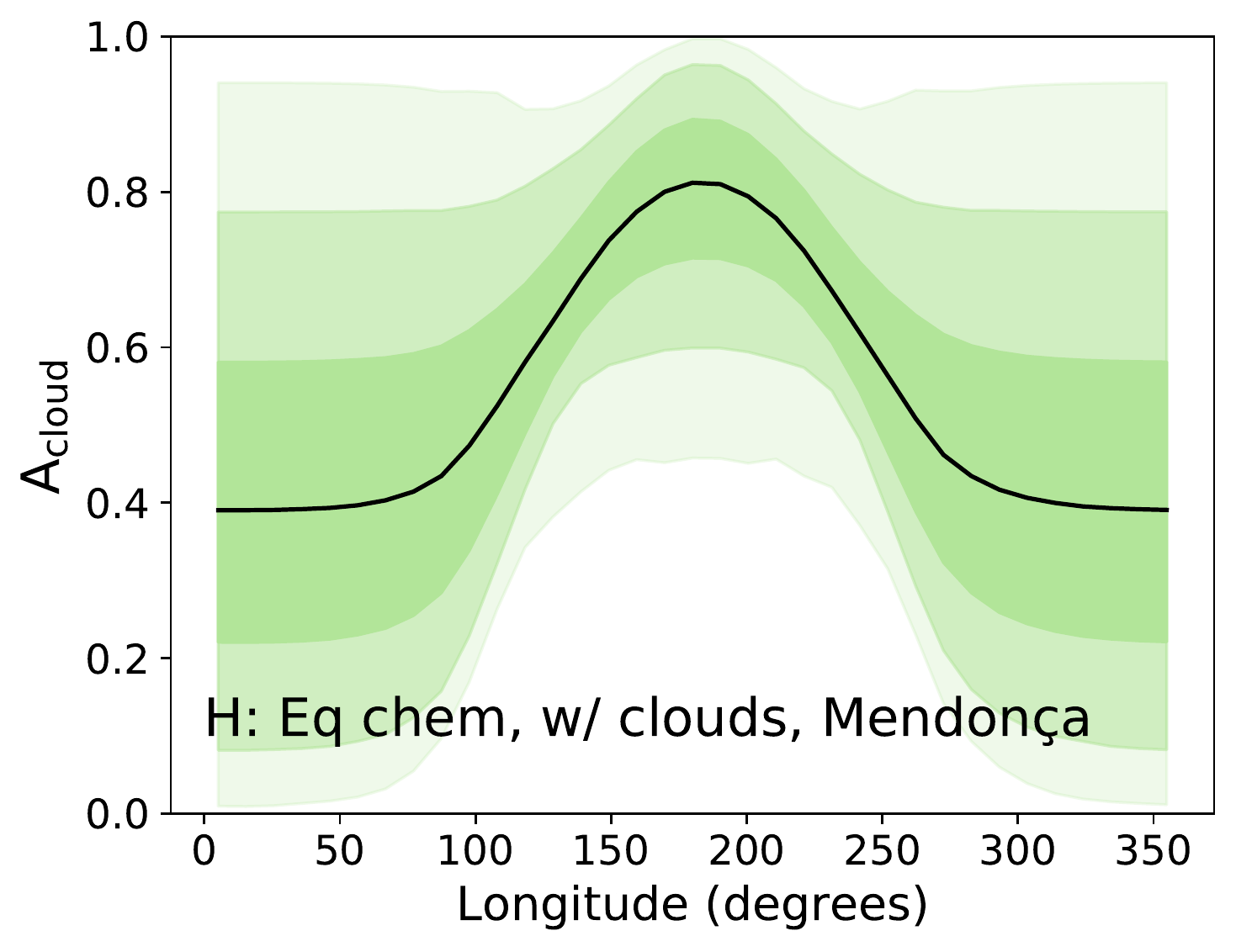}
                        \includegraphics[width=0.3\textwidth]{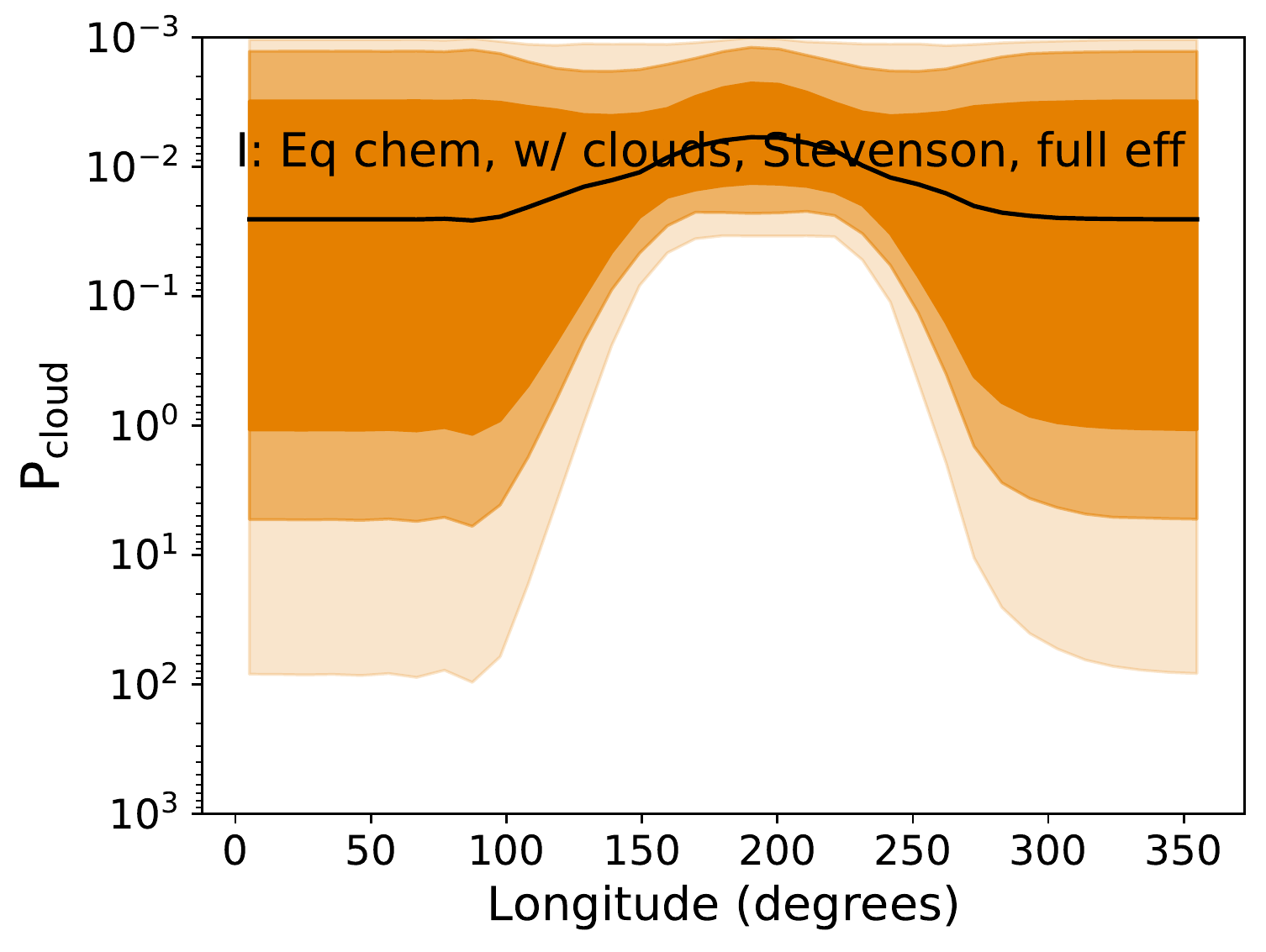}
                        \includegraphics[width=0.3\textwidth]{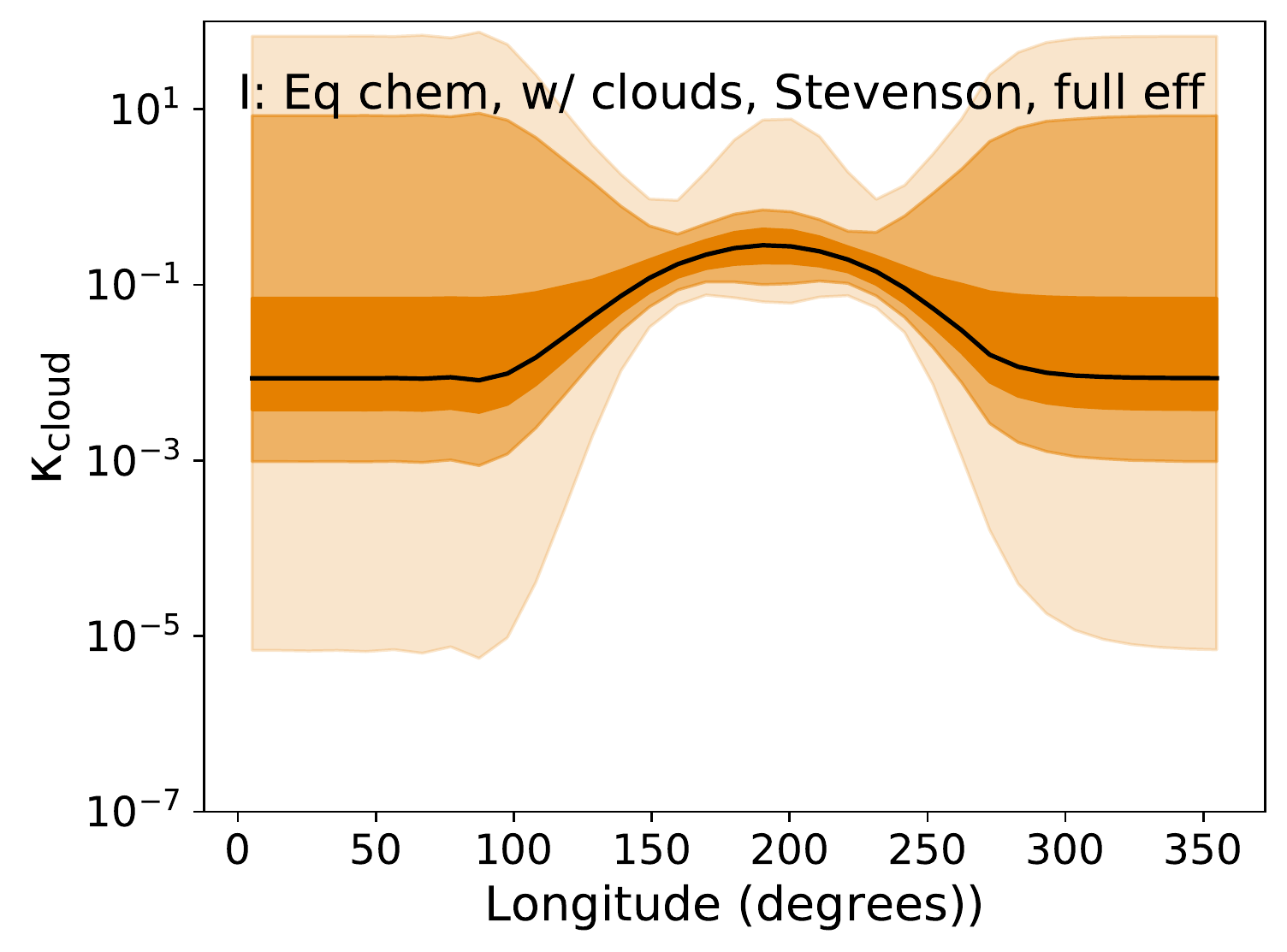}
                        \includegraphics[width=0.3\textwidth]{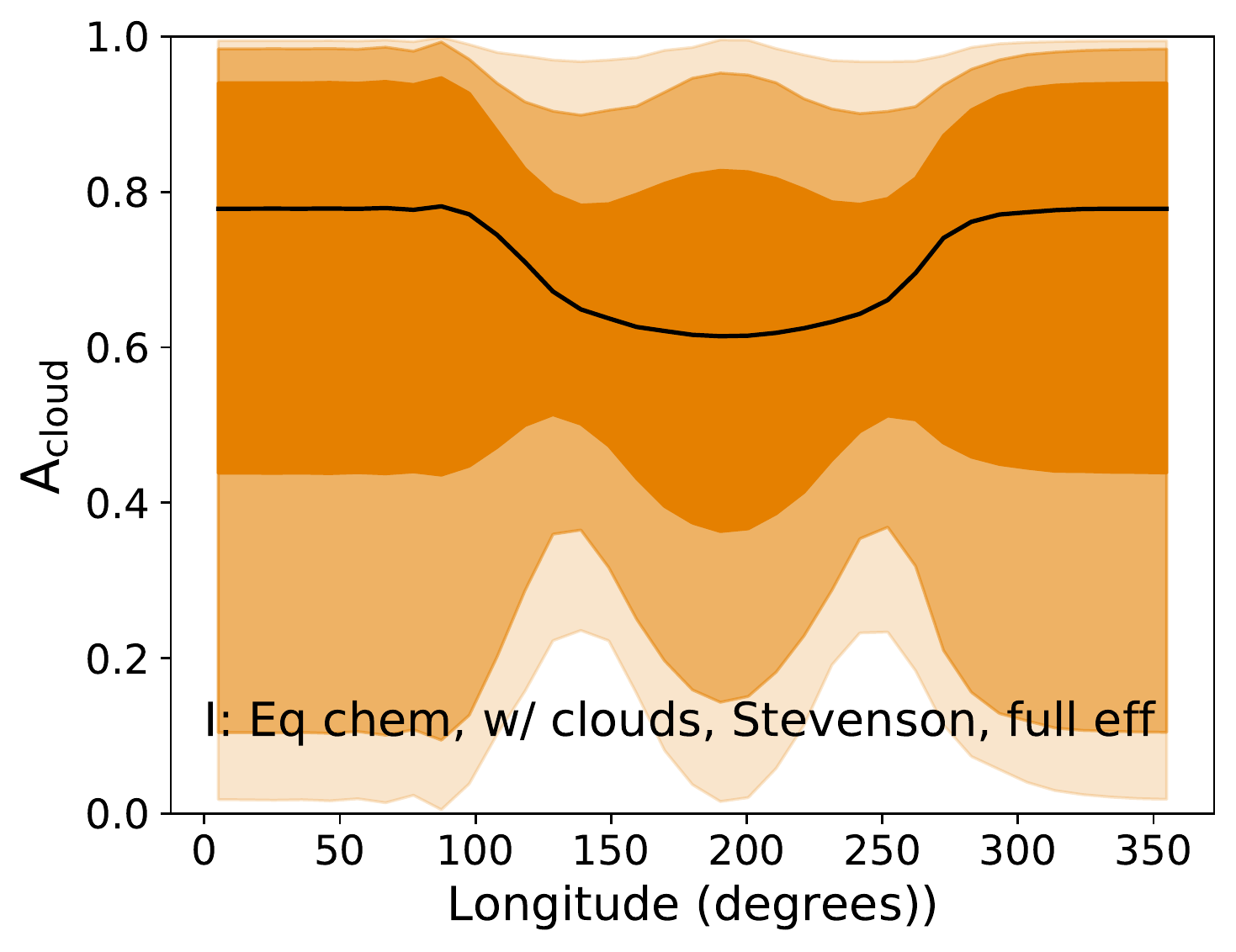}
                        \caption{Variation of retrieved cloud parameters as a function of longitude ($\degree$).  The first column gives P$_{\rm cloud}$ (cloud-top pressure), the second column $\kappa_{\rm cloud}$ (cloud opacity), and the third column $A_{\rm cloud}$ (cloud albedo). Each row is for a  different retrieval setup that include clouds, as detailed in each panel. Stevenson and Mendon{\c{c}}a refer to the use of Spitzer data analysed by \cite{17StLiBe.wasp43b} and \cite{18MeMaDe.wasp43b}, respectively.}\label{fig:cloud_all}
                \end{figure}    
            
            \twocolumn
                
                \section{Emission benchmark}\label{sec:emis_benchmark}
                
                We have tested the emission spectra as computed by ARCiS against the benchmark computations presented in \cite{2017ApJ...850..150B}.
                Here we present both the results for the computations with fixed temperature structure, as well as those where the temperature structure is computed self-consistently. The computations presented in \cite{2017ApJ...850..150B} are for homogeneous planets (so 1D structures only). We therefore run ARCiS in the mode where $f_{red}=1$ and $v_{\Lambda}=0$. This results in a homogeneous planet. We adjust the other parameters to reproduce the benchmark planets presented in that study.
                
                \subsection{Fixed PT profiles}
                
\begin{figure}[!tp]
                        \centerline{\resizebox{\hsize}{!}{\includegraphics{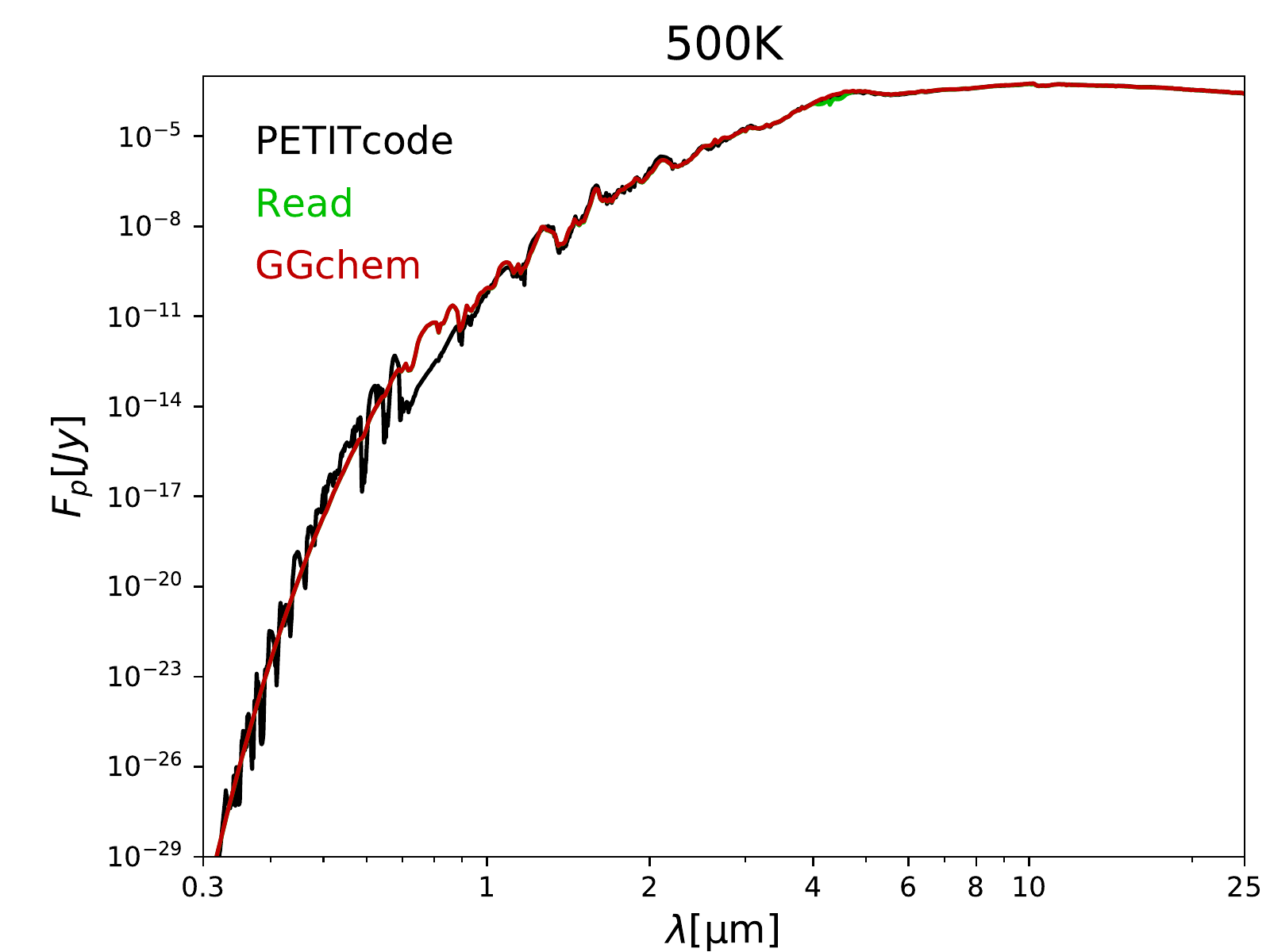}\includegraphics{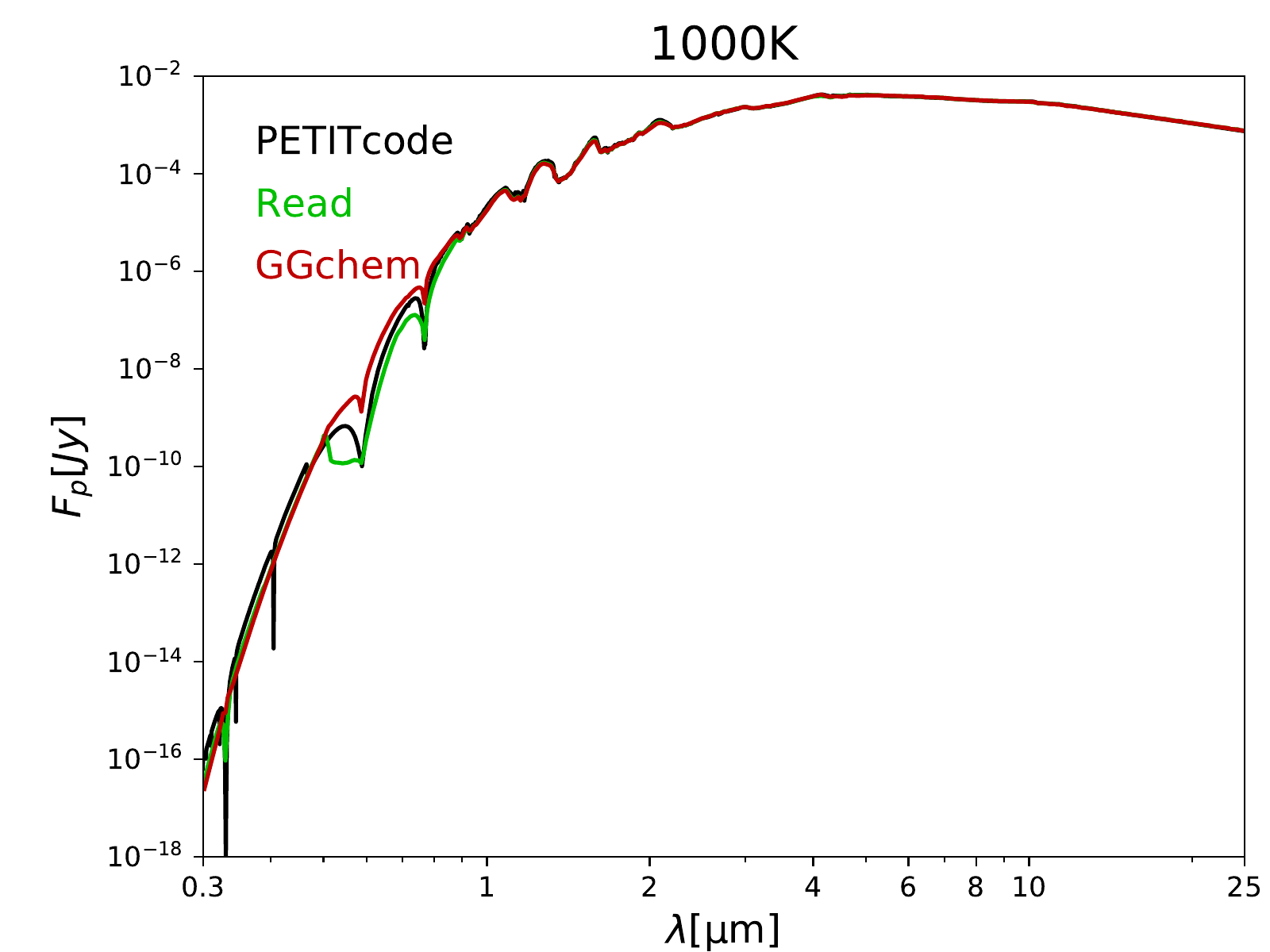}}}
                        \centerline{\resizebox{\hsize}{!}{\includegraphics{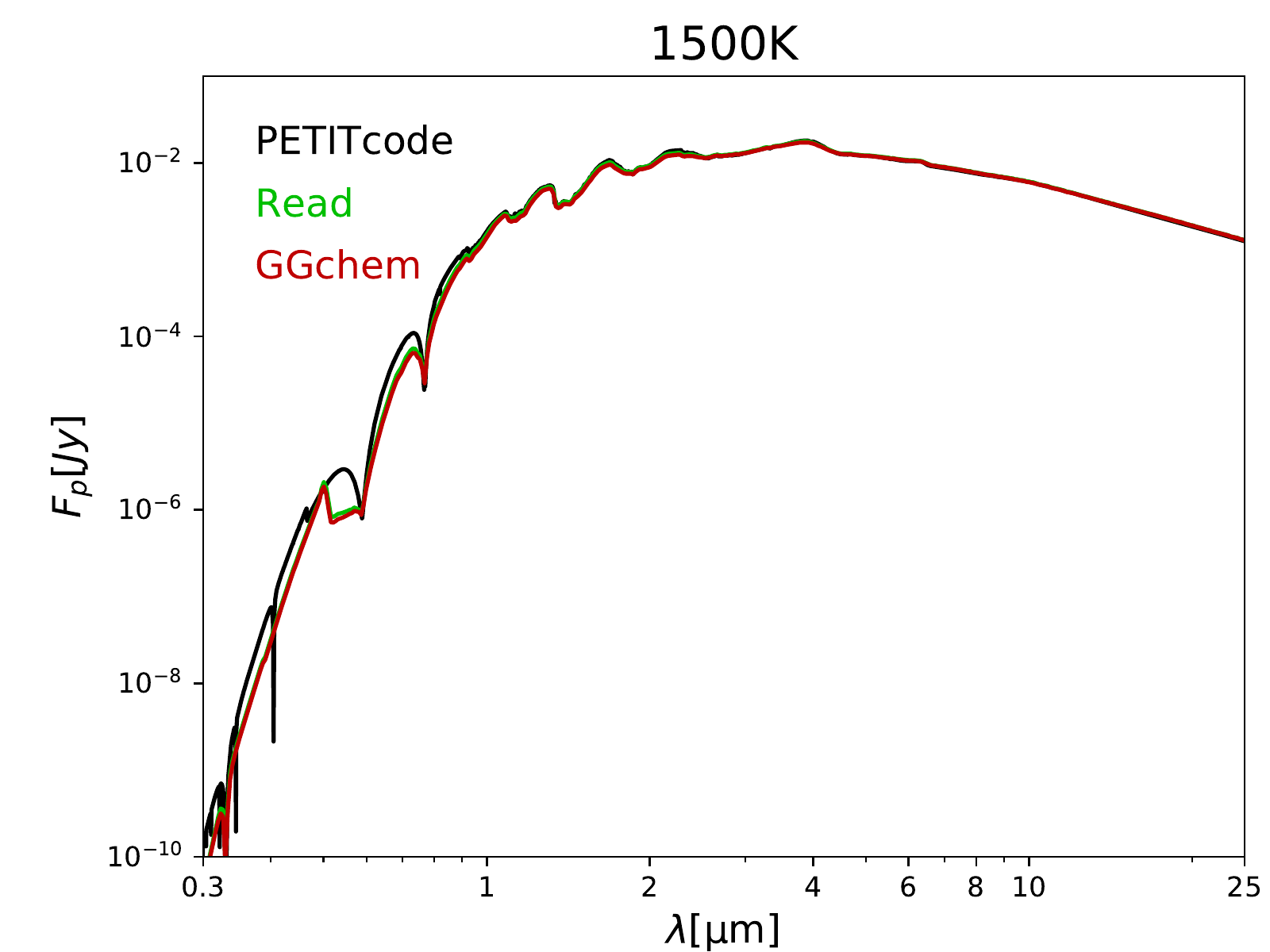}\includegraphics{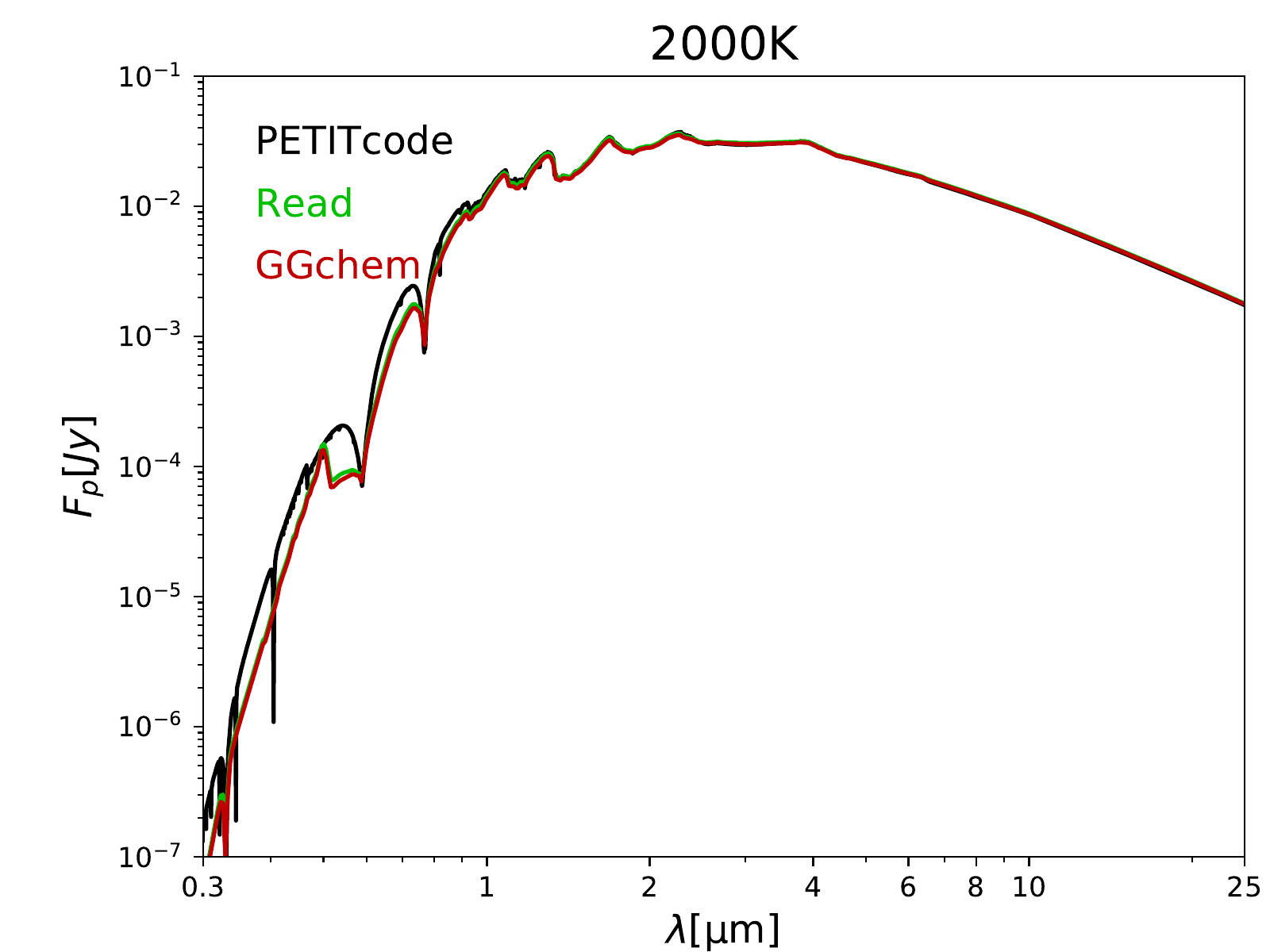}}}
                        \centerline{\resizebox{0.5\hsize}{!}{\includegraphics{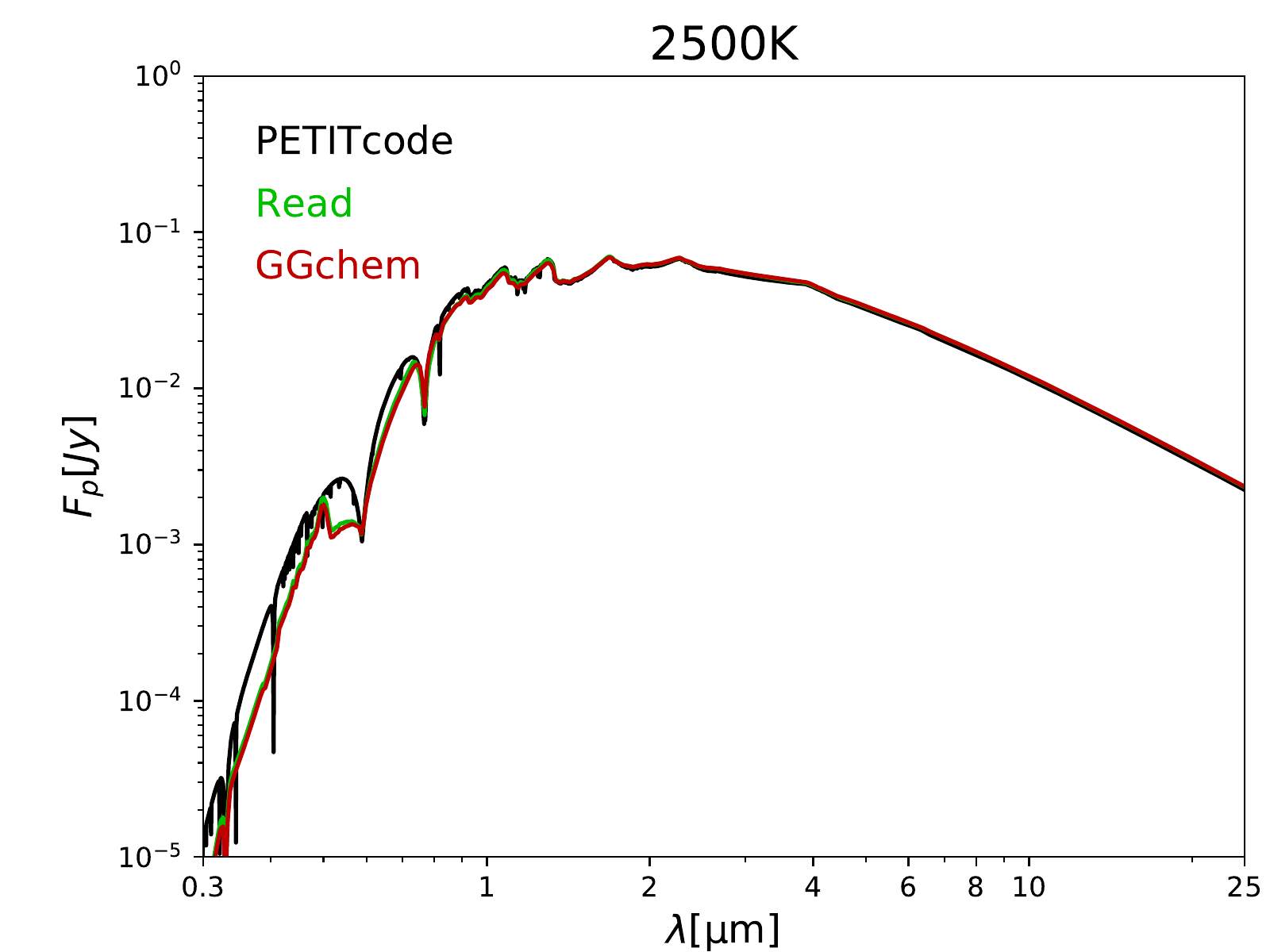}}}
                        \caption{Comparison of the emission spectra of the benchmark planets with fixed pressure temperature structures and different effective temperatures. Shown are the spectra from PETITcode (black), ARCiS reading the precomputed molecular opacities (green), and ARCiS when molecular opacities are computed using GGchem (red).}
                        \label{fig:fixedPT}
                \end{figure}

                Five cases are presented with a fixed pressure temperature profile. Figure \ref{fig:fixedPT} shows the resulting emission spectra compared to the petitCODE~\citep{15MoBoDu.petitCODE,17MoBoBo.exo} spectra presented in \cite{2017ApJ...850..150B}. We show the spectra computed with ARCiS in the mode where the chemistry is computed using GGchem and in the mode where the chemical abundances are read from the ones provided by \cite{2017ApJ...850..150B}. We see that there are small differences in the spectra. Some of them can be assigned to differences in the computed chemical abundances, others cannot. It is most likely that these small remaining differences are caused by different molecular opacities; it can be seen that the line lists used by petitCODE, given in Table 1 of\cite{2017ApJ...850..150B}, are all different to those used in the present study (see Section~\ref{sec:eq_chem}). In particular, the Na and K lines in the optical show significant differences. It is important to note that in all five of these cases, the parameters for the central star were not given in the paper, and so we assume that scattering from the central star was also not taken into account in the spectrum from the planet.
                
                \subsection{Self consistent PT profiles}
                
\begin{figure}[!tp]
                        \centerline{\resizebox{0.75\hsize}{!}{\includegraphics{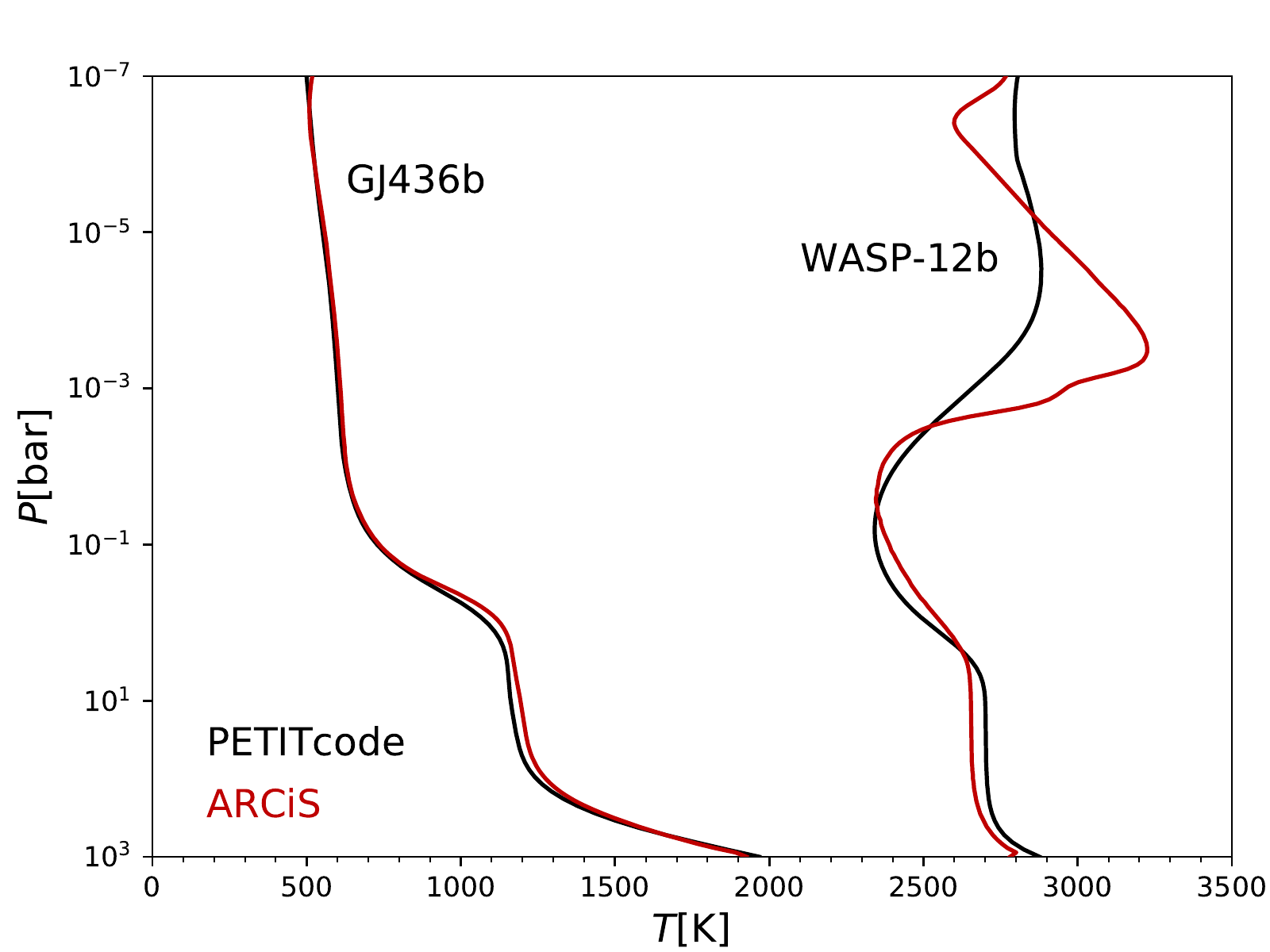}}}
                        \caption{Self-consistent pressure temperature structures as computed using PETITcode \cite{2017ApJ...850..150B} and ARCiS.}
                        \label{fig:consistentPT}
                \end{figure}

\begin{figure}[!tp]
                        \centerline{\resizebox{\hsize}{!}{\includegraphics{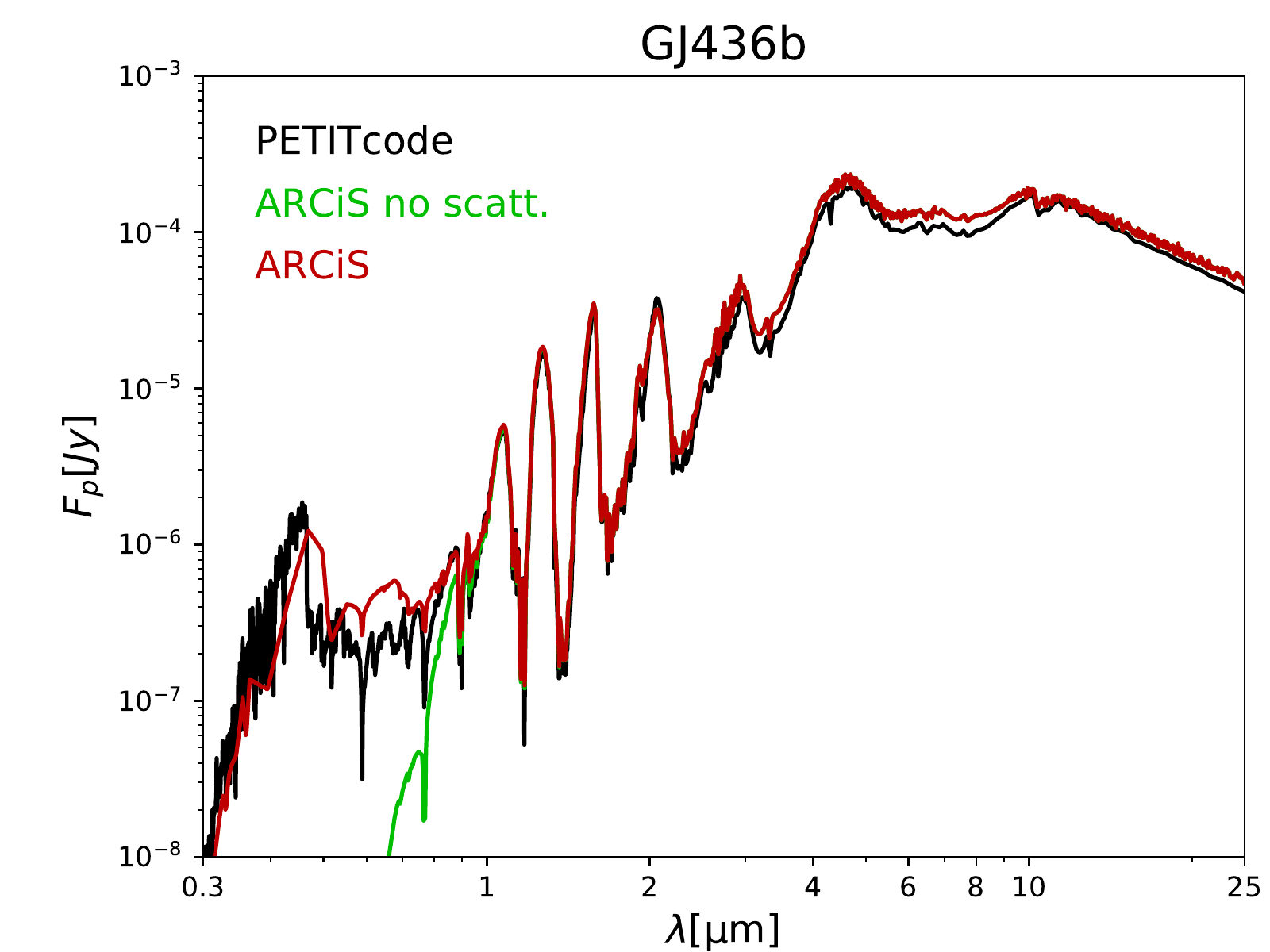}\includegraphics{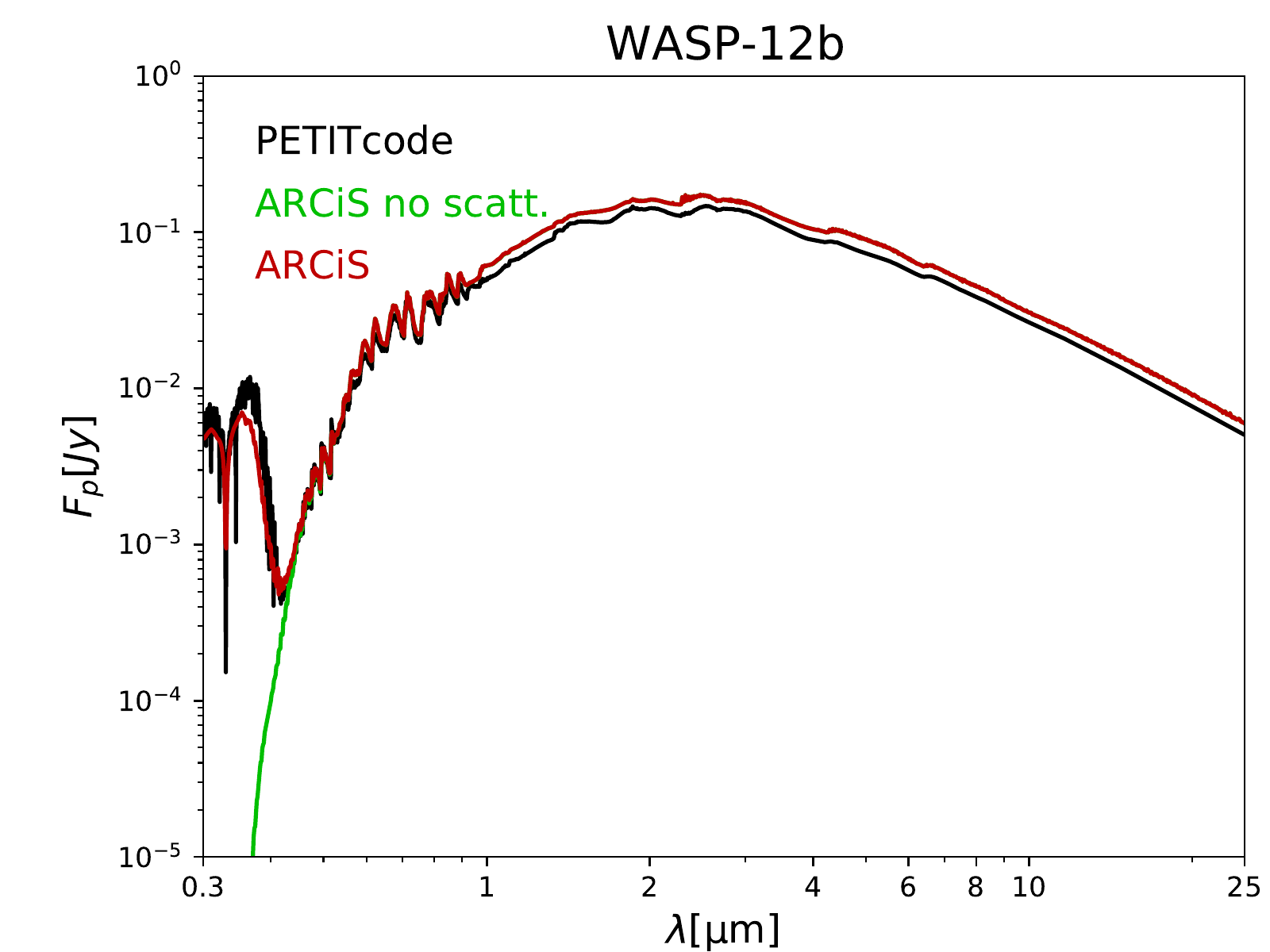}}}
                        \caption{Spectra from the self-consistent models using PETITcode \cite{2017ApJ...850..150B} and ARCiS. In green we also show the spectra as computed using ARCiS but without the scattered light component.}
                        \label{fig:consistentSpec}
                \end{figure}

                In \cite{2017ApJ...850..150B} cases are also presented where the pressure temperature profile is computed self-consistently. We also implemented this mode in the ARCiS code. In Fig.~\ref{fig:consistentPT} we show the resulting PT profiles for two irradiated planets from \cite{2017ApJ...850..150B}. We see that the structure largely agrees. For WASP-12b we see significant differences. These were also found by \cite{2019AJ....157..170M} for the same object. We conclude that likely the chemistry in the upper atmosphere is in a very delicate balance and is disturbed by the slightest difference in the opacities or the way the chemistry is computed. Nevertheless, the overall structure is still similar. In Fig.~\ref{fig:consistentSpec} we plot the resulting spectra for these two cases. It is also important to note the computed scattered light part for these two. This is significant, as can be seen by the difference when we switch the scattering part off, and agrees well with the scattered light computed by PETITcode.
                
                We conclude that, overall, ARCiS successfully reproduces the benchmark results from \cite{2017ApJ...850..150B}. Any differences are likely caused by differences in chemistry and molecular opacities.
                
        \end{appendix}

\end{document}